\documentclass{aa}  
\usepackage{adjustbox}
\usepackage{caption}
\usepackage{float}
\usepackage{graphicx}
\usepackage{xcolor}
\usepackage{txfonts}
\usepackage[]{hyperref}
\hypersetup{
    colorlinks=true,
    linkcolor=blue,
    citecolor=blue,
    filecolor=magenta,      
    urlcolor=magenta,
    }
\usepackage{morefloats}
\extrafloats{100}
\usepackage{soul} 
\usepackage{subcaption}

\defcitealias{2021MNRAS.507L...6C}{Paper I}

\begin{document}

   \title{SMILE: Discriminating milli-lens systems in a VLBI pilot project}

   \subtitle{}

   \author{F. M. P{\"o}tzl\inst{1,2}
          \and
          C. Casadio\inst{1,2}
          \and
          G. Kalaitzidakis\inst{1,2}
          \and
          D. \'Alvarez-Ortega\inst{1,2}
          \and
          A. Kumar\inst{1,2}
          \and
          V. Missaglia\inst{1,2}
          \and
          D. Blinov\inst{1,2}
          \and
          M. Janssen\inst{3,4}
          \and
          N. Loudas\inst{5}
          \and
          V. Pavlidou\inst{1,2}
          \and
          A. C. S. Readhead\inst{1,6}
          \and
          K. Tassis\inst{1,2}
          \and
          P. N. Wilkinson\inst{7}
          \and
          J. A. Zensus\inst{4}
          }
   \institute{
           Institute of Astrophysics, Foundation for Research and Technology – Hellas, N. Plastira 100, Voutes GR-70013, Heraklion, Greece
           \email{fpoetzl@ia.forth.gr}
       \and
           University of Crete, Department of Physics \& Institute of Theoretical \& Computational Physics, 70013 Heraklion, Greece
       \and
           Department of Astrophysics, Institute for Mathematics, Astrophysics and Particle Physics (IMAPP), Radboud University, P.O. Box 9010, 6500 GL Nijmegen, The Netherlands
       \and
           Max-Planck-Institut f{\"u}r Radioastronomie, Auf dem H{\"u}gel 69, D-53121 Bonn, Germany
       \and
           Department of Astrophysical Sciences, Peyton Hall, Princeton University, Princeton, NJ 08544, USA
       \and
           Owens Valley Radio Observatory, California Institute of Technology,  Pasadena, CA 91125, USA
       \and
           Jodrell Bank Observatory, University of Manchester, Nr. Macclesfield, Cheshire SK11 9DL, UK
   }
   
   \date{Received 22 September 2024; accepted 17 January 2025}

 
  \abstract
   {Dark matter (DM) remains poorly probed on critical sub-galactic scales, where predictions from different models diverge in terms of abundance and density profiles of halos. Gravitational lens systems on milli-arcsecond scales (milli-lenses) are expected for a population of dense DM halos (free-floating or sub-halos) and free-floating supermassive black holes (SMBHs) in the mass range of $10^6$ to $10^9\,M_\odot$ that might partly be comprised of primordial black holes (PBHs).}
   {In this paper, we aim to look for possible milli-lens systems via a systematic search in a large sample of radio-loud active galactic nuclei (AGN) observed with very long baseline interferometry (VLBI). We present the observational strategy to discriminate milli-lens systems from contaminant objects mimicking a milli-lens morphology.}
   {In a pilot project, we have investigated VLBI images from 13,828 sources from the Astrogeo VLBI image database and reduced the number of lens candidates to 40 in a first step. We present here the images and analysis of new sensitive follow-up observations with the European VLBI network at 5 and 22\,GHz and streamline our analysis to reject milli-lens candidates. By using constraints such as the surface brightness ratio, conservation of spectral shape, stability of flux ratios over time, and changes in morphology at higher frequencies, we can confidently discriminate between milli-lenses and contaminant objects that mimick them.}
   {Using the above constraints, we ruled out 31 of our initial 40 candidates of milli-lens systems, demonstrating the power of our approach. In addition, we found many new candidate compact symmetric objects (CSOs), which are thought to be primarily short-lived jetted radio sources.}
   {Additional observations of the remaining candidates will be necessary to confirm or reject their nature as milli-lenses or CSOs. This study serves as a pathfinder for the final sample used for the Search for MIlli-LEnses (SMILE) project, which will allow DM models to be constrained by comparing the results to theoretical predictions. This SMILE sample will consist of $\sim$5,000 sources based on the VLA CLASS survey and will include many observations obtained for this project specifically.}


   \maketitle
%

\section{Introduction}

The nature of dark matter (DM) remains elusive. The current and most widely accepted cosmological paradigm, the $\Lambda$CDM model \citep[see, e.g.][for a review]{2012AnP...524..507F}, has been remarkably successful at explaining a great number of observational properties of the Universe, especially observations related to the large-scale structure and the evolution of the Universe \citep{2014MNRAS.444.1518V}.
However, several discrepancies have appeared on smaller scales. These include the problem of missing satellites \citep{1999ApJ...522...82K}, the diversity problem of galaxy rotation curves \citep[see, e.g.][]{2015MNRAS.452.3650O}, and the cusp-core problem \citep[see, e.g.][]{2011AJ....142...24O}. (See also \cite{2017ARA&A..55..343B} for a review of these topics.)

A variety of alternative DM models have been proposed to alleviate these problems, including warm DM \citep[WDM;][]{2005PhRvD..71f3534V}, self-interacting DM \citep[SIDM;][]{2000PhRvL..84.3760S}, fuzzy DM \citep{2022MNRAS.510.1425K}, and primordial black holes \citep[PBHs;][]{2018PDU....22..137C}.
All the proposed DM models make diverse predictions on the abundance and density profiles of DM halos on different mass scales, with CDM predicting a larger number of sub-galactic halos than WDM, for example. The mass range between $10^6$ and $10^9\,M_\odot$ turns out to be an especially crucial discriminating factor between the different models \citep[see, e.g.][]{2020MNRAS.493L..11L}. To include both DM sub-halos and a possible population of (free-floating) supermassive black holes (SMBHs), which might be primordial, we refer to these objects as supermassive compact objects (SMCOs) from hereon. In particular, PBHs have recently gained increased attention with the latest results from the James Webb Space Telescope \citep[\textit{JWST}; see, e.g.][]{2024Natur.627...59M}, but attention was also garnered earlier with the detection of high mass black hole (BH) mergers by the LIGO/Virgo collaboration, such as GW190521 \citep{2020PhRvL.125j1102A}. Because DM halos on the mass scales described above are not expected to form galaxies \citep{2020MNRAS.498.4887B}, they remain largely unprobed.

A unique way to probe SMCOs on these mass scales is through gravitational lensing. Multiple studies have been investigating the effect that different DM sub-halos have on the gravitational lensing potential \citep[see, e.g.][]{2023PhRvD.107j3008G} and thus also on the properties of the multiple-lensed images for the case of strong lensing by galaxy clusters \citep[see, e.g.][for a recent review]{2023arXiv230611781V}. In addition, the lensing effect has been used in fast radio bursts (FRBs) to study the properties of DM on sub-solar mass scales \citep[see, e.g.][]{2022PhRvD.106d3017L} and in searches of lensing signatures in light curves of $\gamma$-ray bursts \citep[GRBs; e.g.][]{2021ApJ...922...77K}.

When considering any DM halo within the mass scales given above along the line of sight, part of the halo should be able to act as a gravitational lens for a background source if that part of the halo is sufficiently dense \citep{1973ApJ...185..397P}. That means it should have a surface mass density above the critical value of $\Sigma_\mathrm{crit}\,{\simeq}\,1.6\,{\times}\,10^3\,\mathrm{M}_\odot\,\mathrm{pc}^{-2}$, assuming lens and background sources are at cosmological distances. SMBHs will naturally fulfil this criterion. For DM halos, it depends crucially on the concentration-mass relation \citep{2020Natur.585...39W}, which dictates what fraction of the actual halo mass can effectively act as a gravitational lens \citep[see also][]{2022A&A...668A.166L}.

For lensing in the mass range between $10^6$ and $10^9\,M_\odot$, the expected angular separation between lensed images should be of the order of ${\sim}\,2\,{\times}\,10^{-6}(M_\mathrm{SMCO}/M_\odot)^{0.5}$\,arcseconds if the source and lens are at cosmological distances \citep{2001PhRvL..86..584W}. For our mass range of interest, that translates to ${\sim}\,1-100$\,mas. These spatial scales can be directly probed by very long baseline interferometry (VLBI) if the lensed background source is a radio-loud, compact active galactic nucleus (AGN). We refer to such systems as milli-lenses from hereon. A previous search in a sample of 300 AGN found no such objects \citep{2001PhRvL..86..584W}, and thus the cosmological number density of uniformly distributed SMCOs in the mass range $10^6$ to $10^8\,\mathrm{M}_\odot$ could be constrained to make up no more than $\sim$1\% of the closure density $\Omega_\mathrm{total}=1$.

More recently, \cite{2019MNRAS.483.2125S} conducted a search for milli-lenses with image separations $>$\,100\,mas at 1.4\,GHz, corresponding to lensing mass scales $>$\,$3\,{\times}\,10^9\,\mathrm{M}_\odot$. Their search resulted in finding one previously unidentified lens candidate. With this, they constrained the lensing rate to $1:(318\,{\pm}\,225)$. Other searches for milli-lenses have also been conducted, but mostly in singular cases. \cite{2022ApJ...927...24P} found evidence for milli-lensing in the long-term light curves of a blazar, most likely caused by an intervening molecular cloud or an undetected dwarf galaxy with a massive BH.

In a previous paper (\citealt{2021MNRAS.507L...6C}, from now on \citetalias{2021MNRAS.507L...6C}), we described the framework in which we aim to constrain the number density of milli-lenses in a large sample of VLBI images of radio sources. With this, \cite{2022A&A...668A.166L} investigated how different DM models can be constrained based on the number of observed milli-lenses. We found that with a large enough sample with known redshifts, one can have significant constraining power over some viable DM models. This is one of the goals of the Search for MIlli-LEnses (SMILE) project.

The Cosmic Lens All-Sky Survey \citep[CLASS;][]{2003MNRAS.341....1M}, conducted with the Very Large Array (VLA), provides the flux-limited sample as the basis for the SMILE project. The survey includes all sources within [0; $+75^\circ$] declination and galactic latitude $|b|\,{\geq}\,10\,^\circ$, with $S_{5\,\mathrm{GHz}}\,{\geq}\,30$\,mJy and spectral index $\alpha\leq0.5$ between the flux density of the NRAO VLA Sky Survey \citep[NVSS;][]{1998AJ....115.1693C} at 1.4\,GHz and GB6 \citep{1996ApJS..103..427G} at 5\,GHz, and it contains 11,685 sources. The CLASS catalogue is an ideal starting point for the SMILE project because of the flat spectrum selection criterion, which maximises chances of detection of radio nuclei, and the available VLA 8\,GHz flux density measurements, which serve as a hint for the expected Very Long Baseline Array (VLBA) 4-8\,GHz flux density. From the CLASS sample, we chose a subsample of 4,968 sources with $S_{8\,{\mathrm{GHz}}}\,{\geq}\,50$\,mJy to increase the probability of detection with VLBI. For the project, we will mostly use archival data from the NRAO archive and combine them with new observations conducted with the VLBA (e.g. Project ID: BP267; PI: Felix P\"otzl).

In a pilot project described in \citetalias{2021MNRAS.507L...6C}, we searched for milli-lens candidates in a different sample of 13,828 sources taken from the Astrogeo VLBI image database\footnote{\url{http://astrogeo.org/vlbi_images/}} \citep{2025ApJS..276...38P}. Serving as a pathfinder for the SMILE project, in this paper we present a thorough investigation of the nature of the 40 best candidates presented in  \citetalias{2021MNRAS.507L...6C}. We combine the available archival data with new observations obtained with the European VLBI network (EVN), which we describe in Sect.~\ref{sec:data}.
In the following, we assume a standard cosmology according to \cite{2020A&A...641A...6P}, with $\Omega_\mathrm{m}=0.315$, $\Omega_\Lambda=0.685$, and $H_0=67.4$\,km/s/Mpc.

\begin{figure*}[h]
    \centering
    \includegraphics[width=0.3\textwidth]{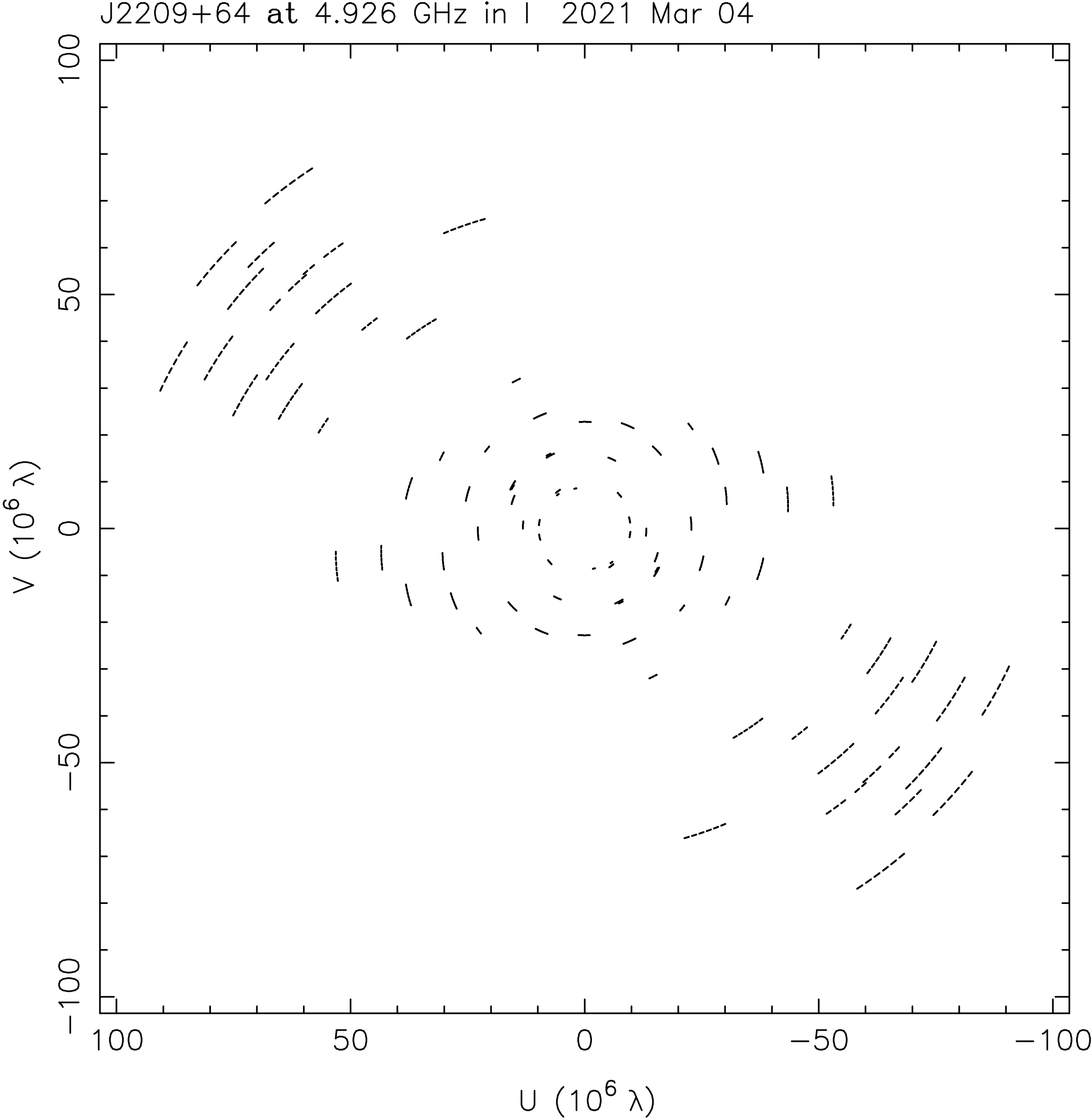}
    \hspace{0.5cm}
    \includegraphics[width=0.3\textwidth]{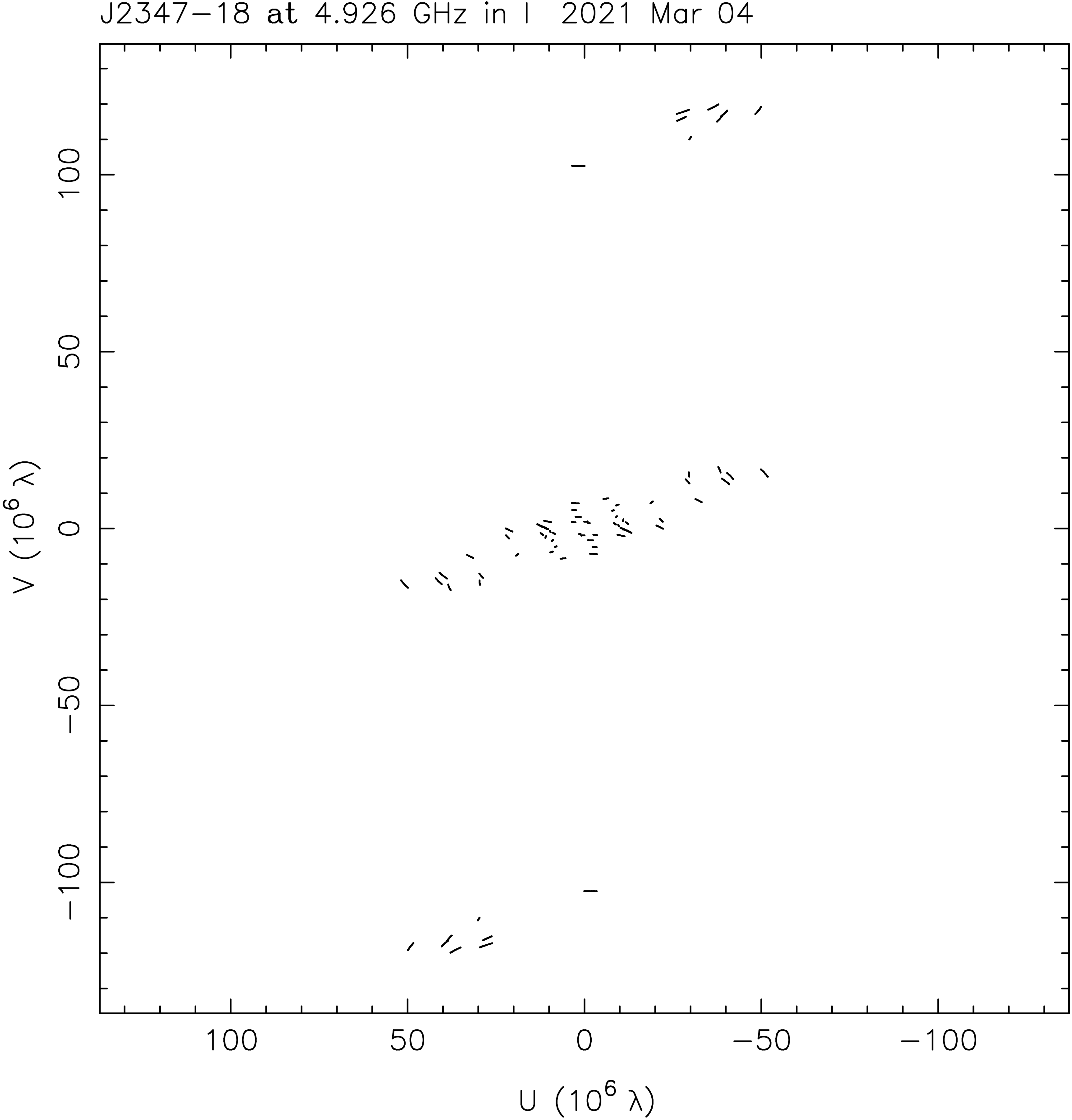}
    \caption{Two examples of $(u,v)$ coverages for the EVN observations at C-band. \textit{Left:} J2209+6442 (high declination target). \textit{Right:} J2347-1856 (low declination target)}
    \label{fig:uv_cov}
\end{figure*}

\section{Data}\label{sec:data}

The 40 lens candidates presented here have been chosen from images of 13,828 unique sources in the Astrogeo database at 2.3 (S-band), 5 (C-band) and 8\,GHz (X-band). The details of the selection process are described in \citetalias{2021MNRAS.507L...6C}, which also lists the coordinates of each source. The basic criteria for the selection were i) a minimum of two compact components were found in the Astrogeo images and ii) the surface brightness ratio (SBR) of the two components, based on Gaussian model-fitting, should be $<$\,7. We chose this more conservative threshold in \citetalias{2021MNRAS.507L...6C} to account for the very heterogeneous Astrogeo sample with partly poor $(u,v)$ coverage and short integration times. Based on our improved error analysis (see Sect.~\ref{sec:modelfit}), we revise this threshold in Sect.~\ref{sec:criteria}. Of the 40 candidates, 38 had available X-band data, and 35 of those were in the declination limit for the EVN. For those sources, we then obtained follow-up observations with the EVN at 5 and 22\,GHz (K-band) in phase-referencing (PR) mode, to help detect the possibly very weak sources and to retain the relative positions of the components between the two frequencies with respect to a compact calibrator. The sources were observed in five sessions, as summarised in Table~\ref{tab:obs}. At C-band, calibrator and target sources were observed in cycles with scans of 90\,s and 210\,s length, respectively. At K-band, scans of 120\,s for calibrators and 90\,s for targets were chosen to capture the faster atmospheric variability at higher frequencies. Typical on-source integration times were 20-30\,minutes at C-band and 60-70\,minutes at K-band for the science targets. Two representative $(u,v)$ coverages are shown in Fig.~\ref{fig:uv_cov}. Together with our new data, for 16 sources, we have data for more than two epochs; for 19 sources, we have data for exactly two epochs at C- or X-band.

\begin{table}[htbp]
    \caption{European VLBI network observing run.}
    \vspace*{1mm}
    \adjustbox{width=0.48\textwidth}{%
    \label{tab:obs}
    \centering
    \begin{tabular}{ c | c | c } 
        \hline
        Obs. session & 5\,GHz & 22\,GHz  \\ [0.5ex] 
        \hline\hline
            1 & 28 Oct 2019 (5 hours)      & 18-19 Oct 2019 (24 hours)    \\
            2 & 26-27 Feb 2020 (5 hours)   & 04-05 Mar 2020 (6+18 hours)  \\
            3 & 29 Oct 2020 (5 hours)      & 22-25 Oct 2020 (12+12 hours) \\
            4 & 04 Mar 2021 (6 hours)      & 10-11 Mar 2021 (24 hours)    \\
            5 & 12+17 Jun 2021 (3+4 hours) & 07-09 Jun 2021 (32 hours)    \\
        \hline
    \end{tabular}
    }
\end{table}

\section{Calibration and imaging}

\begin{figure*}[h]
    \centering
    \includegraphics[width=0.4\textwidth]{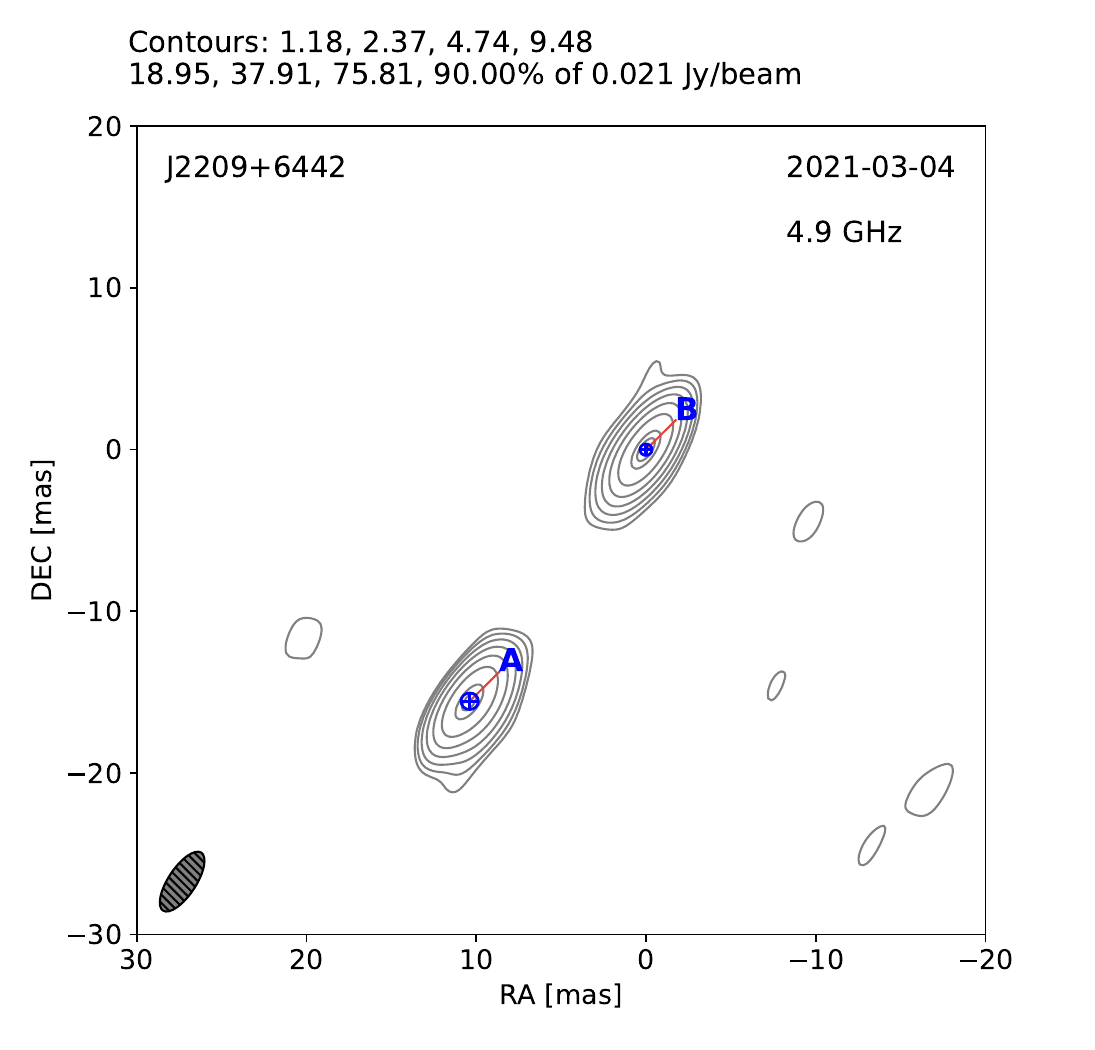}
    \includegraphics[width=0.4\textwidth]{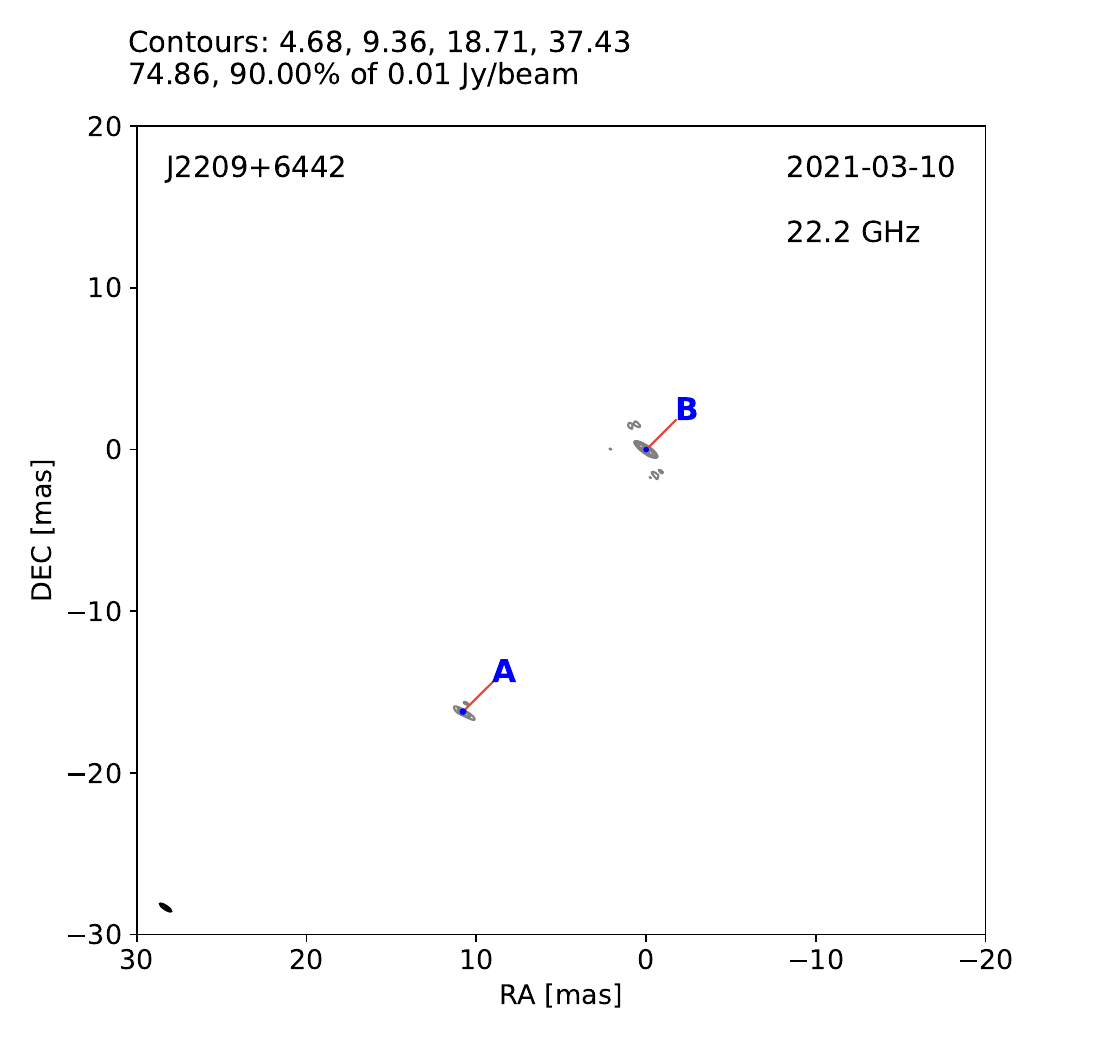}
    \caption{J2209+6442, a source that has both components detected at 22\,GHz and has been rejected as a milli-lens but is retained as a CSO candidate. \textit{Left:} EVN 4.9\,GHz data. The contours start at four times the image rms noise of 0.061\,mJy/beam and increase by factors of two. The restoring beam size is 1.65$\times$4.24\,mas at $-31.7$\,$^\circ$ PA. \textit{Right:} EVN 22.2\,GHz data. The contours start at four times the image rms noise of 0.118\,mJy/beam and increase by factors of two. The restoring beam size is 0.23$\times$0.81\,mas at 55.2\,$^\circ$ PA.}
    \label{fig:J2209}
\end{figure*}

The new EVN data were calibrated with the \texttt{AIPS} package \citep{2003ASSL..285..109G} in a standard manner. First, the a priori amplitude calibration and the parallactic angle correction has been applied by copying the respective calibration tables from the EVN pipeline, as is recommended by the EVN. Only a few stations provide opacity-corrected data or weather information (important at higher frequencies) for the EVN, so no further corrections have been applied. Uncertainties in the amplitude calibration are included in the gain factor in Eq.~\ref{eq:sigma_tot}. In the next step, corrections for the ionospheric dispersive delay were applied to the data at C-band, while it is negligible at K-band. Following that, we removed the instrumental delays by doing a fringe-fit with the task \texttt{FRING} in \texttt{AIPS} on a short scan of a bright calibrator. Then, we applied a global fringe-fit with \texttt{FRING} on the calibrators, as well as a complex bandpass correction. The calibrators were then subsequently imaged in \texttt{difmap} \citep{1997ASPC..125...77S}, with careful self-calibration first in phase only, and then also in amplitudes. We chose a natural weighting scheme for the mostly compact sources for increased sensitivity (\texttt{uvweight\,0,-1} in \texttt{difmap}). The obtained images were then re-read into \texttt{AIPS} to obtain improved self-calibration solutions in the task \texttt{CALIB}. Those solutions were then applied to the target sources, which were subsequently imaged as well. This procedure removes residual gain errors as well as the effects of residual source structure in the calibrators. Since the application of self-calibration on a noisy data set can introduce false flux and structure and even create significant flux from pure noise \citep[see, e.g.][]{2008A&A...480..289M}, we did not attempt self-calibration in images with $\mathrm{S/N}\,{\lesssim}\,5$ in the dirty image. An example of final images at both 4.9\,GHz and 22.2\,GHz is shown in Fig.~\ref{fig:J2209} for the source J2209+6442. In total, we detected 13 out of 35 sources that were observed with the EVN at 22.2\,GHz (see Table~\ref{tab:summary}). Of the 13 detections, seven have both of the components detected that were originally identified as possible lensed images at lower frequencies.

\begin{table*}[h]
    \caption{Derived quantities for J2209+6442.}
    \vspace*{3mm}
    \adjustbox{width=1\textwidth}{%
    \label{tab:J2209}
    \centering
    \begin{tabular}{|| c c c c c | c c | c c | c c | c c | c c | c ||} 
        \hline
        (1)    & (2)       & (3)       & (4)   & (5)    & (6)   & (7)   & (8)   & (9)   & (10)                                & (11)   & (12)  & (13)  & (14)     & (15)  & (16)           \\
        Epochs & Frequency & Component & Flux  & Error  & Flux  & Error & FWHM  & Error & $\log(T_\mathrm{b,obs}/\mathrm{K})$ & Error  & SB    & Error & Distance & Error & Spectral Index \\ [0.5ex] 
               & [GHz]     &           & [mJy] & [mJy]  & Ratio &       & [mas] & [mas] &                                     &        & Ratio &       & [mas]    & [mas] &                \\
        \hline\hline
2016-07-24 & 4.3 & A & 22.6 & 3.2 & 1.21 & 0.18 & 1.14 & 0.09 & 9.10 & 0.09 & 0.99 & 0.27 & 18.80 & 0.34 & -1.38 \\
- & -   & B & 18.6 & 2.8 & -   & -   & 1.03 & 0.09 & 9.11 & 0.10 & -   & -   & -   & -   & -0.31 \\
        \hline
2016-07-24 & 7.6 & A & 10.4 & 2.1 & 0.66 & 0.15 & 1.09 & 0.16 & 8.32 & 0.15 & 0.05 & 0.02 & 19.19 & 0.35 & - \\
- & -   & B & 15.6 & 2.7 & -   & -   & 0.28 & 0.03 & 9.66 & 0.11 & -   & -   & -   & -   & - \\
        \hline\hline
2021-03-04 & 4.9 & A & 23.8 & 2.9 & 1.04 & 0.11 & 1.03 & 0.06 & 9.11 & 0.07 & 0.50 & 0.09 & 18.71 & 0.19 & -1.60 \\
- & -   & B & 23.0 & 2.8 & -   & -   & 0.71 & 0.04 & 9.41 & 0.07 & -   & -   & -   & -   & -0.57 \\
        \hline
2021-03-10 & 22.2 & A & 2.1 & 0.8 & 0.22 & 0.09 & 0.24 & 0.07 & 8.02 & 0.30 & 0.11 & 0.08 & 19.47 & 0.15 & - \\
- & -   & B & 9.7 & 1.8 & -   & -   & 0.17 & 0.02 & 8.98 & 0.13 & -   & -   & -   & -   & - \\
    \hline \noalign{\medskip}
    \multicolumn{16}{p{1.35\textwidth}}{\textsc{Notes:} Columns designate the (1): observation epoch; (2): observing frequency in GHz; (3): studied component (A and B denote the sum over all components in a region); (4): total flux in a component based on the Gaussian model-fit in mJy; (5): respective uncertainty; (6): calculated flux ratio (FR) between components. If multiple components are in one region, we also list the FR of the brightest component over the whole of the other region for comparison; (7): respective uncertainty; (8): FWHM of Gaussian component from model-fitting in mas. If multiple components comprise one region, this is left out; (9): respective uncertainty; (10): observed surface brightness expressed as the logarithm of the brightness temperature (Eq.\ref{eq:Tb}); (11): respective uncertainty; (12): calculated surface brightness ratio (SBR). Again, if multiple components are in one region, we also list the SBR of the brightest component over the whole of the other region for comparison; (13): respective uncertainty; (14): distance calculated between the brightest components in regions A and B in mas; (15): respective uncertainty; (16): spectral index calculated between the total flux densities of the two regions.
    }
    \end{tabular}
    }
\end{table*}

\section{Analysis}

\subsection{Model fitting}\label{sec:modelfit}

To assess the total flux density $S_\mathrm{tot}$, position (radial coordinates $r$ and $\phi$), and FWHM (full width at half maximum) size $\theta$ of components, we fit the visibilities directly with Gaussian brightness templates with the built-in \texttt{modelfit} function of \texttt{difmap}. To find the optimal number of components that describe our data, we tested if the reduced $\chi^2$ of the fit significantly improves adding new components, as described in \cite{2012A&A...537A..70S}. Furthermore, we checked if adding a new component significantly increased the fraction of the total flux recovered (as estimated from the clean component image), that is, adding more than $\sim$10\,\% of the total flux. If either of these criteria was fulfilled, we deemed it a significant improvement of our model; otherwise, additional components were discarded.

We identify the two distinct areas in the images with region A and region B, which may consist of multiple Gaussian components. We denote these components with the region name followed by numbers, starting at 0, which designates the brightest component. Additional, unassociated components are named in alphabetical succession. We note that the components are not necessarily cross-identified between all frequencies and epochs; we simply named them in order for each dataset. Should the parameters for regions A and B turn out to be consistent with a lensing system, we also investigate the other regions in comparison, as they could potentially be associated with a higher number of lensed images. To assess the error budget of our fits, we utilised the following formulas:

\begin{align}
    \mathrm{S/N} &= S_\mathrm{p}/\sigma_\mathrm{rms}\\
    \sigma_\mathrm{p} &= \sigma_\mathrm{rms}\cdot\sqrt{1+\mathrm{S/N}}\\
    (\mathrm{S/N})_\mathrm{p} &= S_\mathrm{p}/\sigma_\mathrm{p}\\
    \sigma_\mathrm{t} &= \sigma_\mathrm{p}\cdot\sqrt{1+(S_\mathrm{tot}^2/S_\mathrm{p}^2)}\nonumber\\
    \sigma_\mathrm{tot} &= \sqrt{\sigma_\mathrm{t}^2+(\Delta{g}S_\mathrm{tot})^2}\label{eq:sigma_tot}\\
    \theta_\mathrm{lim} &= \dfrac{4}{\pi}\sqrt{\pi\log(2)b_\mathrm{maj}b_\mathrm{min}\log\left(\dfrac{\mathrm{S/N}+1}{\mathrm{S/N}}\right)}.
\end{align}

Here, $S_\mathrm{p}$ is the component's peak flux density, as read off in the clean map, $\sigma_\mathrm{rms}$ is the root-mean-square noise measured around the component position in the residual map, $\sigma_\mathrm{p}$ is the error of the peak flux density, $\Delta{g}$ are the errors of the absolute flux calibration due to individual telescopes' gain uncertainties and amplitude calibration, $\sigma_\mathrm{tot}$ is the error on the total flux density, and $\theta_\mathrm{lim}$ is the minimum size of a Gaussian brightness template that the interferometer with given beam ($b_\mathrm{maj}$ and $b_\mathrm{min}$ for beam major and minor axis, respectively) and signal-to-noise ratio $\mathrm{S/N}$ can resolve \citep{2015A&A...574A..84L}. Then,

\begin{align}
    \theta &= \max(\theta_\mathrm{fit}, \theta_\mathrm{lim})\\
    \sigma_\theta &= \theta/(\mathrm{S/N})_\mathrm{p}\\
    \sigma_r &= \sqrt{b_\mathrm{maj}b_\mathrm{min} + \theta^2}/(\mathrm{S/N})_\mathrm{p}\\
    \sigma_\phi &= \arctan(\sigma_r/r){\cdot}180/\pi.
\end{align}

In case $\theta$ of a Gaussian component is smaller than $\theta_\mathrm{lim}$, the size $\theta_\mathrm{lim}$ will become an upper limit on the actual size. In the following, we describe how we use all these measured quantities.

\subsection{Criteria for milli-lenses}\label{sec:criteria}

A variety of different sources could in principle show two or more compact components, which we discuss in more detail below. However, gravitational lensing conserves certain properties of the source. So, a potential milli-lens system should fulfil a set of criteria that we define in the following:

\begin{enumerate}
    \item\label{item1}
        The morphology of the sources should show two (or more) compact components at all frequencies and epochs studied. Exactly two compact images are expected for the core of a background AGN being lensed by an SMCO when the SMCO acts as a point-mass lens. If the components show extended emission or an elongation, it should be tangential to the critical curve along the Einstein radius $\theta_E$ \citep{1995PhDT.......188H}. If any elongation is present along the line of separation between images, it has to be small compared to $\theta_E$. Such an elongation can represent the source-intrinsic structure and is consistent with the lensing scenario \citep[as is the case with the curious lensed blazar JVAS B0218+357][]{1993MNRAS.261..435P, 2003MNRAS.338..599B, 2016MNRAS.457.2263S}.
    \item\label{item2}
        The surface brightness ratio (SBR) between components in region A and B fulfils $0.25<\mathrm{SBR}<4$ in any epoch studied. These are conservative limits that take into account un-modelled systematic uncertainties in imaging and model-fitting \citep[see also][]{2003MNRAS.341...13B}. In principle, the SBR should be close to unity for a milli-lens.
        We tighten our constraints to $0.25<\mathrm{SBR}<4$ compared to \citetalias{2021MNRAS.507L...6C}, where we used a threshold of $1/7<\mathrm{SBR}<7$, because of our improved error analysis and better understanding of the data properties in this work.
    \item\label{item3}
        The flux ratio (FR) between regions at a specific observing frequency should not deviate more than the 2-$\sigma$ combined error between different observing epochs. This reflects the fact that since lensing time-delays are small for small image separations $\Delta\theta$ \citep[${\Delta}t\,{\sim}\,5\left(\Delta\theta/\mathrm{mas}\right)^2\,\mathrm{seconds}$,][]{1973ApJ...185..397P} compared to VLBI observing times, any time variability should be seen in both regions in the same way.
    \item\label{item4}
        The spectral index $\alpha$ (defined as $S_{\nu}\propto\nu^\alpha$) between the components should be comparable ($\Delta\alpha<0.46$). It should be noted that this criterion corresponds to the one employed by \cite{2001PhRvL..86..584W}, where they allowed a maximum of 30\,\% deviation between the FR at two different frequencies, considering frequencies at  C- and X-band. Comparing FRs between S- and X-band, and C- and K-band, this corresponds to $\Delta\alpha<0.22$ and $\Delta\alpha<0.17$, respectively.\\
        While gravitational lensing conserves the shape of the spectrum between the images, it is in principle possible that light rays, travelling along different paths to produce multiple images, cross different plasma regions with different absorption coefficients, leading to differences in their spectra. While this is an important consideration at larger scales, we believe that at scales of $<$\,100\,mas considered here, this should be negligible, as it is highly unlikely to have an absorbing plasma screen with size $\ll$\,100\,mas traverse the line of sight, affecting only one image.
    \item\label{item5}
         There should be no significant radial proper motion between components detected ($<$\,2-$\sigma$ deviation considering the combined error between epochs). Apparent proper motion between putative lensed images can be caused either by motion of the lens or the background source \citep[see, e.g.][for the latter case]{1997VA.....41..281G,2019A&A...630A.108S}. For the first case, \cite{2001PhRvL..86..584W} calculated that for milli-lensing the expected change in image separation by a putative lens moving with 1000\,km/s transverse to the line of sight is $\sim$\,0.01\,$\mu$as/yr and thus negligible. For the second case, we have to consider two possibilities.\\
         First, the background source has a core+knot structure, leading to elongated images \citep[as shown in][]{1995PhDT.......188H}. In this case, one might expect source-intrinsic motion of the knot with respect to the core, which then leads to respective shifts within both elongated images. Since we calculate the distance only between the brightest components in each epoch, which we assume to be the stationary core, there should be no radial proper motion detected for a lens system. We note that \cite{2019A&A...630A.108S} use a detailed lens model to investigate the proper motion seen in multiple images, which is beyond the scope of this work.\\
         Second, the core itself could not be stationary. This is the case if a jet component moving close to or through the core is unresolved, which then leads to an observed shift of the core position \citep{2019MNRAS.485.1822P}. With the magnitude of this shift of the order of 0.2\,mas/yr \citep{2019MNRAS.485.1822P}, we estimate that any radial proper motion between the lensed images would be negligible. This is because the separation between two lensed images in the point-mass lens approximation does not strongly depend on the lensing impact parameter, and is always of the order of $2\theta_\mathrm{E}$. This observed motion of the core is also not expected to last longer than typical flares in AGN jets \citep{2019MNRAS.485.1822P}, and should rather manifest as a stochastic jitter of the core position. That means, if significant radial proper motion is detected between components over timescales $\gtrsim$\,1\,yr, it is against the lensing scenario.\\
         Here, we focus on criteria with which we can safely reject systems as milli-lenses. We note that proper motion can also potentially be used to confirm a system as such, if the observed proper motion reflects a specific symmetry that is related to the lensing configuration. A detailed analysis of this is dedicated to future work on the remaining lens candidates.
\end{enumerate}

Any such system that does not fulfil any one of the above criteria can be discarded as a milli-lens system. We may introduce more criteria in the future to discard any source listed here that is not already discarded by criteria \ref{item1}) - \ref{item5}).

For calculating the surface brightness of a region, we sum up the flux densities and areas of all respective components in that region. The area of a Gaussian component is calculated as $A=\theta^2$. The parameters are then calculated as follows:

\begin{align}
    \mathrm{FR} &= \dfrac{\sum_i S_\mathrm{tot,A,i}}{\sum_i S_\mathrm{tot,B,i}}\,,\\
    \mathrm{SBR} &= \dfrac{\sum_i S_\mathrm{tot,A,i}/\sum_i\theta_\mathrm{A,i}^2}{\sum_i S_\mathrm{tot,B,i}/\sum_i\theta_\mathrm{B,i}^2}\,,\\
    d &= \sqrt{r_A^2+r_B^2-2r_Ar_B\cos(\phi_1-\phi_2)}\,,\\
    \alpha &= \dfrac{\log(S_{\nu_2}/S_{\nu_1})}{\log(\nu_2/\nu_1)}\,,
\end{align}

where FR denotes the flux ratio, SBR the surface brightness ratio, $d$ the distance between components, and $S_{\nu_1}$ and $S_{\nu_2}$ are the sum of the total flux densities of all components in a region at frequencies $\nu_1$ and $\nu_2$, respectively. Here, we define the spectral index $\alpha$ as $S_{\nu}\,{\propto}\,\nu^\alpha$. To calculate the distance $d$ between regions in case of multiple components making up one region, we calculated the distance between the brightest components. All errors on these quantities are calculated with standard error propagation. We did not propagate the gain error in Eq.~\ref{eq:sigma_tot} for the absolute flux calibration in the uncertainties for FR and SBR, since they should scale the same for both components and the ratios should be unaffected. The observed brightness temperature of the components and regions has then been calculated as

\begin{align}
    T_\mathrm{b,obs} = \dfrac{c^2}{2k_\mathrm{B}\nu_\mathrm{obs}^2}\sum_i S_\mathrm{tot,i}/\sum_i\theta_\mathrm{i}^2\,\,\,,\label{eq:Tb}
\end{align}

where $c$ is the vacuum speed of light, $k_\mathrm{B}$ the Boltzmann constant, and $\nu_\mathrm{obs}$ is the observing frequency.

The most common contaminant object mimicking a potential milli-lens system is expected to be an AGN core-jet structure, which would reveal itself by not following criteria \ref{item1}), \ref{item2}) and \ref{item3}). In addition, criterion \ref{item4}) will not be fulfilled, as the spectral indices of the two components would be different because the expanding jet is expected to be more optically thin ($\alpha\lesssim-1$) and the core optically thick ($\alpha\gtrsim-0.5$).

Furthermore, the sources studied here could fall under the class of compact symmetric objects \citep[CSOs; see][]{1993AAS...182.5307R,1994ApJ...432L..87W}, which are, as with our milli-lenses, also morphologically defined as having two (compact) components detected on VLBI-scales. These components might also be straddling a central core. The observed symmetry in these objects, combined with their low abundance among radio-loud AGN as well as their lack of evidence for relativistic beaming made previous authors conclude that these are either very slowly expanding old objects \citep[see, e.g. the reviews by][]{2016AN....337....9O,2021A&ARv..29....3O} or very short-lived objects, with new evidence pointing to the latter scenario. This has been found in the most complete study of CSOs to date, conducted in a series of papers \citep{2024ApJ...961..240K,2024ApJ...961..241K,2024ApJ...961..242R,2024ApJ...977..195D}. \cite{2024ApJ...961..240K} have set out the following criteria to define CSOs, complementing and streamlining earlier definitions: i) no projected radio structure larger than 1\,kpc; ii) evidence of emission on both sides of the centre of activity; iii) no fractional variability greater than 20\,\%/yr; iv) no evidence for apparent superluminal motion in any jet component in excess of $v_\mathrm{app}=2.5c$. These criteria apply broadly to how we defined milli-lens systems above. There is no apparent reason that CSOs should fulfil criterion \ref{item2}) for a given epoch and observing frequency; however, if they do, they will most likely fulfil this criterion for all epochs due to their lack of strong variability. In this case, the spectra of the components can become a discriminating factor, as CSOs are likely to have a gigahertz peaked spectrum \citep[GPS; 82\,\% of sources in the sample of][]{2016MNRAS.459..820T}, and expected to steepen at frequencies above a few gigahertz that we are investigating here. In contrast, milli-lenses are expected to be two images of the same flat-spectrum core. However, in a new study, \cite{2024ApJ...977..195D} find the full range of spectral types in a complete sample of CSOs, with about half of the sample of 17 sources showing a flat-spectrum between 5 and 8\,GHz ($\alpha>-0.5$), and the other half a steep spectrum ($\alpha<-0.5$). Following this, they argue that CSOs and milli-lenses can only be distinguished for a putative milli-lens having $\alpha>-0.3$. In addition, CSOs can show significant proper motion, as long as their apparent speed stays below $v_\mathrm{app} = 2.5c$, while for milli-lenses, no proper motion is expected \citep[][see again \ref{item5}]{2001PhRvL..86..584W}. A prime example of a good CSO candidate found in our study can be seen in Fig.~\ref{fig:J0616}.

Initially, we restricted ourselves to flat-spectrum components as expected for milli-lenses, but as shown in \cite{2024ApJ...977..195D}, CSOs can also show flat ($\alpha>-0.5$) spectra at frequencies well above the spectral peak. Although it seems highly unlikely to observe a steep-spectrum compact component that is isolated, and then gravitationally lensed, it is not impossible if the core of the lensed background AGN is strongly self-absorbed. Currently, studies are underway to carry out a large survey of steep-spectrum CSOs, covering the unlikely scenario of the hotspot of a CSO being lensed as well.

In addition, steep-spectrum sources like CSOs are predominantly found at redshifts below one \citep{2024ApJ...961..240K},  while lensed sources on arcsecond-scales typically have $z>1$ \citep{2003MNRAS.341...13B}. This remains indicative, and we do not rule out lens candidates based on their redshifts at this point.

A third alternative is supermassive black hole binaries (SMBHBs). These objects are expected as the result of galaxy mergers, where the central SMBHs of both galaxies will spiral towards one another \citep{2005LRR.....8....8M}. If both are accreting and possibly also producing extended jets, we can expect to see the radio emission of both of them. The only known example of an SMBHB discovered this way with a separation of 7\,pc is the case of 0402+379, described in \cite{2006ApJ...646...49R}. Another example of a close dual AGN system detected with VLBI is described in \cite{2014Natur.511...57D} with a separation of 26\,mas (138\,pc). For SMBHBs, criterion \ref{item4}) will likely be fulfilled, as both components are expected to show an optically thick radio core. However, there is no reason that our other criteria should all be fulfilled. \cite{2004ApJ...602..123M} demonstrated that monitoring over several years can constrain the variability of two flat-spectrum components to discriminate between the SMBHB and the lensing scenario. In addition, to confirm (or reject) a system as SMBHB, broad-band characterisation is necessary. For example, in an SMBHB, double-peaked emission lines are expected \citep{2014Natur.511...57D}.

In conclusion, we believe that our criteria create a fine enough sieve to find and distinguish milli-lens systems. We categorise all our candidates based on the applicability of these criteria into: being still a viable milli-lens candidate; core-jet; CSO; SMBHB candidate; or other.

\section{Results}

\begin{figure}[ht]
    \centering
    \includegraphics[width=0.45\textwidth]{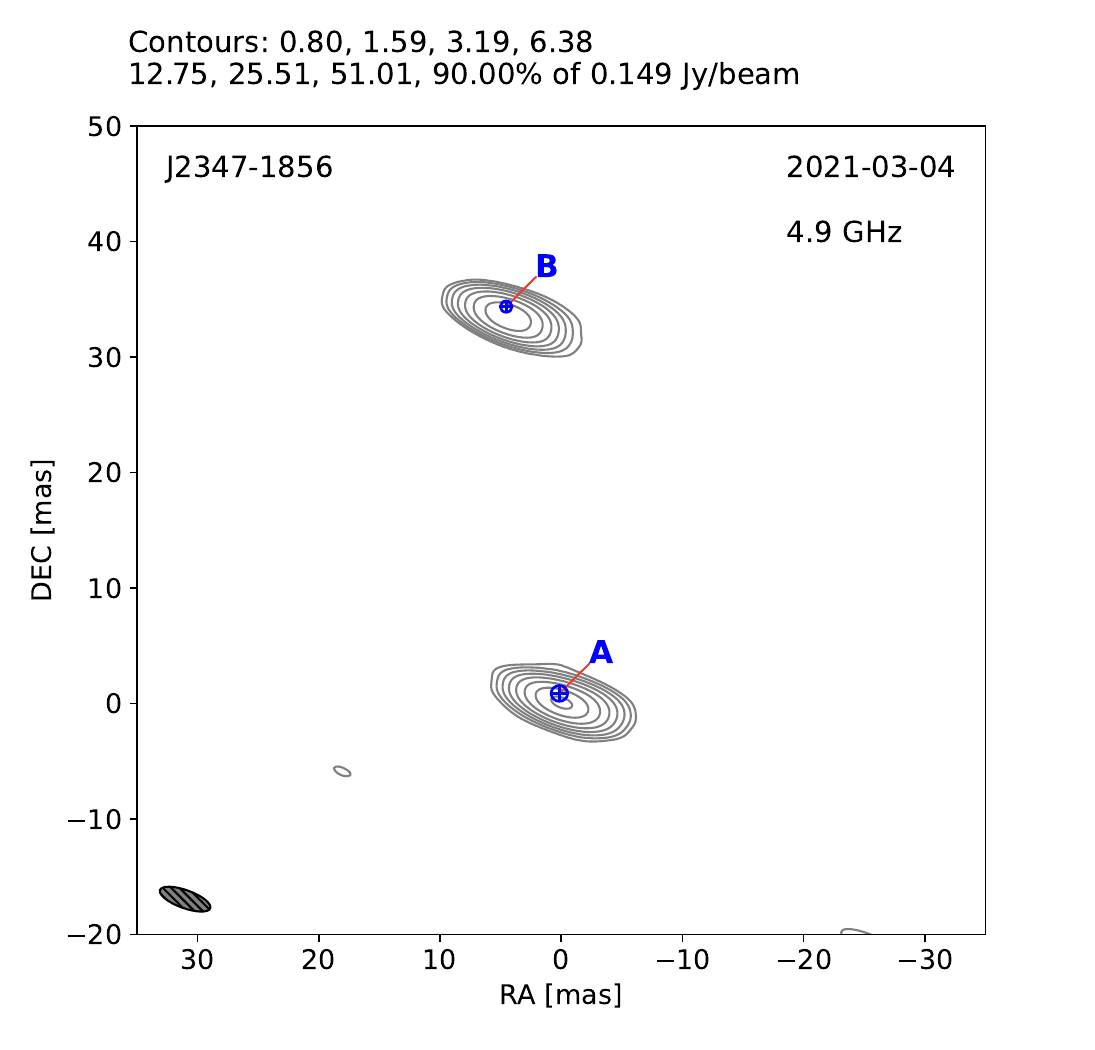}
    \caption{J2347-1856 EVN 4.9\,GHz data. The source is not yet rejected as a milli-lens but has strong indications to be a CSO. The contours start at four times the image rms noise of 0.296\,mJy/beam and increase by factors of two. The restoring beam size is 1.55$\times$4.44\,mas at $68.91$\,$^\circ$ PA. The source is not detected at 22.2\,GHz.}
    \label{fig:J2347}
\end{figure}

Here we present the results of our study. Figure \ref{fig:J2347} shows the 4.9\,GHz image for the source J2347-1856, as an example of a source that we cannot discard as a milli-lens candidate at this stage. Table~\ref{tab:J2347} lists all the derived properties with their respective errors. The images and derived quantities for the other sources are listed in appendix \ref{sec:app1}.

\begin{table*}[ht]
    \caption{Derived quantities for J2347-1856. See Table~\ref{tab:J2209} for a detailed description of the columns.}
    \vspace*{3mm}
    \adjustbox{width=1\textwidth}{%
    \label{tab:J2347}
    \centering
    \begin{tabular}{|| c c c c c | c c | c c | c c | c c | c c | c ||} 
        \hline
        Epochs & Frequency & Component & Flux  & Error  & Flux  & Error & FWHM  & Error & $\log(T_\mathrm{b,obs}/\mathrm{K})$ & Error  & SB    & Error & Distance & Error & Spectral Index \\ [0.5ex] 
               & [GHz]     &           & [mJy] & [mJy]  & Ratio &       & [mas] & [mas] &                                     &        & Ratio &       & [mas]    & [mas] &                \\
        \hline\hline
1997-07-02 & 2.3 & A & 493.4 & 60.8 & 1.48 & 0.17 & 1.98 & 0.11 & 10.53 & 0.07 & 0.43 & 0.09 & 33.31 & 0.47 & -0.85 \\
- & -   & B & 333.2 & 44.2 & -   & -   & 1.06 & 0.07 & 10.90 & 0.08 & -   & -   & -   & -   & -0.72 \\
        \hline
1997-07-02 & 8.3 & A & 163.1 & 35.7 & 1.25 & 0.38 & 1.11 & 0.18 & 9.42 & 0.17 & 1.76 & 1.07 & 21.85 & 0.48 & - \\
- & -   & B & 130.8 & 32.8 & -   & -   & 1.32 & 0.27 & 9.18 & 0.21 & -   & -   & -   & -   & - \\
        \hline\hline
1997-08-27 & 2.3 & A & 273.0 & 60.8 & 1.35 & 0.41 & $<$1.52 & - & $>$10.50 & - & $\sim$1.82 & - & 32.98 & 2.26 & -0.40 \\
- & -   & B & 201.7 & 51.0 & -   & -   & $<$1.77 & - & $>$10.24 & - & -   & -   & -   & -   & -0.72 \\
        \hline
1997-08-27 & 8.3 & A & 163.2 & 56.7 & 2.07 & 1.13 & 1.89 & 0.57 & 8.96 & 0.30 & 0.48 & 0.51 & 21.87 & 1.03 & - \\
- & -   & B & 78.8 & 35.0 & -   & -   & 0.91 & 0.31 & 9.27 & 0.35 & -   & -   & -   & -   & - \\
        \hline\hline
2014-06-09 & 2.3 & A & 372.7 & 42.5 & 1.50 & 0.13 & 2.12 & 0.09 & 10.34 & 0.06 & 0.83 & 0.13 & 33.32 & 0.47 & -0.74 \\
- & -   & B & 248.8 & 30.0 & -   & -   & 1.57 & 0.08 & 10.42 & 0.07 & -   & -   & -   & -   & -0.52 \\
        \hline
2014-06-09 & 8.7 & A & 138.7 & 16.6 & 1.12 & 0.11 & 0.98 & 0.05 & 9.43 & 0.07 & 0.95 & 0.17 & 33.65 & 0.15 & - \\
- & -   & B & 123.5 & 15.0 & -   & -   & 0.90 & 0.05 & 9.45 & 0.07 & -   & -   & -   & -   & - \\
        \hline\hline
2017-05-01 & 2.2 & A & 428.7 & 46.5 & 1.43 & 0.09 & 2.04 & 0.07 & 10.45 & 0.05 & 0.67 & 0.08 & 33.53 & 0.25 & -0.89 \\
- & -   & B & 299.5 & 33.4 & -   & -   & 1.39 & 0.05 & 10.62 & 0.06 & -   & -   & -   & -   & -0.60 \\
        \hline
2017-05-01 & 8.7 & A & 128.9 & 15.0 & 0.97 & 0.08 & 0.82 & 0.04 & 9.55 & 0.07 & 1.43 & 0.23 & 33.61 & 0.11 & - \\
- & -   & B & 133.1 & 15.5 & -   & -   & 1.00 & 0.05 & 9.39 & 0.07 & -   & -   & -   & -   & - \\
        \hline\hline
2018-04-08 & 2.2 & A & 434.1 & 46.6 & 1.41 & 0.09 & 2.09 & 0.06 & 10.44 & 0.05 & 0.67 & 0.07 & 33.44 & 0.24 & -0.88 \\
- & -   & B & 307.3 & 33.9 & -   & -   & 1.44 & 0.05 & 10.61 & 0.06 & -   & -   & -   & -   & -0.59 \\
        \hline
2018-04-08 & 8.7 & A & 133.3 & 15.5 & 0.96 & 0.08 & 0.78 & 0.04 & 9.61 & 0.07 & 1.78 & 0.30 & 33.49 & 0.11 & - \\
- & -   & B & 138.9 & 16.3 & -   & -   & 1.06 & 0.06 & 9.36 & 0.07 & -   & -   & -   & -   & - \\
        \hline\hline
2021-03-04 & 4.9 & A & 211.2 & 24.1 & 1.14 & 0.09 & 1.36 & 0.06 & 9.81 & 0.06 & 0.58 & 0.09 & 33.78 & 0.18 & - \\
- & -   & B & 185.2 & 21.3 & -   & -   & 0.97 & 0.04 & 10.05 & 0.06 & -   & -   & -   & -   & - \\
        \hline
    \end{tabular}
    }
\end{table*}

\begin{figure*}[h]
    \centering
    \includegraphics[width=0.4\textwidth]{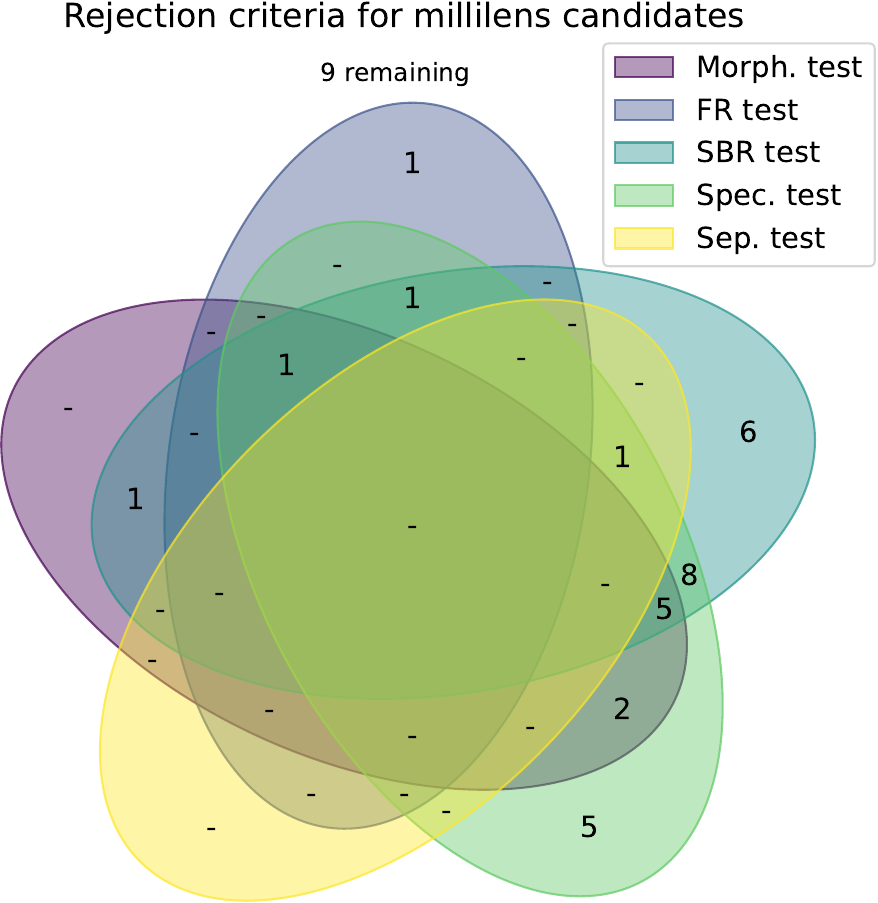}
    \hspace{0.5cm}
    \includegraphics[width=0.52\textwidth]{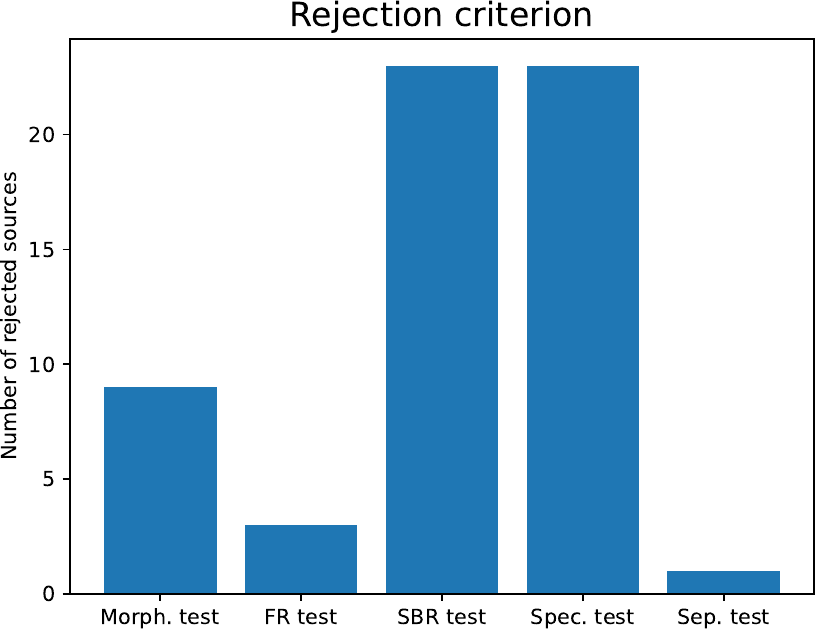}
    \caption{Summary charts for milli-lens rejection criteria. \textit{Left:} Venn diagram showing which combination of criteria lead to source rejection. \textit{Right:} Bar diagram showing the total number of sources rejected for each criterion alone.}
    \label{fig:summary_chart_V2}
\end{figure*}

We summarise our categorisation of lens candidates in Table \ref{tab:summary}. In total, we rule out 31 sources as milli-lens systems from our initial 40 candidates. We discuss the sources rejected as milli-lens candidates in more detail in Sect.~\ref{sec:discussion}. As can be seen from Fig.~\ref{fig:summary_chart_V2}, the most discriminating factors in rejecting milli-lens systems are the spectrum test (23 sources rejected), the SBR test (23 sources rejected) and the morphology test (nine sources rejected). Less discriminating are the FR test (three sources rejected) and the separation test (one source rejected). Investigating both variability (FR test) and proper motion (separation test) is expected to be more discriminating with the addition of more epochs of observations, as we have only two epochs for 19 of our sources. In total, 19 out of 31 rejected lens candidates fail more than one of our criteria.

\section{Discussion}\label{sec:discussion}

We searched the literature and various databases for additional data on our 40 candidates. For 25 of the 40 sources, we found optical counterparts with available redshifts within 1\,\arcsec. Optical images are presented in Figs.~\ref{fig:optical1} and \ref{fig:optical2}. Six of those sources have spectroscopic redshifts available. For another six sources, we found optical counterparts in \textit{Gaia} DR3 (\citealt{2016A&A...595A...1G}; \citealt{2023A&A...674A...1G}).

\subsection{Gaia-VLBI offsets}

To get the offsets of the \textit{Gaia} photo-centre to the VLBI images, we compared the J2000 source positions in the Radio Fundamental Catalog\footnote{\url{http://astrogeo.org/sol/rfc/rfc_2020c/}} (RFC), the International Celestial Reference Frame 3 \citep[ICRF3;][]{2020A&A...644A.159C} and the VLBI Celestial Reference Frame Solution OPA2024a\footnote{\url{https://ivsopar.obspm.fr/24h/index.html}} close to our observing epochs with the \textit{Gaia} coordinates. The \textit{Gaia} coordinates are given in J2016, which differs from J2000 only if significant proper motion is observed. While this correction is mainly necessary for stars, some extragalactic sources also show significant proper motion, which may be due to the motion of bright jet components compared to the jet base. Since we want to compare the positions of the VLBI core with the jet base seen in the optical, we do not apply any corrections to the J2016 coordinates taken from \textit{Gaia} to convert them to J2000.

\textit{Gaia}-VLBI offsets have been systematically studied in various works. \cite{2019ApJ...871..143P} studied the offset in over 1000 AGN, comparing VLBI positions from the RFC and \textit{Gaia} DR2. They find a significant trend of the offset being either upstream or downstream of the jet direction, supporting the fact that the optical emission is either dominated by the accretion disc (upstream offset) or the jet (downstream offset). \cite{2024A&A...684A.202L} find a similar trend, comparing model-fit positions from the MOJAVE survey to the photo centres from \textit{Gaia} DR3. Both studies find a trend for downstream offsets to occur more often in flat spectrum radio quasars (FSRQs) compared to BL Lac objects, supporting the fact that the former have more powerful accretion discs. Variability in the optical jets could cause a jitter in the VLBI-\textit{Gaia} offset \citep{2017MNRAS.471.3775P} and may be used to constrain the location of optical flares. Proper motion of jet components in the optical may be traced with future data releases from \textit{Gaia}. Offsets between VLBI and \textit{Gaia} positions may also be used to investigate possible SMBHB systems. \cite{2023ApJ...958...29C} identified a few SMBHB candidates with VLBI follow-up from an initial selection of candidates based on variability-induced astrometric jitter observed with \textit{Gaia} \citep{2020ApJ...888...73H}.

Out of the six sources with measurable VLBI-\textit{Gaia} offsets in this work, three (J0132+5211, J0237+1116 and JJ2114+4036) show offsets clearly along the direction of the radio jet. These systems are also more likely to classify as core-jet systems instead of being CSOs, and are rejected as milli-lens systems (see Sect.~\ref{sec:notes}). The other three sources show different offsets, and their interpretation is beyond the scope of this paper. Aside from being caused by potential dual or offset AGN systems, the VLBI-\textit{Gaia} offsets could be caused by a variety of effects, including contamination by the emission of the host galaxy, dust obscuration affecting the observed photo-centres, or even \textit{Gaia} observing foreground stars.

\subsection{Notes on individual sources}\label{sec:notes}

\subsubsection{J0010-0740 (0008-079)}
The source (Fig.~\ref{fig:J0010}) is unclassified in the literature. We discarded this source as a milli-lens because of the different spectra of the components and the high SBR of 6.82 at C-band and 22.17 at X-band, respectively. The rejection as a lens candidate and classification as a CSO is also affirmed by the detection of a central component C between regions A and B. The spectra of the two regions A and B are notably different, with A showing a rather flat spectrum, while B has a steep spectrum. Since component C is not identified at the 2015 C-band epoch, we cannot calculate its spectrum to see if it could be a flat-spectrum core straddled by components A and B. The source is a good CSO candidate, the VLBI fractional flux variability being $<$\,1\,\%/yr. J0010-0740 has an optical counterpart classified as a galaxy with 22.17 optical r-band magnitude and photometric redshift of $z=0.31\,{\pm}\,0.09$ in the Sloan Digital Sky Survey \citep[SDSS DR17;][]{2022ApJS..259...35A}. The linear distance between components A and B at C-band is then $\sim$\,61\,pc, consistent with the CSO definition.

\subsubsection{J0011+3443 (0008+344)}
The source (Fig.~\ref{fig:J0011}) has an optical counterpart with 23.2\,mag at r-band with a spectroscopic redshift of $z=0.89$ \citep{2010MNRAS.401.1709V} and is associated with a compact, steep spectrum (CSS) radio source. This source was discarded as a lens candidate because of the low SBR at C-band of 0.09 in the 2020 epoch. The fractional variability in our data is about 9\,\%/yr, consistent with a CSO. The distance between components B and A0 also does not change over a period of more than 4 years. The spectra are remarkably similar, and steep. The linear distance between the components is $\sim$\,315\,pc. All of this favours this source to be a CSO.

\subsubsection{J0024-4202 (0022-423)}
The source (Table~\ref{tab:J0024}) has previously been classified as a GPS radio galaxy \citep[e.g.][]{2007AA...463...97L}, with the radio spectrum peaking at 1.6\,GHz. GPS sources are likely to fall into the CSO category as well \citep{2016MNRAS.459..820T}. The optical counterpart \citep[20.3 r-band magnitude;][]{1995A&AS..114..259D} has a spectroscopic redshift $z=0.937$ and is classified as quasi-stellar object \citep[QSO;][]{2007AA...463...97L}. The integrated radio spectrum of the source appears rather flat around frequencies up to a few gigahertz, but it seems to become considerably steeper at higher frequencies \citep{2007ApJS..171...61H}. \cite{2014ApJS..215...14D} classify the source as FSRQ based on matching known $\gamma$-ray blazars with their IR colours measured with the Wide-field Infrared Survey Explorer (\textit{WISE}).

We did not observe this source with the EVN due to its low declination. The epochs we have from Astrogeo show two similar and steep spectra for both components. The variability between the first two X-band epochs is much more than 20\,\%/yr, while for the other epochs, it is within about 10\,\%. We found that, for the early X-band epochs in Astrogeo, the weaker component is found very close to the edge of the field of view in the map provided in Astrogeo, and thus we are cautious in interpreting the observed high variability in these datasets, and still judge this source as a CSO candidate. The linear distance between the components is $\sim$\,230\,pc, consistent with being a CSO. The previous classification as blazar has to be further scrutinised with future observations in terms of radio variability and possible superluminal motion. Ultimately, we do not reject this source as a lens candidate at this point.

\subsubsection{J0044+2858 (0041+287)}
The source (Fig.~\ref{fig:J0044}) is unclassified in the literature. We found an optical counterpart in SDSS classified as a galaxy with 21.57 r-band magnitude and photometric redshift of $z=0.75\,{\pm}\,0.07$. The source has two steep-spectrum components in the VLBI image, with almost no variability ($<$\,4\,\%/yr) observed over 5 years between the two available epochs. We discarded this source as a milli-lens because of the different component spectra. The linear distance between the identified components A and B is $\sim$\,38\,pc. We classify it as a good CSO candidate.

\subsubsection{J0052+1633 (0049+162)}
For this source (Fig.~\ref{fig:J0052}) we found an optical counterpart with 22.3 r-band magnitude in SDSS classified as a galaxy with photometric redshift of $z=0.43\,{\pm}\,0.04$. That means the linear distance between components A and B is $\sim$\,49\,pc. The spectra are notably different, with component B having a much flatter (or even inverted) spectrum compared to component A. This trend continues for the spectrum between 5 and 22\,GHz, although the difference is not as prominent. The source was discarded as a gravitational lens system because of the spectra, but also because of the too-large FR variations of 0.87 compared to the combined error of 0.81 at C-band, and the high SBR of 6.44 observed at K-band. The morphology also does not support the lensing scenario, since component A shows some elongated emission towards component B. For all frequencies and epochs, the variability is maximally about 20\,\%/yr. The source might still classify as a CSO since for a core-jet system, the brighter component would be expected to be the flat-spectrum core and more variability is expected. The component separation of $\sim$\,49\,pc also fits into the CSO scenario.

\subsubsection{J0118+3810 (0115+379)}
The source (Fig.~\ref{fig:J0118}) is unclassified in the literature. There is an optical counterpart with $r=27.3$\,mag from the Blazar Radio and Optical Survey (BROS) catalogue \citep[BROS J0118.1+3810;][]{2020ApJ...901....3I}, but no available redshift. The two steep-spectrum components, as well as the lack of proper motion and variability (${\sim}\,1(1+z)$\,\%/yr) over 4 years, strongly suggest the source to be a CSO. It was ultimately rejected as milli-lens candidate because of the SBR being above the threshold of $4$ within the uncertainties for all epochs.

\subsubsection{J0132+5211 (0129+519)}
The source (Fig.~\ref{fig:J0132}) has been classified as an FSRQ in \cite{2019ApJS..242....4D} based on its IR colours in \textit{WISE}, which were selected to match those of known $\gamma$-ray emitting blazars. An optical counterpart with photometric $z=0.98\,{\pm}\,0.16$ is found in \cite{2015ApJS..219...39R}, and the r-band magnitude listed in SDSS is 19.97. The spectra of both components are steep but significantly different. Also, the SBR fluctuates between being below the threshold of 0.25 and above the threshold of 4, failing the source as a milli-lens candidate. Component A varies maximally by 4\,\%/yr, component B by 21\,\%/yr, and the total flux by 6\,\%/yr. Surprisingly, the distance between the components has decreased more than the 2-$\sigma$ combined error of 1.54\,mas over the course of 4 years. That is also inconsistent with a milli-lens. In addition, the rejection as a milli-lens is supported by the position of the \textit{Gaia} photo-centre (see Fig.~\ref{fig:J0132}), found in \textit{Gaia} DR3 with 19.85 G-band magnitude, whose position is very close to component A (2.5\,mas away). This hints at the core-jet nature of the source, with component A being the core. The linear distance between the components is $\sim$\,1.2\,kpc at the given redshift, which puts the source outside the CSO definition. The source is more likely a core-jet system, although in that case, the large size would still be puzzling, and more variability should be expected. The source remains a curious case and more data will shed light on the nature of this interesting source.

\subsubsection{J0139+0824 (0137+081)}
The source (Fig.~\ref{fig:J0139}) is unclassified in the literature. There is an optical counterpart in SDSS, classified as a galaxy with 21.10\,mag at r-band, and there is a photometric redshift $z=1.05\,{\pm}\,0.11$ from DES DR8 \citep[Dark Energy Survey;][]{2019AJ....157..168D}. In the VLBI image there are two steep-spectrum components, separated by $\sim$\,130\,pc, with some indication of emission towards the centre between them. This and the lack of significant variability ($\sim$\,4\,\%/yr) of this source over four years suggests it to be a CSO. The difference between the spectral indices between components of $\Delta\alpha=0.47$ is larger than our defined threshold and thus does not fit a milli-lens system.

\subsubsection{J0203+3041 (0200+304)}
The source is associated with an optical counterpart classified as a galaxy in SDSS, and there is also a \textit{Fermi} $\gamma$-ray source associated \citep{2015ApJS..217....2M}. The source was discarded as a milli-lens because of the high SBR (${\gg}$\,4) observed at K-band. Also, a new component C between region A and B appears in the new EVN C-band image (Fig.~\ref{fig:J0203}), and regions A and B slightly extend towards each other. This is highly indicative of the source being a CSO, although not enough to reject the source based on criterion \ref{item1}. The redshift listed in \cite{2015ApJS..217....2M} is $z=0.76$, while \cite{2015ApJ...810...14A} list it at $z=0.96$ \citep[spectroscopic, originally in][]{2012ApJ...748...49S} and categorise it as FSRQ based on optical classification. The magnitude listed in SDSS is 21.84 at r-band. The NVSS image shows two sources separated by $\sim$1\,\arcmin. In the MOJAVE survey \citep{2018ApJS..234...12L}, there is a source about 0.7\,\arcmin away from J0203+3041 with the same B1950 source identifier, which is probably why these MOJAVE data are falsely listed on the Astrogeo webpage for J0203+3041. This other source is also listed at $z=0.76$ and is categorised as BL Lac object \citep{2015ApJ...810...14A}. We decided to adopt a value of $z=0.96$ for J0203+3041. Then, the linear distance between components is $\sim$322\,pc. The spectra of both components are steep. The variability is $\gg$\,20\,\%/yr between the two close S-band epochs but is otherwise observed to be between 10 and 30\,\%/yr. While the morphology and the spectra are suggestive of a CSO, the variability is too high to classify as such. The source is more likely a core-jet system, which is also more likely to be $\gamma$-bright. However, we note that in \cite{2024ApJ...961..240K}, three out of the 79 bona-fide CSOs are also detected in $\gamma$-rays, most likely originating from the radio lobes.

\subsubsection{J0210-2213 (0207-224)}
The source (Fig.~\ref{fig:J0210}) was previously classified as a GPS source \citep{2002MNRAS.337..981S}, peaking at 1.5\,GHz. The optical counterpart is located at redshift $z=1.491\,{\pm}\,0.003$ \citep[spectroscopic;][]{2007AA...464..879D} with a magnitude of 23.52 at r-band \citep{2002MNRAS.337..981S}. The linear distance between components is $\sim$\,247\,pc, falling within the CSO category. The two steep component spectra fit well as a CSO, but we also cannot reject the source as a milli-lens at this stage. The fractional variability is just within $\sim$20\,\%/yr, barely passing as a CSO.

\subsubsection{J0213+8717 (0159+870)}
This source (Fig.~\ref{fig:J0213}) has an unclassified optical counterpart with $z=1.5$ and 19.6 r-band magnitude found in Milliquas v7.2. Since it is not listed in Milliquas v8 \citep{2023OJAp....6E..49F} any more, we decided to consider it as an unknown redshift. We rejected the source as a milli-lens candidate due to the high SBR of 19.8 seen in the EVN C-band data, and the very different spectra between components. \cite{2014MNRAS.438.3058R} studied the radio variability of a sample of mostly $\gamma$-ray detected blazars and found a modulation index of 0.077 (defined as the standard deviation of the flux density measurements in units of the mean measured flux density) in the 15\,GHz OVRO light-curve for this source. The maximum fractional variability we found is ${\sim}\,10(1+z)$\,\%/yr. The spectra of the two components are different, with component B having a much flatter, or even inverted spectrum. While not significant within the errors, there is an indication of an increase in separation between the components over time. The observed component brightness temperatures are also partly in excess of $10^{11}$\,K, a typical value for the equipartition brightness temperature $T_\mathrm{eq}$ in AGN (\citealt{1994ApJ...426...51R}; \citealt{2018ApJ...866..137L}). Values above $T_\mathrm{eq}$ indicate Doppler boosting due to the source moving towards the observer close to the line of sight. All this indicates this source to be a core-jet system.

\subsubsection{J0222+0952 (0219+096)}
This source (Fig.~\ref{fig:J0222}) is unclassified in the literature, with no optical counterpart. We observed the source at C-band in two sessions. It shows a peculiar structure with three main regions A, B and C, and some emission between them, that encompasses a significant fraction of the separation between the regions. While the source does not fulfil the milli-lens criteria for morphology, SBR, and spectra, it might still classify as both a CSO or core-jet. However, the variability we see between the two EVN epochs at C-band is much in excess of 20\,\%/yr, excluding the CSO possibility. The source appears point-like on VLA scales with a rather flat spectrum of $\alpha\,{\sim}-0.4$ \citep{2010A&A...511A..53V}, also more suggestive of a core-jet system.

\subsubsection{J0232-3422 (0230-345)}
This source (Table~\ref{tab:J0232}) is unclassified in the literature, with an optical counterpart found in DES DR2 (J023230.02-342203.8) with 24.51\,mag at r-band, $\sim$\,0.3\,\arcsec away from the VLBI position. We did not yet obtain follow-up data for the source due to its low declination which is inaccessible for the EVN. The spectral indices of the two components differ by 0.76, failing the source as a milli-lens system. Since there is only one epoch available, we cannot make further statements about the nature of the source at this point, and further analysis of archival data and new observations will shed light on this.

\subsubsection{J0237+1116 (0234+110)}
This source (Fig.~\ref{fig:J0237}) was already briefly described and ruled out as a lens in \citetalias{2021MNRAS.507L...6C}. It is associated with an optical counterpart in \textit{Gaia} with 17.91 G-band magnitude, and we found a spectroscopic redshift of $z=0.985\,{\pm}\,0.067$ listed in \cite{2024ApJ...964...69S}. The \textit{Gaia} photo-centre is close to component A with a separation of 6.1\,mas to the south-west. We reported a different offset with a smaller magnitude in \citetalias{2021MNRAS.507L...6C}, where we used the converted J2000 coordinates (see the beginning of Sect.~\ref{sec:discussion}). The difference is due to some proper motion seen between the J2000 and J2016 epochs, which may suggest some moving jet component. The direction of the offset vector is close to the jet direction as estimated from the separation vector between the two components A and B, which further hints at the source being a core-jet system. In addition, we could discard this system as a milli-lens due to its high SBR at X-band of 8.65. The fractional variability of the source is 17\,\%/yr, while both components show flat spectra. The linear distance between components is $\sim$\,41\,pc. This source might be classified as either a core-jet or a CSO, and more observations are needed to study the variability and spectra.

\subsubsection{J0502+1626 (0459+163)}
This source (Fig.~\ref{fig:J0502}) is unclassified in the literature, and we could not find an optical counterpart. The source has been discarded as a milli-lens due to its high SBR at X-band of 8.26 and the significant difference in the spectra of components. No strong variability is observed, but there is an indication of an increase in the distance between the components. The source could be a core-jet or a CSO.

\subsubsection{J0527+1743 (0524+176)}
This source (Fig.~\ref{fig:J0527}) is unclassified in the literature. It has an optical counterpart in SDSS classified as a galaxy with 22.47\,mag at r-band and photometric redshift $z=0.46\,{\pm}\,0.08$. We discarded this source as a milli-lens because of its SBR being $<$\,0.24 at X-band. The linear distance between components is $\sim$\,107\,pc. The fractional variability is in slight excess of 20\,\%/yr, but we still judge this source as a good CSO candidate. Since both spectra are flat, the source could also qualify as SMBHB candidate.

\subsubsection{J0616-1957 (0613-199)}
\begin{figure}[h]
    \centering
    \includegraphics[width=0.45\textwidth]{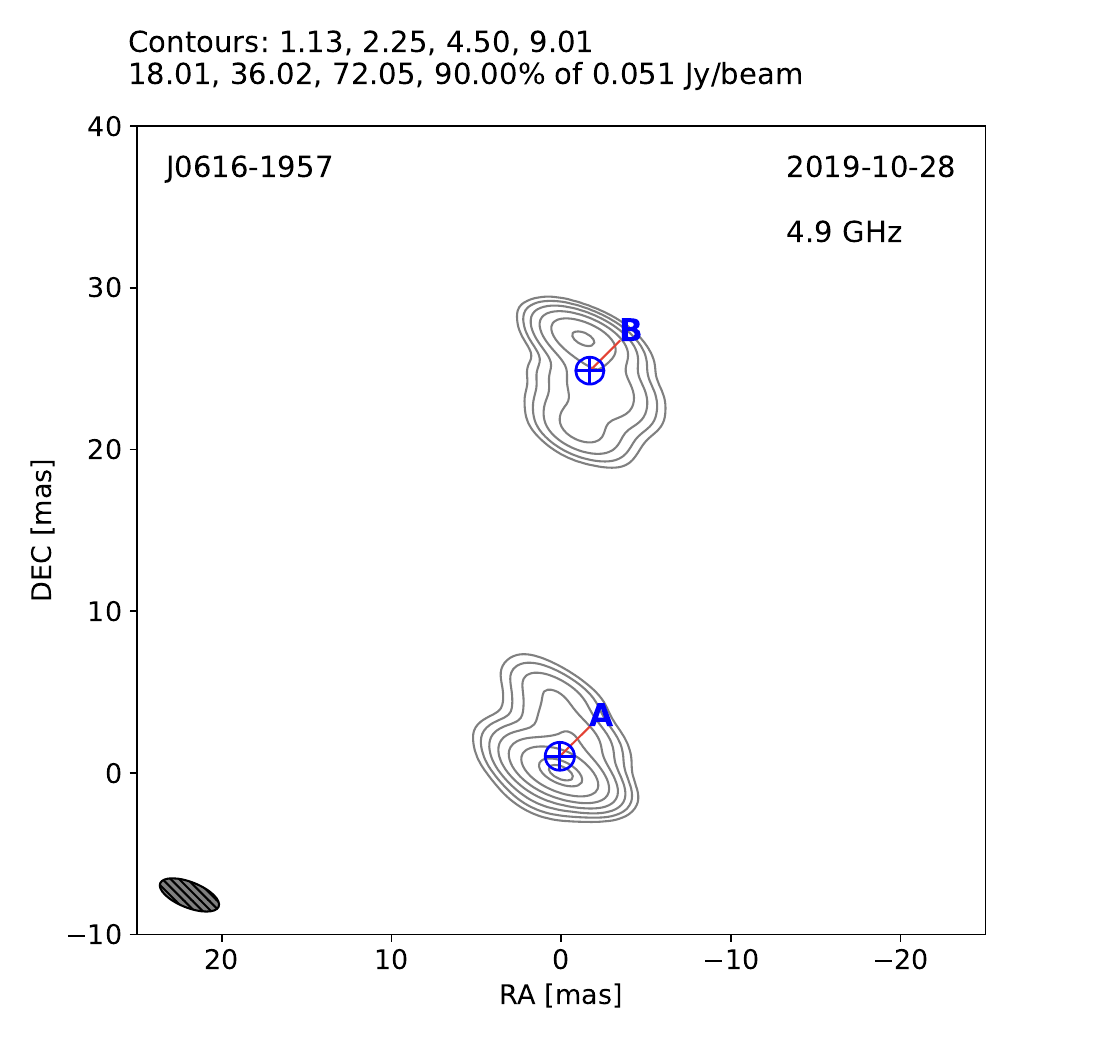}
    \caption{J0616-1957 EVN 4.9\,GHz data. The contours start at four times the image rms noise of 0.142\,mJy/beam and increase by factors of two. The restoring beam size is 1.57$\times$3.74\,mas at 66.9\,$^\circ$ PA. The source is not detected at 22.2\,GHz. Despite being rejected as a milli-lens candidate, it shows clear morphological evidence of being a CSO with its extended radio lobes, which were only revealed through our new EVN observations.}
    \label{fig:J0616}
\end{figure}

This source (Fig.~\ref{fig:J0616}) is unclassified in the literature. We discarded this source as a milli-lens because of the high SBR at X-band of 8.49, the different spectra between components, and the clear extension of both emission regions towards the centre. The steep spectra and the edge-brightening of the two hotspots are typical for CSOs. There is no available optical counterpart and redshift, but at a fiducial redshift of 1, the source would have a separation between components of $\sim$\,200\,pc. The fractional variability of the source is $<2(1+z)$\,\%/yr. We deem this source a CSO.

\begin{table*}[h]
    \caption{Derived quantities for J0616-1957. See Table~\ref{tab:J2209} for a detailed description of the columns.}
    \vspace*{3mm}
    \adjustbox{width=1\textwidth}{%
    \label{tab:J0616}
    \centering
    \begin{tabular}{|| c c c c c | c c | c c | c c | c c | c c | c ||} 
        \hline
        Epochs & Frequency & Component & Flux  & Error  & Flux  & Error & FWHM  & Error & $\log(T_\mathrm{b,obs}/\mathrm{K})$ & Error  & SB    & Error & Distance & Error & Spectral Index \\ [0.5ex] 
               & [GHz]     &           & [mJy] & [mJy]  & Ratio &       & [mas] & [mas] &                                     &        & Ratio &       & [mas]    & [mas] &               \\
        \hline\hline
2016-01-20 & 4.3 & A & 84.8 & 7.8 & 1.62 & 0.15 & -   & -   & 9.30 & 0.05 & 1.57 & 0.28 & -   & -   & -1.68 \\
- & -   & A0 & 53.0 & 6.5 & 1.02 & 0.10 & 0.78 & 0.04 & 9.80 & 0.03 & 4.96 & 0.91 & 24.35 & 0.24 & - \\
- & -   & B & 52.2 & 6.5 & -   & -   & 1.73 & 0.11 & 9.11 & 0.08 & -   & -   & -   & -   & -2.44 \\
        \hline
2016-01-20 & 7.6 & A & 33.0 & 4.4 & 2.50 & 0.45 & 0.89 & 0.06 & 8.99 & 0.08 & 8.49 & 3.08 & 24.31 & 0.34 & - \\
- & -   & B & 13.2 & 2.4 & -   & -   & 1.65 & 0.23 & 8.06 & 0.15 & -   & -   & -   & -   & - \\
        \hline\hline
2019-10-28 & 4.9 & A & 86.5 & 10.1 & 1.77 & 0.18 & 1.73 & 0.09 & 9.22 & 0.07 & 1.64 & 0.33 & 23.91 & 0.25 & - \\
- & -   & B & 48.8 & 6.3 & -   & -   & 1.66 & 0.11 & 9.00 & 0.08 & -   & -   & -   & -   & - \\
        \hline
    \end{tabular}
    }
\end{table*}

\subsubsection{J0732+6023 (0728+605)}
This source (Fig.~\ref{fig:J0732}) is classified as a flat-spectrum radio source measured around gigahertz frequencies \citep{2014ApJS..213....3M}. Our image shows an extended and bright region A, that is model-fitted with three components, and only component B comprising region B. We discarded the source as a milli-lens based on the very different spectra, where region A is very steep and component B has a very flat spectrum. It also did not pass our SBR test at X-band, being with $>$\,23 much in excess of our defined threshold. The maximum fractional variability of the source that we found is $\sim$7\,\%/yr. This source is a CSO candidate, but could also still be a core-jet system. The linear distance between the components is $\sim$163\,pc, based on the photometric redshift of the optical counterpart of $1.064\,{\pm}\,0.162$ found in DES DR8.

\subsubsection{J0923-3435 (0921-343)}
This source is unclassified in the literature, and we did not obtain new data due to the source's low declination which is inaccessible to the EVN. The integrated radio spectrum on VLA-scales peaks at around 1\,GHz \citep{2010A&A...511A..53V}, which is consistent with the data from Astrogeo (Table~\ref{tab:J0923}). The source could be classified as a CSO, but more observations are needed to study the possible proper motion and variability of the components. Without additional data, we cannot reject the source as a milli-lens at this stage. Analysing more archival data as well as possible new observations will shed light on the nature of this source.

\subsubsection{J1132+5100 (1130+512)}
We discarded this source (Fig.~\ref{fig:J1132}) as a milli-lens due to its SBR being 0.17 and 0.12 at two different epochs, the different spectra, and the extension of emission region A towards B, which is revealed in the new EVN C-band image and encompasses a significant fraction of the distance between the two regions. With a photometric redshift of $z=1.152\,{\pm}\,0.303$ (DES DR8), the linear distance between components is $\sim$\,116\,pc, consistent with a CSO. This is also supported by the rather low variability of $\sim$\,7\,\%/yr.

\subsubsection{J1143+1843 (1140+188)}
This source (Fig.~\ref{fig:J1143}) has previously been reported to be a CSO \citep{2011A&A...535A..24S} and an SMBHB candidate \citep{2016MNRAS.459..820T}, which we confirm with our analysis. The rich amount of archival available datasets together with newly obtained data for this source will be discussed in an upcoming, dedicated publication (\'Alvarez-Ortega et al., in prep.). At this stage, we are not rejecting this source as a lens candidate.

\subsubsection{J1218-2159 (1216-217)}
This source (Fig.~\ref{fig:J1218}) is unclassified in the literature. There is no optical counterpart. This source is still a viable milli-lens candidate, passing all our current tests. With the overall variability being maximally $\sim$\,8$(1+z)$\,\%/yr, the source is also a good CSO candidate. This source will be studied in more detail concerning spectra and variability with future, more sensitive observations.

\subsubsection{J1306+0341 (1303+039)}
This source (Fig.~\ref{fig:J1306}) has no optical counterpart. We discarded the source as milli-lens because of the high SBR of 8.38 and 7.42 in two epochs. The source morphology, spectra and the $\sim$\,8$(1+z)$\,\%/yr fractional variability are very indicative of the CSO nature.

\subsubsection{J1340-0335 (1337-033)}
This source (Fig.~\ref{fig:J1340}) has an optical counterpart and associated photometric redshift of $z=0.977\,{\pm}\,0.489$ in DES DR8. The source has been identified to peak at low radio frequencies below 1\,GHz \citep{2017ApJ...836..174C}. We discarded the source as milli-lens because of the high SBR of 9.14 and 12.1 in two X-band epochs and the new EVN epoch having a high SBR of 12.09. The source morphology, showing three emitting regions that are clearly connected, also does not fit a gravitational lens system. The spectra support this source being a CSO, while the variability is only in slight excess of $20$\,\%/yr, so we retain is as a CSO candidate. The linear distance between components, which is $\sim$\,58\,pc, also fits the CSO scenario.

\subsubsection{J1344-1739 (1341-174)}
This source (Fig.~\ref{fig:J1344}) has also been identified to peak at low radio frequencies below 1\,GHz \citep{2017ApJ...836..174C}, and has no optical counterpart. This source failed our milli-lens criteria for the spectra, whose spectral indices differ by 0.53, and the fact that region A shows a clear eastward extension towards region B in both the EVN C-band image and different Astrogeo images. With the variability being less than $10(1+z)$\,\%/yr, this source is a good CSO candidate.

\subsubsection{J1632+3547 (1630+358)}
\cite{2016MNRAS.459..820T} discarded this source (Table\,\ref{tab:J1632}) as a CSO candidate, but they still propose it as SMBHB candidate. The source has a photometric redshift of $z=1.081\,{\pm}\,0.082$ (DES DR8) and an r-band magnitude of 23.70 in the BROS catalogue. The linear distance between components is $\sim$\,91\,pc. Since we did not obtain new data for this source, we cannot make any more definitive statements at this point. Because of the lack of multi-epoch observations, we can also not rule out the source as a milli-lens yet. Analysing more archival data and proposing for new observations should provide more clarity.

\subsubsection{J1653+3503 (1652+351)}
This source (Table\,\ref{tab:J1653}) has been discarded as both CSO and SMBHB candidate \citep{2016MNRAS.459..820T}. It has an optical counterpart in SDSS with 22.85 r-band magnitude and photometric redshift $z=0.81\,{\pm}\,0.10$. The linear distance between components is $\sim$\,78\,pc. As for J1632+3547, we did not obtain any data from follow-up observations and cannot discard this source as a milli-lens. A deeper analysis of archival data and new observations will constrain the nature of this source.

\subsubsection{J1721+5207 (1720+521)}
This source (Fig.~\ref{fig:J1721}) has an optical counterpart in DES DR8 with photometric redshift $z=1.302\,{\pm}\,0.229$. It has a steep integrated spectrum at frequencies around 1\,GHz \citep{2014ApJS..213....3M}, which continues to steepen at higher frequencies, as is confirmed by our data. We rejected this source as milli-lens because of its high SBR of 12.47 for our new EVN data, as well as the spectral indices differing by 0.64 in the 2014 epoch, above our threshold. At the given redshift, the linear distance between components is $\sim$\,408\,pc. The variability is within 7\,\%/yr, making this source a good CSO candidate.

\subsubsection{J1805-0438 (1802-046)}
This source (Fig.~\ref{fig:J1805}) has an IR and optical counterpart with 24.4\,mag at r-band listed in the OCARS catalogue \citep{2018ApJS..239...20M}. It is listed as BL\,Lac object in \cite{2019ApJS..242....4D} according to its IR colours in \textit{WISE}, which were selected to match those of known $\gamma$-ray emitting blazars. Our VLBI images show three components at lower frequencies, but for some epochs from Astrogeo at X-band, only a subset of any of the components A, B, and C is detected in the image, making the interpretation more difficult. Our new EVN image at 4.9\,GHz shows all three components detected, while the 22.2\,GHz image shows only component C. Component B seems undetected, which is not surprising given the steep spectrum and the partial non-detection of component B in earlier epochs. The separation vectors between components A and B and components A and C are almost perpendicular to each other, giving the source a peculiar morphology. Because of the uncertain component identification across epochs, the source passes the FR test, although some epochs at S-band deviate more than the uncertainties allow. We attribute this to uncertainties introduced by blending effects due to the low resolution at S-band. The discriminating factors to discard this source as a lens candidate would be the clearly different spectra of components A and B. As an additional test, we have considered the possibility that components A and B comprise one elongated image, and component C the other, less elongated image of a milli-lens system. In this case, the morphology would match the expectations, since the elongations would be along the critical curve. Then, the corresponding values for SBR, FR, and the spectrum are all consistent with our criteria within the uncertainties. For the spectrum, only the 2012 epoch has been considered. Hence, we keep the source in the list of lens candidates at this stage, and we aim at more sensitive X-band observations to shed light on this peculiar source. The small separation between components makes the source likely to be a CSO but, for confirmation, we would need a redshift. The fractional variability in our data is ${\sim}10(1+z)$\,\%/yr.

\subsubsection{J2010+1513 (2007+150)}
The source (Fig.~\ref{fig:J2010}) does not have an optical counterpart. We observed this source in two sessions with the EVN. While the regions seem blended for the March 2021 C-band epoch due to the large elliptical beam, this is resolved in the June epoch, showing clearly two components for each region and their extension towards each other. The source passes all our criteria but the spectrum, which is significantly different between the regions, and the morphology. Only component A is detected at 22.2\,GHz, indicating that the spectrum of region B must steepen even further than what is indicated between 4.3 and 7.6\,GHz. The variability of the total flux between our EVN epochs is ${\lesssim}\,$20$(1+z)$\,\%/yr, barely fitting the CSO definition. Since there is no evidence of proper motion, this source is still likely to be a CSO.

\subsubsection{J2044+6649 (2044+666)}
This source (Fig.~\ref{fig:J2044}) is unclassified and has no optical counterpart in the literature. We discarded the source based on the spectral indices differing by more than 0.5. The FR is remarkably similar between the components across epochs, and the variability is $<$\,4$(1+z)$\,\%/yr. This source is likely to be a CSO.

\subsubsection{J2114+4036 (2112+404)}
This source (Fig.~\ref{fig:J2114}) has an optical counterpart identified as a galaxy with photometric redshift $z=0.117\,{\pm}\,0.015$ \citep{2014ApJS..210....9B}. We discarded this source as a milli-lens system because of the different spectra between the regions in our new 4.9 and 22.2\,GHz data and the appearance of central emission between regions A and B, connecting the two. Region A steepens considerably, while region B stays rather flat, and could be the core of a core-jet system. The flux variability of component A exceeds 30\,\%/yr, making this scenario more likely, although the total flux density varies maximally by $\sim$\,14\,\%/yr. The \textit{Gaia} counterpart with 21.39 G-band magnitude is found very close to component B, with an offset of 25\,mas from the VLBI position, which is centred on the brighter component A. For a milli-lens, we should expect the photo-centre in between the lensed images, but here the \textit{Gaia} position indicates the optical emission to rather come from a jet component, or the accretion disc if component B is identified as the core. The linear distance between the regions is $\sim$\,38\,pc.

\subsubsection{J2209+6442 (2208+644)}
This source (Fig.~\ref{fig:J2209}) has no optical counterpart. It was rejected as milli-lens because of the low SBR of 0.05 at X-band, as well as the different spectra. The flat spectrum of component B up to 22.2\,GHz could indicate it being the core of a core-jet system. The low variability (maximal 4$(1+z)$\,\%/yr) and the lack of detected proper motion also suggest a possible CSO scenario.

\subsubsection{J2214-2521 (2211-256)}
This source (Fig.~\ref{fig:J2214}) has an optical counterpart classified as an extended galaxy in \cite{2014ApJS..210....9B} with a spectroscopic redshift $z=0.0868\,{\pm}\,0.0002$ \citep{2003astro.ph..6581C}. There is an associated \textit{Gaia} counterpart with 20.99\,mag in G-band, the photo-centre being 152\,mas offset to the south-east from component A. We already discussed the optical properties of this source briefly in \citetalias{2021MNRAS.507L...6C}. Finally, we discarded it as milli-lens because of the different spectra in the 2017 epoch and the low SBR of $<$\,0.06 for the 2017 X-band data. The new EVN image shows a new component C compared to the Astrogeo data, and a much lower separation between components A and B than in previous epochs. We deem this an uncertainty in the imaging and model-fitting process due to bad data quality in this case and do not discard the source because of the separation test. In the last two X-band epochs, the source shows fractional variability in slight excess of 20\,\%/yr, making it less likely to be a CSO and rather a core-jet, but we do not exclude the CSO possibility at this stage. More sensitive observations are required to determine the source morphology and variability more clearly.

\subsubsection{J2225+0841 (2223+084)}
This source (Fig.~\ref{fig:J2225}) has an optical counterpart classified as a galaxy in SDSS with 23.16\,mag at r-band and photometric redshift $z=0.73\,{\pm}\,0.10$. The linear distance between components is thus $\sim$\,33\,pc. We rejected this source as a milli-lens based on the components' different spectra, and the high SBR $\gg$\,4 observed in both the EVN C-band and K-band data. The spectral indices for each of the individual components are remarkably similar to what is found from the Astrogeo data between 4.3 and 7.6\,GHz, and the EVN data between 4.9 and 22.2\,GHz. The source is likely a CSO. The maximum fractional variability of $\sim$\,17\,\%/yr also fits this scenario.

\subsubsection{J2259+4037 (2256+403)}
This source (Fig.~\ref{fig:J2259}) has no optical counterpart. It shows a flat spectrum around 1\,GHz for the integrated flux \citep{2014ApJS..213....3M}. We discarded this source as a milli-lens because of its very different spectral indices ($\Delta\alpha>1$) in our new 5 and 22\,GHz data, very low SBR of 0.01 at K-band, and the source morphology which suggests a connecting bridge between the regions in both new C-band images. The source varies maximally by $\sim$\,16$(1+z)$\,\%/yr and might thus qualify as a CSO.

\subsubsection{J2312+0919 (2309+090)}
This source (Fig.~\ref{fig:J2312}) has an optical counterpart classified as a galaxy with 18.22 r-band magnitude and photometric redshift $z=0.23\pm0.08$ (SDSS). It may be associated with the Fanaroff-Riley II (FRII) high excitation radio galaxy, 3C\,456 \citep[e.g.][]{2020MNRAS.493.4355M}, which is located at redshift $z=0.23$ and about 2.5\,\arcsec (8\,kpc at the given redshift) away from J2312+0919. The photo-centre of the \textit{Gaia} counterpart (19.4 G-band magnitude) is significantly offset to the north of the source by 24.7\,mas, not supporting a milli-lens scenario if it is associated with J2312+0919. Ultimately, we discarded this source as a milli-lens because the FR is too variable at C-band, the difference being 0.81 while the 2-$\sigma$ combined error is 0.80. The maximum fractional variability is just within limits for the CSO classification ($\sim$\,19\,\%/yr), and the linear distance between components is $\sim$21\,pc, fitting this scenario. Both components are dimming over time, and there is an indication of an increase in separation. More monitoring of this source would prove valuable to constrain the trend in flux density, the separation, and if a steepening of the spectra with time could be observed.

\subsubsection{J2324-0058 (2321-012)}
This source (Fig.~\ref{fig:J2324}) has an optical counterpart in SDSS with 21.82 r-band magnitude and photometric redshift $z=0.37\,{\pm}\,0.09$ and is classified as a galaxy. The linear distance between components is $\sim$\,26\,pc. The source passes only the separation test, since it is too variable in FR (the maximum difference of 1.24 compared to the combined error of 0.64), and shows an SBR of 7.11 in the new EVN data. The spectra are also very different, with component A having a flat spectrum compared to the steep-spectrum component B. The spectra and the fractional variability of $\sim$\,30\,\%/yr indicate that this is a core-jet system.

\subsubsection{J2337-0622 (2334-066)}
This source (Fig.~\ref{fig:J2337}) has a relatively bright galaxy (18.35\,mag at r-band) as optical counterpart in SDSS with photometric redshift $z=0.18\,{\pm}\,0.02$. There is an associated \textit{Gaia} source with 19.4 G-band magnitude and a photo-centre 36.9\,mas offset from the VLBI position. The linear distance between components is $\sim$\,20\,pc. We discarded the source as a milli-lens due to the different (and variable) spectra, and the high SBR observed in some X-band epochs (41.2 in 2018). The \textit{Gaia} position also does not support the lensing scenario. Interpreting the source as a core-jet system, with A being the bright, flat-spectrum core, and B a knot in the jet, the \textit{Gaia} position could be tracing either an optical jet, as it lies broadly in the direction of the radio jet, or the host galaxy itself. The strong variability ($\gg$\,20\,\%/yr) of both components also suggest the source to be most likely a core-jet system instead.

\subsubsection{J2347-1856 (2344-192)}
This source (Fig.~\ref{fig:J2347}) was previously classified as a CSO (\citealt{2003ApJ...597..157T}; \citealt{2011A&A...535A..24S}). While it is also claimed as a bona fide CSO in \cite{2024ApJ...961..240K}, it remains the only one out of their 79 bona fide CSOs that does not show compelling morphological evidence of being one, lacking both a clearly identified nucleus or extensions of the two-sided jet. There is an indication of a jet connecting the two components at S-band for some epochs found in Astrogeo, but this still needs to be confirmed by more sensitive observations. We keep the source among the milli-lens candidates for now. We found an optical counterpart in SDSS classified as galaxy with 22.68\,mag at r-band and photometric redshift $z=0.22\,{\pm}\,0.13$. Then, the linear distance between components is $\sim$\,125\,pc. Excluding the first two epochs with uncertain calibration, the fractional variability is within a 7\,\%/yr, so we can confirm this source as a possible CSO. However, at this stage, it also passes all our criteria to be a milli-lens. Studying the spectra in more detail with new observations could constrain this source better. \cite{2003ApJ...597..157T} found no polarised flux stronger than 1.5\,mJy in this source, which would make it difficult to use polarisation observations to study potential differences between the two components. We note, however, that the lack of polarisation makes it less likely to be two lensed images of a radio nucleus, and more likely to be a CSO.

\subsection{Further considerations}

In total, we found nine remaining milli-lens candidates that fulfil all our criteria. For three of them, we did not yet obtain any follow-up observations. We deem it likely that we will reject most or all of the remaining candidates with sufficient follow-up, especially with high-cadence observations, which will reveal proper motion and variability. Even in the unlikely case, that all nine sources were to be confirmed as milli-lenses, we cannot make statistical statements about the population of SMCOs in this study. This is because the parent sample from Astrogeo, that we are using here, while largely unbiased, does not have redshifts available for all the sources. The lensing probability depends crucially on the redshift distribution of the parent sample studied \citep{2022A&A...668A.166L}. For the SMILE sample, we are able to make predictions based on the redshift distribution and expect to find a number of lenses ranging from zero up to the order of a few, depending on the DM model studied \citep{2022A&A...668A.166L}. In that sense, some milli-lenses could indeed be expected to be found in our study here, but no statistical conclusions can be drawn from that.

So far, we have not made predictions about other SMCOs acting as lenses, like PBHs or the centres of globular clusters. This study, as well as a calculation of the expected number of lensing events derived from simulations, is currently being conducted.

As was to be expected, many of the objects in our sample are either CSOs or CSO candidates. In total, we found 34 viable CSO candidates whose data are consistent with the CSO definition at this point. We note that confirming these objects as CSOs will require analysis of more VLBI data and monitoring of their variability, which we leave to a future work (Kumar et al., in prep.). It is interesting to note that none of these objects, which have a spectroscopic or photometric redshift, has a size larger than 500 pc, in good agreement with the findings of \citet{2024ApJ...961..241K} that there is a cut-off in CSO sizes at around 500 pc. Since our sample selected from Astrogeo is not complete it cannot be combined with that of \citet{2024ApJ...961..241K} to make an improved estimate of the statistical significance of this size cut-off. As pointed out by \citet{2024ApJ...977..195D}, quantitative comparisons between samples should only be made using complete samples that do not use a spectral index filter, for example. Figures~\ref{fig:VLBI_spectra_1} and \ref{fig:VLBI_spectra_2} show the spectra for all 40 initial milli-lens candidates using the total fluxes from all available VLBI images used in this work. It is important to note that these data come from different VLBI arrays, and no correction has been made for different resolution or $(u,v)$ range in these observations. Thus, care needs to be taken when interpreting these spectra, and they serve illustrative purposes only. We see primarily steep spectra, with two sources (J1143+1834 and J0052+1633) showing a peak around gigahertz frequencies, as expected for GPS sources, that are likely to be CSOs as well. While J1143+1834 has already been classified as a CSO in previous works, for J0052+1633, it serves as additional new evidence for the CSO nature. A detailed investigation of the spectra of all sources will be dedicated to future work (Kumar et al., in prep.).

\section{Summary and conclusions}

We have presented an investigation of the nature of 40 milli-lens candidates based on a number of VLBI images from the Astrogeo database and sensitive follow-up observations with the EVN at 5 and 22\,GHz. Based on a set of criteria, we ruled out 31 of these candidates as milli-lens systems. Hence, we showed the feasibility of ruling out most milli-lens candidates with observational follow-up. The rejected sources may still be interesting science targets, especially with regard to adding more objects to the class of CSOs, and our preliminary classification here will help future studies.

The remaining nine objects that could not be ruled out will be further scrutinised. We applied for high-frequency observations with the Korean VLBI network (Project ID: KVN-24A-392; PI: Jae-Young Kim) to test the detectability of 20 out of the 40 original candidates and will compute their spectra if detected. This and further monitoring at several frequencies with high-sensitivity observations will prove crucial to ruling out or confirming the remainder of the candidates as milli-lenses. Out of the nine remaining candidates, only one (J1143+1834) was detected in our 22.2\,GHz observations with the EVN. A multi-epoch, multi-frequency observing proposal is planned for the VLBA, for which all nine sources will be visible, in contrast to the EVN. This will put tight constraints on spectra and flux variability, distinguishing core-jet systems, CSOs, and possible milli-lenses. Our classification of sources as CSOs or SMBHBs is necessarily incomplete at this stage since, for example, we only ruled out CSOs if their fractional variability is much over 20\,\%/yr in our data. In addition to new VLBI observations, we will investigate archival data from single-dish and connected-interferometer observations to obtain integrated spectra and constrain the flux variability of sources in much more detail (\'Alvarez-Ortega et al., in prep.). Future VLBI observatories, such as the ngVLA Long Baseline Array (ngVLA-LONG) and Square Kilometer Array (SKA)-VLBI, are expected to help in distinguishing milli-lens systems by providing very sensitive observations that can also constrain the emission at compact and large scales at the same time. This will also enable the study of even larger samples when looking for potential milli-lens systems.

This pilot study is a crucial step in the SMILE project, which aims to discriminate DM models based on an investigation of $\sim$\,5000 sources from the VLA CLASS survey. This flux-complete sample of flat-spectrum sources will largely have redshifts available, thus allowing for accurate predictions of the DM model dependent number of observed milli-lenses, as previously presented in \cite{2022A&A...668A.166L}. It is important to note here that our search for milli-lenses does not require a complete sample as a basis. Rather, what is needed is a sample with known redshifts that gives as many lines of sight as possible to constrain the number density of milli-lenses. We refer to such a sample as a `sufficient sample', meaning a sample that is both clearly defined and unbiased for the purpose of the particular study in question, which in our case is the search for milli-lens systems. (See \cite{2024ApJ...977..195D} for more details on the choosing of appropriate samples.)

\newpage
\begin{table*}[ht]
    \caption{Thirty-one sources clearly rejected as milli-lens systems so far.}
    \vspace*{3mm}
    \adjustbox{width=1\textwidth}{%
    \label{tab:summary}
    \centering
    \begin{tabular}{|| c c c c c c c c c c ||} 
        \hline
        (1)    & (2) & (3)                     & (4)                      & (5)        & (6)     & (7)      & (8)           & (9)             & (10)  \\
        Source & $z$ & $S_{4.9\,\mathrm{GHz}}$ & $S_{22.2\,\mathrm{GHz}}$ & Morphology & FR test & SBR test & Spectrum test & Separation test & Class \\
               &     & [mJy]                   & [mJy]                    &            &         &          &               &                 &       \\
        \hline\hline
        J0010$-$0740 & 0.31$^{1,\mathrm{p}}$  &  29 & -  & FAIL & PASS & FAIL & FAIL & PASS & CSO                    \\
        J0011+3443   & 0.89$^{2,\mathrm{s}}$  & 126 & -  & PASS & PASS & FAIL & PASS & PASS & CSO                    \\
        J0024$-$4202 & 0.94$^{3,\mathrm{s}}$  & -   & -  & PASS & PASS & PASS & PASS & PASS & Milli-lens/CSO/Other   \\
        J0044+2858   & 0.75$^{1,\mathrm{p}}$  &  51 & 13 & PASS & PASS & PASS & FAIL & PASS & CSO                    \\
        J0052+1633   & 0.43$^{1,\mathrm{p}}$  & 155 & 36 & FAIL & FAIL & FAIL & FAIL & PASS & CSO                    \\
        J0118+3810   & -                      & 124 & -  & PASS & PASS & FAIL & PASS & PASS & CSO                    \\
        J0132+5211   & 0.98$^{4,\mathrm{p}}$  &  47 & -  & PASS & PASS & FAIL & FAIL & FAIL & Core-jet               \\
        J0139+0824   & 1.05$^{5,\mathrm{p}}$  &  91 & -  & PASS & PASS & PASS & FAIL & PASS & CSO                    \\
        J0203+3041   & 0.96$^{6,\mathrm{s}}$  & 257 & 32 & PASS & PASS & FAIL & PASS & PASS & Core-jet/CSO           \\
        J0210$-$2213 & 1.49$^{7,\mathrm{s}}$  & 282 & -  & PASS & PASS & PASS & PASS & PASS & Milli-lens/CSO         \\
        J0213+8717   & -                      & 141 & 57 & PASS & PASS & FAIL & FAIL & PASS & Core-jet/CSO           \\
        J0222+0952   & -                      & 148 & -  & FAIL & PASS & FAIL & FAIL & PASS & Core-jet/Other         \\
        J0232$-$3422 & -                      & -   & -  & PASS & PASS & PASS & FAIL & PASS & Core-jet/CSO/Other     \\
        J0237+1116   & 0.99$^{8,\mathrm{s}}$  &  67 & -  & PASS & PASS & FAIL & PASS & PASS & Core-jet/CSO           \\
        J0502+1626   & -                      &  37 & -  & PASS & PASS & FAIL & FAIL & PASS & Core-jet/CSO           \\
        J0527+1743   & 0.46$^{1,\mathrm{p}}$  &  49 & -  & PASS & PASS & FAIL & PASS & PASS & CSO/SMBHB              \\
        J0616$-$1957 & -                      & 135 & -  & FAIL & PASS & FAIL & FAIL & PASS & CSO                    \\
        J0732+6023   & 1.06$^{5,\mathrm{p}}$  &  76 & -  & PASS & PASS & FAIL & FAIL & PASS & CSO/Core-jet           \\
        J0923$-$3435 & -                      & -   & -  & PASS & PASS & PASS & PASS & PASS & Milli-lens/CSO         \\
        J1132+5100   & 1.15$^{5,\mathrm{p}}$  &  95 & -  & FAIL & PASS & FAIL & FAIL & PASS & Core-jet/CSO/Other     \\
        J1143+1834   & 1.10$^{9,\mathrm{p}}$  & 334 & 57 & PASS & PASS & PASS & PASS & PASS & Milli-lens/CSO/SMBHB   \\
        J1218$-$2159 & -                      & 112 & -  & PASS & PASS & PASS & PASS & PASS & Milli-lens/CSO         \\
        J1306+0341   & -                      & 109 & -  & PASS & PASS & FAIL & PASS & PASS & CSO                    \\
        J1340$-$0335 & 0.98$^{5,\mathrm{p}}$  & 238 & -  & FAIL & PASS & FAIL & PASS & PASS & CSO/Other              \\
        J1344$-$1739 & -                      &  77 & 11 & PASS & PASS & PASS & FAIL & PASS & CSO                    \\
        J1632+3547   & 1.08$^{5,\mathrm{p}}$  & -   & -  & PASS & PASS & PASS & PASS & PASS & Milli-lens/SMBHB/Other \\
        J1653+3503   & 0.81$^{1,\mathrm{p}}$  & -   & -  & PASS & PASS & PASS & PASS & PASS & Milli-lens/Other       \\
        J1721+5207   & 1.30$^{5,\mathrm{p}}$  &  71 & -  & PASS & PASS & FAIL & FAIL & PASS & CSO                    \\
        J1805$-$0438 & -                      & 103 & 21 & PASS & PASS & PASS & PASS & PASS & Milli-lens/CSO/Other              \\
        J2010+1513   & -                      & 152 & 19 & FAIL & PASS & PASS & FAIL & PASS & CSO                    \\
        J2044+6649   & -                      & 107 & -  & PASS & PASS & PASS & FAIL & PASS & CSO                    \\
        J2114+4036   & 0.12$^{10,\mathrm{p}}$ & 101 & 21 & FAIL & PASS & PASS & FAIL & PASS & Core-jet/CSO           \\
        J2209+6442   & -                      &  48 & 15 & PASS & PASS & FAIL & FAIL & PASS & Core-jet/CSO           \\
        J2214$-$2521 & 0.09$^{11,\mathrm{s}}$ & 166 & -  & PASS & PASS & FAIL & FAIL & PASS & Core-jet/CSO           \\
        J2225+0841   & 0.73$^{1,\mathrm{p}}$  & 104 & 21 & PASS & PASS & FAIL & FAIL & PASS & CSO                    \\
        J2259+4037   & -                      & 116 & 23 & FAIL & PASS & FAIL & FAIL & PASS & CSO                    \\
        J2312+0919   & 0.23$^{1,\mathrm{p}}$  &  19 & -  & PASS & FAIL & PASS & PASS & PASS & CSO                    \\
        J2324$-$0058 & 0.37$^{1,\mathrm{p}}$  &  56 & 31 & PASS & FAIL & FAIL & FAIL & PASS & Core-jet               \\
        J2337$-$0622 & 0.18$^{1,\mathrm{p}}$  &  45 & -  & PASS & PASS & FAIL & FAIL & PASS & Core-jet               \\
        J2347$-$1856 & 0.22$^{1,\mathrm{p}}$  & 408 & -  & PASS & PASS & PASS & PASS & PASS & Milli-lens/CSO         \\
    \hline \noalign{\medskip}
    \multicolumn{10}{p{1.15\textwidth}}{\textsc{Notes:} Column (1) denotes the source J2000 identifier; (2) the source redshift; (3) the total clean map flux density from our new EVN data at 4.9\,GHz; (4) the total clean map flux density from our new EVN data at 22.2\,GHz; we denote in column (5)-(9) if a source has passed or failed one of the five criteria outlined in Section\,\ref{sec:criteria}; column (11) denotes our source class designation, following our discussion of the results in Section\,\ref{sec:discussion}. References for redshifts: 1 SDSS \citep[DR17,][]{2022ApJS..259...35A}; 2 \cite{2010MNRAS.401.1709V}; 3 \cite{2007AA...463...97L}; 4 \cite{2015ApJS..219...39R}; 5 DES DR8; 6 \cite{2012ApJ...748...49S}; 7 \cite{2007AA...464..879D}; 8 \cite{2024ApJ...964...69S}; 9 DES DR10; 10 \cite{2014ApJS..210....9B}; 11 \cite{2003astro.ph..6581C}. The superscript p denotes photometric and s spectroscopic redshift.}
    \end{tabular}
    }
\end{table*}



\begin{acknowledgements}
We thank the anonymous referee for the very useful suggestions and comments on the manuscript.
F.P., C.C., D.A., A.K., V.M. \& D.B. acknowledge support from the European Research Council (ERC) under the Horizon ERC Grants 2021 programme under grant agreement No. 101040021.\\
We thank Rocco Lico for the useful discussions.\\
We would also like to thank Leonid Petrov for maintaining the Astrogeo VLBI FITS image database and A. Bertarini, L. Vega Garcia, N. Corey, Y. Cui, L. Gurvits, X. He, T. Koryukova, Y. Y. Kovalev, S.-S. Lee, R. Lico, E. Liuzzo, A. Marscher, S. Jorstad, C. Marvin, D. Homan, J.-L. Gomez, M. Lister, A. Pushkarev, E. Ros, T. Savolainen, L. Petrov, A. Popkov, K. Sokolovski, A. Tao, G. Taylor, A. de Witt, M. Xu, and B. Zhang for making VLBI images they produced publicly available. We found that the practice of uploading VLBI images to a publicly available database brings great benefits to the scientific community.\\
The European VLBI Network is a joint facility of independent European, African, Asian, and North American radio astronomy institutes. Scientific results from data presented in this publication are derived from the following EVN project code(s): EC071.\\
This work has made use of data from the European Space Agency (ESA) mission {\it Gaia} (\url{https://www.cosmos.esa.int/gaia}), processed by the {\it Gaia} Data Processing and Analysis Consortium (DPAC, \url{https://www.cosmos.esa.int/web/gaia/dpac/consortium}). Funding for the DPAC has been provided by national institutions, in particular the institutions participating in the {\it Gaia} Multilateral Agreement.\\
Funding for the Sloan Digital Sky Survey V has been provided by the Alfred P. Sloan Foundation, the Heising-Simons Foundation, the National Science Foundation, and the Participating Institutions. SDSS acknowledges support and resources from the Center for High-Performance Computing at the University of Utah. SDSS telescopes are located at Apache Point Observatory, funded by the Astrophysical Research Consortium and operated by New Mexico State University, and at Las Campanas Observatory, operated by the Carnegie Institution for Science. The SDSS web site is \url{www.sdss.org}.
SDSS is managed by the Astrophysical Research Consortium for the Participating Institutions of the SDSS Collaboration, including Caltech, The Carnegie Institution for Science, Chilean National Time Allocation Committee (CNTAC) ratified researchers, The Flatiron Institute, the Gotham Participation Group, Harvard University, Heidelberg University, The Johns Hopkins University, L’Ecole polytechnique f\'{e}d\'{e}rale de Lausanne (EPFL), Leibniz-Institut f{\"u}r Astrophysik Potsdam (AIP), Max-Planck-Institut f{\"u}r Astronomie (MPIA Heidelberg), Max-Planck-Institut f{\"u}r Extraterrestrische Physik (MPE), Nanjing University, National Astronomical Observatories of China (NAOC), New Mexico State University, The Ohio State University, Pennsylvania State University, Smithsonian Astrophysical Observatory, Space Telescope Science Institute (STScI), the Stellar Astrophysics Participation Group, Universidad Nacional Aut\'{o}noma de M\'{e}xico, University of Arizona, University of Colorado Boulder, University of Illinois at Urbana-Champaign, University of Toronto, University of Utah, University of Virginia, Yale University, and Yunnan University.
\end{acknowledgements}

\bibliographystyle{aa} 
\bibliography{mybib.bib}

\appendix
\section{Individual sources}\label{sec:app1}


\begin{figure*}[h!]
    \centering
    \includegraphics[width=0.4\textwidth]{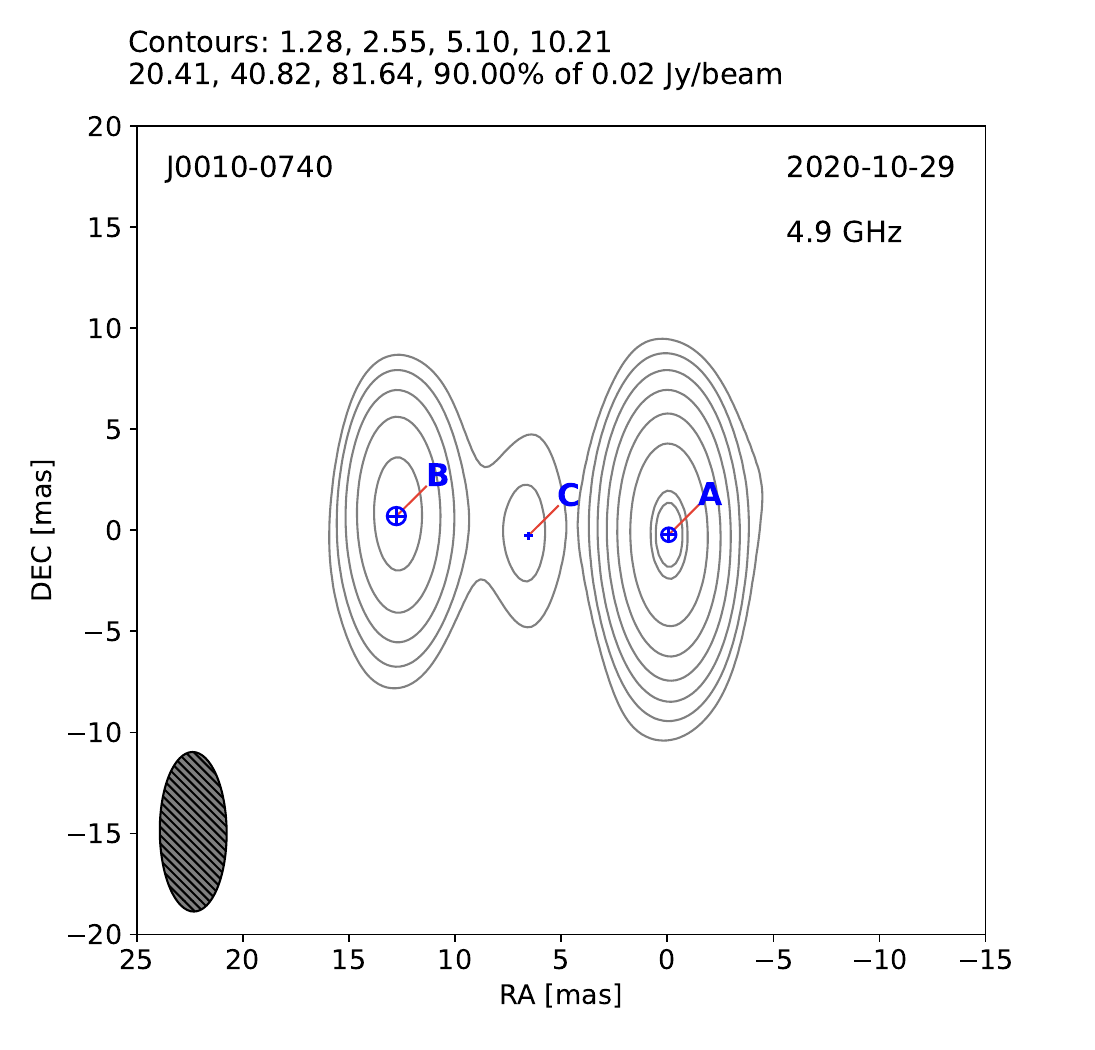}
    \caption{J0010-0740 EVN 4.9\,GHz data. The contours start at four times the image rms noise of 0.065\,mJy/beam and increase by factors of two. The restoring beam size is 3.15$\times$7.90\,mas at 0.5\,$^\circ$ PA. The source is not detected at 22.2\,GHz.}
    \label{fig:J0010}
\end{figure*}

\begin{table*}[h!]
    \caption{Derived quantities for J0010-0740. See Table~\ref{tab:J2209} for a detailed description of the columns.}
    \vspace*{3mm}
    \adjustbox{width=1\textwidth}{%
    \label{tab:J0010}
    \centering
    \begin{tabular}{|| c c c c c | c c | c c | c c | c c | c c | c ||} 
        \hline
        Epochs & Frequency & Component & Flux  & Error  & Flux  & Error & FWHM  & Error & $\log(T_\mathrm{b,obs}/\mathrm{K})$ & Error  & SB    & Error & Distance & Error & Spectral Index  \\ [0.5ex] 
               & [GHz]     &           & [mJy] & [mJy]  & Ratio &       & [mas] & [mas] &                                     &        & Ratio &       & [mas]    & [mas] &                 \\
        \hline\hline
2015-12-27 & 4.3 & A & 21.8 & 3.0 & 3.50 & 0.72 & 0.92 & 0.07 & 9.28 & 0.09 & 6.82 & 2.67 & 12.87 & 0.48 & -0.46 \\
- & -   & B & 6.2 & 1.3 & -   & -   & 1.28 & 0.19 & 8.45 & 0.16 & -   & -   & -   & -   & -1.51 \\
        \hline
2015-12-27 & 7.6 & A & 16.8 & 2.6 & 6.32 & 2.06 & 0.41 & 0.04 & 9.37 & 0.10 & 22.17 & 13.49 & 13.19 & 0.44 & - \\
- & -   & B & 2.7 & 0.9 & -   & -   & 0.77 & 0.19 & 8.03 & 0.25 & -   & -   & -   & -   & - \\
        \hline\hline
2020-10-29 & 4.9 & A & 20.9 & 2.7 & 3.41 & 0.57 & 0.69 & 0.04 & 9.40 & 0.07 & 5.47 & 1.60 & 12.87 & 0.60 & - \\
- & -   & B & 6.1 & 1.1 & -   & -   & 0.87 & 0.09 & 8.66 & 0.12 & -   & -   & -   & -   & - \\
        \hline
        \end{tabular}
    }
\end{table*}


\begin{figure*}[h]
    \centering
    \includegraphics[width=0.4\textwidth]{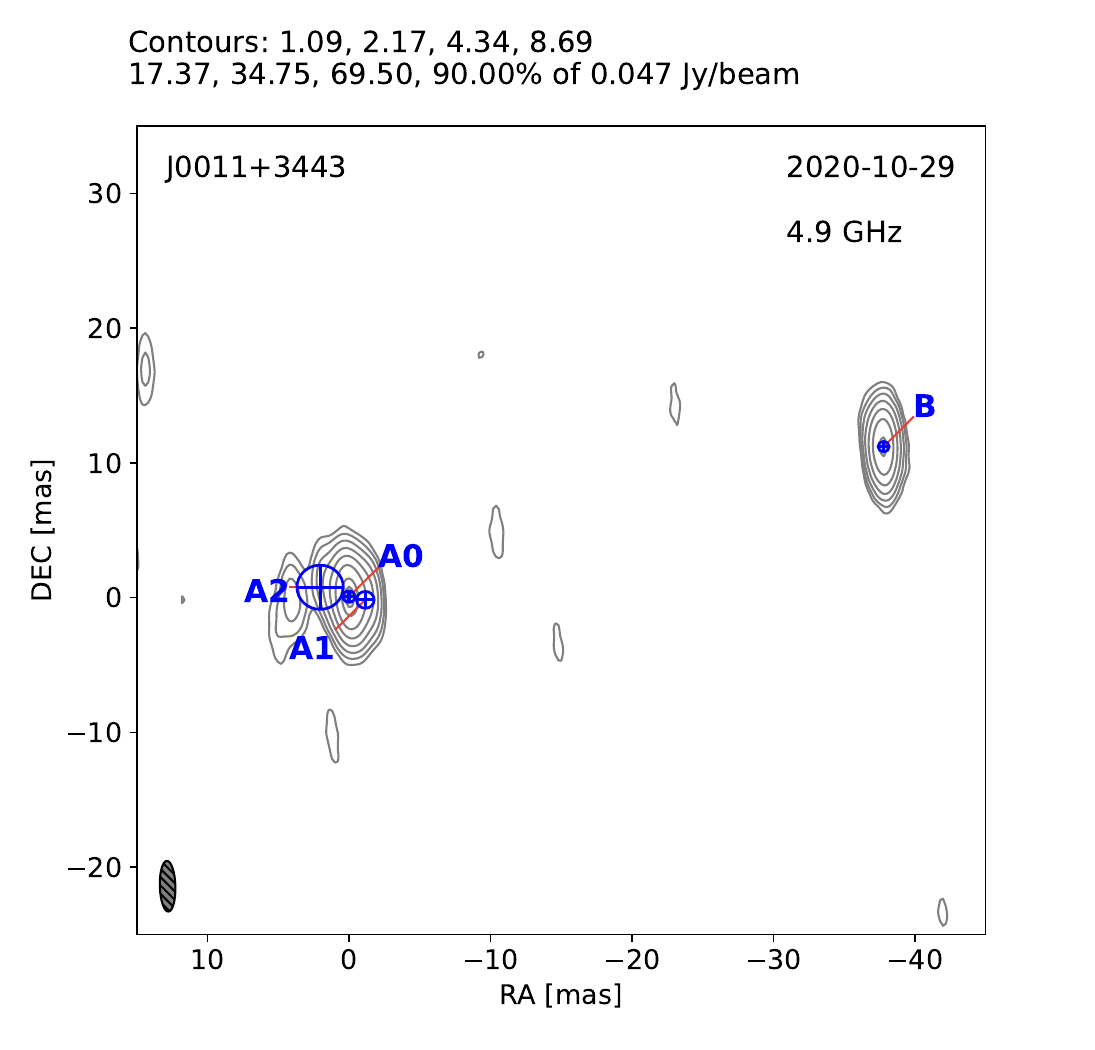}
    \caption{J0011+3443 EVN 4.9\,GHz data. The contours start at four times the image rms noise of 0.126\,mJy/beam and increase by factors of two. The restoring beam size is 1.10$\times$3.76\,mas at 1.6\,$^\circ$ PA. The source is not detected at 22.2\,GHz.}
    \label{fig:J0011}
\end{figure*}

\begin{table*}[h]
    \caption{Derived quantities for J0011+3443. See Table~\ref{tab:J2209} for a detailed description of the columns.}
    \vspace*{3mm}
    \adjustbox{width=1\textwidth}{%
    \label{tab:J0011}
    \centering
    \begin{tabular}{|| c c c c c | c c | c c | c c | c c | c c | c ||} 
        \hline
        Epochs & Frequency & Component & Flux  & Error  & Flux  & Error & FWHM  & Error & $\log(T_\mathrm{b,obs}/\mathrm{K})$ & Error  & SB    & Error & Distance & Error & Spectral Index \\ [0.5ex] 
               & [GHz]     &           & [mJy] & [mJy]  & Ratio &       & [mas] & [mas] &                                     &        & Ratio &       & [mas]    & [mas] &                \\
        \hline\hline
2016-05-30 & 4.3 & A & 103.7 & 10.7 & 1.86 & 0.15 & -   & -   & 9.15 & 0.08 & 0.32 & 0.06 & -   & -   & -1.74 \\
- & -   & A0 & 94.4 & 10.6 & 1.69 & 0.14 & 1.53 & 0.06 & 9.47 & 0.02 & 0.67 & 0.10 & 39.38 & 0.19 & -   \\
- & -   & B & 55.7 & 6.6 & -   & -   & 0.96 & 0.05 & 9.65 & 0.07 & -   & -   & -   & -   & -1.75 \\
        \hline
2016-05-30 & 7.6 & A & 39.0 & 4.9 & 1.88 & 0.24 & 1.37 & 0.09 & 8.69 & 0.08 & 0.62 & 0.15 & 39.30 & 0.22 & - \\
- & -   & B & 20.8 & 3.0 & -   & -   & 0.79 & 0.06 & 8.90 & 0.09 & -   & -   & -   & -   & -  \\
        \hline\hline
2020-10-29 & 4.9 & A & 87.0 & 8.0 & 1.90 & 0.18 & -   & -   & 8.58 & 0.15 & 0.09 & 0.03 & -   & -   & - \\
- & -   & A0 & 55.4 & 6.7 & 1.21 & 0.12 & 0.85 & 0.04 & 9.64 & 0.03 & 1.03 & 0.19 & 39.40 & 0.17 & - \\
- & -   & B & 45.8 & 5.7 & -   & -   & 0.79 & 0.05 & 9.62 & 0.08 & -   & -   & -   & -   & - \\
        \hline
    \end{tabular}
    }
\end{table*}


\begin{table*}[h]
    \caption{Derived quantities for J0024-4202. See Table~\ref{tab:J2209} for a detailed description of the columns.}
    \vspace*{3mm}
    \adjustbox{width=1\textwidth}{%
    \label{tab:J0024}
    \centering
    \begin{tabular}{|| c c c c c | c c | c c | c c | c c | c c | c ||} 
        \hline
        Epochs & Frequency & Component & Flux  & Error  & Flux  & Error & FWHM  & Error & $\log(T_\mathrm{b,obs}/\mathrm{K})$ & Error  & SB    & Error & Distance & Error & Spectral Index \\ [0.5ex] 
               & [GHz]     &           & [mJy] & [mJy]  & Ratio &       & [mas] & [mas] &                                     &        & Ratio &       & [mas]    & [mas] &                \\
        \hline\hline
2002-01-31 & 8.7 & A & 462.4 & 85.3 & 3.01 & 0.86 & 3.17 & 0.48 & 8.93 & 0.15 & 1.79 & 1.13 & 27.78 & 0.75 & - \\
- & -   & B & 153.7 & 40.2 & -   & -   & 2.45 & 0.58 & 8.68 & 0.23 & -   & -   & -   & -   & - \\
        \hline\hline
2002-05-14 & 2.3 & A & 1924.2 & 237.8 & 2.99 & 0.43 & 3.33 & 0.21 & 10.66 & 0.08 & 2.08 & 0.58 & 28.52 & 0.64 & -0.85 \\
- & -   & B & 644.1 & 102.1 & -   & -   & 2.78 & 0.28 & 10.34 & 0.11 & -   & -   & -   & -   & -0.82 \\
        \hline
2002-05-14 & 8.7 & A & 624.8 & 115.0 & 2.89 & 1.14 & 3.61 & 0.55 & 8.95 & 0.16 & 3.08 & 2.70 & 27.75 & 1.49 & - \\
- & -   & B & 216.3 & 81.3 & -   & -   & 3.72 & 1.35 & 8.46 & 0.35 & -   & -   & -   & -   & - \\
        \hline\hline
2017-01-16 & 2.2 & A & 1826.4 & 204.2 & 3.66 & 0.38 & 3.48 & 0.15 & 10.62 & 0.06 & 1.83 & 0.37 & 28.81 & 0.45 & -1.10 \\
- & -   & B & 499.3 & 67.9 & -   & -   & 2.46 & 0.18 & 10.35 & 0.09 & -   & -   & -   & -   & -1.45 \\
        \hline
2017-01-16 & 8.7 & A & 414.7 & 60.2 & 5.82 & 1.44 & 2.91 & 0.30 & 8.96 & 0.11 & 2.13 & 1.14 & 28.18 & 0.53 & - \\
- & -   & B & 71.2 & 17.5 & -   & -   & 1.76 & 0.38 & 8.63 & 0.21 & -   & -   & -   & -   & - \\
        \hline
        \end{tabular}
    }
\end{table*}


\begin{figure*}[h]
    \centering
    \includegraphics[width=0.4\textwidth]{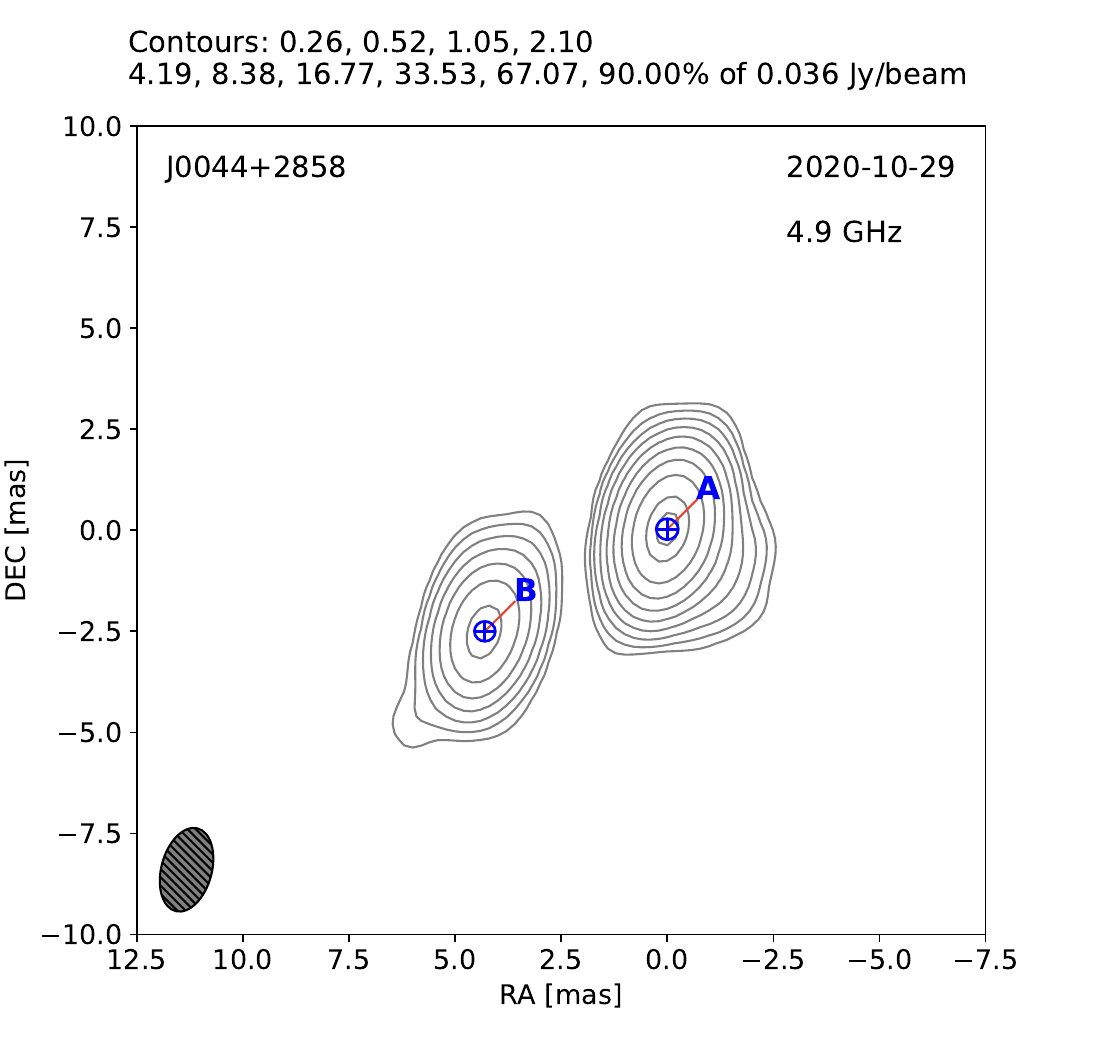}
    \includegraphics[width=0.4\textwidth]{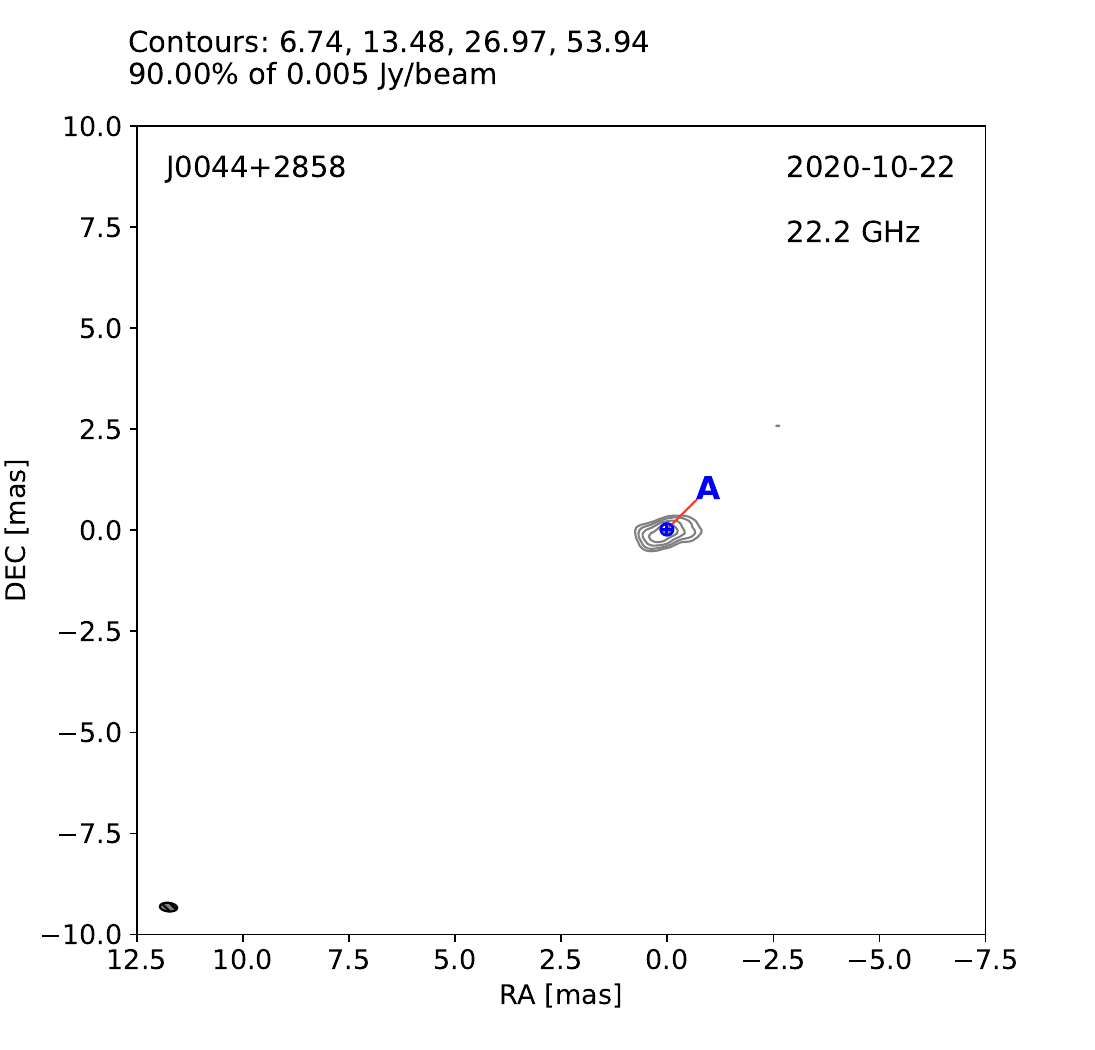}
    \caption{J0044+2858. \textit{Left:} EVN 4.9\,GHz data. The contours start at four times the image rms noise of 0.023\,mJy/beam and increase by factors of two. The restoring beam size is 1.19$\times$2.11\,mas at $-13.7$\,$^\circ$ PA. \textit{Right:} EVN 22.2\,GHz data. The contours start at four times the image rms noise of 0.078\,mJy/beam and increase by factors of two. The restoring beam size is 0.22$\times$0.41\,mas at 83.8\,$^\circ$ PA.}
    \label{fig:J0044}
\end{figure*}

\begin{table*}[h]
    \caption{Derived quantities for J0044+2858. See Table~\ref{tab:J2209} for a detailed description of the columns.}
    \vspace*{3mm}
    \adjustbox{width=1\textwidth}{%
    \label{tab:J0044}
    \centering
    \begin{tabular}{|| c c c c c | c c | c c | c c | c c | c c | c ||} 
        \hline
        Epochs & Frequency & Component & Flux  & Error  & Flux  & Error & FWHM  & Error & $\log(T_\mathrm{b,obs}/\mathrm{K})$ & Error  & SB    & Error & Distance & Error & Spectral Index \\ [0.5ex] 
               & [GHz]     &           & [mJy] & [mJy]  & Ratio &       & [mas] & [mas] &                                     &        & Ratio &       & [mas]    & [mas] &                \\
        \hline\hline
2015-12-12 & 4.3 & A & 43.6 & 5.4 & 4.89 & 0.85 & 0.82 & 0.04 & 9.68 & 0.07 & $<$0.54 & - & 4.89 & 0.32 & -1.15 \\
- & -   & B & 8.9 & 1.7 & -   & -   & $<$0.27 & - & $>$9.94 & - & -   & -   & -   & -   & -1.88 \\
        \hline
2015-12-12 & 7.6 & A & 22.8 & 3.3 & 7.37 & 2.25 & 0.51 & 0.04 & 9.32 & 0.09 & 2.84 & 1.51 & 4.91 & 0.33 & - \\
- & -   & B & 3.1 & 0.9 & -   & -   & 0.32 & 0.06 & 8.86 & 0.22 & -   & -   & -   & -   & - \\
        \hline\hline
2020-10-29 & 4.9 & A & 39.9 & 4.2 & 4.75 & 0.39 & 0.52 & 0.01 & 9.93 & 0.05 & 4.28 & 0.64 & 4.99 & 0.10 & -0.78 \\
- & -   & B & 8.4 & 1.0 & -   & -   & 0.49 & 0.03 & 9.29 & 0.07 & -   & -   & -   & -   & - \\
        \hline
2020-10-22 & 22.2 & A & 12.2 & 2.0 & - & - & 0.29 & 0.03 & 8.61 & 0.12 & - & - & - & - & - \\
- & -   & B & - & - & -   & -   & - & - & - & - & -   & -   & -   & -   & - \\
        \hline
    \end{tabular}
    }
\end{table*}


\begin{figure*}[h]
    \centering
    \includegraphics[width=0.4\textwidth]{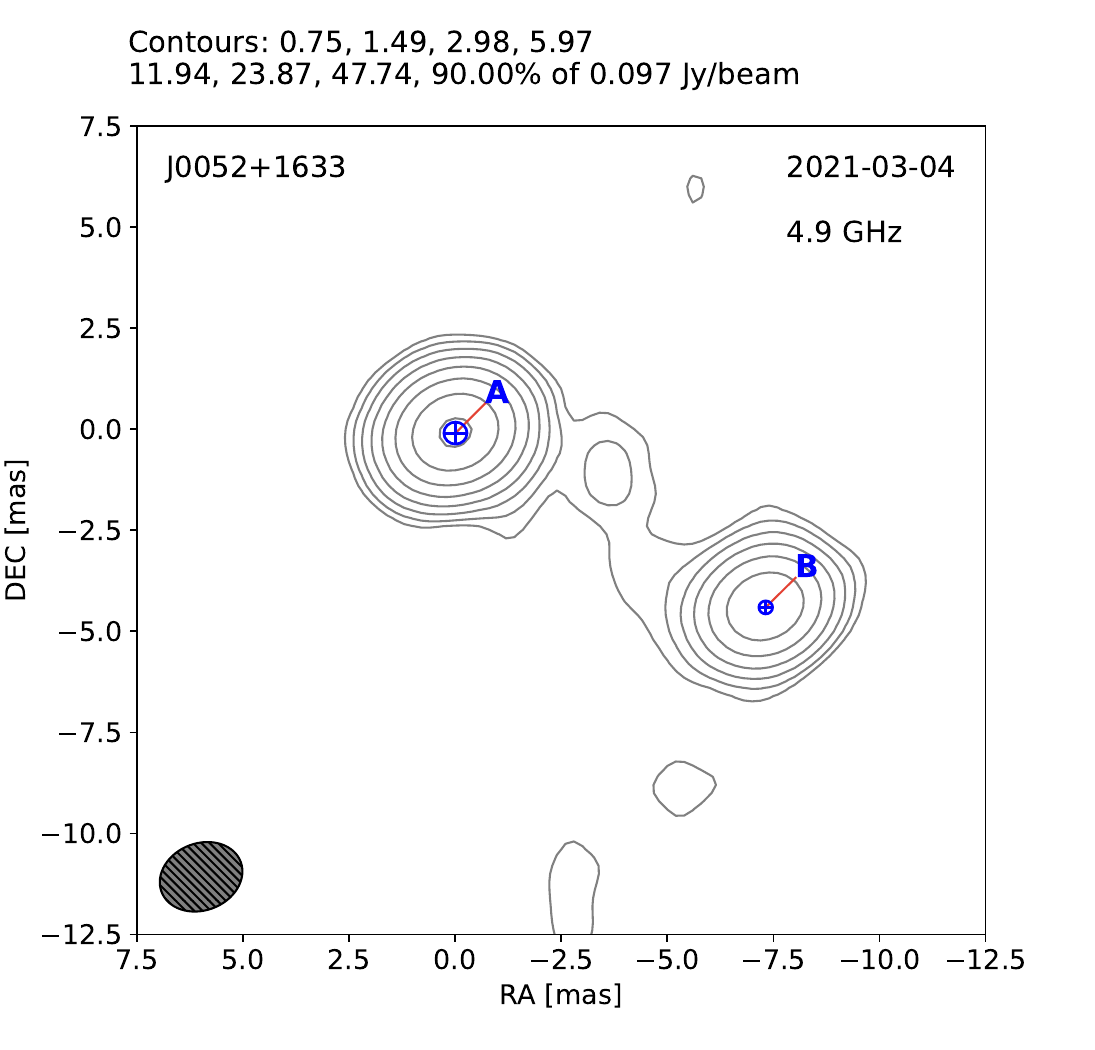}
    \includegraphics[width=0.4\textwidth]{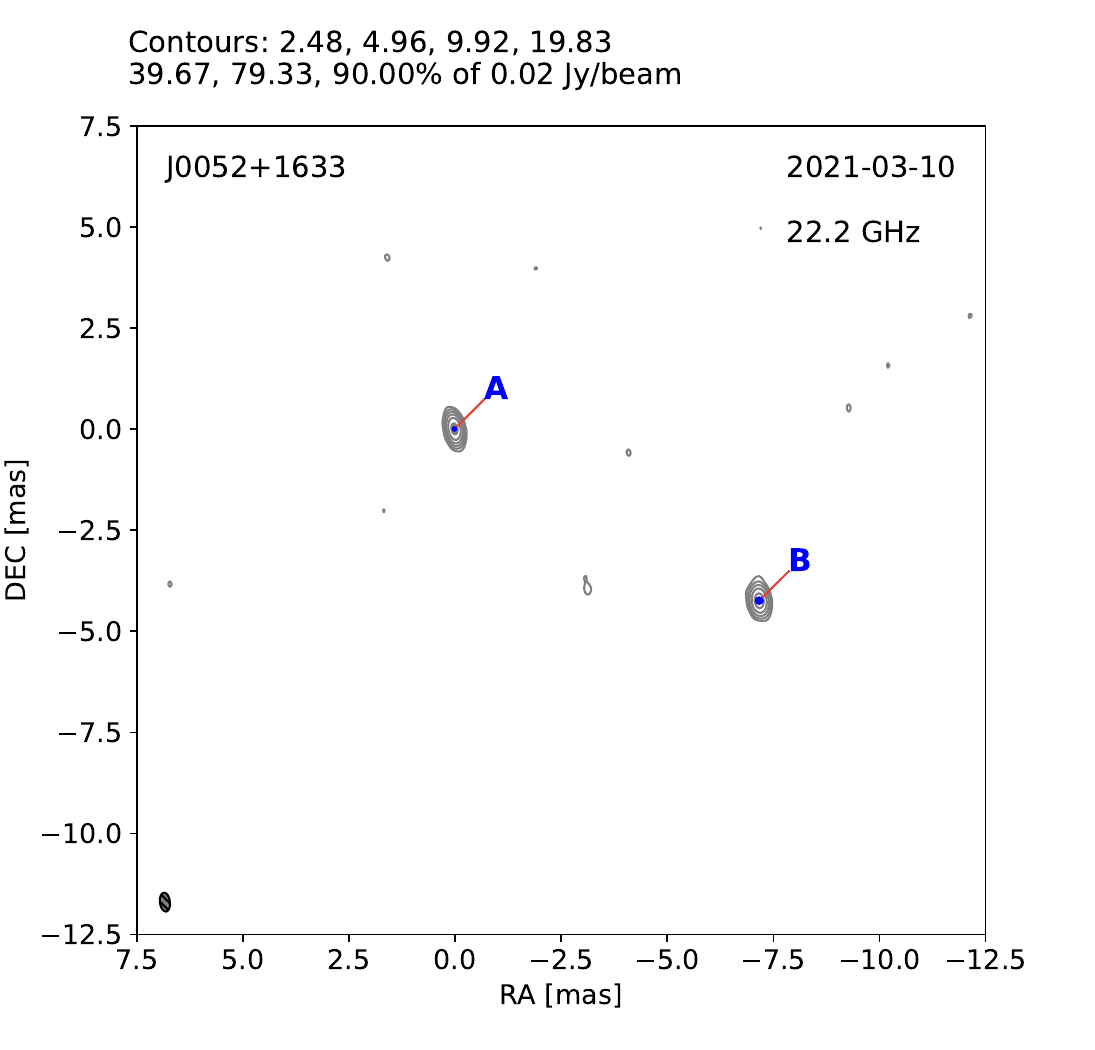}
    \caption{J0052+1633. \textit{Left:} EVN 4.9\,GHz data. The contours start at four times the image rms noise of 0.182\,mJy/beam and increase by factors of two. The restoring beam size is 1.65$\times$2.01\,mas at $-64.3$\,$^\circ$ PA. \textit{Right:} EVN 22.2\,GHz data. The contours start at four times the image rms noise of 0.122\,mJy/beam and increase by factors of two. The restoring beam size is 0.24$\times$0.47\,mas at 6.6\,$^\circ$ PA.}
    \label{fig:J0052}
\end{figure*}

\begin{table*}[h]
    \caption{Derived quantities for J0052+1633. See Table~\ref{tab:J2209} for a detailed description of the columns.}
    \vspace*{3mm}
    \adjustbox{width=1\textwidth}{%
    \label{tab:J0052}
    \centering
    \begin{tabular}{|| c c c c c | c c | c c | c c | c c | c c | c ||} 
        \hline
        Epochs & Frequency & Component & Flux  & Error  & Flux  & Error & FWHM  & Error & $\log(T_\mathrm{b,obs}/\mathrm{K})$ & Error  & SB    & Error & Distance & Error & Spectral Index \\ [0.5ex] 
               & [GHz]     &           & [mJy] & [mJy]  & Ratio &       & [mas] & [mas] &                                     &        & Ratio &       & [mas]    & [mas] &                \\
        \hline\hline
2013-07-17 & 4.3 & A & 130.6 & 14.4 & 3.55 & 0.35 & 0.71 & 0.02 & 10.28 & 0.06 & 1.02 & 0.18 & 8.50 & 0.19 & -0.86 \\
- & -   & B & 36.9 & 4.9 & -   & -   & 0.38 & 0.02 & 10.27 & 0.08 & -   & -   & -   & -   & 0.47 \\
        \hline
2013-07-17 & 7.6 & A & 80.8 & 9.2 & 1.68 & 0.15 & 0.41 & 0.02 & 10.06 & 0.06 & 1.65 & 0.27 & 8.45 & 0.10 & - \\
- & -   & B & 48.1 & 5.9 & -   & -   & 0.41 & 0.02 & 9.84 & 0.07 & -   & -   & -   & -   & - \\
        \hline\hline
2019-07-29 & 4.3 & A & 136.8 & 14.8 & 2.68 & 0.21 & 1.04 & 0.03 & 9.97 & 0.05 & 0.99 & 0.14 & 8.37 & 0.24 & -0.79 \\
- & -   & B & 51.0 & 6.1 & -   & -   & 0.63 & 0.03 & 9.98 & 0.07 & -   & -   & -   & -   & -0.07 \\
        \hline
2019-07-29 & 7.6 & A & 87.8 & 9.6 & 1.79 & 0.13 & 0.80 & 0.03 & 9.52 & 0.06 & 1.02 & 0.13 & 8.42 & 0.14 & - \\
- & -   & B & 49.1 & 5.7 & -   & -   & 0.60 & 0.03 & 9.51 & 0.06 & -   & -   & -   & -   & - \\
        \hline\hline
2020-03-10 & 2.3 & A & 93.0 & 12.9 & 2.87 & 0.54 & 1.03 & 0.07 & 10.37 & 0.08 & 39.55 & 12.83 & 7.89 & 1.02 & - \\
- & -   & B & 32.4 & 6.2 & -   & -   & 3.81 & 0.43 & 8.77 & 0.13 & -   & -   & -   & -   & -   \\
        \hline\hline
2021-03-04 & 4.9 & A & 106.9 & 12.4 & 2.36 & 0.25 & 0.55 & 0.02 & 10.31 & 0.06 & 0.82 & 0.16 & 8.49 & 0.14 & -1.10 \\
- & -   & B & 45.4 & 6.1 & -   & -   & 0.32 & 0.02 & 10.40 & 0.08 & -   & -   & -   & -   & -0.78 \\
        \hline
2021-03-10 & 22.2 & A & 20.4 & 3.0 & 1.46 & 0.25 & 0.07 & 0.01 & 10.10 & 0.09 & 6.44 & 2.01 & 8.34 & 0.05 & - \\
- & -   & B & 14.0 & 2.3 & -   & -   & 0.14 & 0.01 & 9.29 & 0.12 & -   & -   & -   & -   & - \\
        \hline
    \end{tabular}
    }
\end{table*}


\begin{figure*}[h]
    \centering
    \includegraphics[width=0.4\textwidth]{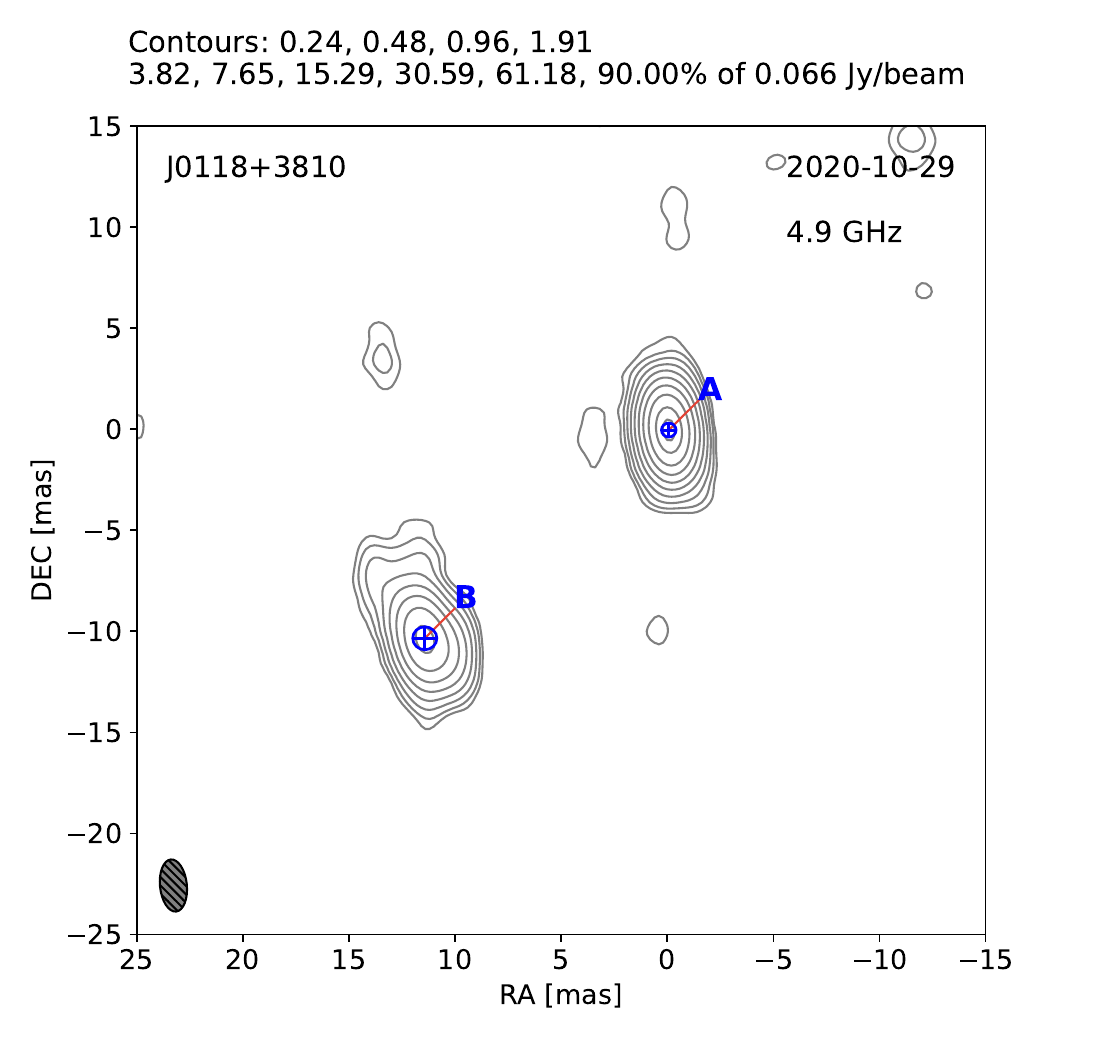}
    \caption{J0118+3810 EVN 4.9\,GHz data. The contours start at four times the image rms noise of 0.040\,mJy/beam and increase by factors of two. The restoring beam size is 1.27$\times$2.59\,mas at 5.5\,$^\circ$ PA. The source is not detected at 22.2\,GHz.}
    \label{fig:J0118}
\end{figure*}

\begin{table*}[h]
    \caption{Derived quantities for J0118+3810. See Table~\ref{tab:J2209} for a detailed description of the columns.}
    \vspace*{3mm}
    \adjustbox{width=1\textwidth}{%
    \label{tab:J0118}
    \centering
    \begin{tabular}{|| c c c c c | c c | c c | c c | c c | c c | c ||} 
        \hline
        Epochs & Frequency & Component & Flux  & Error  & Flux  & Error & FWHM  & Error & $\log(T_\mathrm{b,obs}/\mathrm{K})$ & Error  & SB    & Error & Distance & Error & Spectral Index \\ [0.5ex] 
               & [GHz]     &           & [mJy] & [mJy]  & Ratio &       & [mas] & [mas] &                                     &        & Ratio &       & [mas]    & [mas] &                \\
        \hline\hline
2016-08-05 & 4.3 & A & 74.1 & 8.6 & 1.99 & 0.21 & 0.78 & 0.03 & 9.95 & 0.06 & 6.72 & 1.32 & 15.44 & 0.25 & -1.01 \\
- & -   & B & 37.2 & 4.9 & -   & -   & 1.44 & 0.10 & 9.12 & 0.08 & -   & -   & -   & -   & -1.47 \\
        \hline
2016-08-05 & 7.6 & A & 41.9 & 5.2 & 2.56 & 0.36 & 0.54 & 0.03 & 9.54 & 0.07 & 10.20 & 2.77 & 15.58 & 0.21 & - \\
- & -   & B & 16.3 & 2.6 & -   & -   & 1.07 & 0.11 & 8.53 & 0.11 & -   & -   & -   & -   & - \\
        \hline\hline
2020-10-29 & 4.9 & A & 79.0 & 8.3 & 2.14 & 0.12 & 0.69 & 0.02 & 9.98 & 0.05 & 5.83 & 0.64 & 15.44 & 0.10 & - \\
- & -   & B & 36.9 & 4.1 & -   & -   & 1.14 & 0.05 & 9.21 & 0.06 & -   & -   & -   & -   & - \\
        \hline
    \end{tabular}
    }
\end{table*}


\begin{figure*}[h]
    \centering
    \includegraphics[width=0.4\textwidth]{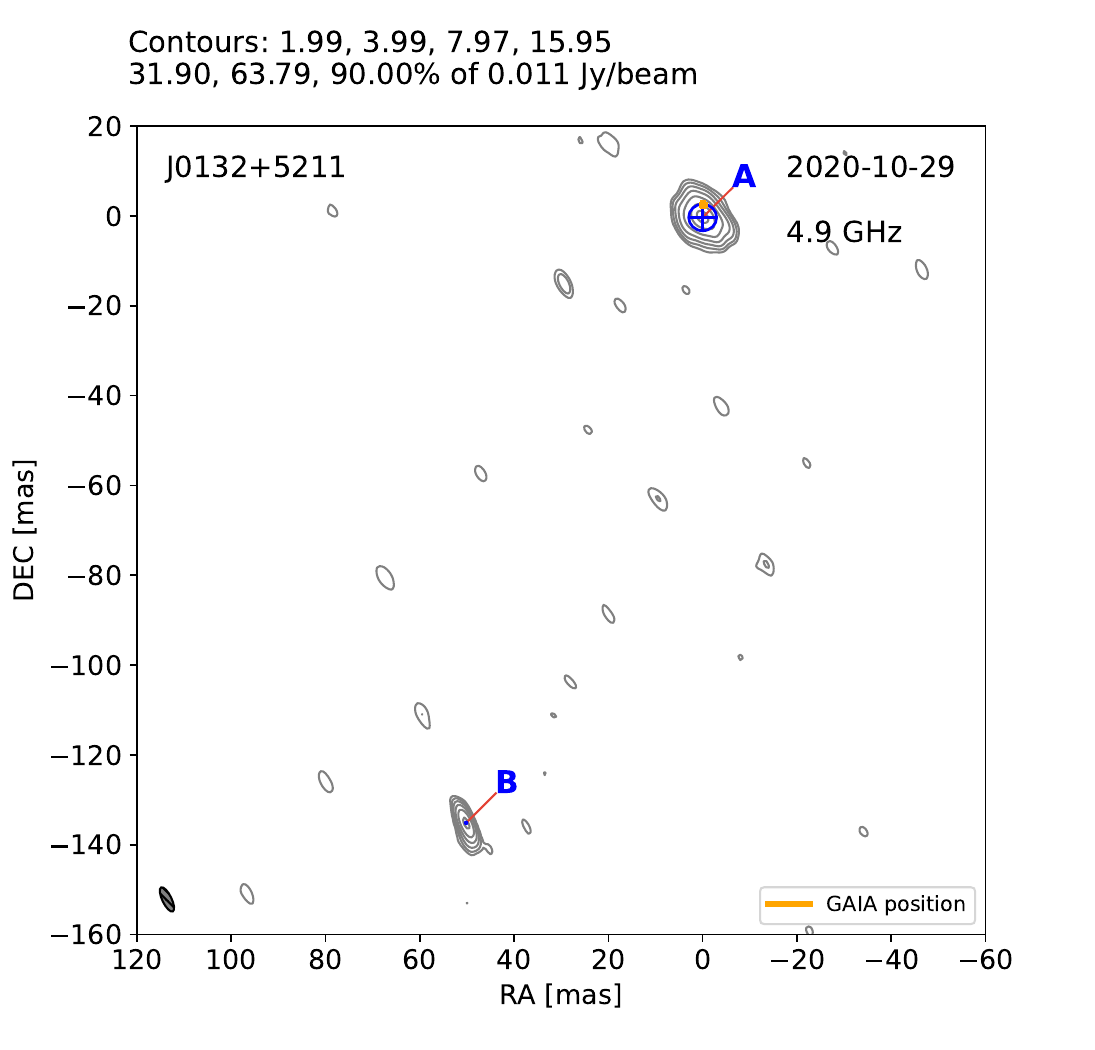}
    \caption{EVN 4.9\,GHz data with GAIA position (orange cross). The contours start at four times the image rms noise of 0.056\,mJy/beam and increase by factors of two. The restoring beam size is 1.97$\times$5.88\,mas at 24.6\,$^\circ$ PA. The source is not detected at 22.2\,GHz.}
    \label{fig:J0132}
\end{figure*}

\begin{table*}[h]
    \caption{Derived quantities for J0132+5211. See Table~\ref{tab:J2209} for a detailed description of the columns.}
    \vspace*{3mm}
    \adjustbox{width=1\textwidth}{%
    \label{tab:J0132}
    \centering
    \begin{tabular}{|| c c c c c | c c | c c | c c | c c | c c | c ||} 
        \hline
        Epochs & Frequency & Component & Flux  & Error  & Flux  & Error & FWHM  & Error & $\log(T_\mathrm{b,obs}/\mathrm{K})$ & Error  & SB    & Error & Distance & Error & Spectral Index \\ [0.5ex] 
               & [GHz]     &           & [mJy] & [mJy]  & Ratio &       & [mas] & [mas] &                                     &        & Ratio &       & [mas]    & [mas] &                \\
        \hline\hline
2016-05-15 & 4.3 & A & 48.3 & 6.4 & 3.71 & 0.60 & 4.62 & 0.38 & 8.22 & 0.09 & 0.57 & 0.19 & 145.41 & 0.51 & -1.91 \\
- & -   & B & 13.0 & 2.2 & -   & -   & 1.81 & 0.21 & 8.46 & 0.12 & -   & -   & -   & -   & -1.19 \\
        \hline
2016-05-15 & 7.6 & A & 16.5 & 2.4 & 2.48 & 0.68 & 0.53 & 0.04 & 9.15 & 0.10 & 108.05 & 63.34 & 145.71 & 0.99 & - \\
- & -   & B & 6.6 & 1.8 & -   & -   & 3.47 & 0.85 & 7.12 & 0.24 & -   & -   & -   & -   & - \\
        \hline\hline
2020-10-29 & 4.9 & A & 44.7 & 5.7 & 5.11 & 0.70 & 5.92 & 0.45 & 7.86 & 0.09 & 0.02 & 0.01 & 143.83 & 0.58 & - \\
- & -   & B & 8.8 & 1.3 & -   & -   & 0.38 & 0.03 & 9.55 & 0.10 & -   & -   & -   & -   & - \\
        \hline
    \end{tabular}
    }
\end{table*}


\begin{figure*}[h]
    \centering
    \includegraphics[width=0.4\textwidth]{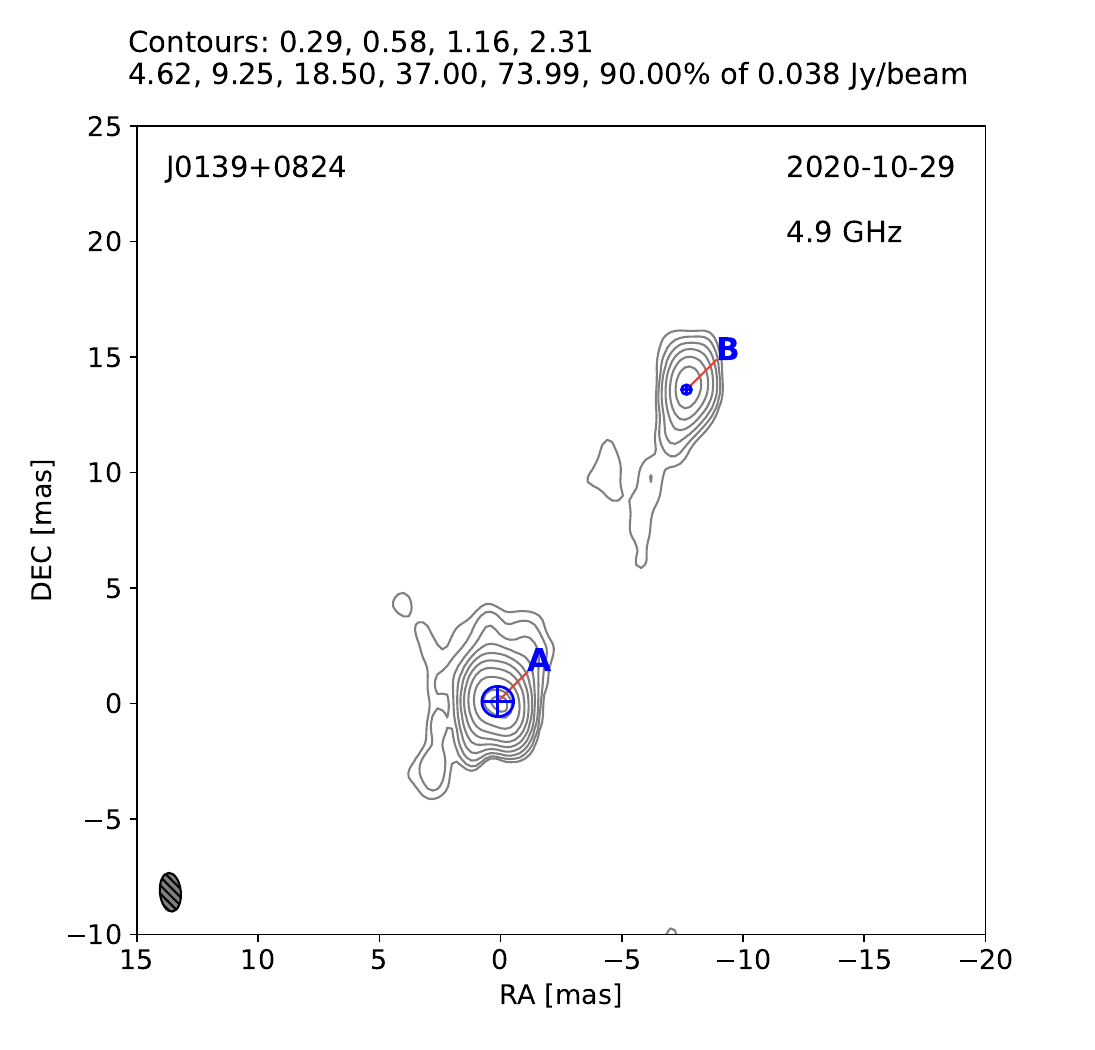}
    \caption{EVN 4.9\,GHz data. The contours start at four times the image rms noise of 0.028\,mJy/beam and increase by factors of two. The restoring beam size is 0.86$\times$1.68\,mas at 5.3\,$^\circ$ PA. The source is not detected at 22.2\,GHz.}
    \label{fig:J0139}
\end{figure*}

\begin{table*}[h]
    \caption{Derived quantities for J0139+0824. See Table~\ref{tab:J2209} for a detailed description of the columns.}
    \vspace*{3mm}
    \adjustbox{width=1\textwidth}{%
    \label{tab:J0139}
    \centering
    \begin{tabular}{|| c c c c c | c c | c c | c c | c c | c c | c ||} 
        \hline
        Epochs & Frequency & Component & Flux  & Error  & Flux  & Error & FWHM  & Error & $\log(T_\mathrm{b,obs}/\mathrm{K})$ & Error  & SB    & Error & Distance & Error & Spectral Index \\ [0.5ex] 
               & [GHz]     &           & [mJy] & [mJy]  & Ratio &       & [mas] & [mas] &                                     &        & Ratio &       & [mas]    & [mas] &                \\
        \hline\hline
2016-08-29 & 4.3 & A & 85.0 & 9.5 & 7.23 & 1.03 & 1.25 & 0.05 & 9.60 & 0.06 & 2.98 & 0.75 & 15.57 & 0.39 & -1.01 \\
- & -   & B & 11.8 & 2.0 & -   & -   & 0.80 & 0.08 & 9.13 & 0.11 & -   & -   & -   & -   & -1.48 \\
        \hline
2016-08-29 & 7.6 & A & 48.2 & 5.6 & 9.40 & 1.78 & 1.23 & 0.06 & 8.88 & 0.07 & 4.31 & 1.52 & 15.53 & 0.34 & - \\
- & -   & B & 5.1 & 1.1 & -   & -   & 0.83 & 0.12 & 8.25 & 0.15 & -   & -   & -   & -   & - \\
        \hline\hline
2020-10-29 & 4.9 & A & 79.4 & 8.3 & 8.28 & 0.69 & 1.31 & 0.04 & 9.42 & 0.05 & 0.84 & 0.13 & 15.58 & 0.09 & - \\
- & -   & B & 9.6 & 1.2 & -   & -   & 0.42 & 0.02 & 9.50 & 0.08 & -   & -   & -   & -   & - \\
        \hline
    \end{tabular}
    }
\end{table*}


\begin{figure*}[h]
    \centering
    \includegraphics[width=0.4\textwidth]{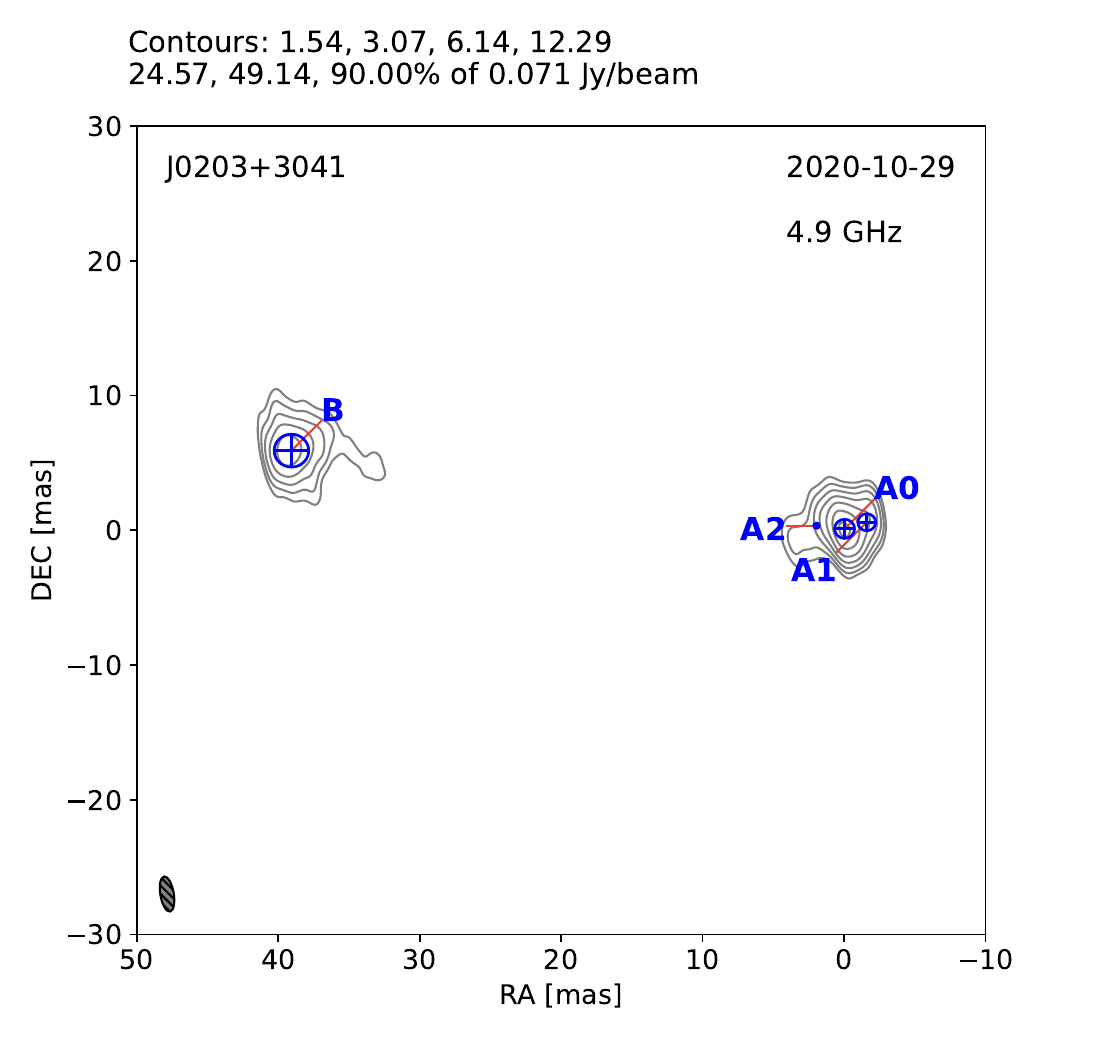}
    \includegraphics[width=0.4\textwidth]{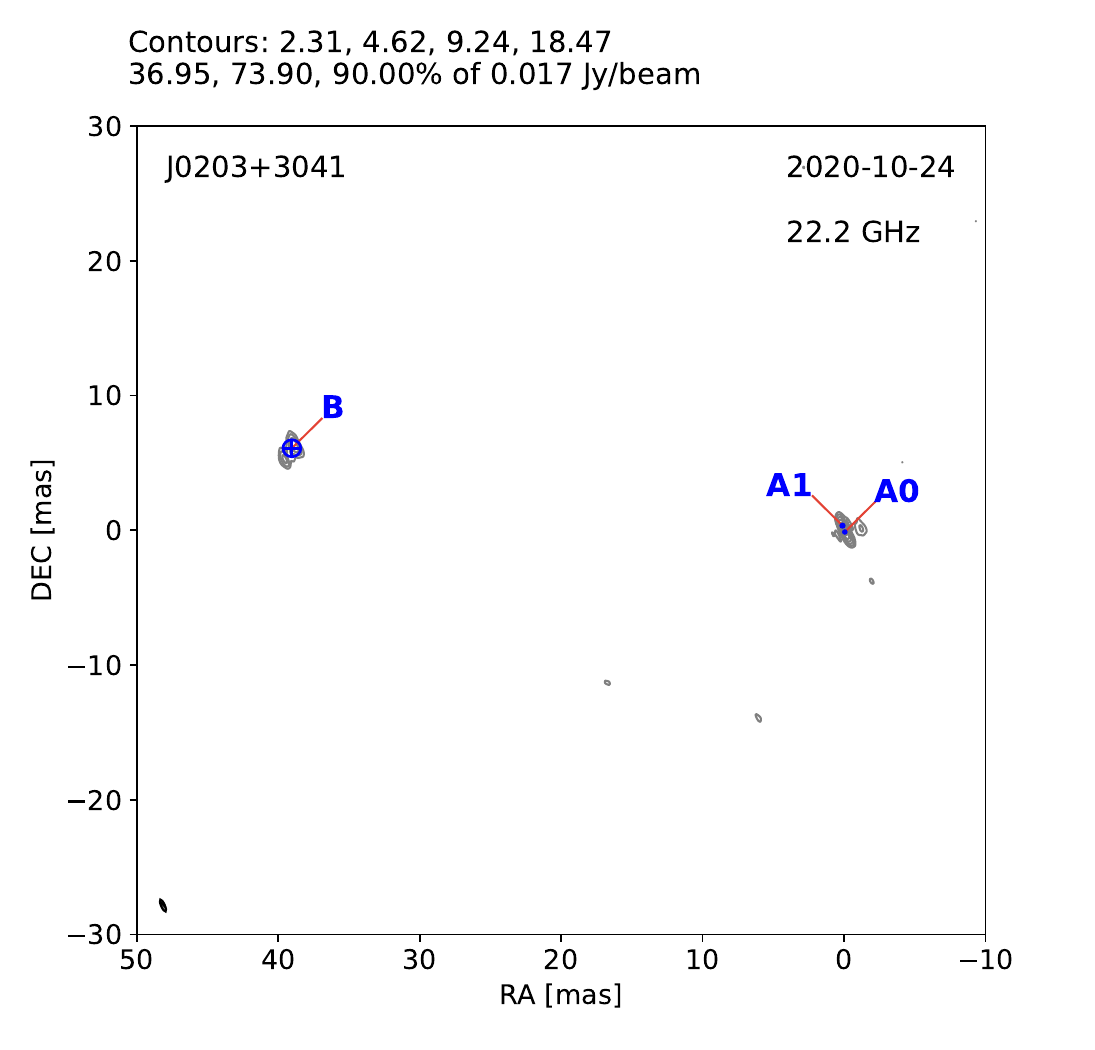}
    \caption{J0203+3041. \textit{Left:} EVN 4.9\,GHz data. The contours start at four times the image rms noise of 0.272\,mJy/beam and increase by factors of two. The restoring beam size is 0.95$\times$2.63\,mas at 9.1\,$^\circ$ PA. \textit{Right:} EVN 22.2\,GHz data. The contours start at four times the image rms noise of 0.096\,mJy/beam and increase by factors of two. The restoring beam size is 0.27$\times$0.97\,mas at 22.8\,$^\circ$ PA.}
    \label{fig:J0203}
\end{figure*}

\begin{table*}[h]
    \caption{Derived quantities for J0203+3041. See Table~\ref{tab:J2209} for a detailed description of the columns.}
    \vspace*{3mm}
    \adjustbox{width=1\textwidth}{%
    \label{tab:J0203}
    \centering
    \begin{tabular}{|| c c c c c | c c | c c | c c | c c | c c | c ||} 
        \hline
        Epochs & Frequency & Component & Flux  & Error  & Flux  & Error & FWHM  & Error & $\log(T_\mathrm{b,obs}/\mathrm{K})$ & Error  & SB    & Error & Distance & Error & Spectral Index \\ [0.5ex] 
               & [GHz]     &           & [mJy] & [mJy]  & Ratio &       & [mas] & [mas] &                                     &        & Ratio &       & [mas]    & [mas] &                \\
        \hline\hline
1996-05-15 & 2.3 & A & 359.5 & 65.1 & 2.16 & 0.59 & 1.98 & 0.22 & 10.39 & 0.13 & 3.41 & 1.67 & 39.75 & 1.08 & -0.88 \\
- & -   & B & 166.0 & 40.8 & -   & -   & 2.48 & 0.43 & 9.86 & 0.18 & -   & -   & -   & -   & - \\
        \hline
1996-05-15 & 8.3 & A & 113.9 & 24.7 & - & - & 1.78 & 0.32 & 8.86 & 0.18 & - & - & - & - & - \\
- & -   & B & - & - & -   & -   & - & - & - & - & -   & -   & -   & -   & - \\
        \hline\hline
2011-05-13 & 8.3 & A & 120.4 & 13.9 & 1.77 & 0.18 & 1.60 & 0.08 & 8.97 & 0.07 & 2.48 & 0.52 & 39.85 & 0.20 & - \\
- & -   & B & 68.1 & 8.8 & -   & -   & 1.89 & 0.15 & 8.58 & 0.09 & -   & -   & -   & -   & - \\
        \hline\hline
2011-06-28 & 2.3 & A & 369.3 & 36.1 & 1.92 & 0.14 & -   & -   & 9.77 & 0.06 & 4.07 & 1.00 & -   & -   & - \\
- & -   & A0 & 323.5 & 35.3 & 1.68 & 0.13 & 1.99 & 0.07 & 10.32 & 0.02 & 14.51 & 3.37 & 39.95 & 0.28 & - \\
- & -   & B & 192.1 & 18.8 & -   & -   & -   & -   & 9.16 & 0.10 & -   & -   & -   & -   & - \\
- & -   & B0 & 138.8 & 16.7 & 2.66 & 0.21 & 1.93 & 0.10 & 9.98 & 0.03 & 0.62 & 0.10 & -   & -   & - \\
        \hline\hline
2017-06-15 & 2.2 & A & 359.4 & 41.4 & 2.23 & 0.32 & -   & -   & 9.73 & 0.08 & 1.14 & 0.31 & -   & -   & -1.05 \\
- & -   & A0 & 289.5 & 38.3 & 1.79 & 0.26 & 2.08 & 0.13 & 10.26 & 0.04 & 3.83 & 1.06 & 40.02 & 0.64 & - \\
- & -   & B & 161.3 & 25.1 & -   & -   & 3.04 & 0.30 & 9.68 & 0.11 & -   & -   & -   & -   & -0.97 \\
        \hline
2017-06-15 & 8.7 & A & 88.0 & 10.7 & 2.03 & 0.24 & 1.37 & 0.08 & 8.94 & 0.08 & 1.55 & 0.37 & 39.90 & 0.19 & - \\
- & -   & B & 43.4 & 6.0 & -   & -   & 1.19 & 0.10 & 8.75 & 0.09 & -   & -   & -   & -   & - \\
        \hline\hline
2017-07-09 & 2.3 & A & 381.5 & 34.6 & 1.93 & 0.11 & -   & -   & 9.81 & 0.05 & 4.72 & 0.80 & -   & -   & -0.98 \\
- & -   & A0 & 313.4 & 33.3 & 1.59 & 0.09 & 1.88 & 0.05 & 10.37 & 0.02 & 16.94 & 2.87 & 39.97 & 0.26 & - \\
- & -   & B & 197.6 & 17.7 & -   & -   & -   & -   & 9.14 & 0.08 & -   & -   & -   & -   & -1.55 \\
- & -   & B0 & 138.8 & 15.8 & 2.75 & 0.18 & 2.05 & 0.08 & 9.94 & 0.02 & 0.74 & 0.09 & -   & -   & - \\
        \hline\hline
2018-11-03 & 2.2 & A & 408.3 & 40.1 & 2.12 & 0.31 & -   & -   & 9.26 & 0.07 & 3.31 & 1.02 & -   & -   & -1.26 \\
- & -   & A0 & 219.3 & 30.0 & 1.14 & 0.18 & 5.73 & 0.47 & 9.26 & 0.04 & 3.29 & 1.10 & 38.63 & 1.57 & - \\
- & -   & B & 192.7 & 31.4 & -   & -   & 9.73 & 1.20 & 8.75 & 0.13 & -   & -   & -   & -   & -1.30 \\
        \hline
2018-11-03 & 8.7 & A & 74.3 & 10.6 & 2.21 & 0.44 & 2.67 & 0.24 & 8.28 & 0.10 & 3.48 & 1.45 & 38.79 & 0.74 & - \\
- & -   & B & 33.6 & 6.7 & -   & -   & 3.35 & 0.54 & 7.74 & 0.16 & -   & -   & -   & -   & - \\
        \hline\hline
2020-10-29 & 4.9 & A & 171.7 & 17.6 & 2.10 & 0.28 & -   & -   & 9.41 & 0.07 & 3.33 & 0.96 & -   & -   & -1.32 \\
- & -   & A0 & 136.9 & 16.8 & 1.67 & 0.23 & 1.42 & 0.09 & 9.59 & 0.03 & 4.96 & 1.45 & 39.54 & 0.36 & - \\
- & -   & B & 81.7 & 12.6 & -   & -   & 2.44 & 0.28 & 8.89 & 0.12 & -   & -   & -   & -   & -1.35 \\
        \hline
2020-10-24 & 22.2 & A & 23.3 & 2.9 & 2.19 & 0.63 & -   & -   & 8.84 & 0.08 & 36.55 & 23.03 & -   & -   & - \\
- & -   & A0 & 18.2 & 2.6 & 1.71 & 0.50 & 0.19 & 0.02 & 9.13 & 0.04 & 71.78 & 45.74 & 39.59 & 0.37 & - \\
- & -   & B & 10.7 & 3.1 & -   & -   & 1.26 & 0.34 & 7.27 & 0.27 & -   & -   & -   & -   & - \\
        \hline
    \end{tabular}
    }
\end{table*}


\begin{figure*}[h]
    \centering
    \includegraphics[width=0.4\textwidth]{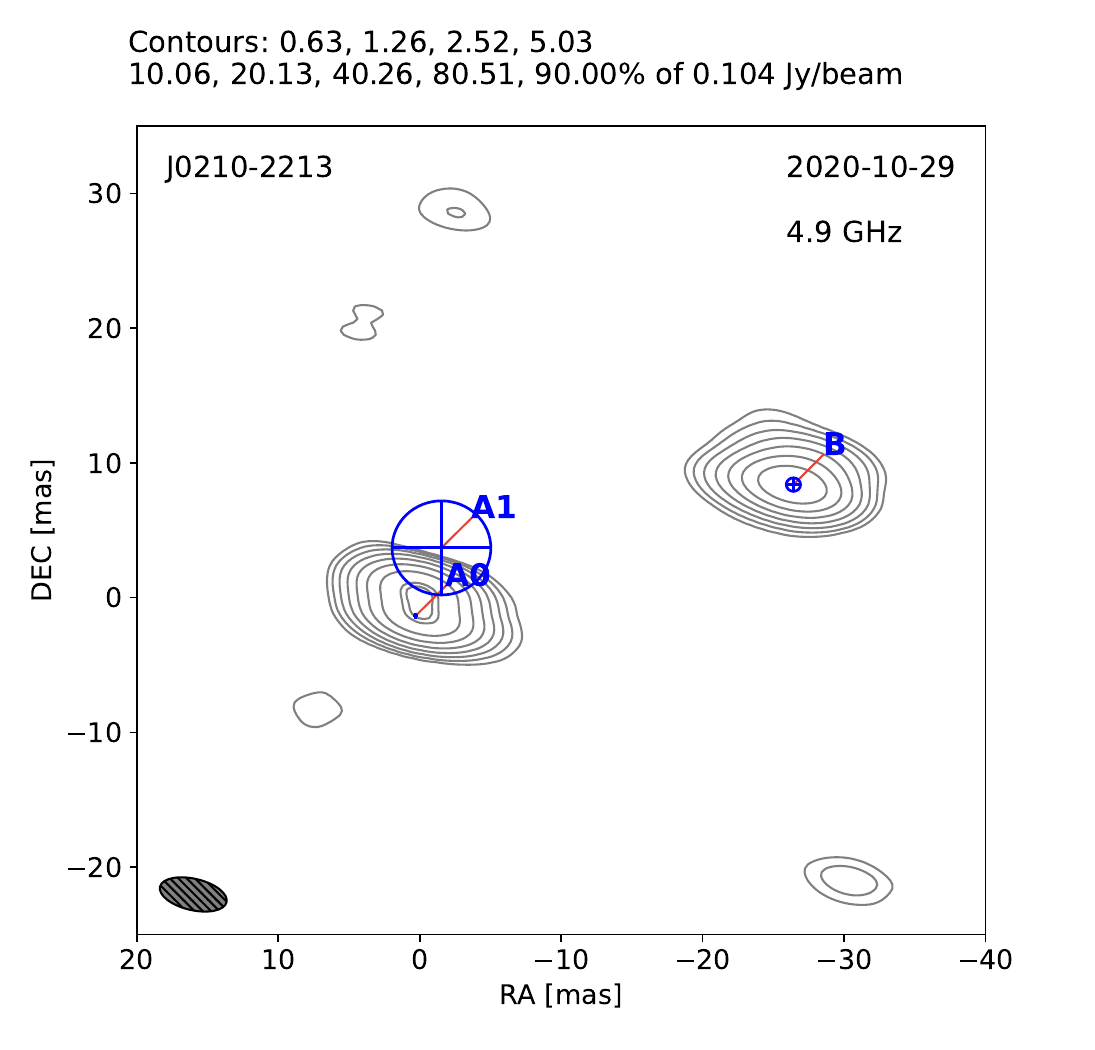}
    \caption{J0210-2213 EVN 4.9\,GHz data. The contours start at four times the image rms noise of 0.163\,mJy/beam and increase by factors of two. The restoring beam size is 2.32$\times$4.85\,mas at 75.9\,$^\circ$ PA. The source is not detected at 22.2\,GHz.}
    \label{fig:J0210}
\end{figure*}

\begin{table*}[h]
    \caption{Derived quantities for J0210-2213. See Table~\ref{tab:J2209} for a detailed description of the columns.}
    \vspace*{3mm}
    \adjustbox{width=1\textwidth}{%
    \label{tab:J0210}
    \centering
    \begin{tabular}{|| c c c c c | c c | c c | c c | c c | c c | c ||} 
        \hline
        Epochs & Frequency & Component & Flux  & Error  & Flux  & Error & FWHM  & Error & $\log(T_\mathrm{b,obs}/\mathrm{K})$ & Error  & SB    & Error & Distance & Error & Spectral Index \\ [0.5ex] 
               & [GHz]     &           & [mJy] & [mJy]  & Ratio &       & [mas] & [mas] &                                     &        & Ratio &       & [mas]    & [mas] &                \\
        \hline\hline
2015-08-31 & 4.3 & A & 251.0 & 27.3 & 1.66 & 0.11 & 1.31 & 0.04 & 10.03 & 0.06 & 1.39 & 0.18 & 27.78 & 0.15 & -1.31 \\
- & -   & B & 151.6 & 17.3 & -   & -   & 1.20 & 0.05 & 9.89 & 0.06 & -   & -   & -   & -   & -1.55 \\
        \hline
2015-08-31 & 7.6 & A & 120.3 & 14.0 & 1.89 & 0.19 & 1.06 & 0.05 & 9.41 & 0.07 & 1.00 & 0.19 & 27.56 & 0.14 & - \\
- & -   & B & 63.6 & 8.2 & -   & -   & 0.77 & 0.05 & 9.41 & 0.08 & -   & -   & -   & -   & - \\
        \hline\hline
2020-10-29 & 4.9 & A & 183.9 & 17.7 & 1.78 & 0.13 & -   & -   & $>$8.33 & - & $>$0.04 & - & -   & -   & - \\
- & -   & A0 & 151.4 & 16.7 & 1.47 & 0.11 & $<$0.10 & - & $>$11.91 & - & 140.32 & 18.11 & 28.44 & 0.18 & -  \\
- & -   & B & 103.3 & 11.9 & -   & -   & 1.01 & 0.04 & 9.76 & 0.06 & -   & -   & -   & -   & - \\
        \hline
    \end{tabular}
    }
\end{table*}


\begin{figure*}[h]
    \centering
    \includegraphics[width=0.4\textwidth]{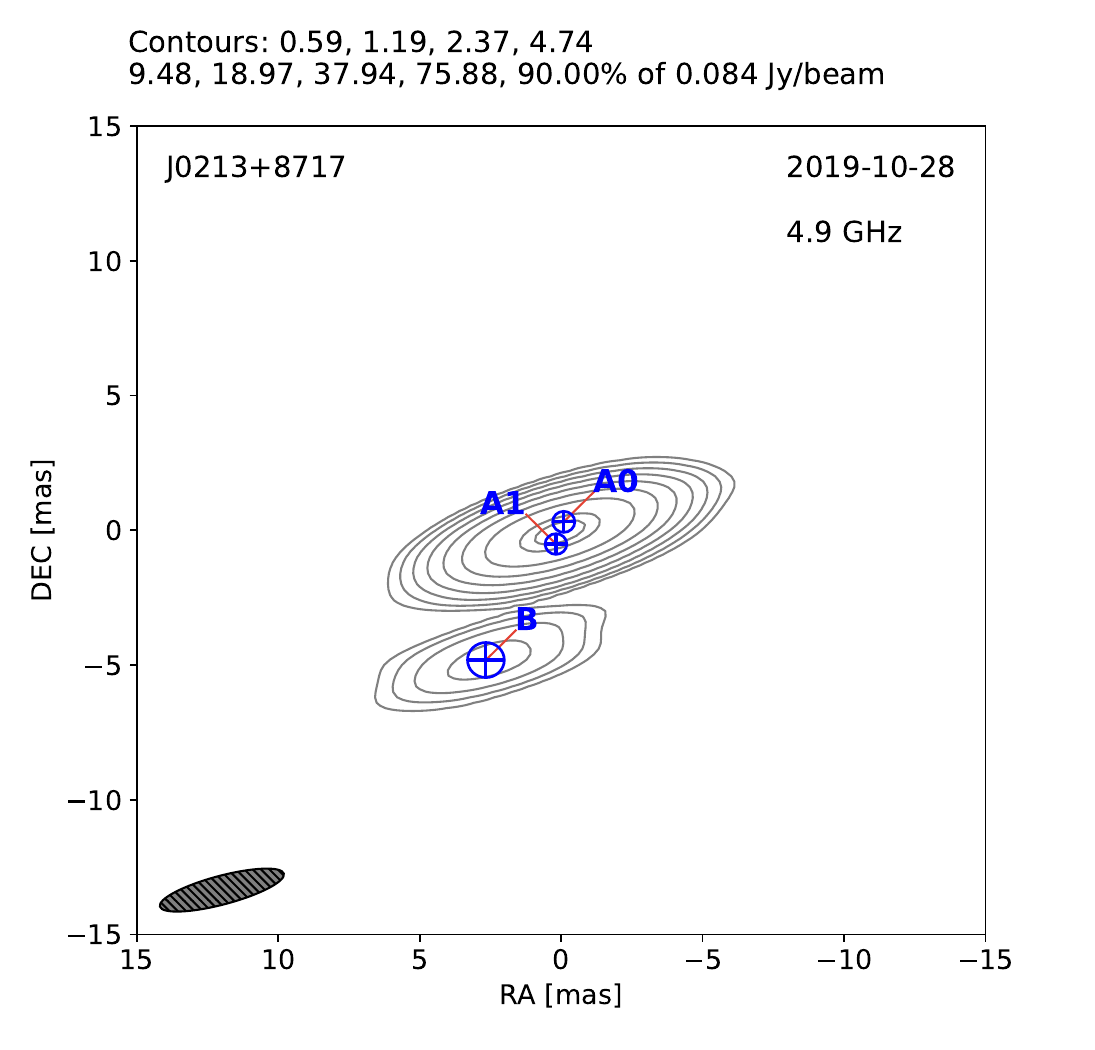}
    \includegraphics[width=0.4\textwidth]{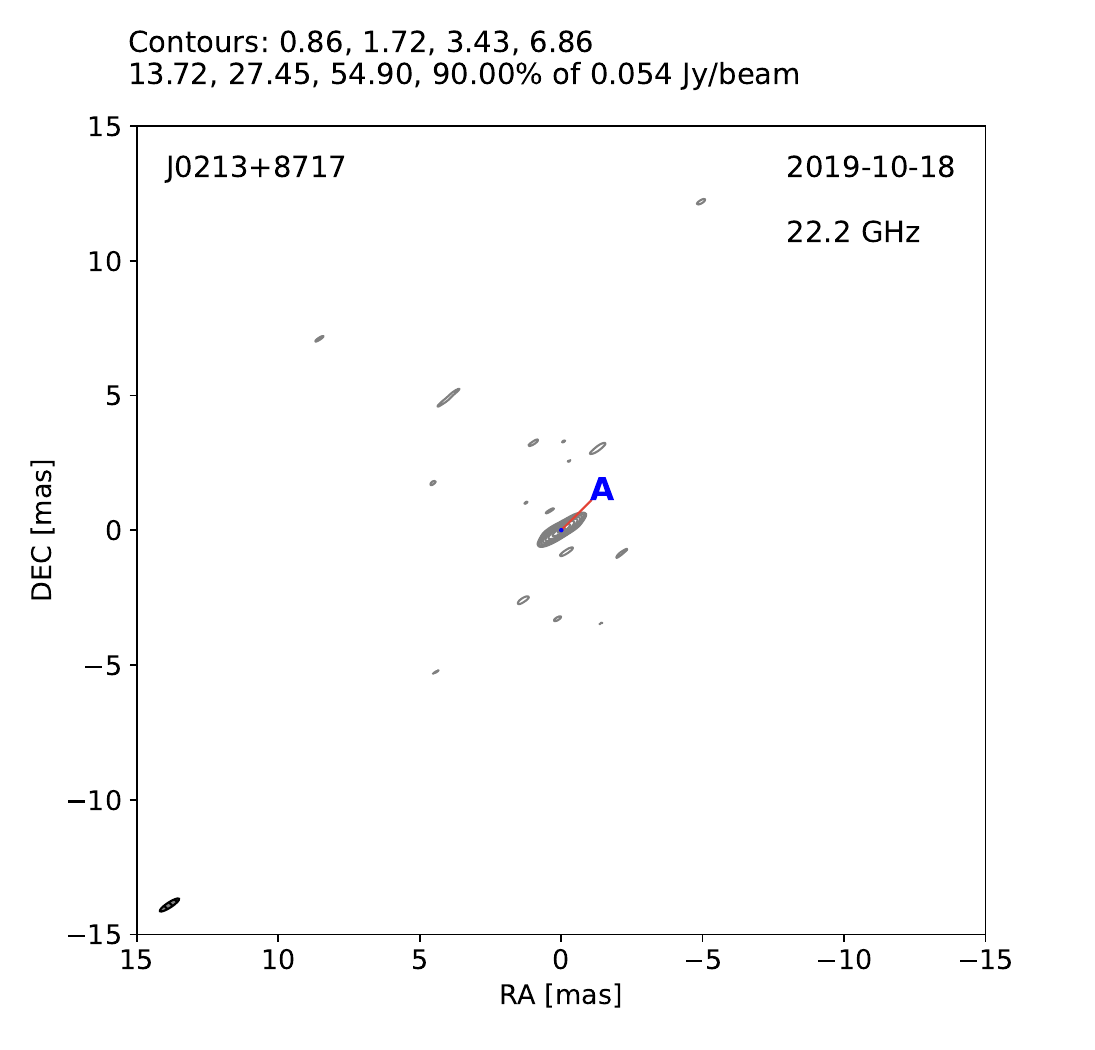}
    \caption{J0213+8717. \textit{Left:} EVN 4.9\,GHz data. The contours start at four times the image rms noise 0.124\,mJy/beam and increase by factors of two. The restoring beam size is 1.06$\times$4.55\,mas at $-74.4$\,$^\circ$ PA. \textit{Right:} EVN 22.2\,GHz data. The contours start at four times the image rms noise of 0.115\,mJy/beam and increase by factors of two. The restoring beam size is 0.19$\times$0.82\,mas at $-55.6$\,$^\circ$ PA.}
    \label{fig:J0213}
\end{figure*}

\begin{table*}[h]
    \caption{Derived quantities for J0213+8717. See Table~\ref{tab:J2209} for a detailed description of the columns.}
    \vspace*{3mm}
    \adjustbox{width=1\textwidth}{%
    \label{tab:J0213}
    \centering
    \begin{tabular}{|| c c c c c | c c | c c | c c | c c | c c | c ||} 
        \hline
        Epochs & Frequency & Component & Flux  & Error  & Flux  & Error & FWHM  & Error & $\log(T_\mathrm{b,obs}/\mathrm{K})$ & Error  & SB    & Error & Distance & Error & Spectral Index \\ [0.5ex] 
               & [GHz]     &           & [mJy] & [mJy]  & Ratio &       & [mas] & [mas] &                                     &        & Ratio &       & [mas]    & [mas] &                \\
        \hline\hline
2004-04-30 & 2.3 & A & 179.0 & 22.1 & 33.53 & 30.62 & 0.35 & 0.02 & 11.59 & 0.07 & 1456.61 & 1368.62 & 3.33 & 0.50 & -0.34 \\
- & -   & B & 5.3 & 4.9 & -   & -   & 2.29 & 0.22 & 8.42 & 0.41 & -   & -   & -   & -   & 0.81 \\
        \hline
2004-04-30 & 8.7 & A & 115.2 & 15.5 & 7.37 & 1.94 & 0.39 & 0.03 & 10.15 & 0.08 & $<$1.81 & - & 4.88 & 0.22 & - \\
- & -   & B & 15.6 & 4.2 & -   & -   & $<$0.19 & - & $>$9.90 & - & -   & -   & -   & -   & - \\
        \hline\hline
2014-05-31 & 2.3 & A & 176.6 & 19.5 & 17.84 & 6.59 & 0.50 & 0.02 & 11.27 & 0.06 & 1578.94 & 616.50 & 3.34 & 0.45 & -0.51 \\
- & -   & B & 9.9 & 3.8 & -   & -   & 4.73 & 0.25 & 8.07 & 0.17 & -   & -   & -   & -   & -0.19 \\
        \hline
2014-05-31 & 8.7 & A & 90.0 & 10.6 & 11.67 & 2.63 & 0.42 & 0.02 & 9.97 & 0.07 & $<$2.76 & - & 5.18 & 0.23 & - \\
- & -   & B & 7.7 & 1.8 & -   & -   & $<$0.21 & - & $>$9.53 & - & -   & -   & -   & -   & - \\
        \hline\hline
2017-03-23 & 2.2 & A & 228.1 & 25.7 & 29.21 & 10.07 & 0.98 & 0.04 & 10.81 & 0.06 & 145.08 & 63.35 & 4.71 & 0.67 & -0.76 \\
- & -   & B & 7.8 & 2.8 & -   & -   & 2.19 & 0.28 & 8.65 & 0.19 & -   & -   & -   & -   & -0.17 \\
        \hline
2017-03-23 & 8.7 & A & 81.5 & 9.6 & 13.08 & 3.02 & 0.58 & 0.03 & 9.65 & 0.07 & 3.03 & 1.24 & 5.24 & 0.21 & - \\
- & -   & B & 6.2 & 1.5 & -   & -   & 0.28 & 0.05 & 9.17 & 0.18 & -   & -   & -   & -   & - \\
        \hline\hline
2019-10-28 & 4.9 & A & 132.9 & 11.0 & 14.04 & 2.51 & -   & -   & 9.80 & 0.04 & 19.80 & 6.92 & -   & -   & -0.57 \\
- & -   & A0 & 69.3 & 8.1 & 7.32 & 1.34 & 0.78 & 0.03 & 9.81 & 0.03 & 20.27 & 7.24 & 5.82 & 0.39 & - \\
- & -   & B & 9.5 & 1.9 & -   & -   & 1.30 & 0.19 & 8.50 & 0.15 & -   & -   & -   & -   & - \\
        \hline
2019-10-18 & 22.2 & A & 56.2 & 6.7 & - & - & 0.06 & 0.00 & 10.67 & 0.07 & - & - & - & - & - \\
- & -   & B & - & - & -   & -   & - & - & - & - & -   & -   & -   & -   & - \\
        \hline
    \end{tabular}
    }
\end{table*}


\begin{figure*}[h]
    \centering
    \includegraphics[width=0.4\textwidth]{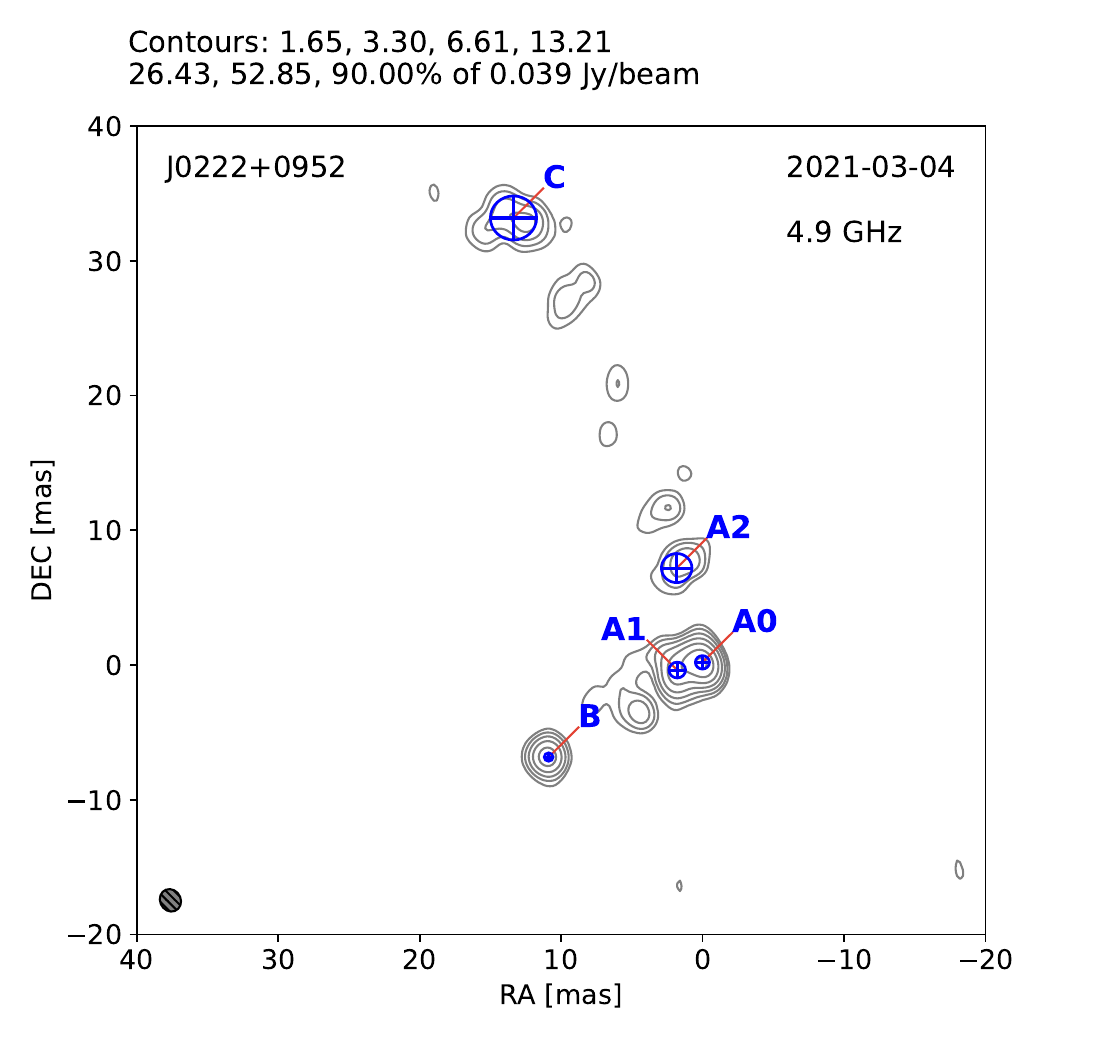}
    \includegraphics[width=0.4\textwidth]{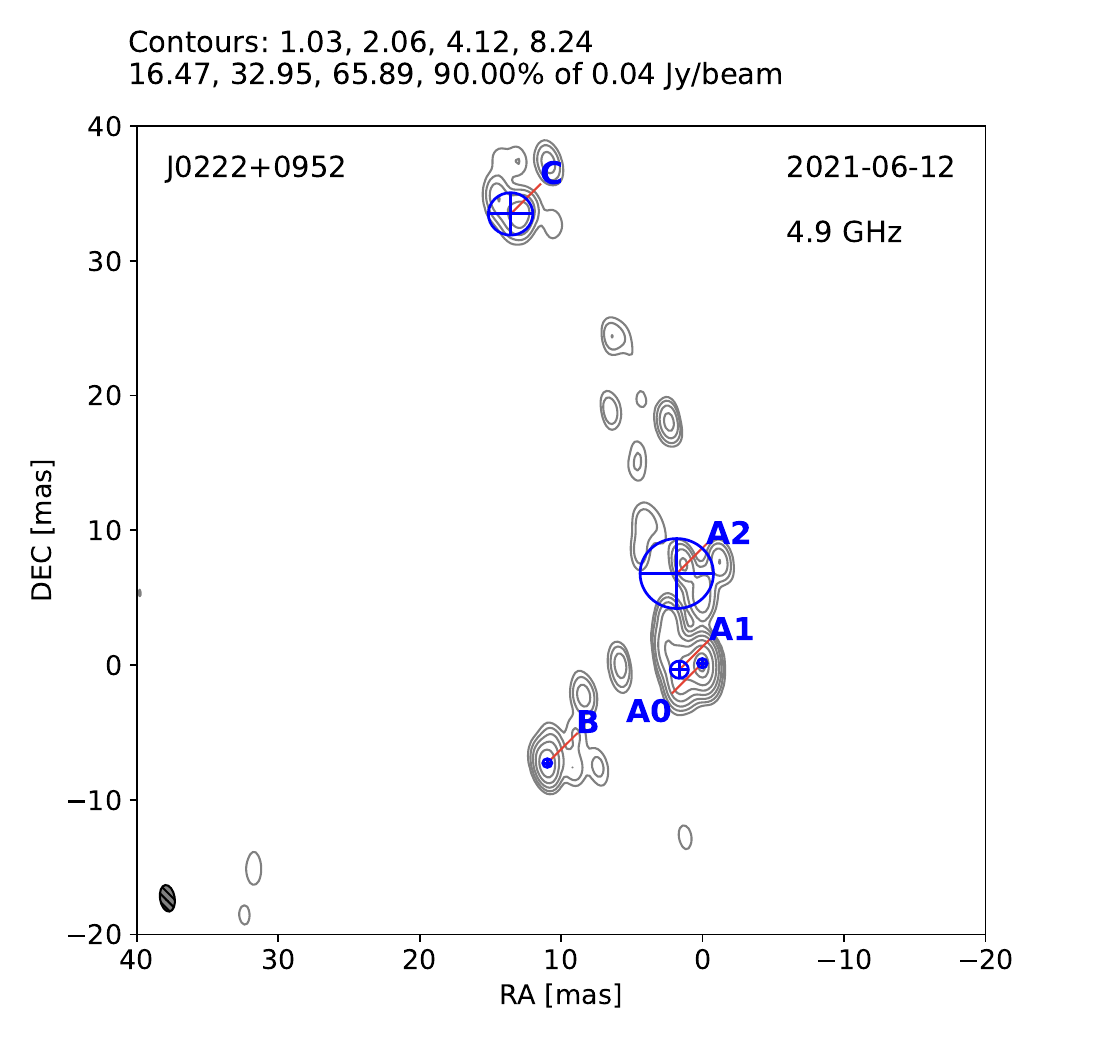}
    \caption{J0222+0952. \textit{Left:} EVN 4.9\,GHz data from session 4. The contours start at four times the image rms noise of 0.164\,mJy/beam and increase by factors of two. The restoring beam size is 1.47$\times$1.70\,mas at 22.0\,$^\circ$ PA. \textit{Right:} EVN 4.9\,GHz data from session 5. The contours start at four times the image rms noise 0.104\,mJy/beam and increase by factors of two. The restoring beam size is 1.04$\times$1.99\,mas at 8.6\,$^\circ$ PA. The source is not detected at 22.2\,GHz.}
    \label{fig:J0222}
\end{figure*}

\begin{table*}[h]
    \caption{Derived quantities for J0222+0952. See Table~\ref{tab:J2209} for a detailed description of the columns.}
    \vspace*{3mm}
    \adjustbox{width=1\textwidth}{%
    \label{tab:J0222}
    \centering
    \begin{tabular}{|| c c c c c | c c | c c | c c | c c | c c | c ||} 
        \hline
        Epochs & Frequency & Component & Flux  & Error  & Flux  & Error & FWHM  & Error & $\log(T_\mathrm{b,obs}/\mathrm{K})$ & Error  & SB    & Error & Distance & Error & Spectral Index \\ [0.5ex] 
               & [GHz]     &           & [mJy] & [mJy]  & Ratio &       & [mas] & [mas] &                                     &        & Ratio &       & [mas]    & [mas] &                \\
        \hline\hline
2014-02-23 & 4.3 & A & 124.5 & 14.5 & 6.63 & 1.35 & -   & -   & 8.61 & 0.18 & 0.14 & 0.07 & -   & -   & -0.95 \\
- & -   & A0 & 103.0 & 13.4 & 5.48 & 1.12 & 1.93 & 0.13 & 9.31 & 0.04 & 0.71 & 0.26 & 12.58 & 0.45 & -  \\
- & -   & B & 18.8 & 4.0 & -   & -   & 0.69 & 0.09 & 9.46 & 0.15 & -   & -   & -   & -   & -0.07 \\
        \hline
2014-02-23 & 7.6 & A & 72.8 & 8.5 & 4.02 & 0.78 & -   & -   & 9.02 & 0.08 & 0.82 & 0.30 & -   & -   & - \\
- & -   & A0 & 55.0 & 7.7 & 3.04 & 0.61 & 0.97 & 0.07 & 9.15 & 0.04 & 1.11 & 0.41 & 12.95 & 0.28 & - \\
- & -   & B & 18.1 & 3.6 & -   & -   & 0.58 & 0.08 & 9.11 & 0.15 & -   & -   & -   & -   & - \\
        \hline\hline
2018-12-29 & 2.2 & A & 138.1 & 14.9 & 7.81 & 1.55 & -   & -   & 9.68 & 0.06 & 5.18 & 1.83 & -   & -   & -0.60 \\
- & -   & A0 & 115.2 & 14.2 & 6.52 & 1.30 & 2.23 & 0.12 & 9.80 & 0.03 & 6.82 & 2.44 & 11.38 & 1.08 & - \\
- & -   & B & 17.7 & 3.7 & -   & -   & 2.28 & 0.31 & 8.97 & 0.15 & -   & -   & -   & -   & -0.13 \\
        \hline
2018-12-29 & 8.7 & A & 61.7 & 6.1 & 4.15 & 0.55 & -   & -   & 8.76 & 0.08 & 0.50 & 0.13 & -   & -   & - \\
- & -   & A0 & 45.0 & 5.5 & 3.02 & 0.42 & 0.78 & 0.04 & 9.14 & 0.03 & 1.19 & 0.30 & 13.21 & 0.17 & - \\
- & -   & B & 14.9 & 2.3 & -   & -   & 0.49 & 0.04 & 9.06 & 0.10 & -   & -   & -   & -   & - \\
        \hline\hline
2021-03-04 & 4.9 & A & 97.1 & 8.7 & 6.21 & 0.98 & -   & -   & 8.90 & 0.12 & 0.34 & 0.13 & -   & -   & - \\
- & -   & A0 & 53.9 & 6.9 & 3.44 & 0.57 & 1.00 & 0.06 & 9.49 & 0.03 & 1.35 & 0.40 & 12.95 & 0.21 & - \\
- & -   & B & 15.6 & 2.8 & -   & -   & 0.62 & 0.07 & 9.36 & 0.12 & -   & -   & -   & -   & - \\
        \hline\hline
2021-06-12 & 4.9 & A & 113.8 & 11.0 & 5.74 & 0.73 & -   & -   & 8.35 & 0.29 & 0.08 & 0.06 & -   & -   & - \\
- & -   & A0 & 54.2 & 6.4 & 2.73 & 0.33 & 0.70 & 0.03 & 9.80 & 0.03 & 2.35 & 0.53 & 13.23 & 0.15 & - \\
- & -   & B & 19.8 & 2.9 & -   & -   & 0.65 & 0.05 & 9.43 & 0.09 & -   & -   & -   & -   & - \\
        \hline
    \end{tabular}
    }
\end{table*}


\begin{table*}[h]
    \caption{Derived quantities for J0232-3422. See Table~\ref{tab:J2209} for a detailed description of the columns.}
    \vspace*{3mm}
    \adjustbox{width=1\textwidth}{%
    \label{tab:J0232}
    \centering
    \begin{tabular}{|| c c c c c | c c | c c | c c | c c | c c | c ||} 
        \hline
        Epochs & Frequency & Component & Flux  & Error  & Flux  & Error & FWHM  & Error & $\log(T_\mathrm{b,obs}/\mathrm{K})$ & Error  & SB    & Error & Distance & Error & Spectral Index \\ [0.5ex] 
               & [GHz]     &           & [mJy] & [mJy]  & Ratio &       & [mas] & [mas] &                                     &        & Ratio &       & [mas]    & [mas] &                \\
        \hline\hline
2016-08-02 & 4.3 & A & 49.8 & 8.4 & 3.02 & 0.83 & 0.64 & 0.07 & 9.95 & 0.11 & $<$1.10 & - & 6.35 & 0.47 & -0.88 \\
- & -   & B & 16.5 & 4.3 & -   & -   & $<$0.39 & - & $>$9.90 & - & -   & -   & -   & -   & -0.12 \\
        \hline
2016-08-02 & 7.6 & A & 30.4 & 4.4 & 1.98 & 0.36 & 0.28 & 0.02 & 9.96 & 0.09 & 3.61 & 1.15 & 6.37 & 0.18 & - \\
- & -   & B & 15.4 & 2.7 & -   & -   & 0.38 & 0.04 & 9.40 & 0.12 & -   & -   & -   & -   & - \\
        \hline
        \end{tabular}
    }
\end{table*}


\begin{figure*}[h]
    \centering
    \includegraphics[width=0.4\textwidth]{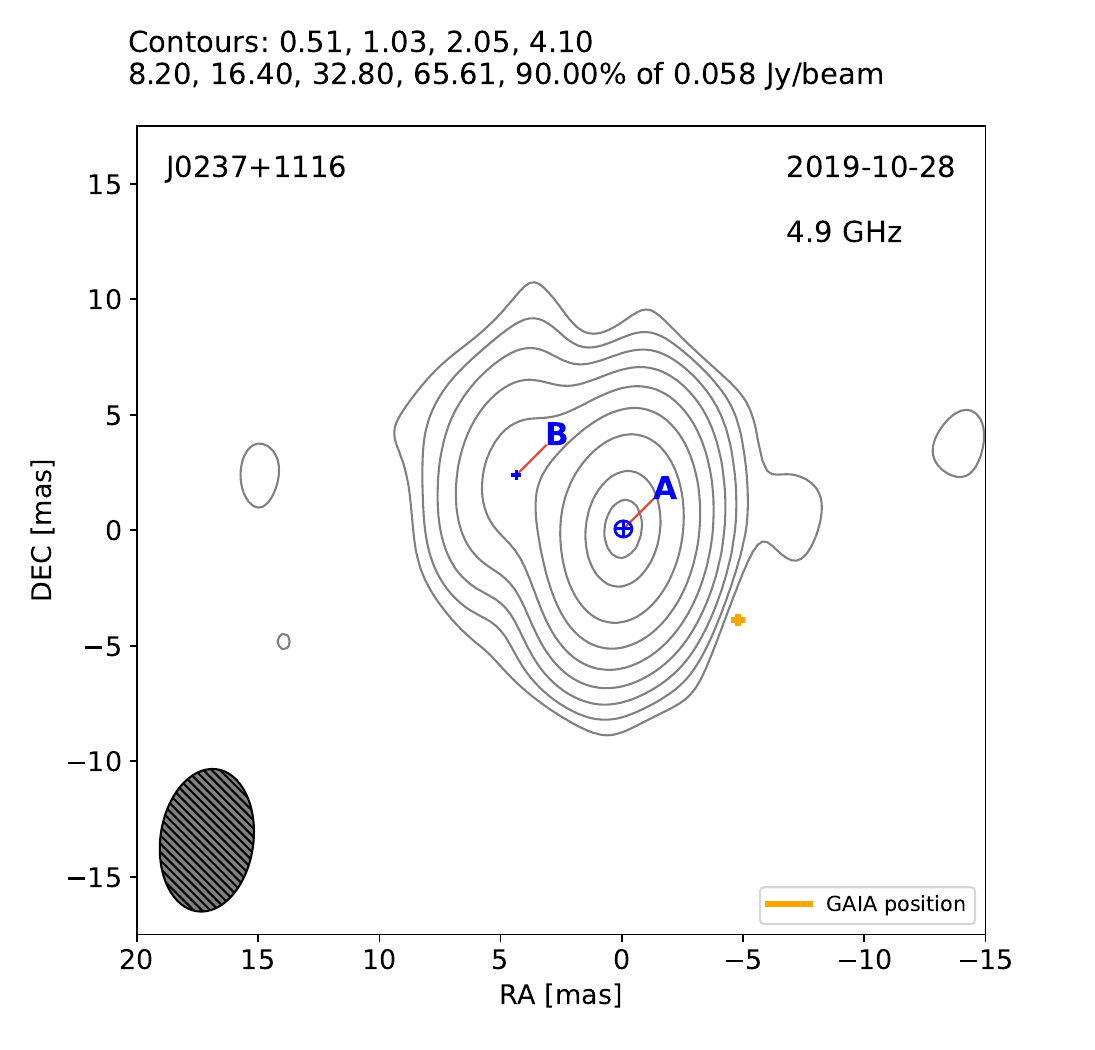}
    \caption{J0237+1116 EVN 4.9\,GHz data with GAIA position (orange cross). The contours start at four times the image rms noise of 0.074\,mJy/beam and increase by factors of two. The restoring beam size is 3.84$\times$6.20\,mas at $-6.95$\,$^\circ$ PA. The source is not detected at 22.2\,GHz.}
    \label{fig:J0237}
\end{figure*}

\begin{table*}[h]
    \caption{Derived quantities for J0237+1116. See Table~\ref{tab:J2209} for a detailed description of the columns.}
    \vspace*{3mm}
    \adjustbox{width=1\textwidth}{%
    \label{tab:J0237}
    \centering
    \begin{tabular}{|| c c c c c | c c | c c | c c | c c | c c | c ||} 
        \hline
        Epochs & Frequency & Component & Flux  & Error  & Flux  & Error & FWHM  & Error & $\log(T_\mathrm{b,obs}/\mathrm{K})$ & Error  & SB    & Error & Distance & Error & Spectral Index \\ [0.5ex] 
               & [GHz]     &           & [mJy] & [mJy]  & Ratio &       & [mas] & [mas] &                                     &        & Ratio &       & [mas]    & [mas] &               \\
        \hline\hline
2016-05-27 & 4.3 & A & 75.5 & 9.1 & 7.58 & 1.49 & 0.70 & 0.03 & 10.05 & 0.07 & 3.04 & 1.05 & 4.97 & 0.38 & -0.34 \\
- & -   & B & 10.0 & 2.1 & -   & -   & 0.45 & 0.06 & 9.57 & 0.15 & -   & -   & -   & -   & -0.42 \\
        \hline
2016-05-27 & 7.6 & A & 62.5 & 7.3 & 7.94 & 1.44 & 0.71 & 0.03 & 9.46 & 0.07 & 8.65 & 2.93 & 4.79 & 0.24 & - \\
- & -   & B & 7.9 & 1.6 & -   & -   & 0.75 & 0.10 & 8.53 & 0.15 & -   & -   & -   & -   & - \\
        \hline\hline
2019-10-28 & 4.9 & A & 59.1 & 6.6 & 8.61 & 1.35 & 0.71 & 0.03 & 9.83 & 0.06 & $<$3.45 & - & 5.00 & 0.51 & - \\
- & -   & B & 6.9 & 1.2 & -   & -   & $<$0.45 & - & $>$9.29 & - & -   & -   & -   & -   & - \\
        \hline
    \end{tabular}
    }
\end{table*}


\begin{figure*}[h]
    \centering
    \includegraphics[width=0.4\textwidth]{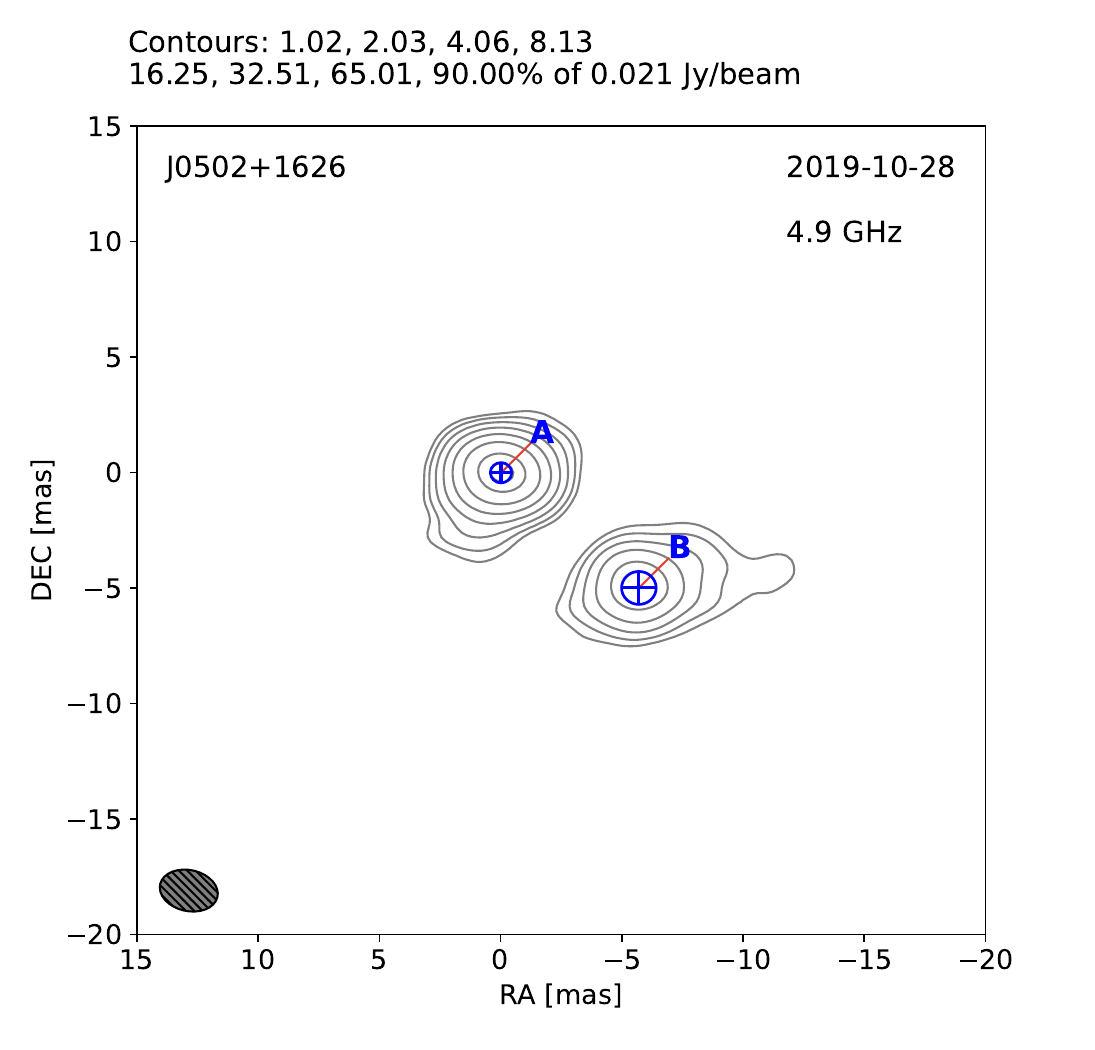}
    \caption{J0502+1626 EVN 4.9\,GHz data. The contours start at four times the image rms noise of 0.054 mJy/beam and increase by factors of two. The restoring beam size is 1.77$\times$2.43\,mas at 75.91\,$^\circ$ PA. The source is not detected at 22.2\,GHz.}
    \label{fig:J0502}

\end{figure*}

\begin{table*}[h]
    \caption{Derived quantities for J0502+1626. See Table~\ref{tab:J2209} for a detailed description of the columns.}
    \vspace*{3mm}
    \adjustbox{width=1\textwidth}{%
    \label{tab:J0502}
    \centering
    \begin{tabular}{|| c c c c c | c c | c c | c c | c c | c c | c ||} 
        \hline
        Epochs & Frequency & Component & Flux  & Error  & Flux  & Error & FWHM  & Error & $\log(T_\mathrm{b,obs}/\mathrm{K})$ & Error  & SB    & Error & Distance & Error & Spectral Index \\ [0.5ex] 
               & [GHz]     &           & [mJy] & [mJy]  & Ratio &       & [mas] & [mas] &                                     &        & Ratio &       & [mas]    & [mas] &                \\
        \hline\hline
2015-12-01 & 4.3 & A & 28.9 & 3.8 & 5.25 & 1.13 & 0.78 & 0.05 & 9.54 & 0.08 & $<$1.38 & - & 7.27 & 0.46 & -1.25 \\
- & -   & B & 5.5 & 1.2 & -   & -   & $<$0.40 & - & $>$9.40 & - & -   & -   & -   & -   & 0.02 \\
        \hline
2015-12-01 & 7.6 & A & 14.3 & 2.3 & 2.57 & 0.60 & 0.32 & 0.03 & 9.52 & 0.10 & 2.41 & 0.99 & 7.27 & 0.29 & - \\
- & -   & B & 5.6 & 1.2 & -   & -   & 0.31 & 0.04 & 9.14 & 0.16 & -   & -   & -   & -   & - \\
        \hline\hline
2019-10-28 & 4.9 & A & 26.8 & 3.2 & 3.14 & 0.42 & 0.88 & 0.04 & 9.30 & 0.07 & 8.26 & 2.07 & 7.56 & 0.26 & - \\
- & -   & B & 8.6 & 1.3 & -   & -   & 1.43 & 0.13 & 8.38 & 0.11 & -   & -   & -   & -   & - \\
        \hline
    \end{tabular}
    }
\end{table*}


\begin{figure*}[h]
    \centering
    \includegraphics[width=0.4\textwidth]{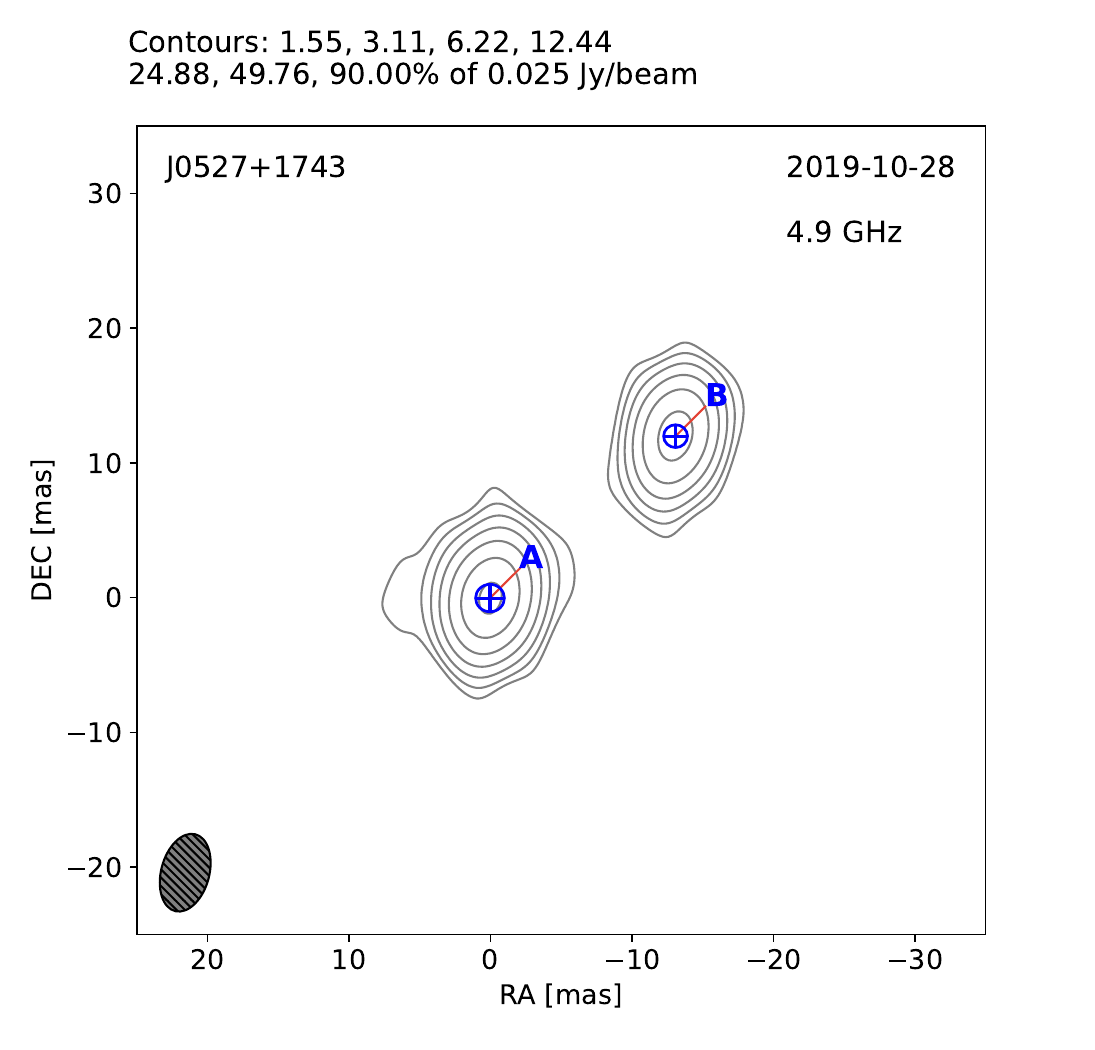}
    \caption{J0527+1743 EVN 4.9\,GHz data. The contours start at four times the image rms noise of 0.097 mJy/beam and increase by factors of two. The restoring beam size is 3.41$\times$5.88\,mas at $-13.63$\,$^\circ$ PA. The source is not detected at 22.2\,GHz.}
    \label{fig:J0527}
\end{figure*}

\begin{table*}[h]
    \caption{Derived quantities for J0527+1743. See Table~\ref{tab:J2209} for a detailed description of the columns.}
    \vspace*{3mm}
    \adjustbox{width=1\textwidth}{%
    \label{tab:J0527}
    \centering
    \begin{tabular}{|| c c c c c | c c | c c | c c | c c | c c | c ||} 
        \hline
        Epochs & Frequency & Component & Flux  & Error  & Flux  & Error & FWHM  & Error & $\log(T_\mathrm{b,obs}/\mathrm{K})$ & Error  & SB    & Error & Distance & Error & Spectral Index \\ [0.5ex] 
               & [GHz]     &           & [mJy] & [mJy]  & Ratio &       & [mas] & [mas] &                                     &        & Ratio &       & [mas]    & [mas] &                \\
        \hline\hline
2015-10-30 & 4.3 & A & 21.8 & 2.9 & 2.35 & 0.39 & 1.06 & 0.07 & 9.16 & 0.08 & 0.72 & 0.21 & 17.63 & 0.43 & -0.38 \\
- & -   & B & 9.3 & 1.6 & -   & -   & 0.58 & 0.06 & 9.30 & 0.12 & -   & -   & -   & -   & -0.33 \\
        \hline
2015-10-30 & 7.6 & A & 17.6 & 2.6 & 2.28 & 0.44 & 0.69 & 0.06 & 8.95 & 0.09 & $<$0.24 & - & 17.47 & 0.28 & - \\
- & -   & B & 7.7 & 1.5 & -   & -   & $<$0.22 & - & $>$9.56 & - & -   & -   & -   & -   & - \\
        \hline\hline
2019-10-28 & 4.9 & A & 30.2 & 3.9 & 1.61 & 0.21 & 2.04 & 0.13 & 8.62 & 0.08 & 1.11 & 0.26 & 17.79 & 0.48 & - \\
- & -   & B & 18.8 & 2.7 & -   & -   & 1.69 & 0.13 & 8.57 & 0.09 & -   & -   & -   & -   & - \\
        \hline
    \end{tabular}
    }
\end{table*}




\begin{figure*}[h]
    \centering
    \includegraphics[width=0.4\textwidth]{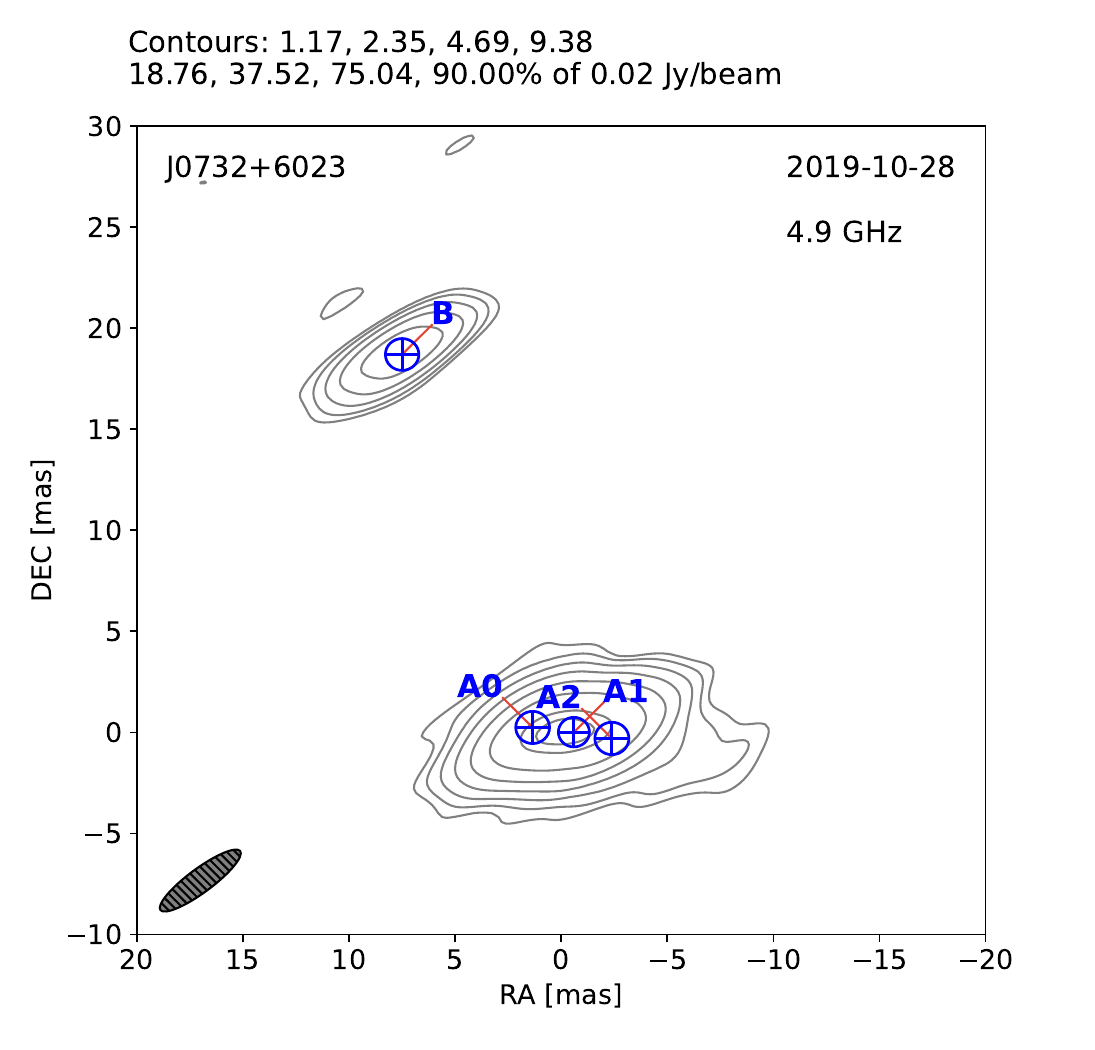}
    \caption{J0732+6023 EVN 4.9\,GHz data. The source is not detected at 22.2\,GHz. The contours start at four times the image rms noise of 0.058\,mJy/beam and increase by factors of two. The restoring beam size is 1.16$\times$4.76\,mas at $-52.0$\,$^\circ$ PA. The source is not detected at 22.2\,GHz.}
    \label{fig:J0732}
\end{figure*}

\begin{table*}[h]
    \caption{Derived quantities for J0732+6023. See Table~\ref{tab:J2209} for a detailed description of the columns.}
    \vspace*{3mm}
    \adjustbox{width=1\textwidth}{%
    \label{tab:J0732}
    \centering
    \begin{tabular}{|| c c c c c | c c | c c | c c | c c | c c | c ||} 
        \hline
        Epochs & Frequency & Component & Flux  & Error  & Flux  & Error & FWHM  & Error & $\log(T_\mathrm{b,obs}/\mathrm{K})$ & Error  & SB    & Error & Distance & Error & Spectral Index \\ [0.5ex] 
               & [GHz]     &           & [mJy] & [mJy]  & Ratio &       & [mas] & [mas] &                                     &        & Ratio &       & [mas]    & [mas] &                \\
        \hline\hline
2016-08-11 & 4.3 & A & 67.0 & 8.3 & 4.75 & 1.10 & -   & -   & 8.99 & 0.09 & 1.55 & 0.68 & -   & -   & -1.97 \\
- & -   & A0 & 47.7 & 7.3 & 3.38 & 0.81 & 1.80 & 0.18 & 9.04 & 0.05 & 1.73 & 0.80 & 19.41 & 0.57 & - \\
- & -   & B & 14.1 & 3.3 & -   & -   & 1.28 & 0.22 & 8.80 & 0.18 & -   & -   & -   & -   & -0.39 \\
        \hline
2016-08-11 & 7.6 & A & 22.1 & 3.1 & 1.95 & 0.42 & -   & -   & $>$9.41 & - & $>$23.31 & - & -   & -   & - \\
- & -   & A0 & 16.2 & 2.7 & 1.43 & 0.32 & 0.38 & 0.04 & 9.42 & 0.06 & 23.60 & 12.24 & 21.74 & 0.55 & - \\
- & -   & B & 11.3 & 2.2 & -   & -   & -   & -   & 8.04 & 0.20 & -   & -   & -   & -   & - \\
- & -   & B1 & 5.8 & 1.7 & 3.80 & 1.14 & 1.41 & 0.36 & 7.84 & 0.12 & 36.98 & 22.81 & -   & -   & - \\
        \hline\hline
2019-10-28 & 4.9 & A & 64.4 & 4.7 & 7.09 & 0.91 & -   & -   & 8.70 & 0.04 & 2.43 & 0.62 & -   & -   & - \\
- & -   & A0 & 24.2 & 3.0 & 2.67 & 0.38 & 1.59 & 0.10 & 8.74 & 0.03 & 2.61 & 0.73 & 19.45 & 0.34 & - \\
- & -   & B & 9.1 & 1.4 & -   & -   & 1.58 & 0.16 & 8.32 & 0.11 & -   & -   & -   & -   & - \\
        \hline
    \end{tabular}
    }
\end{table*}


\begin{table*}[h]
    \caption{Derived quantities for J0923-3435. See Table~\ref{tab:J2209} for a detailed description of the columns.}
    \vspace*{3mm}
    \adjustbox{width=1\textwidth}{%
    \label{tab:J0923}
    \centering
    \begin{tabular}{|| c c c c c | c c | c c | c c | c c | c c | c ||} 
        \hline
        Epochs & Frequency & Component & Flux  & Error  & Flux  & Error & FWHM  & Error & $\log(T_\mathrm{b,obs}/\mathrm{K})$ & Error  & SB    & Error & Distance & Error & Spectral Index \\ [0.5ex] 
               & [GHz]     &           & [mJy] & [mJy]  & Ratio &       & [mas] & [mas] &                                     &        & Ratio &       & [mas]    & [mas] &                \\
        \hline\hline
2013-02-22 & 4.3 & A & 198.1 & 31.8 & 2.42 & 0.55 & 2.47 & 0.26 & 9.38 & 0.11 & 1.29 & 0.56 & 17.53 & 0.89 & -1.11 \\
- & -   & B & 81.9 & 17.7 & -   & -   & 1.81 & 0.27 & 9.27 & 0.16 & -   & -   & -   & -   & -1.49 \\
        \hline
2013-02-22 & 7.6 & A & 105.7 & 14.2 & 2.98 & 0.54 & 1.37 & 0.11 & 9.13 & 0.09 & 3.07 & 1.10 & 17.05 & 0.41 & - \\
- & -   & B & 35.5 & 6.6 & -   & -   & 1.39 & 0.19 & 8.64 & 0.14 & -   & -   & -   & -   & - \\
        \hline
        \end{tabular}
    }
\end{table*}


\begin{figure*}[h]
    \centering
    \includegraphics[width=0.4\textwidth]{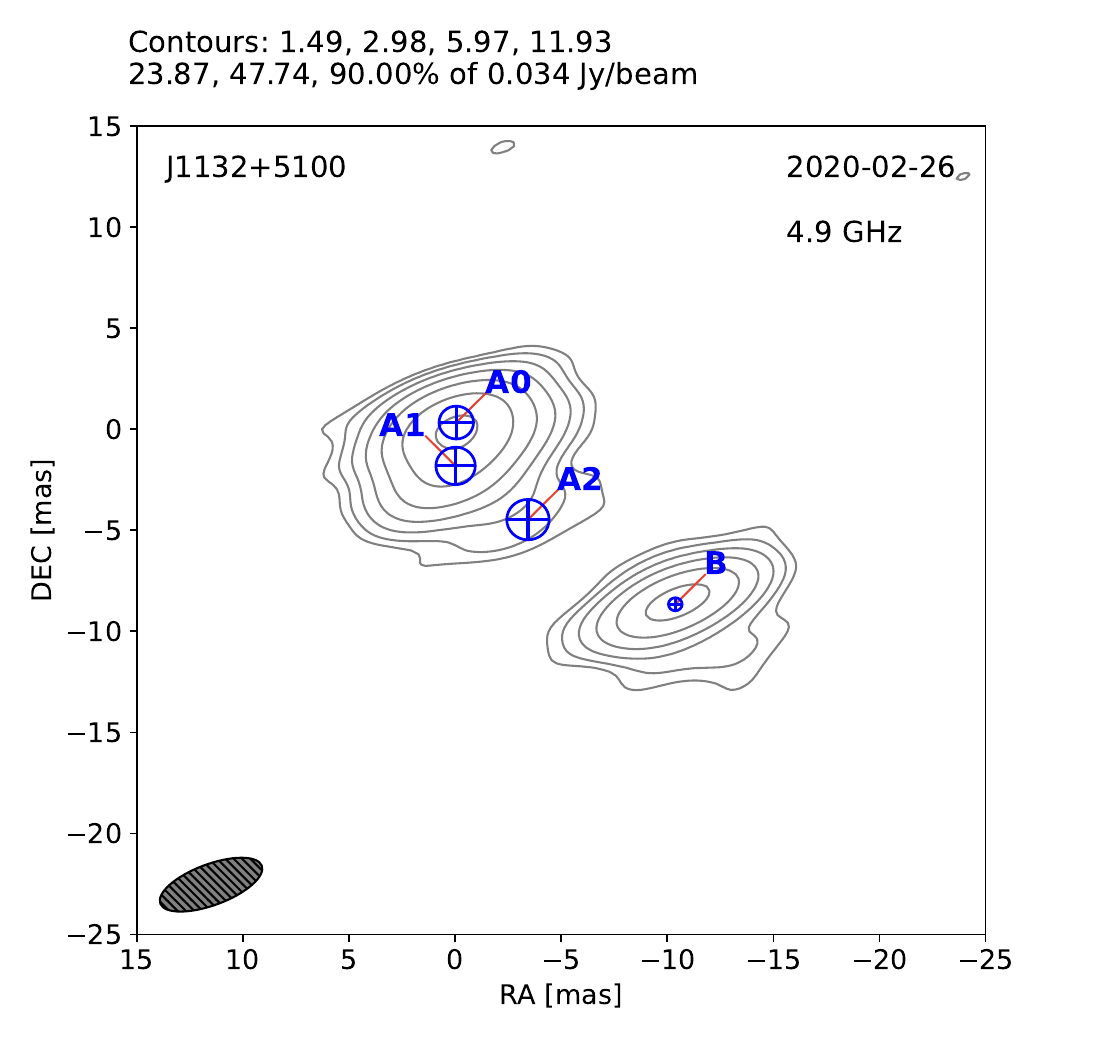}
    \caption{J1132+5100 EVN 4.9\,GHz data. The contours start at four times the image rms noise of 0.128\,mJy/beam and increase by factors of two. The restoring beam size is 2.02$\times$5.14\,mas at $-68.4$\,$^\circ$ PA. The source is not detected at 22.2\,GHz.}
    \label{fig:J1132}
\end{figure*}

\begin{table*}[h]
    \caption{Derived quantities for J1132+5100. See Table~\ref{tab:J2209} for a detailed description of the columns. Note that the 2013 epoch lists A1 as the brightest component. The calculated distance between components A1 and B can not be readily compared to the distance between components A0 and B for the 2020 epoch. In 2013, the distance between components A0 and B is $14.04\,{\pm}\,0.31$\,mas, consistent within the combined error for both epochs.}
    \vspace*{3mm}
    \adjustbox{width=1\textwidth}{%
    \label{tab:J1132}
    \centering
    \begin{tabular}{|| c c c c c | c c | c c | c c | c c | c c | c ||} 
        \hline
        Epochs & Frequency & Component & Flux  & Error  & Flux  & Error & FWHM  & Error & $\log(T_\mathrm{b,obs}/\mathrm{K})$ & Error  & SB    & Error & Distance & Error & Spectral Index \\ [0.5ex] 
               & [GHz]     &           & [mJy] & [mJy]  & Ratio &       & [mas] & [mas] &                                     &        & Ratio &       & [mas]    & [mas] &                \\
        \hline\hline
2013-04-22 & 4.4 & A & 90.0 & 8.4 & 3.51 & 0.45 & -   & -   & 9.27 & 0.05 & 0.59 & 0.13 & -   & -   & -1.17 \\
- & -   & A1 & 46.9 & 6.1 & 1.83 & 0.26 & 1.51 & 0.08 & 9.17 & 0.04 & 0.48 & 0.12 & 12.54 & 0.32 & - \\
- & -   & B & 25.6 & 3.9 & -   & -   & 0.77 & 0.06 & 9.50 & 0.10 & -   & -   & -   & -   & -0.66 \\
        \hline
2013-04-22 & 7.6 & A & 46.7 & 4.8 & 2.65 & 0.38 & -   & -   & 8.75 & 0.07 & 0.17 & 0.04 & -   & -   & - \\
- & -   & A0 & 24.1 & 3.5 & 1.37 & 0.22 & 1.14 & 0.10 & 8.65 & 0.05 & 0.13 & 0.04 & 14.23 & 0.22 & - \\
- & -   & B & 17.7 & 2.8 & -   & -   & 0.35 & 0.03 & 9.53 & 0.10 & -   & -   & -   & -   & - \\
        \hline\hline
2020-02-27 & 4.9 & A & 70.7 & 6.5 & 2.99 & 0.36 & -   & -   & 8.60 & 0.10 & 0.12 & 0.04 & -   & -   & - \\
- & -   & A0 & 37.8 & 4.9 & 1.60 & 0.21 & 1.63 & 0.10 & 8.91 & 0.04 & 0.25 & 0.06 & 13.69 & 0.34 & - \\
- & -   & B & 23.6 & 3.4 & -   & -   & 0.65 & 0.05 & 9.51 & 0.09 & -   & -   & -   & -   & - \\
        \hline
    \end{tabular}
    }
\end{table*}


\begin{figure*}[h]
    \centering
    \includegraphics[width=0.4\textwidth]{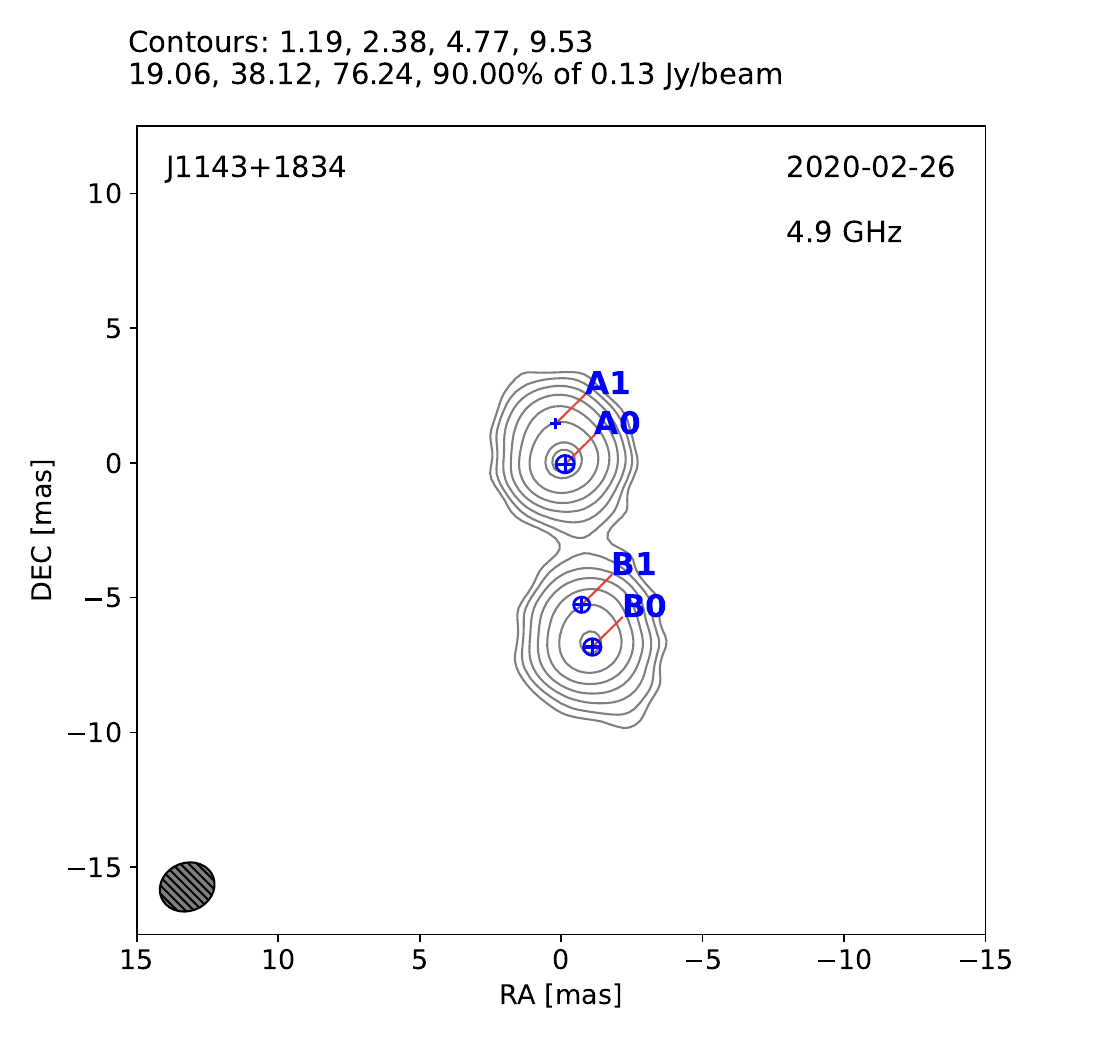}
    \includegraphics[width=0.4\textwidth]{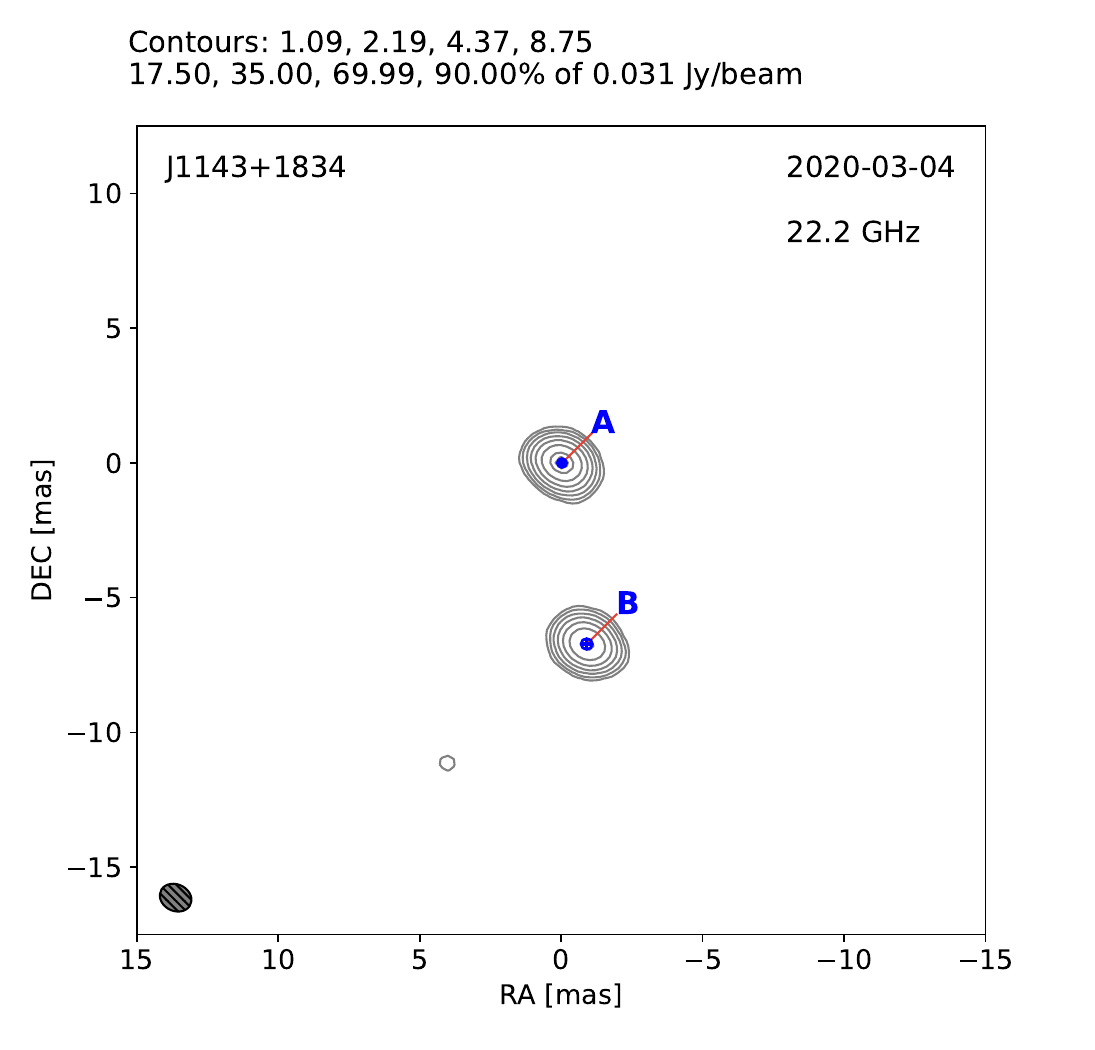}
    \caption{J1143+1843. \textit{Left:} EVN 4.9\,GHz data. The contours start at four times the image rms noise of 0.388\,mJy/beam and increase by factors of two. The restoring beam size is 2.02$\times$5.14\,mas at $-68.4$\,$^\circ$ PA. \textit{Right:} EVN 22.2\,GHz data. The contours start at four times the image rms noise of 0.084\,mJy/beam and increase by factors of two. The restoring beam size is 0.98$\times$1.17\,mas at 58.4\,$^\circ$ PA.}
    \label{fig:J1143}
\end{figure*}

\begin{table*}[h]
    \caption{Derived quantities for J1143+1843. See Table~\ref{tab:J2209} for a detailed description of the columns.}
    \vspace*{3mm}
    \adjustbox{width=1\textwidth}{%
    \label{tab:J1143}
    \centering
    \begin{tabular}{|| c c c c c | c c | c c | c c | c c | c c | c |} 
        \hline
        Epochs & Frequency & Component & Flux  & Error  & Flux  & Error & FWHM  & Error & $\log(T_\mathrm{b,obs}/\mathrm{K})$ & Error  & SB    & Error & Distance & Error & Spectral Index \\ [0.5ex] 
               & [GHz]     &           & [mJy] & [mJy]  & Ratio &       & [mas] & [mas] &                                     &        & Ratio &       & [mas]    & [mas] &                \\
        \hline\hline
1996-01-02 & 2.3 & A & 57.8 & 14.2 & 0.20 & 0.05 & 1.05 & 0.15 & 10.15 & 0.16 & 1.88 & 0.77 & 7.68 & 0.92 & - \\
- & -   & B & 282.3 & 40.3 & -   & -   & 3.18 & 0.26 & 9.88 & 0.09 & -   & -   & -   & -   & - \\
        \hline
1996-01-02 & 8.3 & A & 64.1 & 16.0 & 0.60 & 0.17 & 0.33 & 0.06 & 10.07 & 0.18 & 0.91 & 0.47 & 6.97 & 0.30 & - \\
- & -   & B & 107.0 & 21.7 & -   & -   & 0.41 & 0.05 & 10.11 & 0.14 & -   & -   & -   & -   & - \\
        \hline\hline
1996-03-13 & 2.3 & A & 188.9 & 27.7 & 1.14 & 0.18 & 1.05 & 0.08 & 10.66 & 0.09 & 1.01 & 0.26 & 7.03 & 0.63 & -0.48 \\
- & -   & B & 166.4 & 25.3 & -   & -   & 0.99 & 0.07 & 10.66 & 0.09 & -   & -   & -   & -   & -0.51 \\
        \hline
1996-03-13 & 8.3 & A & 100.6 & 20.7 & 1.17 & 0.31 & 0.44 & 0.06 & 10.02 & 0.15 & 1.61 & 0.78 & 6.89 & 0.29 & - \\
- & -   & B & 85.9 & 18.9 & -   & -   & 0.51 & 0.08 & 9.81 & 0.16 & -   & -   & -   & -   & - \\
        \hline\hline
2006-02-09 & 4.8 & A & 186.3 & 21.3 & 1.16 & 0.10 & 0.31 & 0.01 & 11.05 & 0.06 & 3.58 & 0.51 & 6.84 & 0.14 & - \\
- & -   & B & 160.4 & 18.7 & -   & -   & 0.55 & 0.02 & 10.50 & 0.06 & -   & -   & -   & -   & - \\
        \hline\hline
2008-05-15 & 15.1 & A & 51.2 & 9.2 & 1.15 & 0.25 & 0.21 & 0.02 & 9.84 & 0.12 & 1.72 & 0.67 & 6.83 & 0.14 & - \\
- & -   & B & 44.6 & 8.4 & -   & -   & 0.26 & 0.03 & 9.60 & 0.13 & -   & -   & -   & -   & - \\
        \hline\hline
2012-02-20 & 8.3 & A & 117.9 & 14.7 & 1.20 & 0.13 & 0.43 & 0.02 & 10.11 & 0.07 & 1.69 & 0.34 & 6.79 & 0.13 & - \\
- & -   & B & 98.2 & 12.7 & -   & -   & 0.51 & 0.03 & 9.88 & 0.08 & -   & -   & -   & -   & - \\
        \hline\hline
2015-01-23 & 2.3 & A & 205.5 & 21.1 & 1.32 & 0.10 & -   & -   & 10.14 & 0.06 & 0.66 & 0.09 & -   & -   & -0.45 \\
- & -   & A0 & 180.8 & 20.5 & 1.16 & 0.09 & 1.01 & 0.04 & 10.67 & 0.02 & 2.23 & 0.30 & 6.75 & 0.28 & - \\
- & -   & B & 155.4 & 17.9 & -   & -   & 1.40 & 0.06 & 10.32 & 0.06 & -   & -   & -   & -   & -0.26 \\
        \hline
2015-01-23 & 8.7 & A & 113.3 & 12.6 & 1.04 & 0.07 & 0.37 & 0.01 & 10.19 & 0.06 & 1.52 & 0.19 & 6.84 & 0.07 & - \\
- & -   & B & 109.1 & 12.2 & -   & -   & 0.44 & 0.02 & 10.01 & 0.06 & -   & -   & -   & -   & -  \\
        \hline\hline
2017-02-24 & 2.3 & A & 240.0 & 23.3 & 1.71 & 0.15 & -   & -   & 10.01 & 0.06 & 0.65 & 0.11 & -   & -   & -0.51 \\
- & -   & A0 & 188.9 & 21.9 & 1.35 & 0.12 & 1.04 & 0.04 & 10.67 & 0.03 & 2.96 & 0.45 & 6.57 & 0.33 & - \\
- & -   & B & 140.0 & 16.9 & -   & -   & 1.54 & 0.07 & 10.19 & 0.07 & -   & -   & -   & -   & -0.15 \\
        \hline
2017-02-24 & 8.7 & A & 122.2 & 13.9 & 1.06 & 0.08 & 0.40 & 0.02 & 10.15 & 0.06 & 1.24 & 0.17 & 6.83 & 0.08 & - \\
- & -   & B & 115.2 & 13.2 & -   & -   & 0.43 & 0.02 & 10.06 & 0.06 & -   & -   & -   & -   & - \\
        \hline\hline
2018-05-07 & 2.2 & A & 192.3 & 20.7 & 1.10 & 0.08 & -   & -   & 10.16 & 0.09 & 0.36 & 0.07 & -   & -   & -0.25 \\
- & -   & A0 & 184.9 & 20.6 & 1.06 & 0.07 & 0.77 & 0.03 & 10.94 & 0.02 & 2.17 & 0.26 & 7.04 & 0.26 & - \\
- & -   & B & 175.1 & 19.6 & -   & -   & 1.10 & 0.04 & 10.60 & 0.06 & -   & -   & -   & -   & -0.26 \\
        \hline
2018-05-07 & 8.7 & A & 136.8 & 14.8 & 1.11 & 0.07 & 0.41 & 0.01 & 10.17 & 0.05 & 1.14 & 0.12 & 6.86 & 0.06 & - \\
- & -   & B & 123.2 & 13.4 & -   & -   & 0.42 & 0.01 & 10.12 & 0.05 & -   & -   & -   & -   & - \\
        \hline\hline
2020-02-26 & 4.9 & A & 172.6 & 18.3 & 1.12 & 0.11 & -   & -   & 10.35 & 0.07 & 1.79 & 0.32 & -   & -   & -1.08 \\
- & -   & A0 & 135.2 & 17.0 & 0.87 & 0.09 & 0.65 & 0.04 & 10.26 & 0.03 & 1.47 & 0.27 & 6.85 & 0.16 & - \\
- & -   & B & 154.5 & 16.5 & -   & -   & -   & -   & 10.10 & 0.06 & -   & -   & -   & -   & -1.14  \\
- & -   & B0 & 116.0 & 15.0 & 1.49 & 0.16 & 0.62 & 0.04 & 10.24 & 0.04 & 1.29 & 0.25 & -   & -   & - \\
        \hline
2020-03-04 & 22.2 & A & 33.6 & 4.1 & 1.21 & 0.13 & 0.33 & 0.02 & 8.95 & 0.07 & 1.83 & 0.35 & 6.77 & 0.08 & - \\
- & -   & B & 27.8 & 3.5 & -   & -   & 0.40 & 0.02 & 8.68 & 0.08 & -   & -   & -   & -   & - \\
        \hline
    \end{tabular}
    }
\end{table*}


\begin{figure*}[h]
    \centering
    \includegraphics[width=0.4\textwidth]{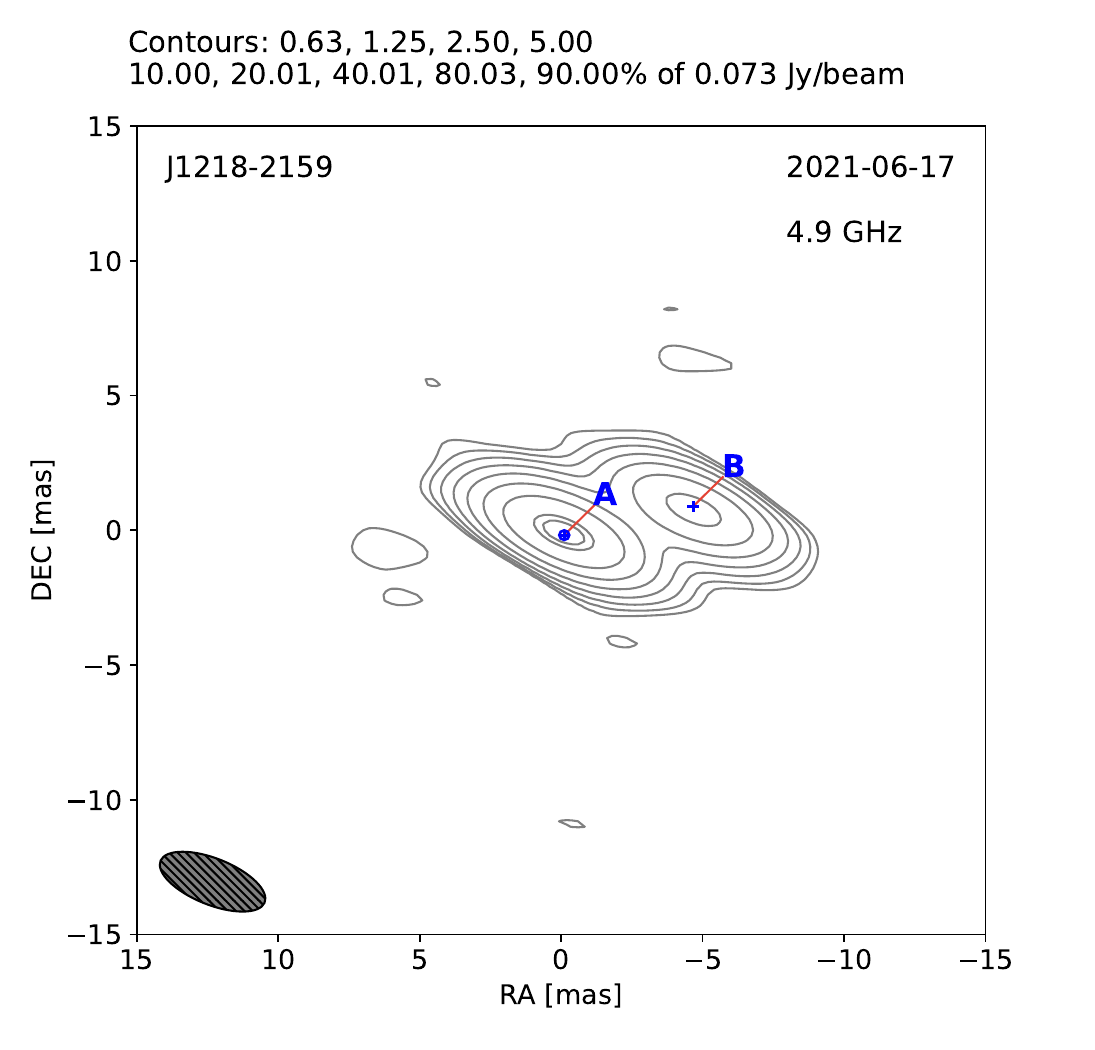}
    \caption{J1218-2159 EVN 4.9\,GHz data. The contours start at four times the image rms noise of 0.115\,mJy/beam and increase by factors of two. The restoring beam size is 1.71$\times$3.99\,mas at 66.8\,$^\circ$ PA. The source is not detected at 22.2\,GHz.}
    \label{fig:J1218}
\end{figure*}

\begin{table*}[h]
    \caption{Derived quantities for J1218-2159. See Table~\ref{tab:J2209} for a detailed description of the columns.}
    \vspace*{3mm}
    \adjustbox{width=1\textwidth}{%
    \label{tab:J1218}
    \centering
    \begin{tabular}{|| c c c c c | c c | c c | c c | c c | c c | c ||} 
        \hline
        Epochs & Frequency & Component & Flux  & Error  & Flux  & Error & FWHM  & Error & $\log(T_\mathrm{b,obs}/\mathrm{K})$ & Error  & SB    & Error & Distance & Error & Spectral Index \\ [0.5ex] 
               & [GHz]     &           & [mJy] & [mJy]  & Ratio &       & [mas] & [mas] &                                     &        & Ratio &       & [mas]    & [mas] &                \\
        \hline\hline
2005-07-20 & 2.3 & A & 149.0 & 23.9 & - & - & 5.87 & 0.56 & 9.05 & 0.11 & - & - & - & - & - \\
- & -   & B & - & - & -   & -   & - & - & - & - & -   & -   & -   & -   & - \\
        \hline\hline
2014-12-20 & 2.3 & A & 100.1 & 13.0 & 2.21 & 0.33 & 1.75 & 0.11 & 9.93 & 0.08 & 1.58 & 0.42 & 4.33 & 0.54 & -0.40 \\
- & -   & B & 45.3 & 7.2 & -   & -   & 1.48 & 0.13 & 9.74 & 0.10 & -   & -   & -   & -   & -0.33 \\
        \hline
2014-12-20 & 8.7 & A & 58.5 & 7.0 & 2.00 & 0.23 & 0.41 & 0.02 & 9.81 & 0.07 & 1.99 & 0.41 & 4.71 & 0.12 & - \\
- & -   & B & 29.3 & 4.0 & -   & -   & 0.41 & 0.03 & 9.51 & 0.09 & -   & -   & -   & -   & - \\
        \hline\hline
2017-03-27 & 2.2 & A & 114.7 & 24.5 & 1.98 & 0.65 & 1.92 & 0.27 & 9.93 & 0.15 & 1.24 & 0.72 & 4.44 & 1.20 & -0.47 \\
- & -   & B & 58.0 & 16.5 & -   & -   & 1.52 & 0.29 & 9.84 & 0.21 & -   & -   & -   & -   & -0.59 \\
        \hline
2017-03-27 & 8.7 & A & 60.9 & 7.3 & 2.33 & 0.28 & 0.44 & 0.02 & 9.76 & 0.07 & 0.62 & 0.13 & 4.75 & 0.11 & - \\
- & -   & B & 26.1 & 3.7 & -   & -   & 0.23 & 0.02 & 9.97 & 0.09 & -   & -   & -   & -   & - \\
        \hline\hline
2018-12-04 & 2.2 & A & 121.8 & 14.8 & 2.62 & 0.34 & 1.70 & 0.09 & 10.06 & 0.07 & 1.06 & 0.24 & 4.45 & 0.48 & -0.52\\
- & -   & B & 46.6 & 6.9 & -   & -   & 1.08 & 0.08 & 10.04 & 0.09 & -   & -   & -   & -   & -0.37 \\
        \hline
2018-12-04 & 8.7 & A & 60.8 & 7.2 & 2.15 & 0.24 & 0.47 & 0.02 & 9.71 & 0.07 & 1.89 & 0.39 & 4.67 & 0.11 & - \\
- & -   & B & 28.2 & 3.9 & -   & -   & 0.44 & 0.03 & 9.44 & 0.08 & -   & -   & -   & -   & - \\
        \hline\hline
2021-06-17 & 4.9 & A & 77.2 & 8.8 & 2.11 & 0.20 & 0.35 & 0.01 & 10.55 & 0.06 & $<$0.32 & - & 4.68 & 0.18 & - \\
- & -   & B & 36.6 & 4.7 & -   & -   & $<$0.14 & 0.01 & $>$11.04 & - & -   & -   & -   & -   & - \\
        \hline
        \end{tabular}
    }
\end{table*}


\begin{figure*}[h]
    \centering
    \includegraphics[width=0.4\textwidth]{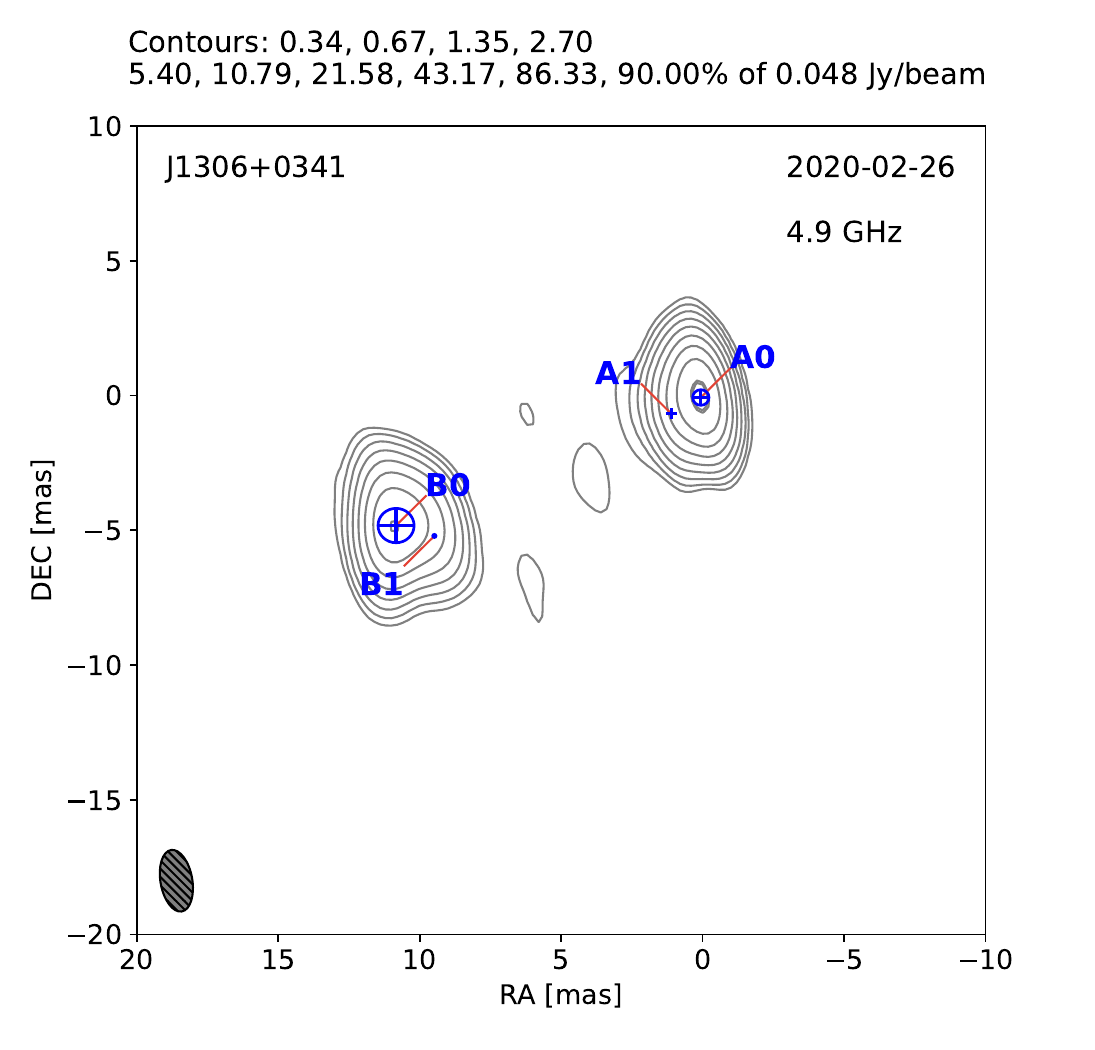}
    \caption{J1306+0341 EVN 4.9\,GHz data. The contours start at four times the image rms noise of 0.040\,mJy/beam and increase by factors of two. The restoring beam size is 1.13$\times$2.31\,mas at 9.3\,$^\circ$ PA. The source is not detected at 22.2\,GHz.}
    \label{fig:J1306}
\end{figure*}

\begin{table*}[h]
    \caption{Derived quantities for J1306+0341. See Table~\ref{tab:J2209} for a detailed description of the columns.}
    \vspace*{3mm}
    \adjustbox{width=1\textwidth}{%
    \label{tab:J1306}
    \centering
    \begin{tabular}{|| c c c c c | c c | c c | c c | c c | c c | c ||} 
        \hline
        Epochs & Frequency & Component & Flux  & Error  & Flux  & Error & FWHM  & Error & $\log(T_\mathrm{b,obs}/\mathrm{K})$ & Error  & SB    & Error & Distance & Error & Spectral Index \\ [0.5ex] 
               & [GHz]     &           & [mJy] & [mJy]  & Ratio &       & [mas] & [mas] &                                     &        & Ratio &       & [mas]    & [mas] &                \\
        \hline\hline
2014-01-27 & 4.3 & A & 97.6 & 10.9 & 1.62 & 0.13 & 0.90 & 0.03 & 9.95 & 0.06 & 4.76 & 0.73 & 11.62 & 0.18 & -1.27 \\
- & -   & B & 60.3 & 7.2 & -   & -   & 1.54 & 0.08 & 9.27 & 0.07 & -   & -   & -   & -   & -1.26 \\
        \hline
2014-01-27 & 7.6 & A & 47.8 & 5.8 & 1.61 & 0.19 & 0.65 & 0.03 & 9.43 & 0.07 & 8.38 & 1.99 & 11.79 & 0.20 & - \\
- & -   & B & 29.7 & 4.2 & -   & -   & 1.49 & 0.13 & 8.51 & 0.10 & -   & -   & -   & -   & - \\
        \hline\hline
2020-02-26 & 4.9 & A & 65.6 & 6.2 & 1.55 & 0.09 & -   & -   & 10.04 & 0.05 & 7.42 & 0.88 & -   & -   & - \\
- & -   & A0 & 56.8 & 6.1 & 1.34 & 0.08 & 0.58 & 0.02 & 9.98 & 0.02 & 6.57 & 0.79 & 11.78 & 0.10 & -  \\
- & -   & B & 42.3 & 4.2 & -   & -   & -   & -   & 9.17 & 0.06 & -   & -   & -   & -   & - \\
- & -   & B0 & 36.4 & 4.1 & 1.80 & 0.11 & 1.28 & 0.06 & 9.11 & 0.02 & 8.55 & 1.04 & -   & -   & - \\
        \hline
    \end{tabular}
    }
\end{table*}


\begin{figure*}[h]
    \centering
    \includegraphics[width=0.4\textwidth]{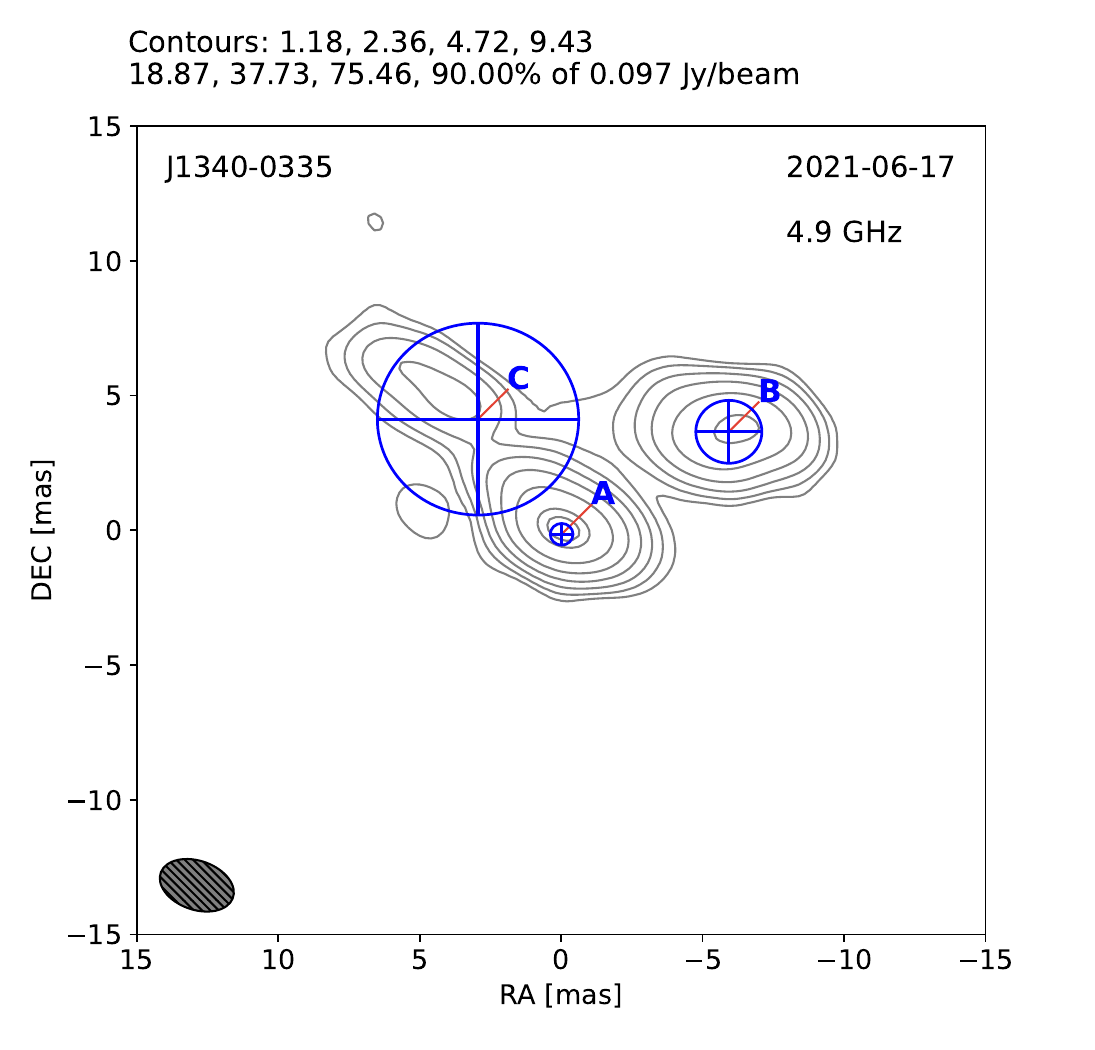}
    \caption{J1340-0335 EVN 4.9\,GHz data. The contours start at four times the image rms noise of 0.286\,mJy/beam and increase by factors of two. The restoring beam size is 1.81$\times$2.72\,mas at 68.8\,$^\circ$ PA. The source is not detected at 22.2\,GHz.}
    \label{fig:J1340}
\end{figure*}

\begin{table*}[h]
    \caption{Derived quantities for J1340-0335. See Table~\ref{tab:J2209} for a detailed description of the columns.}
    \vspace*{3mm}
    \adjustbox{width=1\textwidth}{%
    \label{tab:J1340}
    \centering
    \begin{tabular}{|| c c c c c | c c | c c | c c | c c | c c | c ||} 
        \hline
        Epochs & Frequency & Component & Flux  & Error  & Flux  & Error & FWHM  & Error & $\log(T_\mathrm{b,obs}/\mathrm{K})$ & Error  & SB    & Error & Distance & Error & Spectral Index \\ [0.5ex] 
               & [GHz]     &           & [mJy] & [mJy]  & Ratio &       & [mas] & [mas] &                                     &        & Ratio &       & [mas]    & [mas] &                \\
        \hline\hline
2014-12-20 & 2.3 & A & 316.7 & 43.5 & 1.75 & 0.28 & 2.30 & 0.17 & 10.20 & 0.09 & 2.99 & 0.89 & 6.54 & 0.72 & -1.31 \\
- & -   & B & 180.8 & 29.3 & -   & -   & 3.00 & 0.31 & 9.72 & 0.11 & -   & -   & -   & -   & -1.29 \\
        \hline
2014-12-20 & 8.7 & A & 55.4 & 6.8 & 1.69 & 0.29 & -   & -   & 9.09 & 0.13 & 9.14 & 3.94 & -   & -   & - \\
- & -   & A0 & 49.2 & 6.5 & 1.50 & 0.26 & 0.55 & 0.04 & 9.48 & 0.04 & 22.56 & 8.28 & 7.03 & 0.38 & - \\
- & -   & B & 32.7 & 5.9 & -   & -   & 2.12 & 0.31 & 8.13 & 0.15 & -   & -   & -   & -   & - \\
        \hline\hline
2017-01-16 & 2.2 & A & 327.3 & 35.7 & 1.66 & 0.12 & 2.03 & 0.07 & 10.34 & 0.06 & 4.33 & 0.60 & 6.37 & 0.34 & -1.35 \\
- & -   & B & 197.2 & 22.9 & -   & -   & 3.28 & 0.16 & 9.70 & 0.07 & -   & -   & -   & -   & -1.28 \\
        \hline
2017-01-16 & 8.7 & A & 53.1 & 6.0 & 1.50 & 0.25 & -   & -   & 8.74 & 0.14 & 5.76 & 2.58 & -   & -   & - \\
- & -   & A0 & 44.2 & 5.6 & 1.25 & 0.21 & 0.52 & 0.03 & 9.47 & 0.03 & 31.42 & 11.09 & 7.08 & 0.43 & - \\
- & -   & B & 35.3 & 6.3 & -   & -   & 2.63 & 0.38 & 7.98 & 0.15 & -   & -   & -   & -   & - \\
        \hline\hline
2018-01-25 & 2.2 & A & 354.9 & 45.3 & 1.72 & 0.23 & 1.70 & 0.10 & 10.53 & 0.07 & 3.68 & 0.86 & 6.42 & 0.76 & -1.31 \\
- & -   & B & 206.0 & 29.8 & -   & -   & 2.49 & 0.20 & 9.96 & 0.09 & -   & -   & -   & -   & -1.24 \\
        \hline
2018-01-25 & 8.7 & A & 60.5 & 6.9 & 1.56 & 0.21 & -   & -   & 9.20 & 0.08 & 12.10 & 3.48 & -   & -   & - \\
- & -   & A0 & 52.8 & 6.7 & 1.36 & 0.19 & 0.62 & 0.04 & 9.41 & 0.03 & 19.47 & 5.38 & 7.01 & 0.34 & - \\
- & -   & B & 38.8 & 5.8 & -   & -   & 2.33 & 0.25 & 8.12 & 0.11 & -   & -   & -   & -   & - \\
        \hline\hline
2021-06-17 & 4.9 & A & 116.1 & 14.2 & 1.44 & 0.17 & 0.81 & 0.04 & 10.01 & 0.07 & 12.09 & 2.92 & 7.03 & 0.32 & - \\
- & -   & B & 80.7 & 11.3 & -   & -   & 2.34 & 0.21 & 8.93 & 0.10 & -   & -   & -   & -   & - \\
        \hline
    \end{tabular}
    }
\end{table*}


\begin{figure*}[h]
    \centering
    \includegraphics[width=0.4\textwidth]{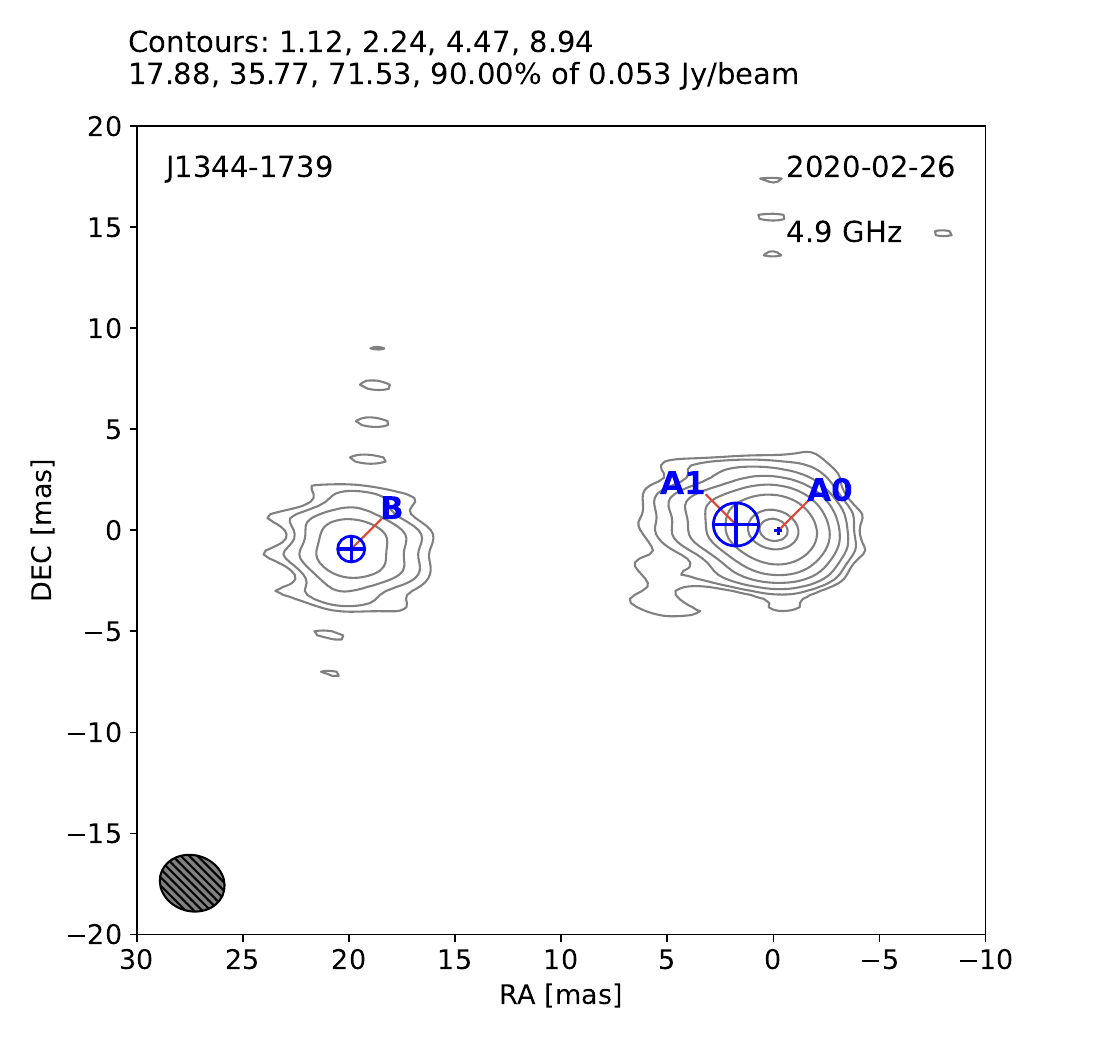}
    \includegraphics[width=0.4\textwidth]{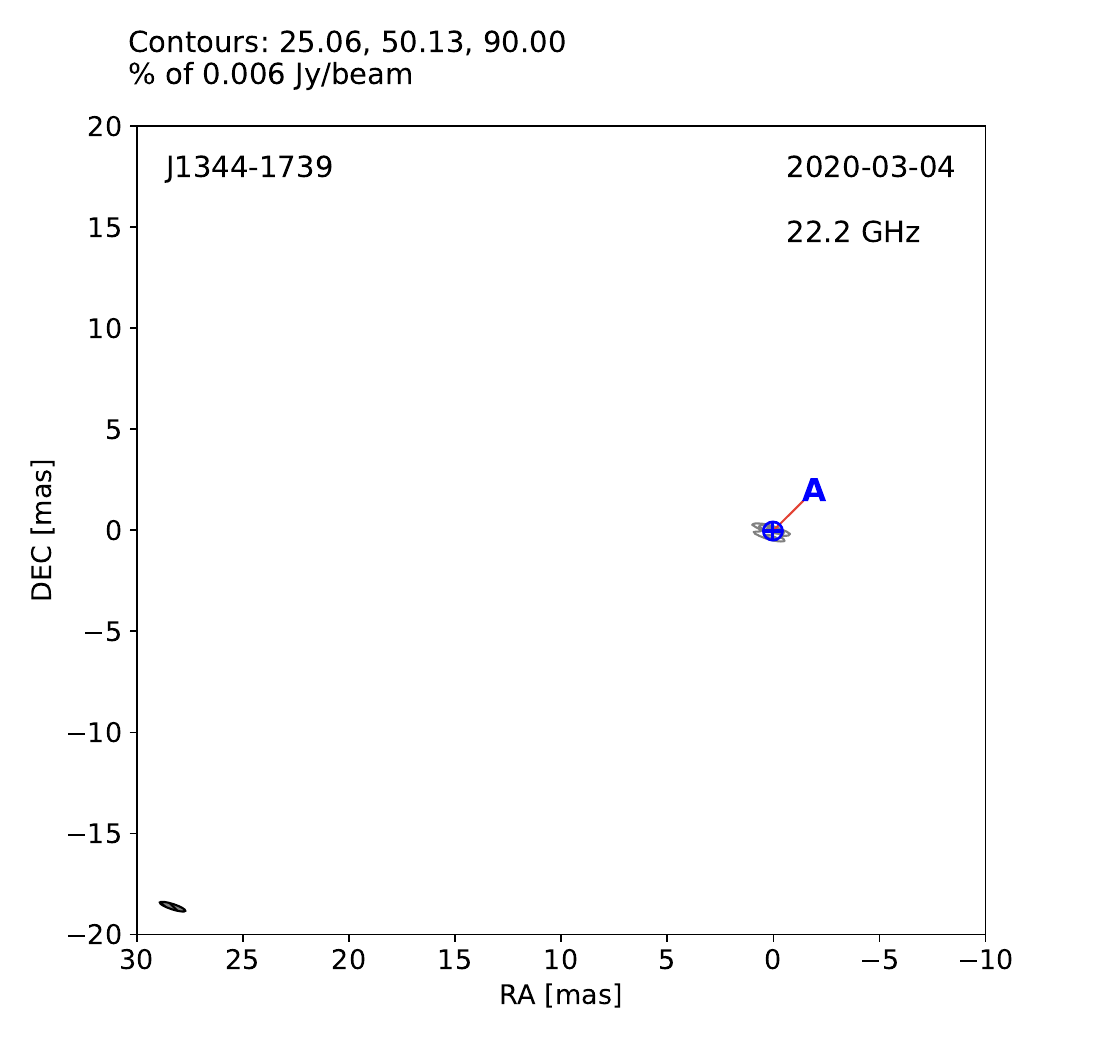}
    \caption{J1344-1739. \textit{Left:} EVN 4.9\,GHz data. The contours start at four times the image rms noise of 0.148\,mJy/beam and increase by factors of two. The restoring beam size is 2.75$\times$3.11\,mas at 66.7\,$^\circ$ PA. \textit{Right:} EVN 22.2\,GHz data. The contours start at four times the image rms noise of 0.392\,mJy/beam and increase by factors of two. The restoring beam size is 0.27$\times$1.27\,mas at 70.9\,$^\circ$ PA.}
    \label{fig:J1344}
\end{figure*}

\begin{table*}[h]
    \caption{Derived quantities for J1344-1739. See Table~\ref{tab:J2209} for a detailed description of the columns.}
    \vspace*{3mm}
    \adjustbox{width=1\textwidth}{%
    \label{tab:J1344}
    \centering
    \begin{tabular}{|| c c c c c | c c | c c | c c | c c | c c | c ||} 
        \hline
        Epochs & Frequency & Component & Flux  & Error  & Flux  & Error & FWHM  & Error & $\log(T_\mathrm{b,obs}/\mathrm{K})$ & Error  & SB    & Error & Distance & Error & Spectral Index \\ [0.5ex] 
               & [GHz]     &           & [mJy] & [mJy]  & Ratio &       & [mas] & [mas] &                                     &        & Ratio &       & [mas]    & [mas] &                \\
        \hline\hline
2014-01-27 & 4.3 & A & 101.8 & 11.7 & 5.66 & 1.09 & -   & -   & 8.83 & 0.18 & 2.73 & 1.48 & -   & -   & -1.10 \\
- & -   & A0 & 89.9 & 11.2 & 5.00 & 0.96 & 1.60 & 0.10 & 9.41 & 0.03 & 10.32 & 4.06 & 20.01 & 0.56 & - \\
- & -   & B & 18.0 & 3.7 & -   & -   & 2.31 & 0.37 & 8.40 & 0.16 & -   & -   & -   & -   & -1.63 \\
        \hline
2014-01-27 & 7.6 & A & 54.7 & 6.5 & 7.61 & 2.28 & -   & -   & 8.58 & 0.09 & 4.64 & 2.95 & -   & -   & - \\
- & -   & A1 & 39.7 & 5.8 & 5.53 & 1.68 & 1.80 & 0.17 & 8.46 & 0.05 & 3.56 & 2.28 & 20.12 & 0.56 & - \\
- & -   & B & 7.2 & 2.2 & -   & -   & 1.45 & 0.38 & 7.91 & 0.27 & -   & -   & -   & -   & - \\
        \hline\hline
2020-02-27 & 4.9 & A & 62.8 & 6.6 & 5.76 & 1.04 & -   & -   & 8.89 & 0.08 & $>$2.05 & - & -   & -   & -1.26 \\
- & -   & A0 & 46.2 & 5.9 & 4.23 & 0.79 & $<$0.15 & - & $>$11.09 & - & 321.64 & 107.26 & 20.14 & 0.44 & - \\
- & -   & B & 10.9 & 2.1 & -   & -   & 1.28 & 0.16 & 8.58 & 0.14 & -   & -   & -   & -   & - \\
        \hline
2020-03-05 & 22.2 & A & 9.3 & 3.5 & - & - & 0.90 & 0.30 & 7.51 & 0.33 & - & - & - & - & - \\
- & -   & B & - & - & -   & -   & - & - & - & - & -   & -   & -   & -   & - \\
        \hline
    \end{tabular}
    }
\end{table*}


\begin{table*}[h]
    \caption{Derived quantities for J1632+3547. See Table~\ref{tab:J2209} for a detailed description of the columns.}
    \vspace*{3mm}
    \adjustbox{width=1\textwidth}{%
    \label{tab:J1632}
    \centering
    \begin{tabular}{|| c c c c c | c c | c c | c c | c c | c c | c ||} 
        \hline
        Epochs & Frequency & Component & Flux  & Error  & Flux  & Error & FWHM  & Error & $\log(T_\mathrm{b,obs}/\mathrm{K})$ & Error  & SB    & Error & Distance & Error & Spectral Index \\ [0.5ex] 
               & [GHz]     &           & [mJy] & [mJy]  & Ratio &       & [mas] & [mas] &                                     &        & Ratio &       & [mas]    & [mas] &                \\
        \hline\hline
2006-01-03 & 4.8 & A & 204.2 & 19.1 & 6.26 & 1.15 & -   & -   & 9.20 & 0.11 & 0.29 & 0.11 & -   & -   & - \\
- & -   & A0 & 107.4 & 14.7 & 3.29 & 0.64 & 0.47 & 0.03 & 10.45 & 0.04 & 5.25 & 1.78 & 10.81 & 0.32 & - \\
- & -   & B & 32.6 & 6.4 & -   & -   & 0.60 & 0.07 & 9.73 & 0.14 & -   & -   & -   & -   & - \\
        \hline
        \end{tabular}
    }
\end{table*}


\begin{table*}[h]
    \caption{Derived quantities for J1653+3503. See Table~\ref{tab:J2209} for a detailed description of the columns.}
    \vspace*{3mm}
    \adjustbox{width=1\textwidth}{%
    \label{tab:J1653}
    \centering
    \begin{tabular}{|| c c c c c | c c | c c | c c | c c | c c | c ||} 
        \hline
        Epochs & Frequency & Component & Flux  & Error  & Flux  & Error & FWHM  & Error & $\log(T_\mathrm{b,obs}/\mathrm{K})$ & Error  & SB    & Error & Distance & Error & Spectral Index \\ [0.5ex] 
               & [GHz]     &           & [mJy] & [mJy]  & Ratio &       & [mas] & [mas] &                                     &        & Ratio &       & [mas]    & [mas] &                \\
        \hline\hline
2006-08-07 & 4.8 & A & 72.9 & 9.8 & 2.57 & 0.44 & 0.91 & 0.06 & 9.71 & 0.08 & 2.85 & 0.89 & 10.08 & 0.32 & - \\
- & -   & B & 28.3 & 5.0 & -   & -   & 0.96 & 0.11 & 9.26 & 0.12 & -   & -   & -   & -   & - \\
        \hline
        \end{tabular}
    }
\end{table*}


\begin{figure*}[h]
    \centering
    \includegraphics[width=0.4\textwidth]{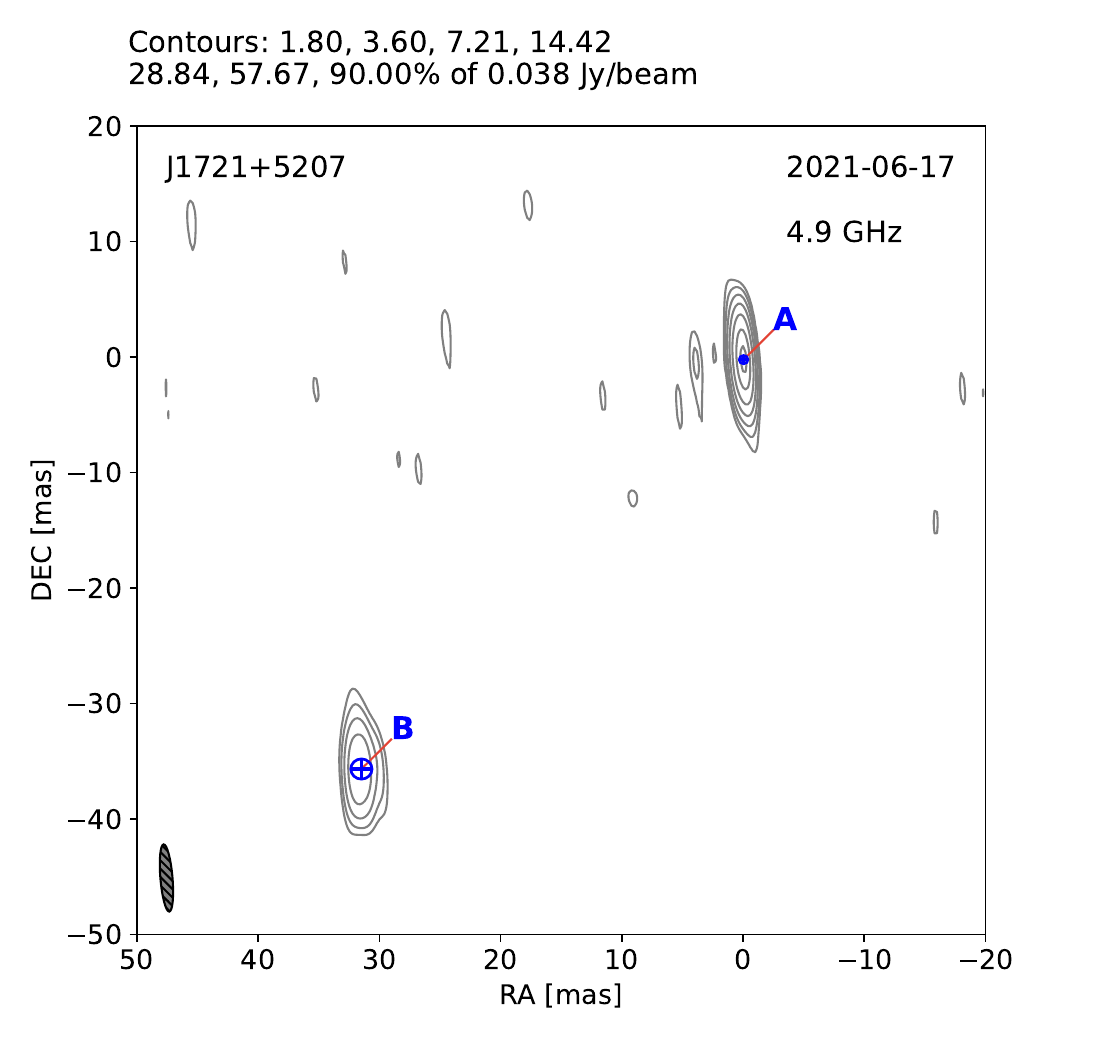}
    \caption{J1721+5207 EVN 4.9\,GHz data. The contours start at four times the image rms noise of 0.171\,mJy/beam and increase by factors of two. The restoring beam size is 1.00$\times$5.84\,mas at 4.6\,$^\circ$ PA. The source is not detected at 22.2\,GHz.}
    \label{fig:J1721}
\end{figure*}

\begin{table*}[h]
    \caption{Derived quantities for J1721+5207. See Table~\ref{tab:J2209} for a detailed description of the columns.}
    \vspace*{3mm}
    \adjustbox{width=1\textwidth}{%
    \label{tab:J1721}
    \centering
    \begin{tabular}{|| c c c c c | c c | c c | c c | c c | c c | c ||} 
        \hline
        Epochs & Frequency & Component & Flux  & Error  & Flux  & Error & FWHM  & Error & $\log(T_\mathrm{b,obs}/\mathrm{K})$ & Error  & SB    & Error & Distance & Error & Spectral Index \\ [0.5ex] 
               & [GHz]     &           & [mJy] & [mJy]  & Ratio &       & [mas] & [mas] &                                     &        & Ratio &       & [mas]    & [mas] &                \\
        \hline\hline
2014-02-18 & 4.3 & A & 59.3 & 7.5 & 2.19 & 0.31 & 1.30 & 0.08 & 9.41 & 0.08 & 2.66 & 0.70 & 47.69 & 0.37 & -1.90 \\
- & -   & B & 27.0 & 4.2 & -   & -   & 1.43 & 0.13 & 8.99 & 0.10 & -   & -   & -   & -   & -2.54 \\
        \hline
2014-02-18 & 7.6 & A & 20.4 & 3.6 & 3.15 & 0.95 & 1.12 & 0.12 & 8.59 & 0.12 & 3.01 & 1.62 & 47.53 & 0.91 & - \\
- & -   & B & 6.5 & 1.8 & -   & -   & 1.09 & 0.21 & 8.11 & 0.21 & -   & -   & -   & -   & - \\
        \hline\hline
2021-06-17 & 4.9 & A & 46.9 & 6.2 & 2.09 & 0.35 & 0.72 & 0.05 & 9.72 & 0.08 & 12.47 & 4.22 & 47.43 & 0.42 & - \\
- & -   & B & 22.4 & 3.9 & -   & -   & 1.75 & 0.23 & 8.62 & 0.14 & -   & -   & -   & -   & - \\
        \hline
        \end{tabular}
    }
\end{table*}


\begin{figure*}[h]
    \centering
    \includegraphics[width=0.4\textwidth]{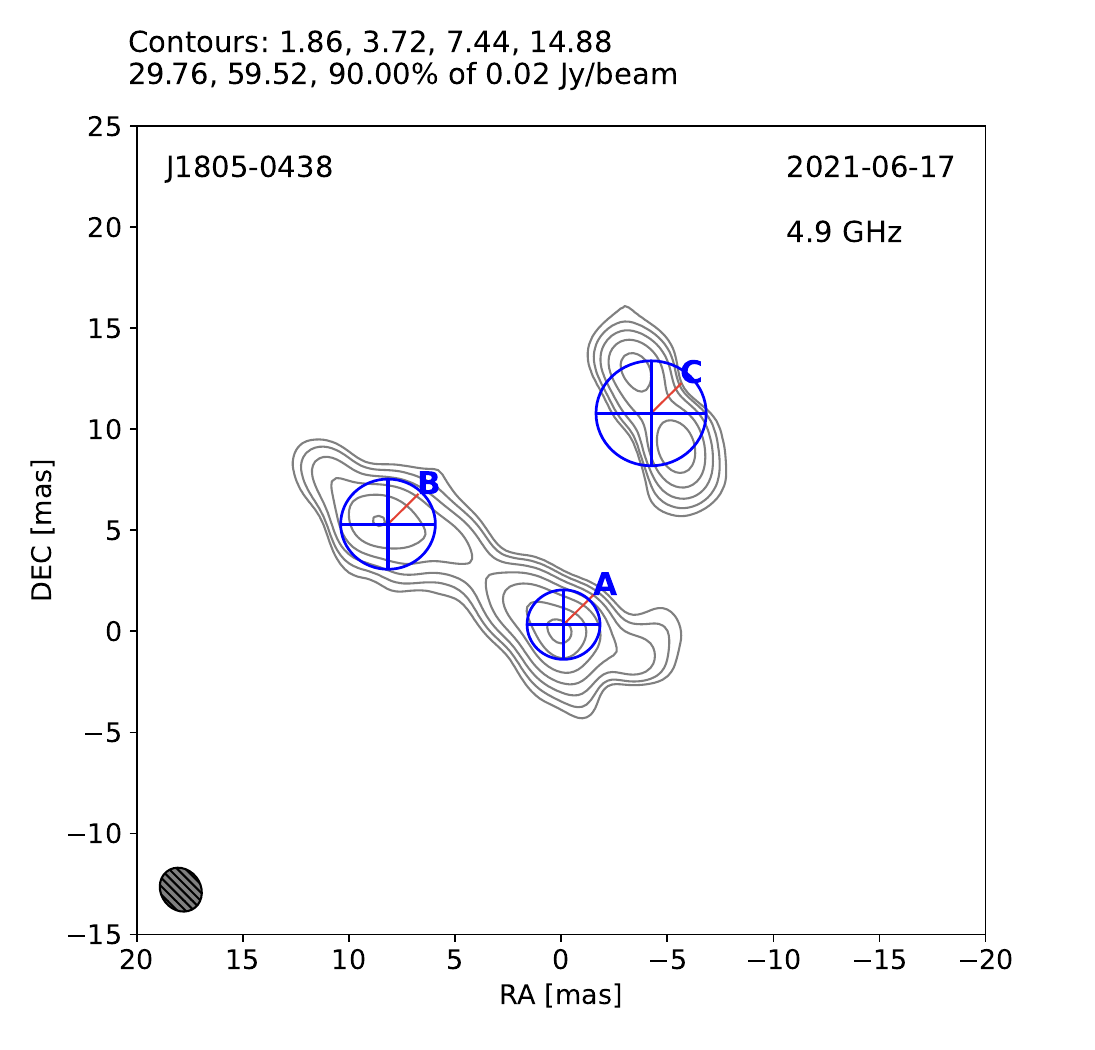}
    \includegraphics[width=0.4\textwidth]{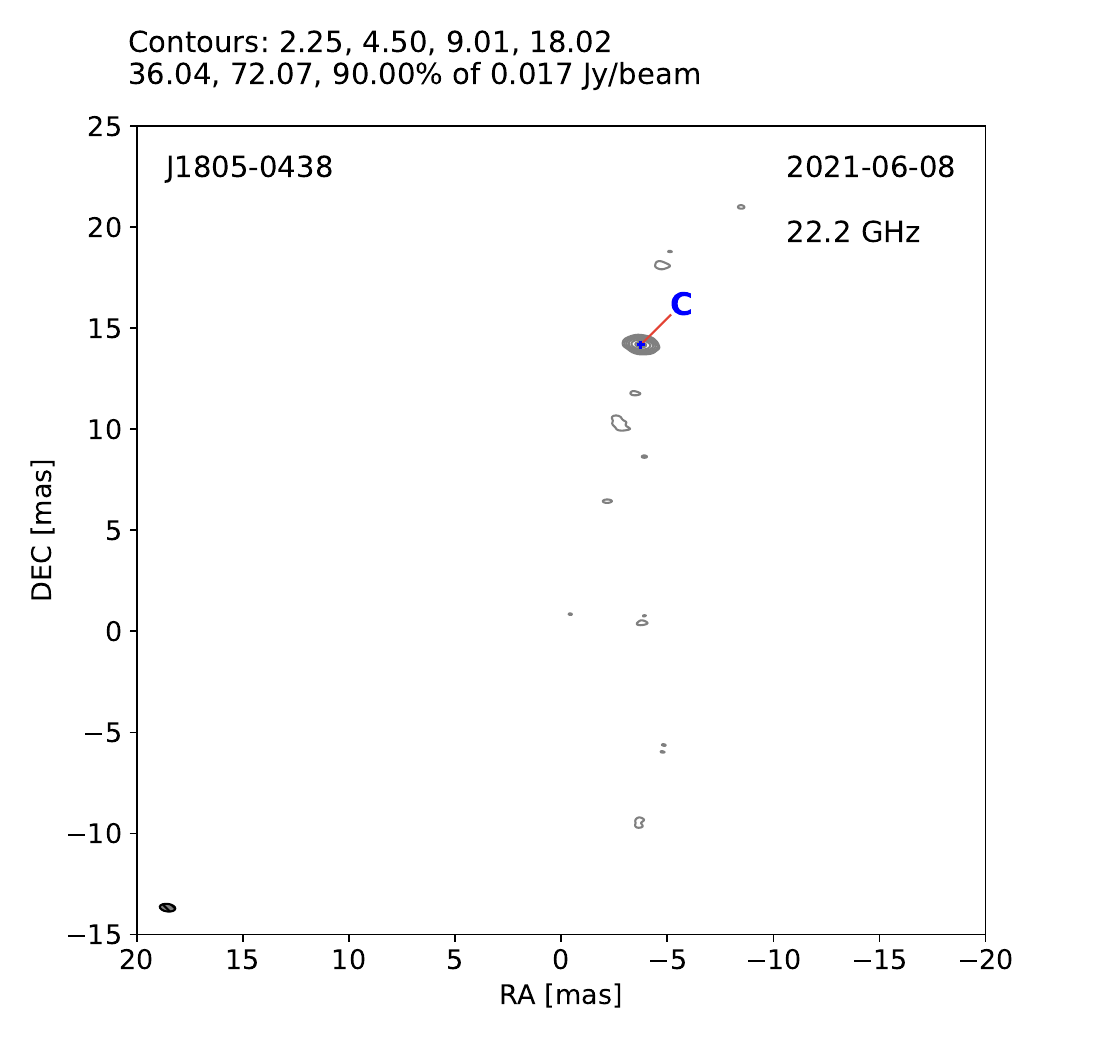}
    \caption{J1805-0438. \textit{Left:} EVN 4.9\,GHz data. The contours start at four times the image rms noise of 0.094\,mJy/beam and increase by factors of two. The restoring beam size is 1.90$\times$2.24\,mas at 27.8\,$^\circ$ PA. \textit{Right:} EVN 22.2\,GHz data. The contours start at four times the image rms noise of 0.058\,mJy/beam and increase by factors of two. The restoring beam size is 0.38$\times$0.74\,mas at 82.1\,$^\circ$ PA. Only component C is reliably detected in the 22\,GHz image, so we cannot compare the quantities with the 5\,GHz image.}
    \label{fig:J1805}
\end{figure*}

\begin{table*}[h]
    \caption{Derived quantities for J1805-0438. See Table~\ref{tab:J2209} for a detailed description of the columns.}
    \vspace*{3mm}
    \adjustbox{width=1\textwidth}{%
    \label{tab:J1805}
    \centering
    \begin{tabular}{|| c c c c c | c c | c c | c c | c c | c c | c ||}
        \hline
        Epochs & Frequency & Component & Flux  & Error  & Flux  & Error & FWHM  & Error & $\log(T_\mathrm{b,obs}/\mathrm{K})$ & Error  & SB    & Error & Distance & Error & Spectral Index \\ [0.5ex] 
               & [GHz]     &           & [mJy] & [mJy]  & Ratio &       & [mas] & [mas] &                                     &        & Ratio &       & [mas]    & [mas] &                \\
        \hline\hline
2004-04-30 & 2.3 & A & 173.4 & 27.5 & 6.63 & 2.30 & 1.70 & 0.13 & 10.19 & 0.10 & 52.71 & 33.10 & 17.43 & 3.95 & - \\
- & -   & B & 26.1 & 8.9 & -   & -   & 4.79 & 1.20 & 8.47 & 0.26 & -   & -   & -   & -   & - \\
        \hline\hline
2012-02-08 & 2.3 & A & 137.2 & 17.4 & 2.29 & 0.33 & 3.24 & 0.21 & 9.53 & 0.08 & 2.68 & 0.74 & 9.95 & 0.74 & -0.69 \\
- & -   & B & 60.0 & 9.3 & -   & -   & 3.51 & 0.35 & 9.10 & 0.11 & -   & -   & -   & -   & -1.66 \\
        \hline
2012-02-08 & 8.6 & A & 55.2 & 7.3 & 8.16 & 2.17 & 0.26 & 0.02 & 10.18 & 0.08 & $<$17.75 & - & 5.73 & 0.44 & - \\
- & -   & B & 6.8 & 1.8 & -   & -   & $<$0.39 & - & $>$8.93 & - & -   & -   & -   & -   & - \\
        \hline\hline
2017-07-09 & 2.2 & A & 151.0 & 17.1 & 1.80 & 0.16 & 3.44 & 0.15 & 9.54 & 0.06 & 2.21 & 0.39 & 9.50 & 0.48 & -0.96 \\
- & -   & B & 83.7 & 10.3 & -   & -   & 3.81 & 0.23 & 9.20 & 0.08 & -   & -   & -   & -   & -2.33 \\
        \hline
2017-07-09 & 8.7 & A & 41.3 & 4.5 & 11.42 & 2.94 & -   & -   & 8.90 & 0.12 & $<$0.63 & - & -   & -   & - \\
- & -   & A0 & 31.1 & 4.0 & 8.59 & 2.24 & 0.17 & 0.01 & 10.29 & 0.04 & 15.47 & 6.96 & 8.16 & 0.26 & - \\
- & -   & B & 3.6 & 1.0 & -   & -   & $<$0.23 & - & $>$9.10 & - & -   & -   & -   & -   & - \\
        \hline\hline
2018-09-02 & 2.3 & A & 137.3 & 15.6 & 1.90 & 0.18 & 3.40 & 0.15 & 9.50 & 0.06 & 2.08 & 0.37 & 9.64 & 0.49 & -0.90 \\
- & -   & B & 72.3 & 9.1 & -   & -   & 3.56 & 0.22 & 9.18 & 0.08 & -   & -   & -   & -   & - \\
        \hline
2018-09-02 & 8.7 & A & 41.1 & 5.4 & - & - & 0.51 & 0.03 & 9.46 & 0.08 & - & - & - & - & - \\
- & -   & B & - & - & -   & -   & - & - & - & - & -   & -   & -   & -   & - \\
        \hline\hline
2021-06-17 & 4.9 & A & 47.7 & 6.3 & 1.42 & 0.22 & 3.44 & 0.28 & 8.36 & 0.09 & 2.39 & 0.80 & 9.65 & 0.69 & -1.87 \\
- & -   & B & 33.6 & 5.4 & -   & -   & 4.46 & 0.55 & 7.98 & 0.13 & -   & -   & -   & -   & - \\
        \hline
    \end{tabular}
    }
\end{table*}


\begin{figure*}[h]
    \centering
    \includegraphics[width=0.33\textwidth]{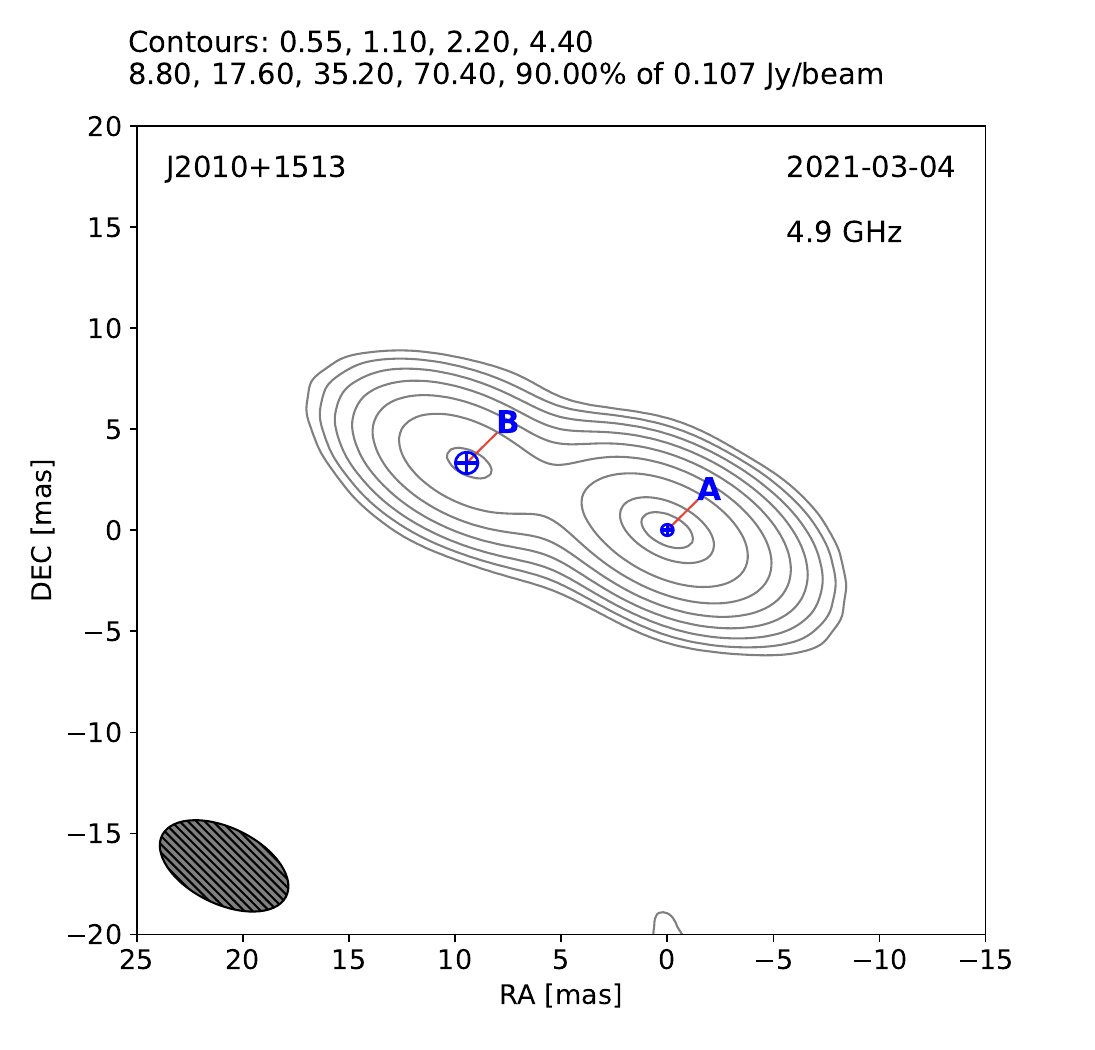}
    \includegraphics[width=0.33\textwidth]{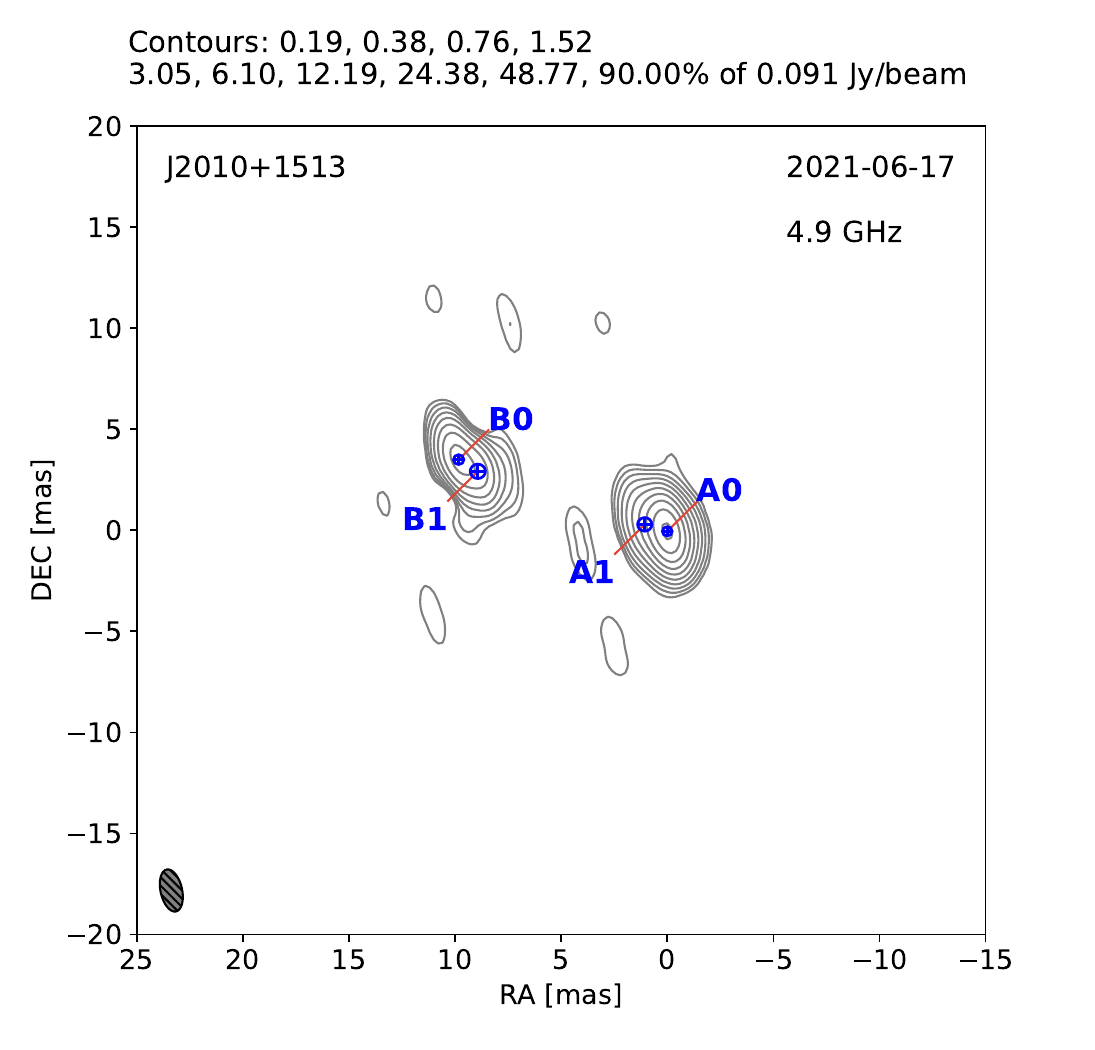}
    \includegraphics[width=0.33\textwidth]{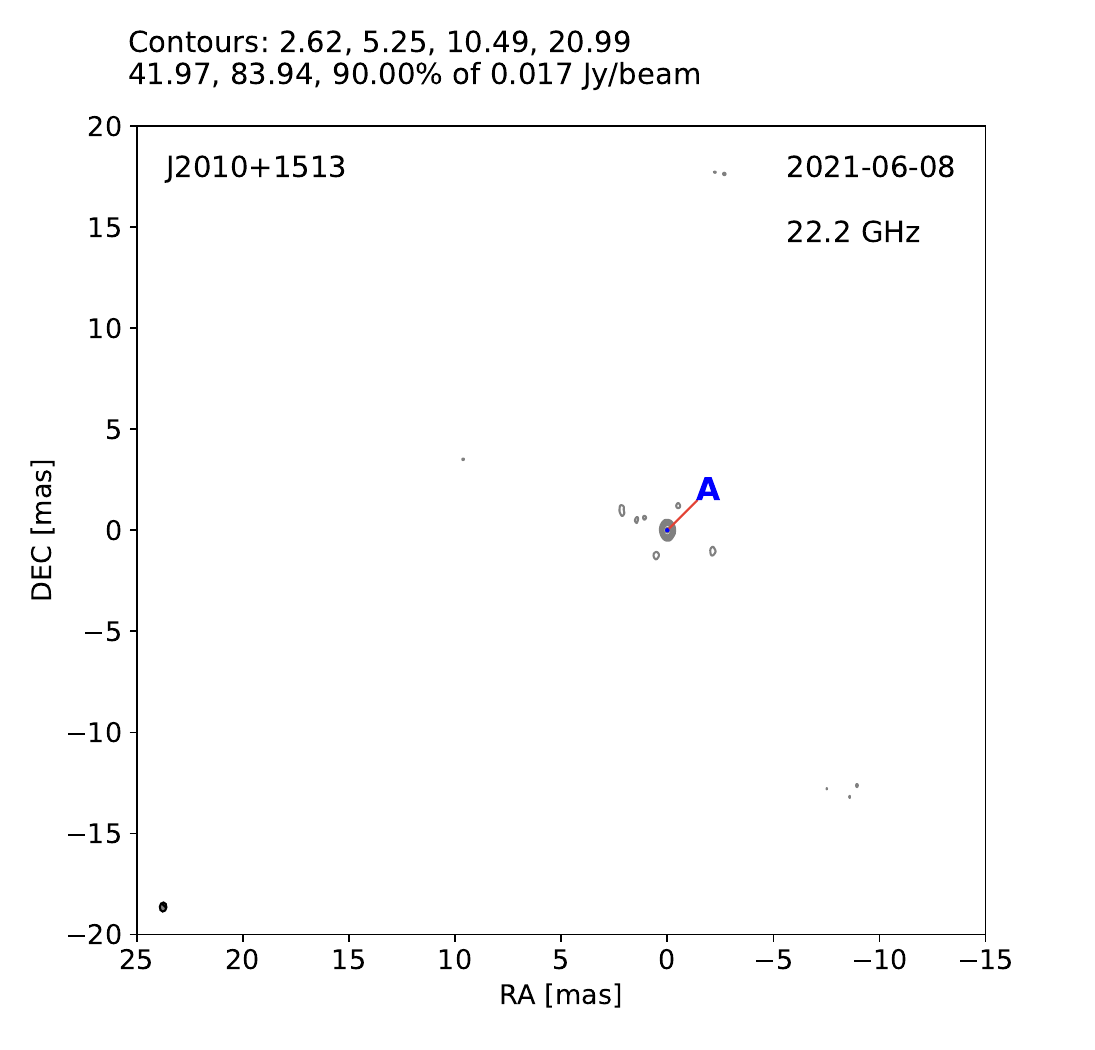}
    \caption{J2010+1513. \textit{Left:} EVN 4.9\,GHz data (session 4). The contours start at four times the image rms noise of 0.181\,mJy/beam and increase by factors of two. The restoring beam size is 3.73$\times$6.58\,mas at 61.8\,$^\circ$ PA. \textit{Centre:} EVN 4.9\,GHz data (session 5). The contours start at four times the image rms noise of 0.043\,mJy/beam and increase by factors of two. The restoring beam size is 1.02$\times$2.12\,mas at 11.7\,$^\circ$ PA. \textit{Right:} EVN 22.2\,GHz data (session 5). The contours start at four times the image rms noise of 0.114\,mJy/beam and increase by factors of two. The restoring beam size is 0.31$\times$0.45\,mas at $-4.8$\,$^\circ$ PA. Only component A is detected.}
    \label{fig:J2010}
\end{figure*}

\begin{table*}[h]
    \caption{Derived quantities for J2010+1513. See Table~\ref{tab:J2209} for a detailed description of the columns.}
    \vspace*{3mm}
    \adjustbox{width=1\textwidth}{%
    \label{tab:J2010}
    \centering
    \begin{tabular}{|| c c c c c | c c | c c | c c | c c | c c | c ||}
        \hline
        Epochs & Frequency & Component & Flux  & Error  & Flux  & Error & FWHM  & Error & $\log(T_\mathrm{b,obs}/\mathrm{K})$ & Error  & SB    & Error & Distance & Error & Spectral Index \\ [0.5ex] 
               & [GHz]     &           & [mJy] & [mJy]  & Ratio &       & [mas] & [mas] &                                     &        & Ratio &       & [mas]    & [mas] &                \\
        \hline\hline
2014-01-29 & 4.3 & A & 129.8 & 14.6 & 2.35 & 0.22 & 0.81 & 0.03 & 10.16 & 0.06 & 3.30 & 0.55 & 10.05 & 0.22 & -0.48 \\
- & -   & B & 55.2 & 7.0 & -   & -   & 0.96 & 0.06 & 9.64 & 0.07 & -   & -   & -   & -   & -1.36 \\
        \hline
2014-01-29 & 7.6 & A & 99.3 & 11.9 & 3.86 & 0.58 & 0.63 & 0.03 & 9.78 & 0.07 & 9.54 & 2.67 & 10.20 & 0.25 & - \\
- & -   & B & 25.7 & 4.3 & -   & -   & 0.99 & 0.11 & 8.80 & 0.12 & -   & -   & -   & -   & - \\
        \hline\hline
2021-03-04 & 4.9 & A & 108.0 & 12.2 & 2.58 & 0.25 & 0.57 & 0.02 & 10.28 & 0.06 & 9.14 & 1.58 & 10.03 & 0.35 & - \\
- & -   & B & 41.9 & 5.5 & -   & -   & 1.07 & 0.06 & 9.32 & 0.08 & -   & -   & -   & -   & - \\
        \hline
2021-06-17 & 4.9 & A & 112.1 & 10.5 & 2.46 & 0.13 & -   & -   & 9.98 & 0.05 & 2.79 & 0.29 & -   & -   & -1.25 \\
- & -   & A0 & 99.7 & 10.4 & 2.18 & 0.12 & 0.46 & 0.01 & 10.43 & 0.01 & 7.76 & 0.75 & 10.45 & 0.07 & - \\
- & -   & B & 45.7 & 3.9 & -   & -   & -   & -   & 9.54 & 0.05 & -   & -   & -   & -   & - \\
- & -   & B0 & 28.7 & 3.3 & 3.91 & 0.24 & 0.49 & 0.02 & 9.83 & 0.02 & 1.43 & 0.17 & -   & -   & - \\
        \hline
2021-06-09 & 22.2 & A & 16.9 & 2.6 & - & - & 0.07 & 0.01 & 9.94 & 0.10 & - & - & - & - & - \\
- & -   & B & - & - & -   & -   & - & - & - & - & -   & -   & -   & -   & - \\
        \hline
    \end{tabular}
    }
\end{table*}


\begin{figure*}[h]
    \centering
    \includegraphics[width=0.4\textwidth]{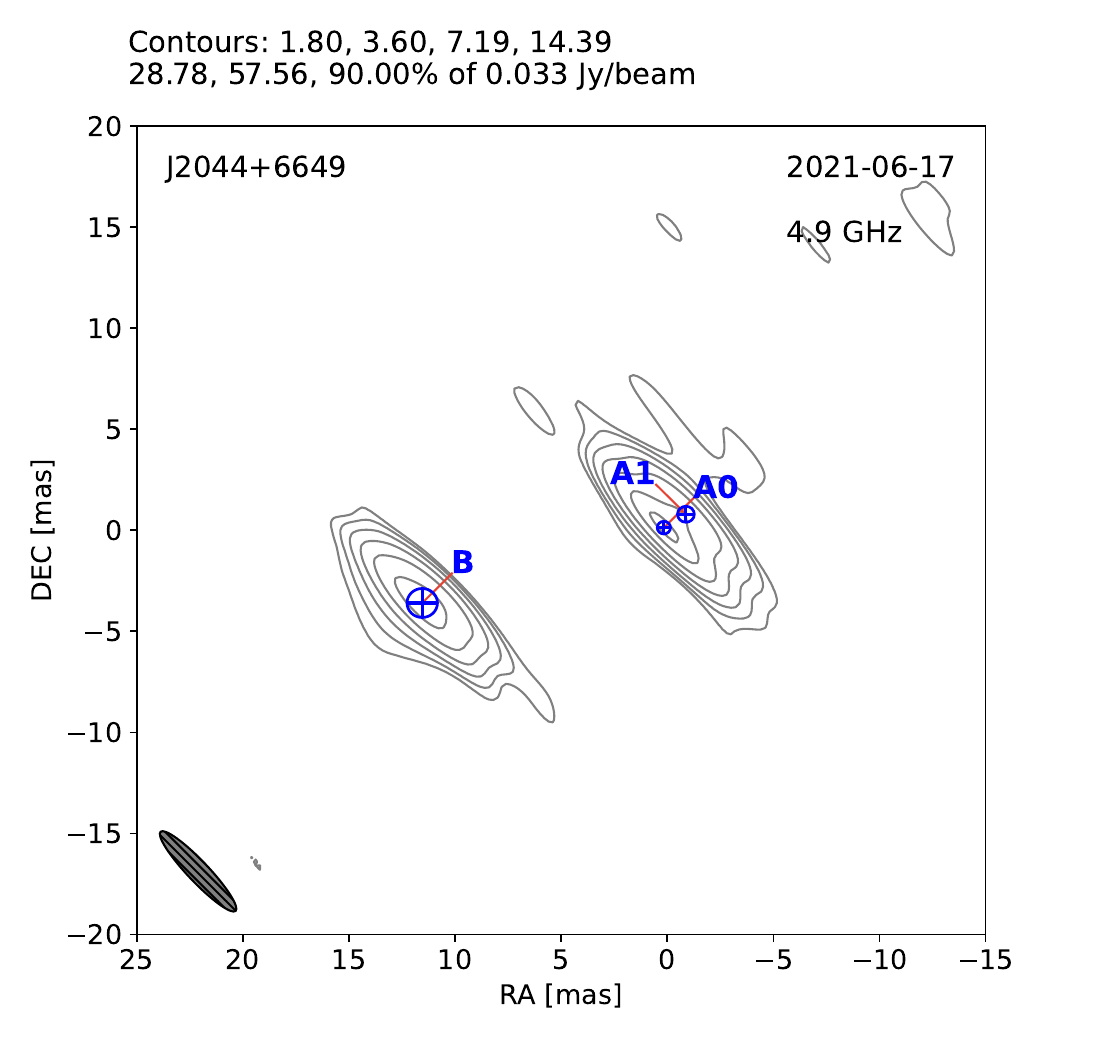}
    \caption{J2044+6649 EVN 4.9\,GHz data. The contours start at four times the image rms noise of 0.150\,mJy/beam and increase by factors of two. The restoring beam size is 0.99$\times$5.28\,mas at 42.0\,$^\circ$ PA. The source is not detected at 22.2\,GHz.}
    \label{fig:J2044}
\end{figure*}

\begin{table*}[h]
    \caption{Derived quantities for J2044+6649. See Table~\ref{tab:J2209} for a detailed description of the columns.}
    \vspace*{3mm}
    \adjustbox{width=1\textwidth}{%
    \label{tab:J2044}
    \centering
    \begin{tabular}{|| c c c c c | c c | c c | c c | c c | c c | c ||}
        \hline
        Epochs & Frequency & Component & Flux  & Error  & Flux  & Error & FWHM  & Error & $\log(T_\mathrm{b,obs}/\mathrm{K})$ & Error  & SB    & Error & Distance & Error & Spectral Index \\ [0.5ex] 
               & [GHz]     &           & [mJy] & [mJy]  & Ratio &       & [mas] & [mas] &                                     &        & Ratio &       & [mas]    & [mas] &                \\
        \hline\hline
2016-08-07 & 4.3 & A & 50.7 & 6.2 & 1.20 & 0.13 & 0.70 & 0.04 & 9.88 & 0.07 & 2.51 & 0.48 & 12.48 & 0.26 & -0.71 \\
- & -   & B & 42.3 & 5.4 & -   & -   & 1.01 & 0.06 & 9.48 & 0.08 & -   & -   & -   & -   & -1.26 \\
        \hline
2016-08-07 & 7.6 & A & 34.0 & 4.6 & 1.63 & 0.25 & 0.67 & 0.05 & 9.26 & 0.09 & 4.27 & 1.27 & 12.05 & 0.25 & - \\
- & -   & B & 20.8 & 3.3 & -   & -   & 1.08 & 0.11 & 8.63 & 0.11 & -   & -   & -   & -   & - \\
        \hline\hline
2021-06-17 & 4.9 & A & 60.4 & 6.1 & 1.26 & 0.14 & -   & -   & 9.53 & 0.07 & 2.55 & 0.57 & -   & -   & -  \\
- & -   & A0 & 37.2 & 5.0 & 0.78 & 0.10 & 0.62 & 0.04 & 9.73 & 0.04 & 4.09 & 0.97 & 11.98 & 0.26 & - \\
- & -   & B & 48.0 & 6.3 & -   & -   & 1.44 & 0.11 & 9.12 & 0.09 & -   & -   & -   & -   & - \\
        \hline
        \end{tabular}
    }
\end{table*}


\begin{figure*}[h]
    \centering
    \includegraphics[width=0.33\textwidth]{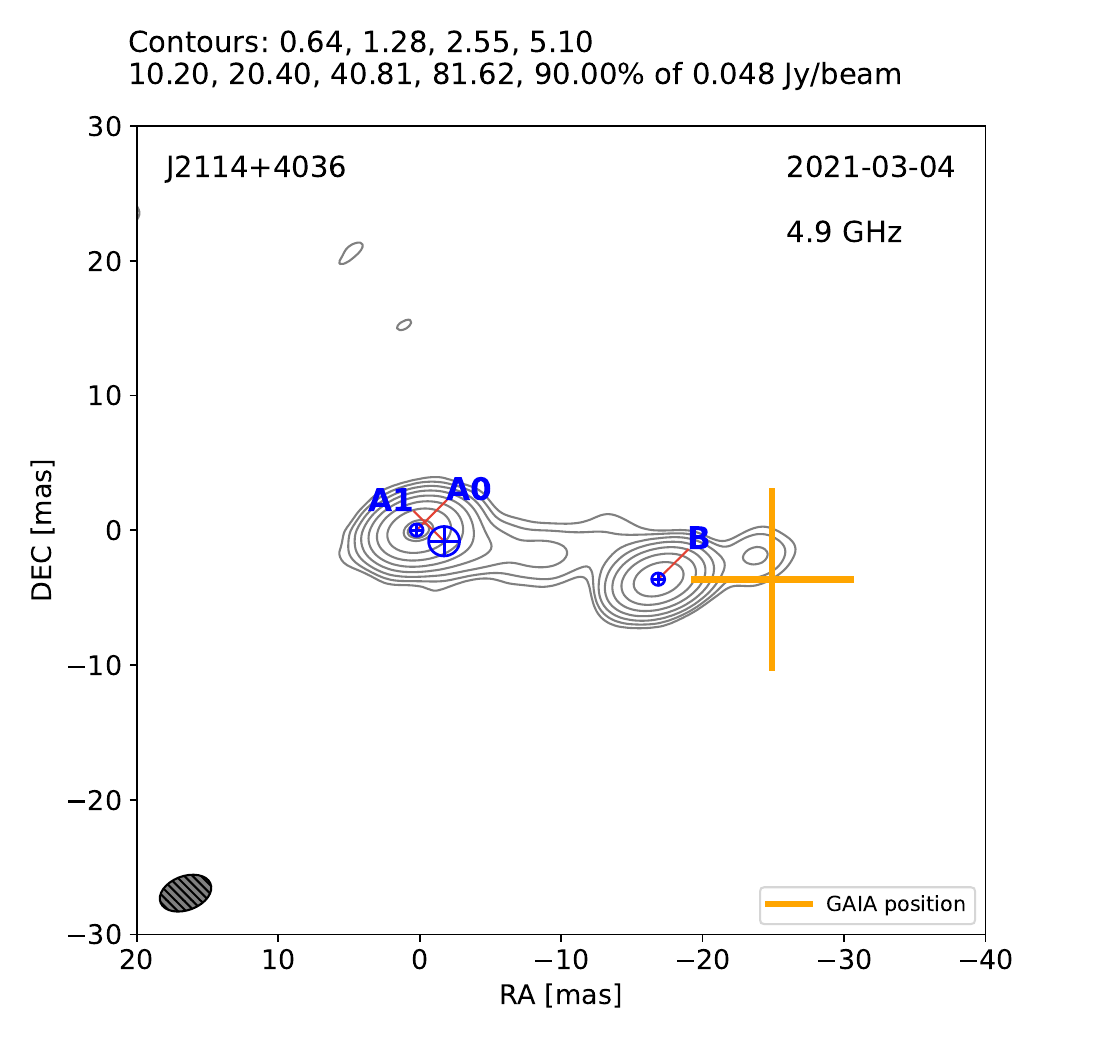}
    \includegraphics[width=0.33\textwidth]{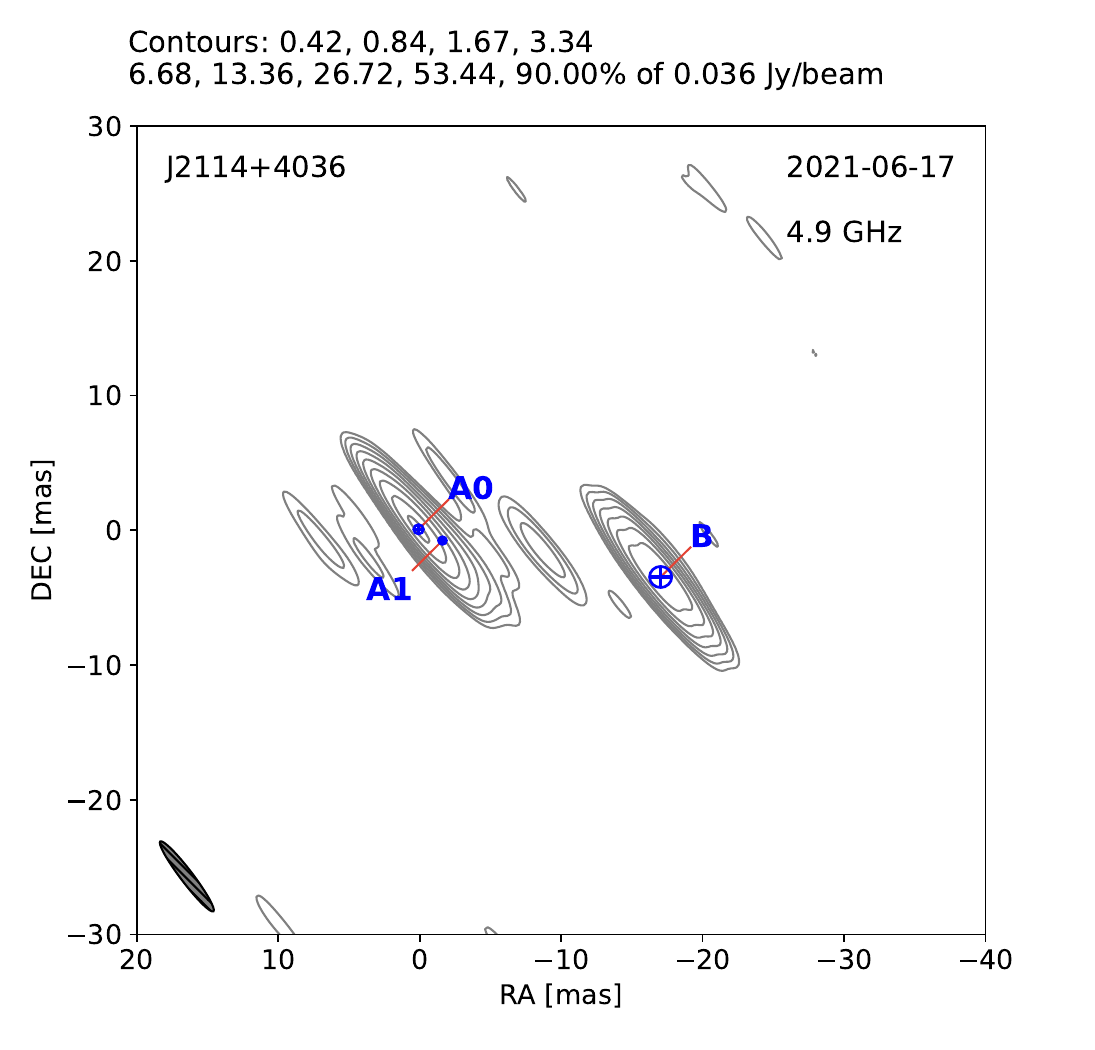}
    \includegraphics[width=0.33\textwidth]{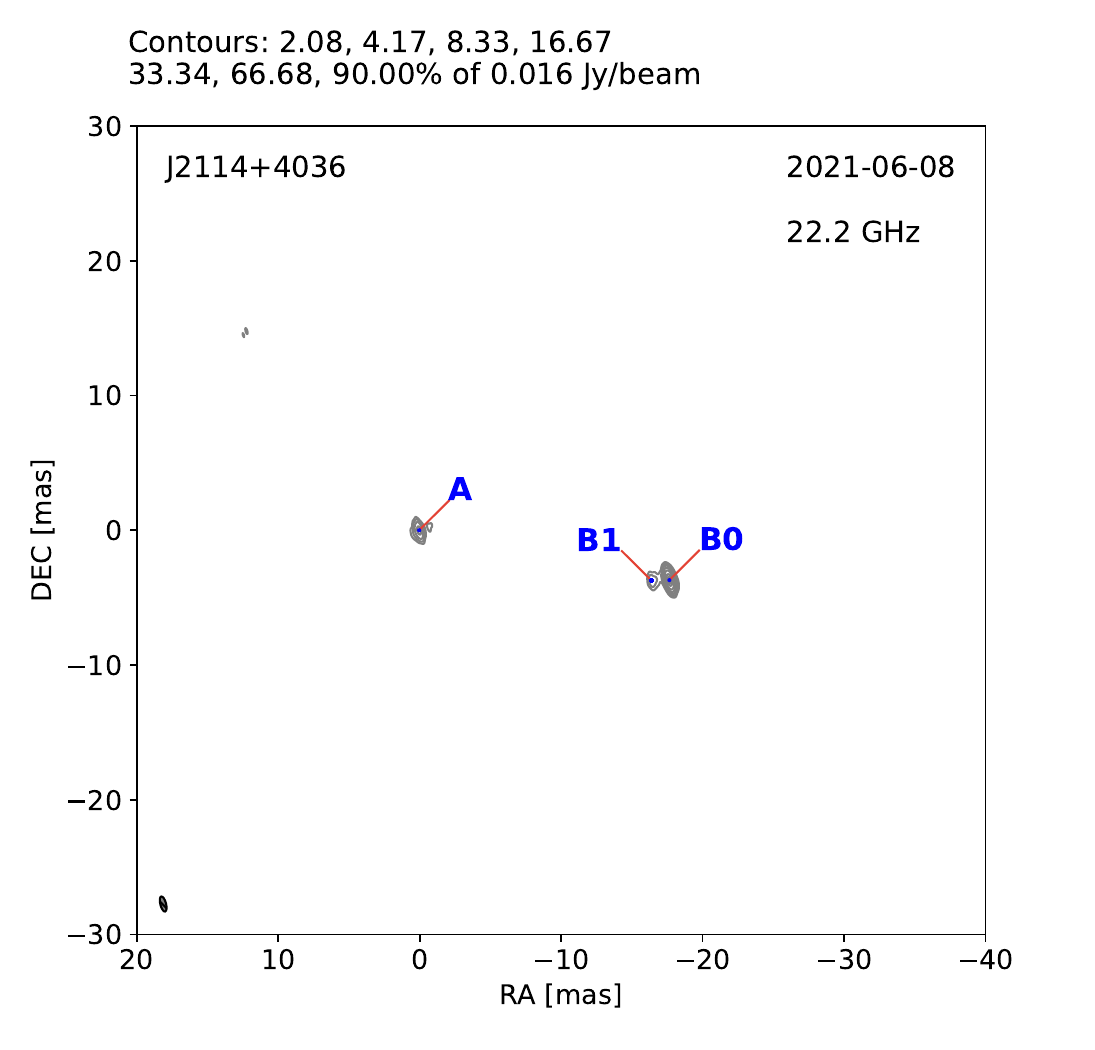}
    \caption{J2114+4036. \textit{Left:} EVN 4.9\,GHz data (session 4) with GAIA position (orange cross). The contours start at four times the image rms noise of 0.077\,mJy/beam and increase by factors of two. The restoring beam size is 2.51$\times$3.80\,mas at $-67.4$\,$^\circ$ PA. \textit{Centre:} EVN 4.9\,GHz data (session 5). The contours start at four times the image rms noise of 0.037\,mJy/beam and increase by factors of two. The restoring beam size is 1.06$\times$5.46\,mas at 35.7\,$^\circ$ PA. \textit{Right:} EVN 22.2\,GHz data (session 5). The contours start at four times the image rms noise of 0.081\,mJy/beam and increase by factors of two. The restoring beam size is 0.41$\times$1.15\,mas at 13.8\,$^\circ$ PA.}
    \label{fig:J2114}
\end{figure*}

\begin{table*}[h]
    \caption{Derived quantities for J2114+4036. See Table~\ref{tab:J2209} for a detailed description of the columns.}
    \vspace*{3mm}
    \adjustbox{width=1\textwidth}{%
    \label{tab:J2114}
    \centering
    \begin{tabular}{|| c c c c c | c c | c c | c c | c c | c c | c ||} 
        \hline
        Epochs & Frequency & Component & Flux  & Error  & Flux  & Error & FWHM  & Error & $\log(T_\mathrm{b,obs}/\mathrm{K})$ & Error  & SB    & Error & Distance & Error & Spectral Index \\ [0.5ex] 
               & [GHz]     &           & [mJy] & [mJy]  & Ratio &       & [mas] & [mas] &                                     &        & Ratio &       & [mas]    & [mas] &                \\
        \hline\hline
2015-09-05 & 4.3 & A & 60.9 & 7.3 & 1.31 & 0.13 & 1.18 & 0.06 & 9.51 & 0.07 & 0.46 & 0.08 & 16.65 & 0.22 & -0.89  \\
- & -   & B & 46.4 & 5.8 & -   & -   & 0.70 & 0.04 & 9.84 & 0.07 & -   & -   & -   & -   & -0.98 \\
        \hline
2015-09-05 & 7.6 & A & 36.9 & 4.8 & 1.38 & 0.17 & 0.87 & 0.06 & 9.07 & 0.08 & 0.35 & 0.08 & 16.77 & 0.16 & - \\
- & -   & B & 26.8 & 3.7 & -   & -   & 0.44 & 0.03 & 9.53 & 0.09 & -   & -   & -   & -   & - \\
        \hline\hline
2021-03-04 & 4.9 & A & 59.6 & 5.7 & 1.60 & 0.13 & -   & -   & 8.78 & 0.06 & 0.24 & 0.04 & -   & -   & - \\
- & -   & A0 & 47.0 & 5.4 & 1.26 & 0.11 & 0.92 & 0.04 & 9.50 & 0.02 & 1.27 & 0.19 & 17.46 & 0.18 & - \\
- & -   & B & 37.3 & 4.4 & -   & -   & 0.92 & 0.04 & 9.40 & 0.07 & -   & -   & -   & -   & - \\
        \hline
2021-06-17 & 4.9 & A & 54.9 & 5.1 & 1.43 & 0.09 & -   & -   & 9.59 & 0.05 & 4.39 & 0.52 & -   & -   & -1.80 \\
- & -   & A0 & 45.2 & 4.9 & 1.18 & 0.07 & 0.69 & 0.02 & 9.73 & 0.02 & 5.99 & 0.74 & 17.48 & 0.15 & - \\
- & -   & B & 38.5 & 4.3 & -   & -   & 1.57 & 0.07 & 8.95 & 0.06 & -   & -   & -   & -   & -0.49 \\
        \hline
2021-06-09 & 22.2 & A & 3.6 & 0.9 & 0.20 & 0.05 & 0.10 & 0.01 & 9.03 & 0.17 & 0.61 & 0.32 & -   & -   & -  \\
- & -   & B & 18.4 & 2.4 & -   & -   & -   & -   & 9.25 & 0.16 & -   & -   & -   & -   & - \\
- & -   & B0 & 16.9 & 2.4 & 0.21 & 0.05 & 0.09 & 0.01 & 9.80 & 0.04 & 0.17 & 0.07 & 18.07 & 0.11 & - \\
        \hline
    \end{tabular}
    }
\end{table*}




\begin{figure*}[h]
    \centering
    \includegraphics[width=0.4\textwidth]{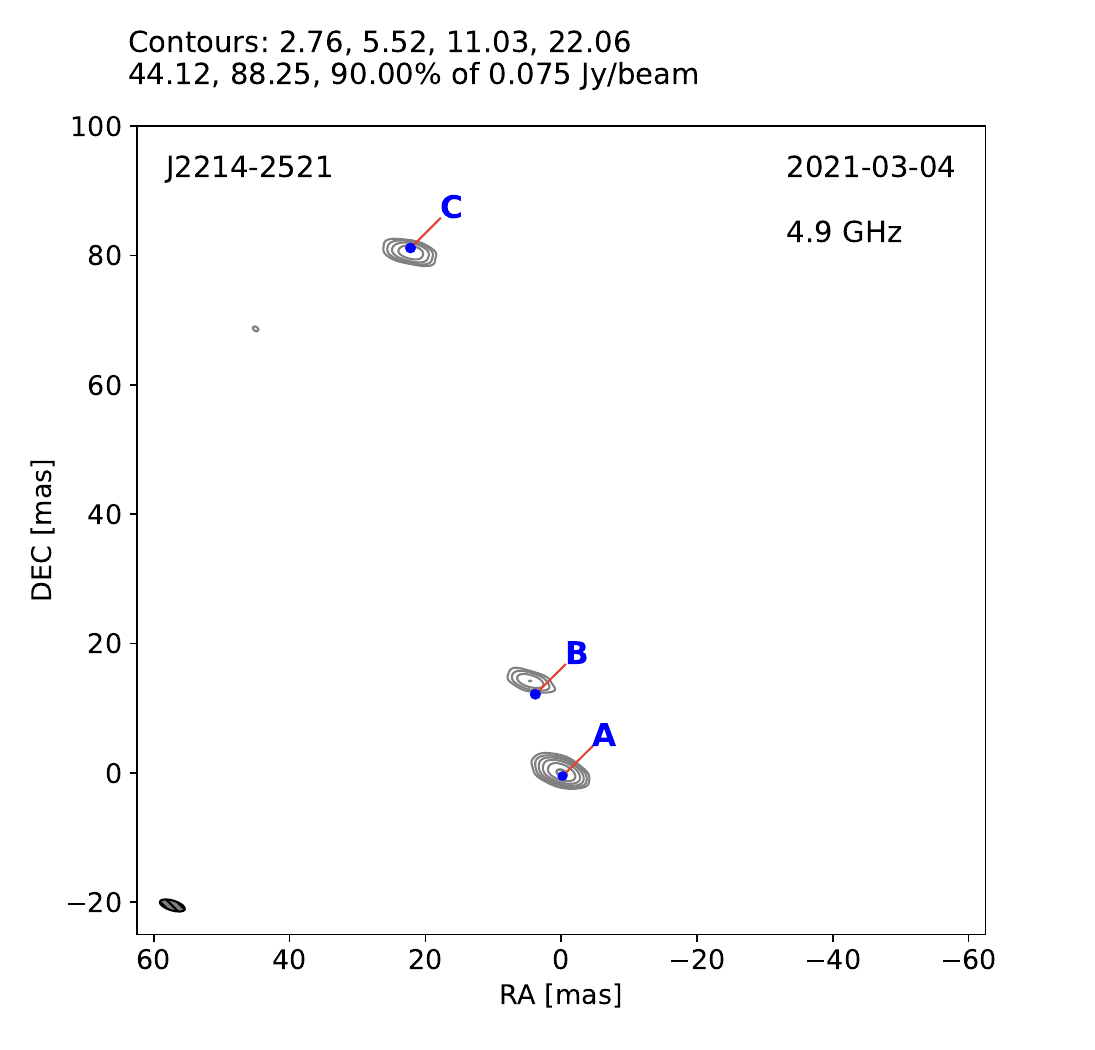}
    \includegraphics[width=0.4\textwidth]{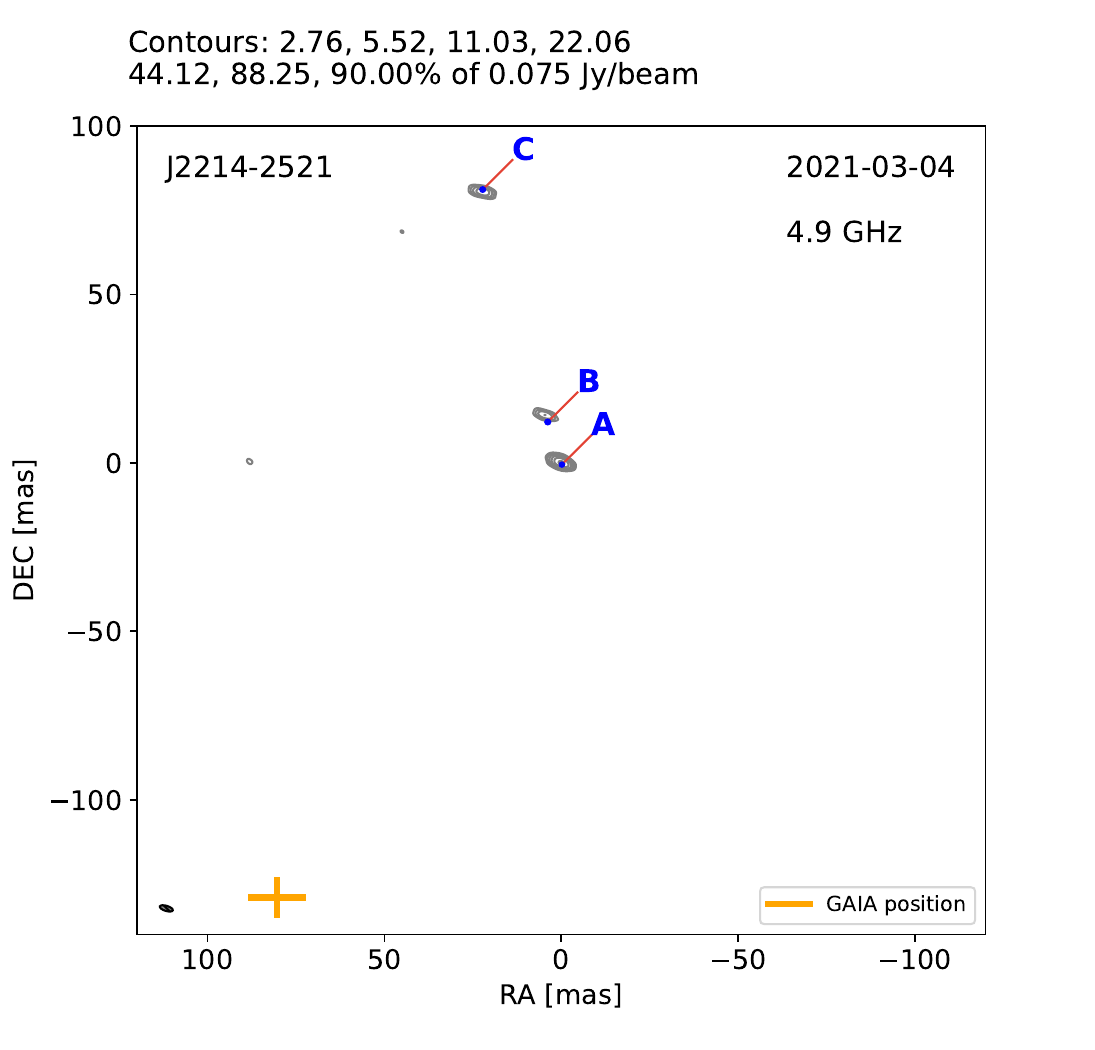}
    \caption{\textit{Left:} J2214-2521 EVN 4.9\,GHz data. The contours start at four times the image rms noise of 0.585\,mJy/beam and increase by factors of two. The restoring beam size is 1.53$\times$3.86\,mas at $71.7$\,$^\circ$ PA. \textit{Right:} EVN 4.9\,GHz data with GAIA position (orange cross). The source is not detected at 22.2\,GHz.}
    \label{fig:J2214}
\end{figure*}

\begin{table*}[h]
    \caption{Derived quantities for J2214-2521. See Table~\ref{tab:J2209} for a detailed description of the columns.}
    \vspace*{3mm}
    \adjustbox{width=1\textwidth}{%
    \label{tab:J2214}
    \centering
    \begin{tabular}{|| c c c c c | c c | c c | c c | c c | c c | c ||} 
        \hline
        Epochs & Frequency & Component & Flux  & Error  & Flux  & Error & FWHM  & Error & $\log(T_\mathrm{b,obs}/\mathrm{K})$ & Error  & SB    & Error & Distance & Error & Spectral Index \\ [0.5ex] 
               & [GHz]     &           & [mJy] & [mJy]  & Ratio &       & [mas] & [mas] &                                     &        & Ratio &       & [mas]    & [mas] &                \\
        \hline\hline
2005-07-20 & 2.3 & A & 291.1 & 36.0 & - & - & 2.02 & 0.11 & 10.27 & 0.07 & - & - & - & - & - \\
- & -   & B & - & - & -   & -   & - & - & - & - & -   & -   & -   & -   & - \\
        \hline
2005-07-20 & 8.7 & A & 113.7 & 18.7 & 7.67 & 2.99 & 0.99 & 0.10 & 9.33 & 0.11 & $<$3.82 & - & 19.43 & 0.84 & - \\
- & -   & B & 14.8 & 5.7 & -   & -   & $<$0.70 & - & $>$8.75 & - & -   & -   & -   & -   & - \\
        \hline\hline
2017-07-09 & 2.2 & A & 300.0 & 25.8 & 12.37 & 1.88 & -   & -   & 9.46 & 0.07 & 0.79 & 0.23 & -   & -   & -0.71 \\
- & -   & A0 & 204.9 & 22.9 & 8.45 & 1.31 & 0.83 & 0.03 & 10.91 & 0.02 & 22.22 & 6.11 & 18.89 & 0.61 & - \\
- & -   & B & 24.3 & 4.3 & -   & -   & 1.35 & 0.15 & 9.56 & 0.12 & -   & -   & -   & -   & -0.07 \\
        \hline
2017-07-09 & 8.7 & A & 114.9 & 13.1 & 5.17 & 0.66 & 1.02 & 0.05 & 9.31 & 0.06 & $<$0.06 & - & 19.31 & 0.14 & - \\
- & -   & B & 22.2 & 3.4 & -   & -   & $<$0.11 & - & $>$10.55 & - & -   & -   & -   & -   & - \\
        \hline\hline
2018-08-10 & 2.2 & A & 269.2 & 25.5 & 11.23 & 1.77 & -   & -   & 9.78 & 0.08 & $<$0.26 & - & -   & -   & -0.75 \\
- & -   & A0 & 219.3 & 24.5 & 9.15 & 1.45 & 1.26 & 0.05 & 10.58 & 0.02 & 1.62 & 0.45 & 18.86 & 0.59 & - \\
- & -   & B & 24.0 & 4.3 & -   & -   & $<$0.53 & - & $>$10.37 & - & -   & -   & -   & -   & -0.43 \\
        \hline
2018-08-10 & 8.7 & A & 97.7 & 11.3 & 7.25 & 1.15 & 0.91 & 0.04 & 9.34 & 0.07 & 1.78 & 0.53 & 18.96 & 0.17 & - \\
- & -   & B & 13.5 & 2.4 & -   & -   & 0.45 & 0.05 & 9.09 & 0.13 & -   & -   & -   & -   & - \\
        \hline\hline
2021-03-04 & 4.9 & A & 109.5 & 15.4 & 4.35 & 1.01 & 1.07 & 0.08 & 9.74 & 0.09 & 5.67 & 2.45 & 13.26 & 0.49  & - \\
- & -   & B & 25.1 & 5.8 & -   & -   & 1.22 & 0.20 & 8.98 & 0.18 & -   & -   & -   & -   & - \\
        \hline
    \end{tabular}
    }
\end{table*}


\begin{figure*}[h]
    \centering
    \includegraphics[width=0.33\textwidth]{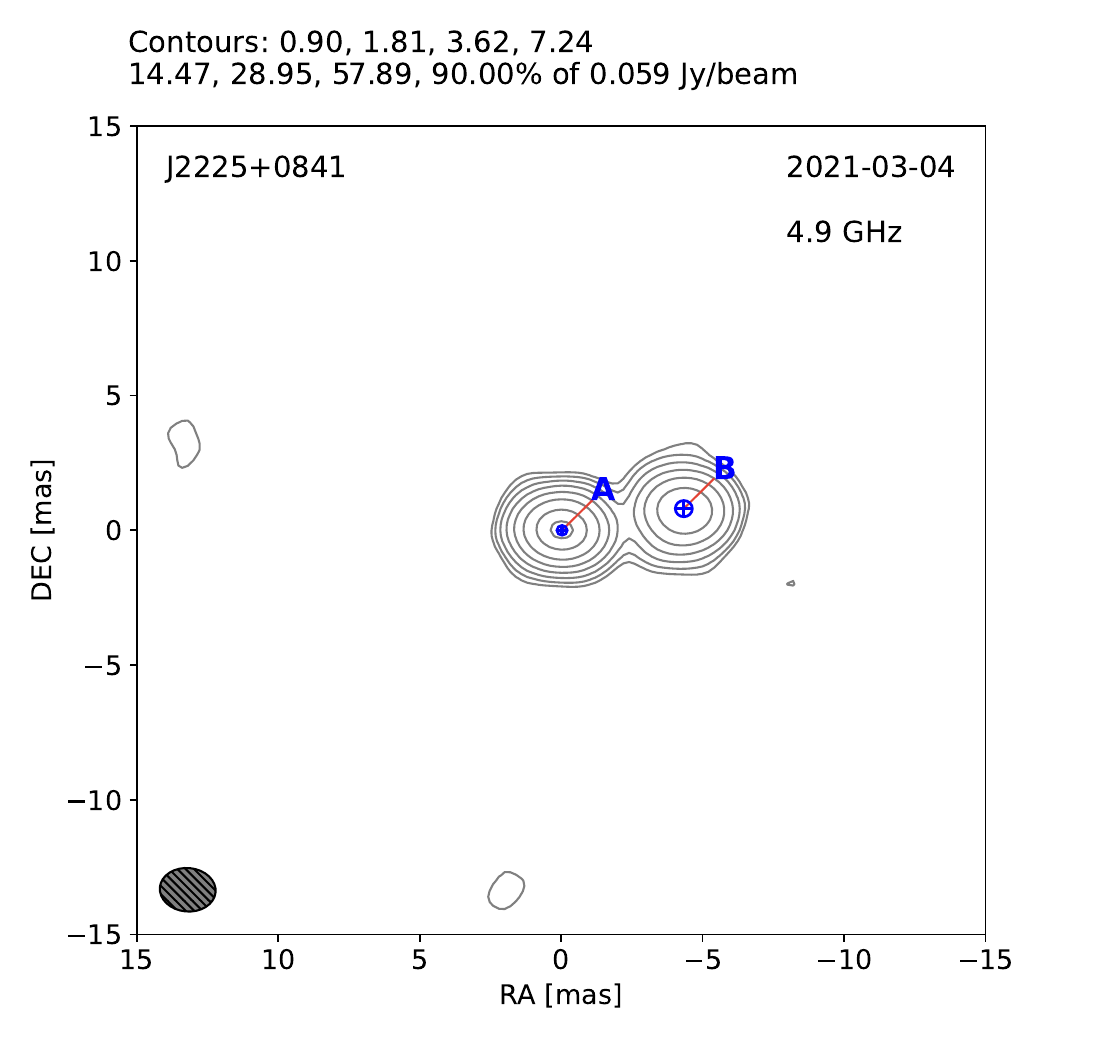}
    \includegraphics[width=0.33\textwidth]{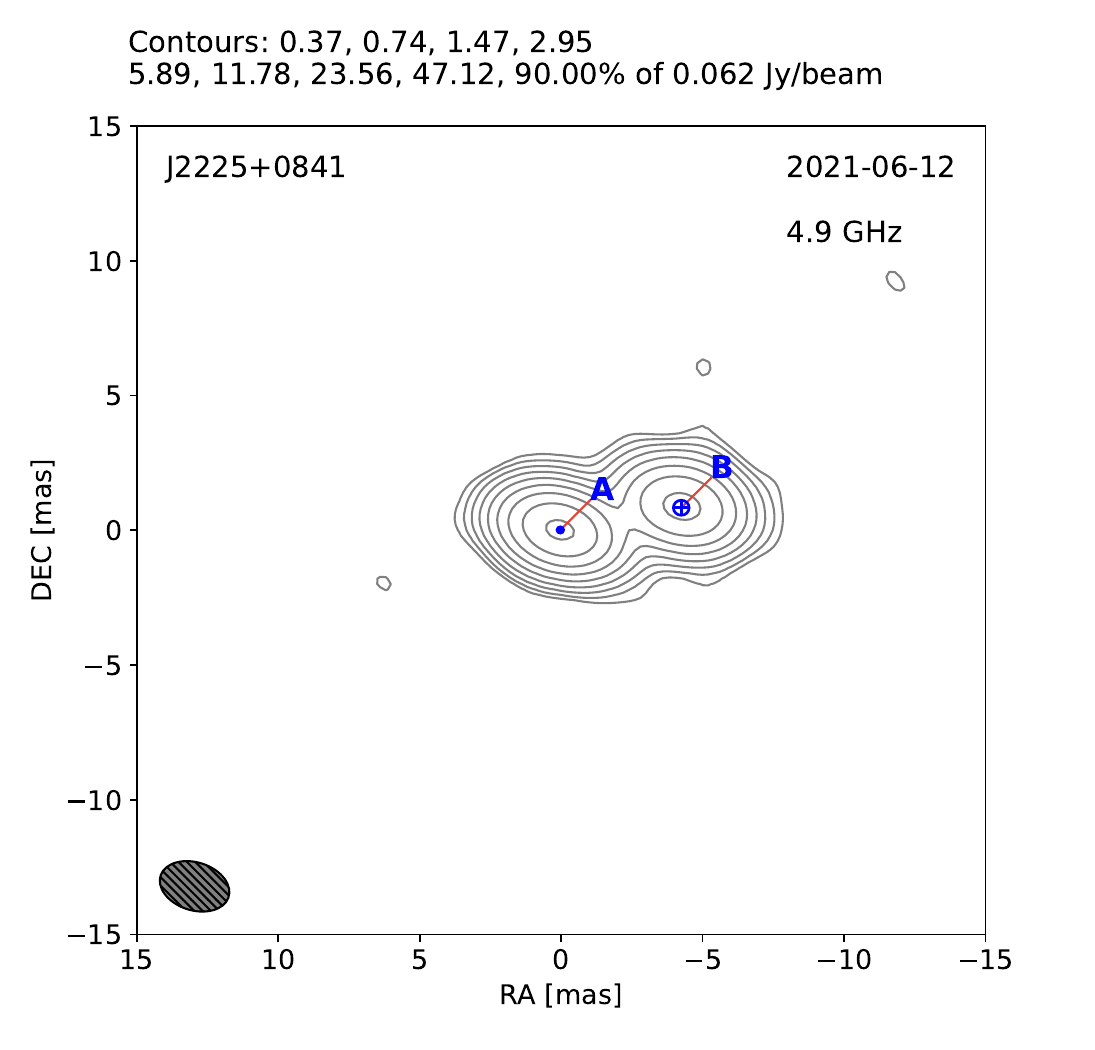}
    \includegraphics[width=0.33\textwidth]{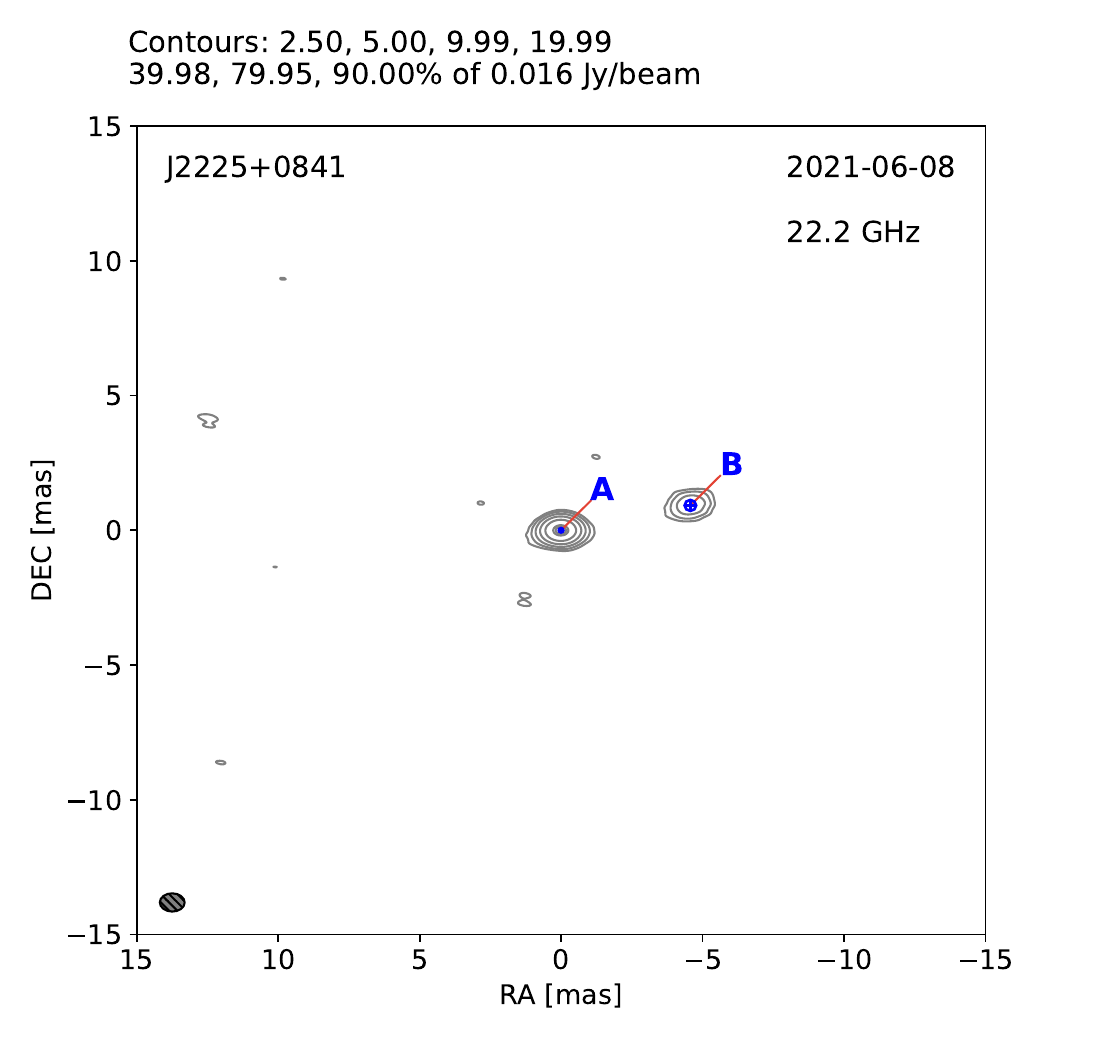}
    \caption{J2225+0841. \textit{Left:} EVN 4.9\,GHz data (session 4). The contours start at four times the image rms noise of 0.134\,mJy/beam and increase by factors of two. The restoring beam size is 1.61$\times$1.98\,mas at $84.1$\,$^\circ$ PA. \textit{Centre:} EVN 4.9\,GHz data (session 5). The contours start at four times the image rms noise of 0.057\,mJy/beam and increase by factors of two. The restoring beam size is 1.77$\times$2.53\,mas at $70.3$\,$^\circ$ PA. \textit{Right:} EVN 22.2\,GHz data (session 5). The contours start at four times the image rms noise of 0.102\,mJy/beam and increase by factors of two. The restoring beam size is 0.68$\times$0.88\,mas at $89.28$\,$^\circ$ PA.}
    \label{fig:J2225}
\end{figure*}

\begin{table*}[h]
    \caption{Derived quantities for J2225+0841. See Table~\ref{tab:J2209} for a detailed description of the columns.}
    \vspace*{3mm}
    \adjustbox{width=1\textwidth}{%
    \label{tab:J2225}
    \centering
    \begin{tabular}{|| c c c c c | c c | c c | c c | c c | c c | c ||} 
        \hline
        Epochs & Frequency & Component & Flux  & Error  & Flux  & Error & FWHM  & Error & $\log(T_\mathrm{b,obs}/\mathrm{K})$ & Error  & SB    & Error & Distance & Error & Spectral Index \\ [0.5ex] 
               & [GHz]     &           & [mJy] & [mJy]  & Ratio &       & [mas] & [mas] &                                     &        & Ratio &       & [mas]    & [mas] &                \\
        \hline\hline
2015-12-12 & 4.3 & A & 54.1 & 6.6 & 1.72 & 0.20 & 0.67 & 0.03 & 9.94 & 0.07 & 3.66 & 0.76 & 4.41 & 0.22 & -0.82 \\
- & -   & B & 31.5 & 4.2 & -   & -   & 0.98 & 0.07 & 9.38 & 0.08 & -   & -   & -   & -   & -1.31 \\
        \hline
2015-12-12 & 7.6 & A & 34.1 & 4.6 & 2.26 & 0.37 & 0.69 & 0.05 & 9.24 & 0.09 & 1.61 & 0.49 & 4.52 & 0.18 & - \\
- & -   & B & 15.1 & 2.5 & -   & -   & 0.58 & 0.06 & 9.03 & 0.12 & -   & -   & -   & -   & - \\
        \hline\hline
2021-03-04 & 4.9 & A & 61.6 & 7.4 & 1.67 & 0.18 & 0.37 & 0.02 & 10.42 & 0.07 & 4.55 & 0.88 & 4.38 & 0.14 & - \\
- & -   & B & 36.8 & 4.9 & -   & -   & 0.60 & 0.04 & 9.76 & 0.08 & -   & -   & -   & -   & - \\
        \hline\hline
2021-06-12 & 4.9 & A & 62.7 & 6.8 & 1.65 & 0.11 & 0.23 & 0.01 & 10.82 & 0.05 & 9.46 & 1.16 & 4.35 & 0.11 & -0.88 \\
- & -   & B & 38.0 & 4.3 & -   & -   & 0.56 & 0.02 & 9.84 & 0.06 & -   & -   & -   & -   & -1.40 \\
        \hline
2021-06-09 & 22.2 & A & 16.5 & 2.5 & 3.59 & 0.88 & 0.16 & 0.01 & 9.25 & 0.10 & 23.15 & 10.54 & 4.66 & 0.16 & - \\
- & -   & B & 4.6 & 1.1 & -   & -   & 0.41 & 0.07 & 7.89 & 0.18 & -   & -   & -   & -   & - \\
        \hline
    \end{tabular}
    }
\end{table*}


\begin{figure*}[h]
    \centering
    \includegraphics[width=0.33\textwidth]{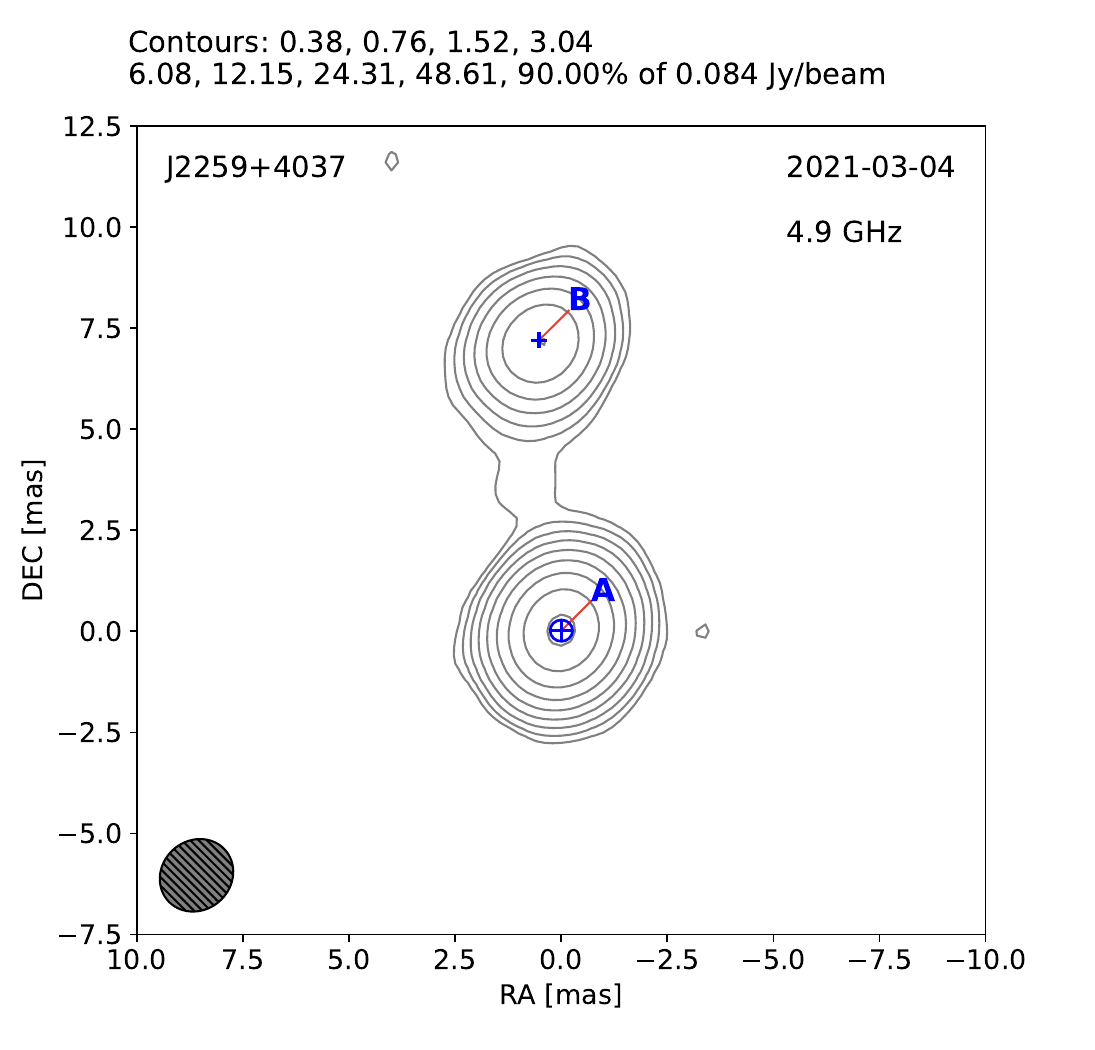}
    \includegraphics[width=0.33\textwidth]{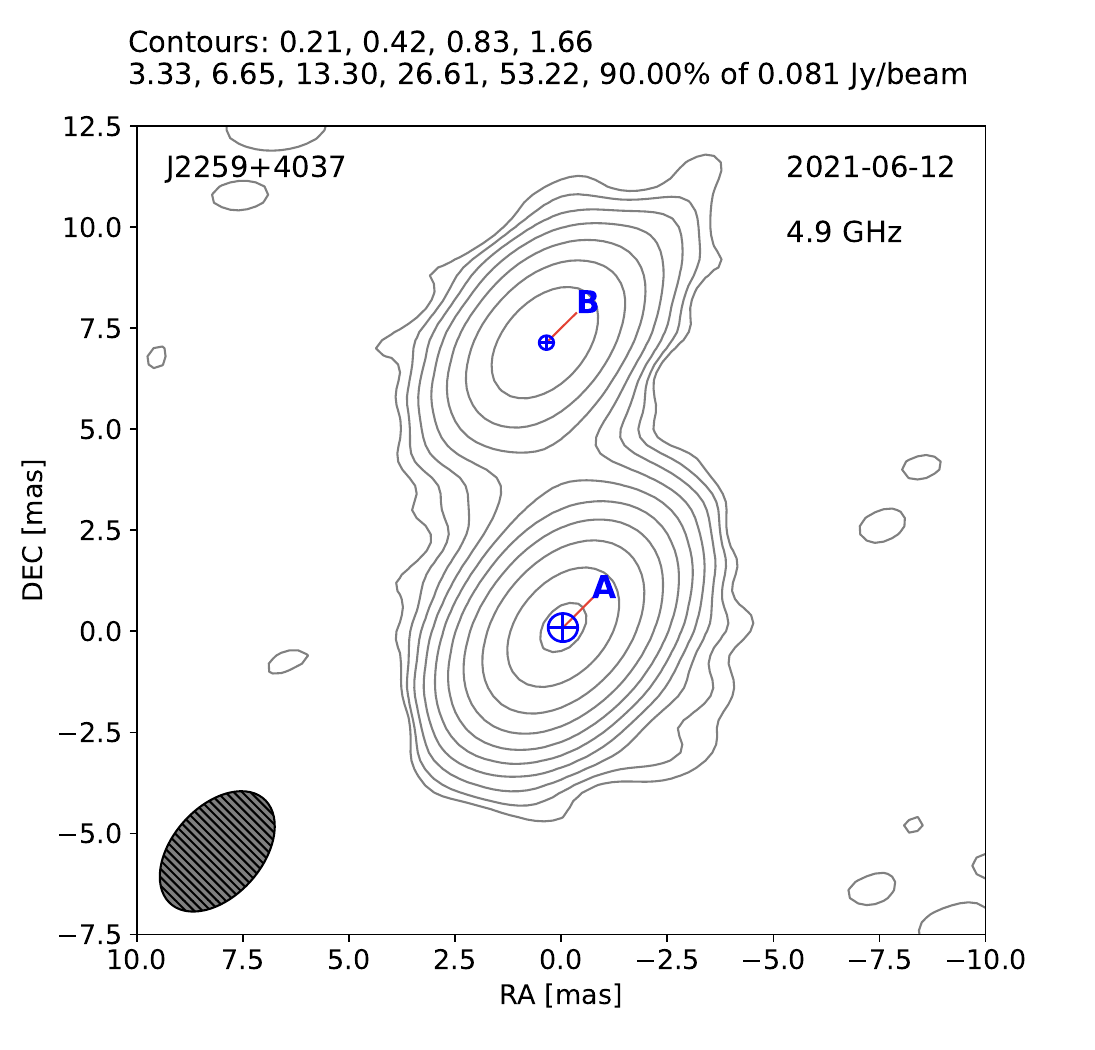}
    \includegraphics[width=0.33\textwidth]{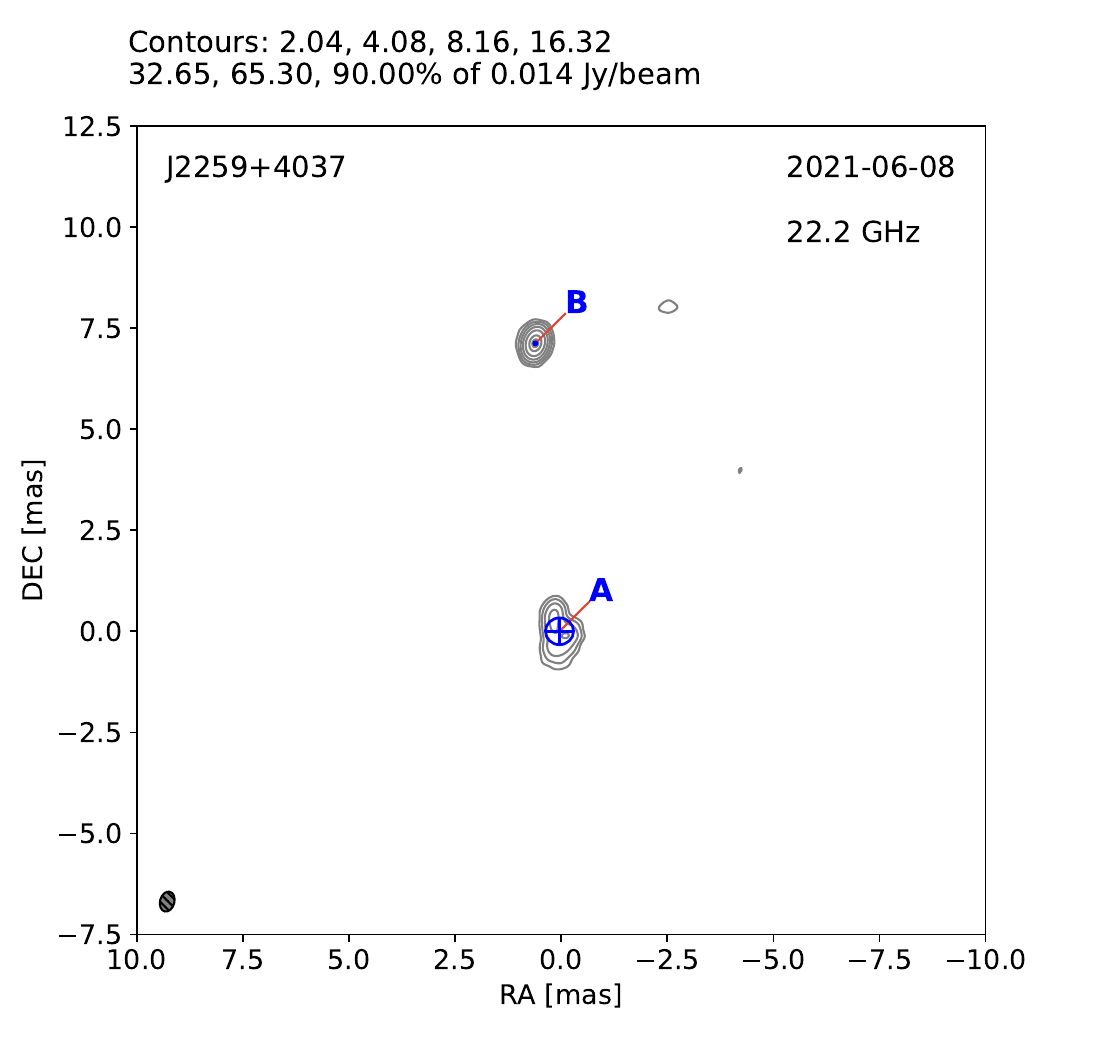}
    \caption{J2259+4037. \textit{Left:} EVN 4.9\,GHz data (session 4). The contours start at four times the image rms noise of 0.800\,mJy/beam and increase by factors of two. The restoring beam size is 1.67$\times$1.85\,mas at $-34.9$\,$^\circ$ PA. \textit{Centre:} EVN 4.9\,GHz data (session 5). The contours start at four times the image rms noise of 0.042\,mJy/beam and increase by factors of two. The restoring beam size is 1.84$\times$2.57\,mas at $-55.9$\,$^\circ$ PA. \textit{Right:} EVN 22.2\,GHz data (session 5). The contours start at four times the image rms noise of 0.074\,mJy/beam and increase by factors of two. The restoring beam size is 0.34$\times$0.48\,mas at $-16.4$\,$^\circ$ PA.}
    \label{fig:J2259}
\end{figure*}

\begin{table*}[h]
    \caption{Derived quantities for J2259+4037. See Table~\ref{tab:J2209} for a detailed description of the columns.}
    \vspace*{3mm}
    \adjustbox{width=1\textwidth}{%
    \label{tab:J2259}
    \centering
    \begin{tabular}{|| c c c c c | c c | c c | c c | c c | c c | c ||} 
        \hline
        Epochs & Frequency & Component & Flux  & Error  & Flux  & Error & FWHM  & Error & $\log(T_\mathrm{b,obs}/\mathrm{K})$ & Error  & SB    & Error & Distance & Error & Spectral Index \\ [0.5ex] 
               & [GHz]     &           & [mJy] & [mJy]  & Ratio &       & [mas] & [mas] &                                     &        & Ratio &       & [mas]    & [mas] &              \\
        \hline\hline
2016-08-20 & 4.3 & A & 84.1 & 9.2 & 4.14 & 0.43 & 0.60 & 0.02 & 10.23 & 0.06 & 0.65 & 0.12 & 6.96 & 0.22 & -1.12 \\
- & -   & B & 20.3 & 2.8 & -   & -   & 0.24 & 0.02 & 10.42 & 0.08 & -   & -   & -   & -   & -0.90 \\
        \hline
2016-08-20 & 7.6 & A & 44.8 & 5.5 & 3.68 & 0.57 & 0.58 & 0.03 & 9.51 & 0.07 & 4.14 & 1.16 & 7.19 & 0.21 & - \\
- & -   & B & 12.2 & 2.1 & -   & -   & 0.61 & 0.06 & 8.89 & 0.12 & -   & -   & -   & -   & - \\
        \hline\hline
2021-03-04 & 4.9 & A & 92.9 & 10.1 & 4.21 & 0.40 & 0.53 & 0.02 & 10.28 & 0.05 & 0.14 & 0.02 & 7.20 & 0.12 & - \\
- & -   & B & 22.1 & 2.9 & -   & -   & 0.10 & 0.01 & 11.13 & 0.08 & -   & -   & -   & -   & - \\
        \hline\hline
2021-06-12 & 4.9 & A & 88.5 & 9.3 & 4.21 & 0.30 & 0.70 & 0.02 & 10.02 & 0.05 & 1.02 & 0.13 & 7.06 & 0.11 & -1.48 \\
- & -   & B & 21.0 & 2.5 & -   & -   & 0.34 & 0.02 & 10.01 & 0.07 & -   & -   & -   & -   & -0.22 \\
        \hline
2021-06-09 & 22.2 & A & 9.4 & 2.0 & 0.63 & 0.14 & 0.66 & 0.12 & 7.78 & 0.19 & 0.01 & 0.00 & 7.15 & 0.15 & - \\
- & -   & B & 15.0 & 2.1 & -   & -   & 0.07 & 0.01 & 9.88 & 0.09 & -   & -   & -   & -   & - \\
        \hline
    \end{tabular}
    }
\end{table*}


\begin{figure*}[h]
    \centering
    \includegraphics[width=0.4\textwidth]{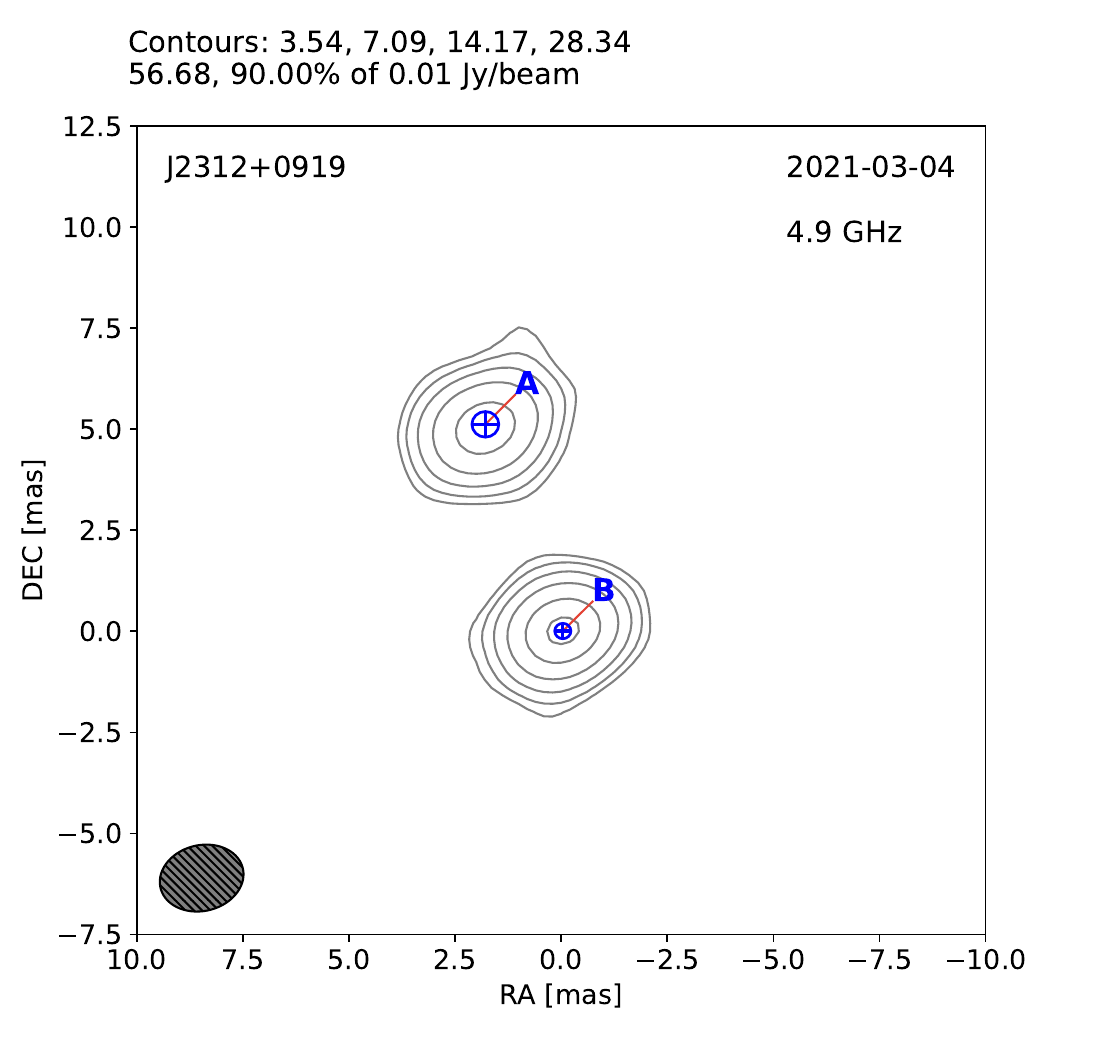}
    \includegraphics[width=0.4\textwidth]{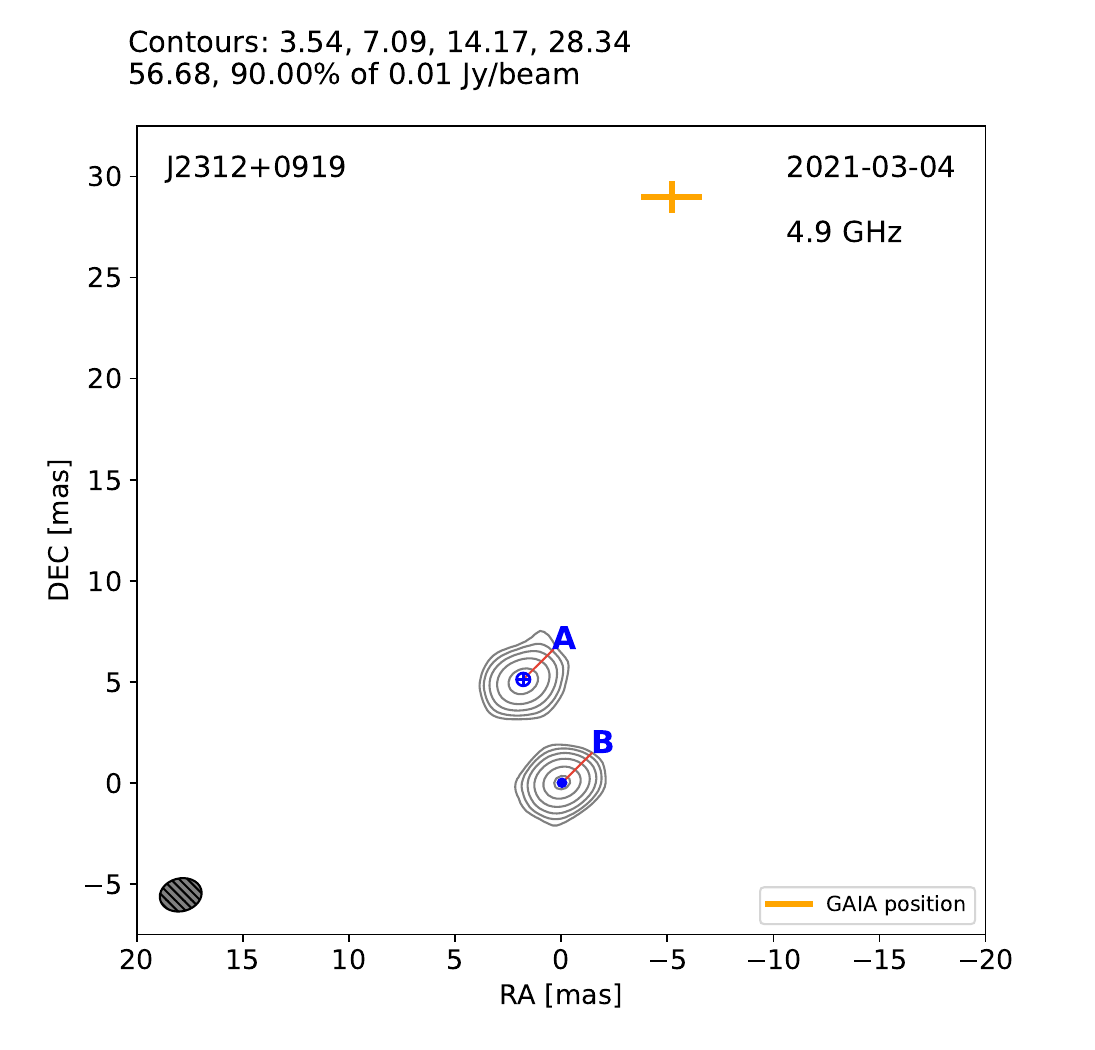}
    \caption{\textit{Left:} J2312+0919 EVN 4.9\,GHz data; \textit{Right:} EVN 4.9\,GHz data with GAIA position (orange cross). The contours start at four times the image rms noise of 0.086\,mJy/beam and increase by factors of two. The restoring beam size is 1.62$\times$2.00\,mas at $-73.5$\,$^\circ$ PA. The source is not detected at 22.2\,GHz.}
    \label{fig:J2312}
\end{figure*}

\begin{table*}[h]
    \caption{Derived quantities for J2312+0919. See Table~\ref{tab:J2209} for a detailed description of the columns.}
    \vspace*{3mm}
    \adjustbox{width=1\textwidth}{%
    \label{tab:J2312}
    \centering
    \begin{tabular}{|| c c c c c | c c | c c | c c | c c | c c | c ||} 
        \hline
        Epochs & Frequency & Component & Flux  & Error  & Flux  & Error & FWHM  & Error & $\log(T_\mathrm{b,obs}/\mathrm{K})$ & Error  & SB    & Error & Distance & Error & Spectral Index \\ [0.5ex] 
               & [GHz]     &           & [mJy] & [mJy]  & Ratio &       & [mas] & [mas] &                                     &        & Ratio &       & [mas]    & [mas] &          \\
        \hline\hline
2015-08-18 & 4.3 & A & 24.2 & 3.4 & 2.17 & 0.37 & 0.24 & 0.02 & 10.49 & 0.08 & 3.47 & 1.03 & 5.33 & 0.30 & -0.23 \\
- & -   & B & 11.2 & 1.9 & -   & -   & 0.31 & 0.03 & 9.94 & 0.12 & -   & -   & -   & -   & -0.33 \\
        \hline
2015-08-18 & 7.6 & A & 21.3 &66 3.0 & 2.29 & 0.41 & 0.25 & 0.02 & 9.90 & 0.09 & 1.48 & 0.46 & 5.45 & 0.18 & - \\
- & -   & B & 9.3 & 1.7 & -   & -   & 0.20 & 0.02 & 9.73 & 0.12 & -   & -   & -   & -   & - \\
        \hline\hline
2021-03-04 & 4.9 & A & 8.8 & 1.5 & 0.85 & 0.16 & 0.63 & 0.07 & 9.10 & 0.12 & 0.31 & 0.10 & 5.43 & 0.26 & - \\
- & -   & B & 10.4 & 1.7 & -   & -   & 0.38 & 0.04 & 9.61 & 0.11 & -   & -   & -   & -   & - \\
        \hline
    \end{tabular}
    }
\end{table*}


\begin{figure*}[h]
    \centering
    \includegraphics[width=0.4\textwidth]{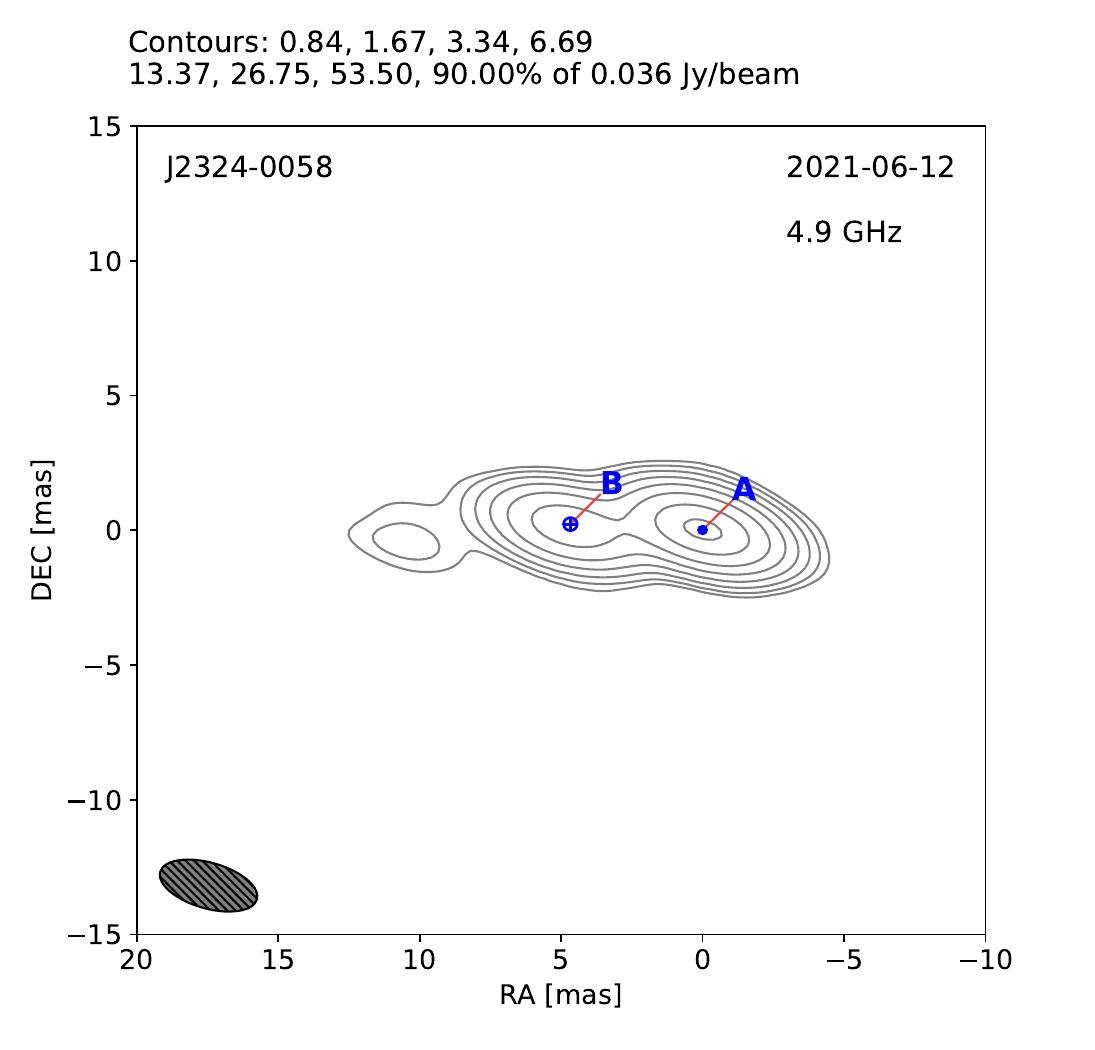}
    \includegraphics[width=0.4\textwidth]{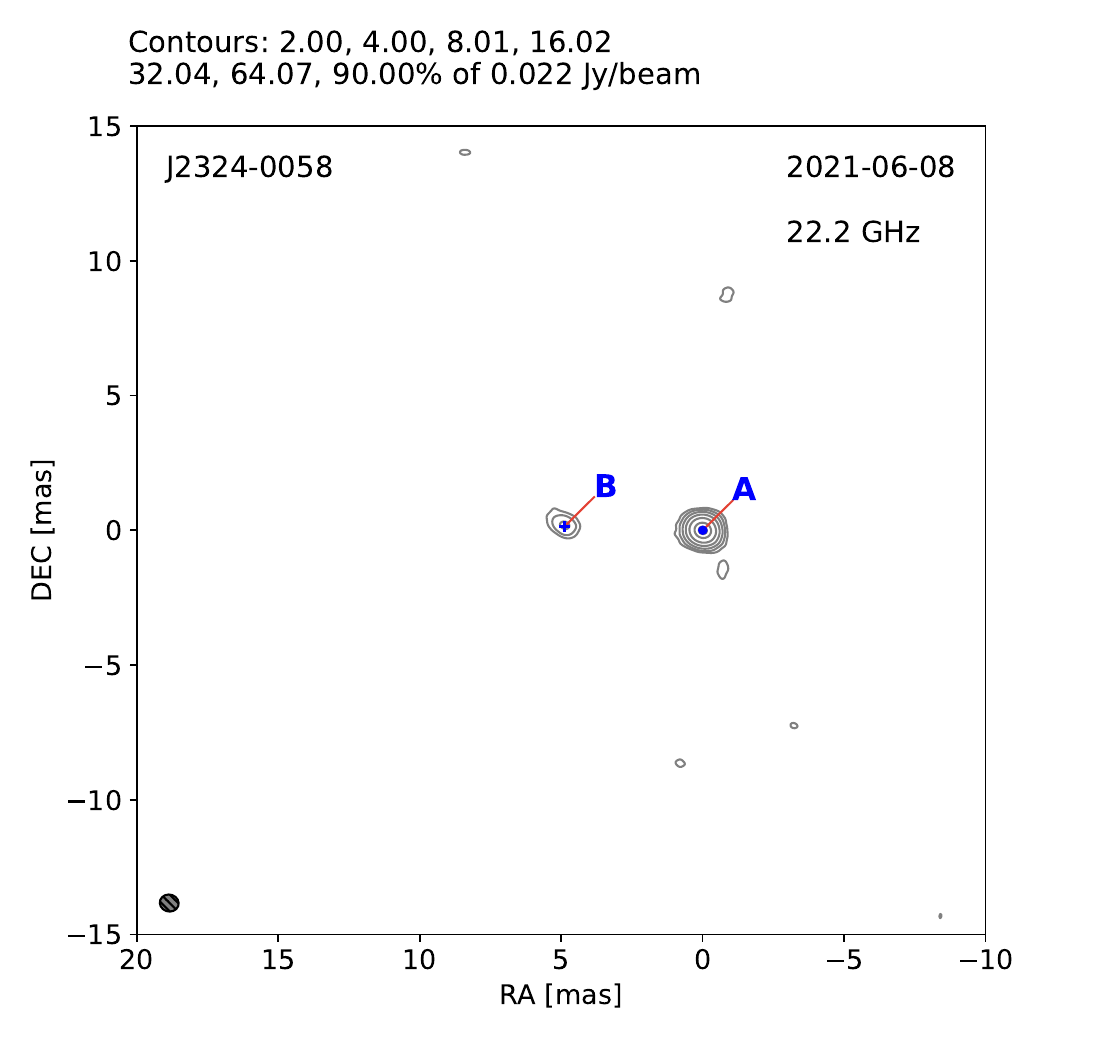}
    \caption{J2324-0058. \textit{Left:} EVN 4.9\,GHz data. The contours start at four times the image rms noise of 0.076\,mJy/beam and increase by factors of two. The restoring beam size is 1.69$\times$3.57\,mas at $72.9$\,$^\circ$ PA. \textit{Right:} EVN 22.2\,GHz data. The contours start at four times the image rms noise of 0.110\,mJy/beam and increase by factors of two. The restoring beam size is 0.63$\times$0.67\,mas at $69.5$\,$^\circ$ PA.}
    \label{fig:J2324}
\end{figure*}

\begin{table*}[h]
    \caption{Derived quantities for J2324-0058. See Table~\ref{tab:J2209} for a detailed description of the columns.}
    \vspace*{3mm}
    \adjustbox{width=1\textwidth}{%
    \label{tab:J2324}
    \centering
    \begin{tabular}{|| c c c c c | c c | c c | c c | c c | c c | c ||} 
        \hline
        Epochs & Frequency & Component & Flux  & Error  & Flux  & Error & FWHM  & Error & $\log(T_\mathrm{b,obs}/\mathrm{K})$ & Error  & SB    & Error & Distance & Error & Spectral Index \\ [0.5ex] 
               & [GHz]     &           & [mJy] & [mJy]  & Ratio &       & [mas] & [mas] &                                     &        & Ratio &       & [mas]    & [mas] &          \\
        \hline\hline
2018-05-25 & 4.3 & A & 47.8 & 5.9 & 1.20 & 0.13 & 0.72 & 0.04 & 9.84 & 0.07 & 1.06 & 0.20 & 4.57 & 0.19 & -0.16 \\
- & -   & B & 39.7 & 5.0 & -   & -   & 0.67 & 0.04 & 9.81 & 0.07 & -   & -   & -   & -   & -1.02 \\
        \hline
2018-05-25 & 7.6 & A & 43.6 & 5.3 & 1.95 & 0.23 & 0.27 & 0.01 & 10.14 & 0.07 & 3.89 & 0.82 & 4.60 & 0.12 & - \\
- & -   & B & 22.4 & 3.1 & -   & -   & 0.39 & 0.03 & 9.55 & 0.09 & -   & -   & -   & -   & - \\
        \hline\hline
2021-06-12 & 4.9 & A & 37.3 & 4.4 & 2.44 & 0.29 & 0.28 & 0.01 & 10.43 & 0.06 & 7.11 & 1.48 & 4.68 & 0.21 & -0.24 \\
- & -   & B & 15.3 & 2.2 & -   & -   & 0.48 & 0.03 & 9.57 & 0.09 & -   & -   & -   & -   & -1.27 \\
        \hline
2021-06-09 & 22.2 & A & 25.8 & 3.5 & 11.51 & 3.85 & 0.24 & 0.02 & 9.10 & 0.08 & 3.92 & 2.34 & 4.89 & 0.16 & - \\
- & -   & B & 2.2 & 0.8 & -   & -   & 0.14 & 0.03 & 8.51 & 0.25 & -   & -   & -   & -   & - \\
        \hline
    \end{tabular}
    }
\end{table*}


\begin{figure*}[h]
    \centering
    \includegraphics[width=0.4\textwidth]{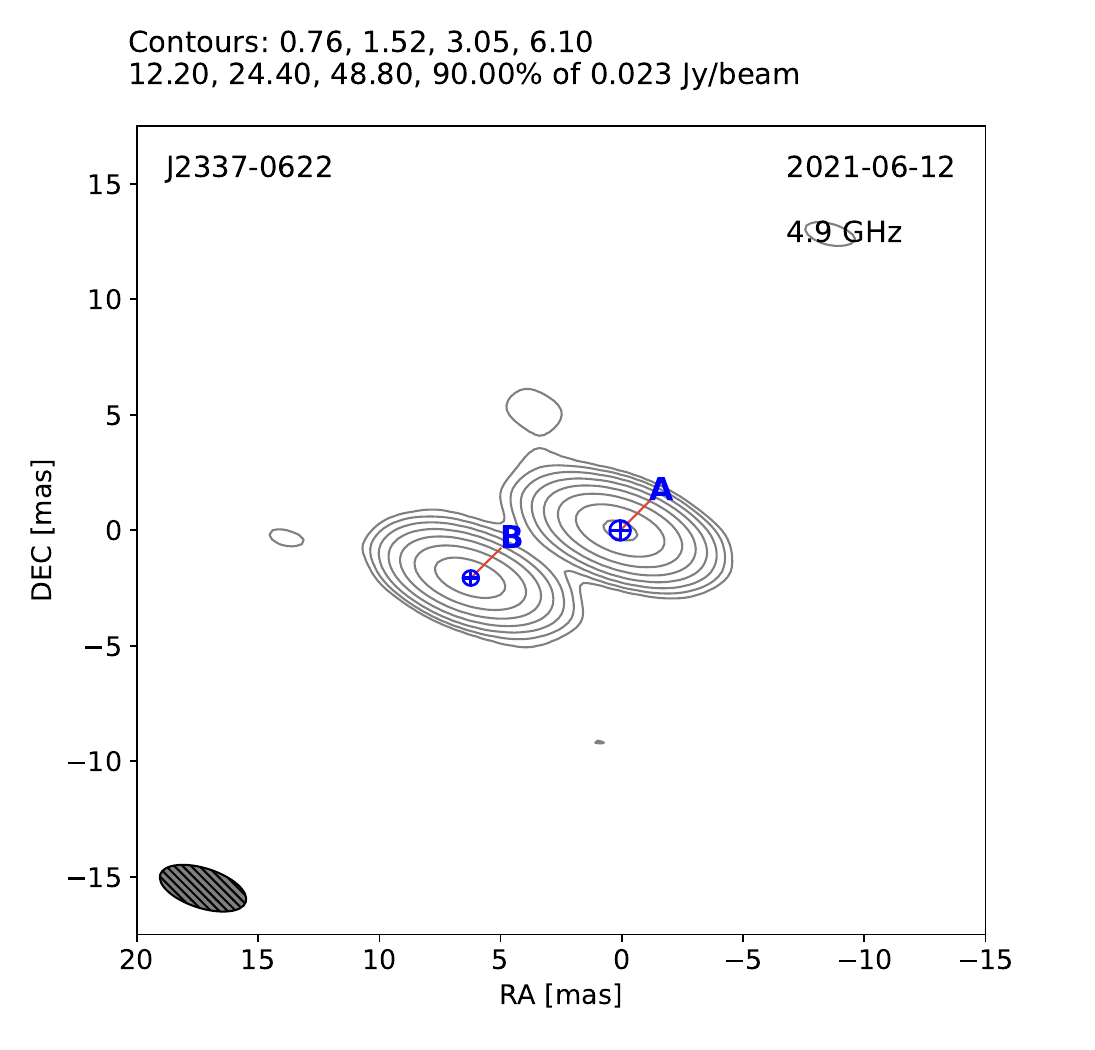}
    \includegraphics[width=0.4\textwidth]{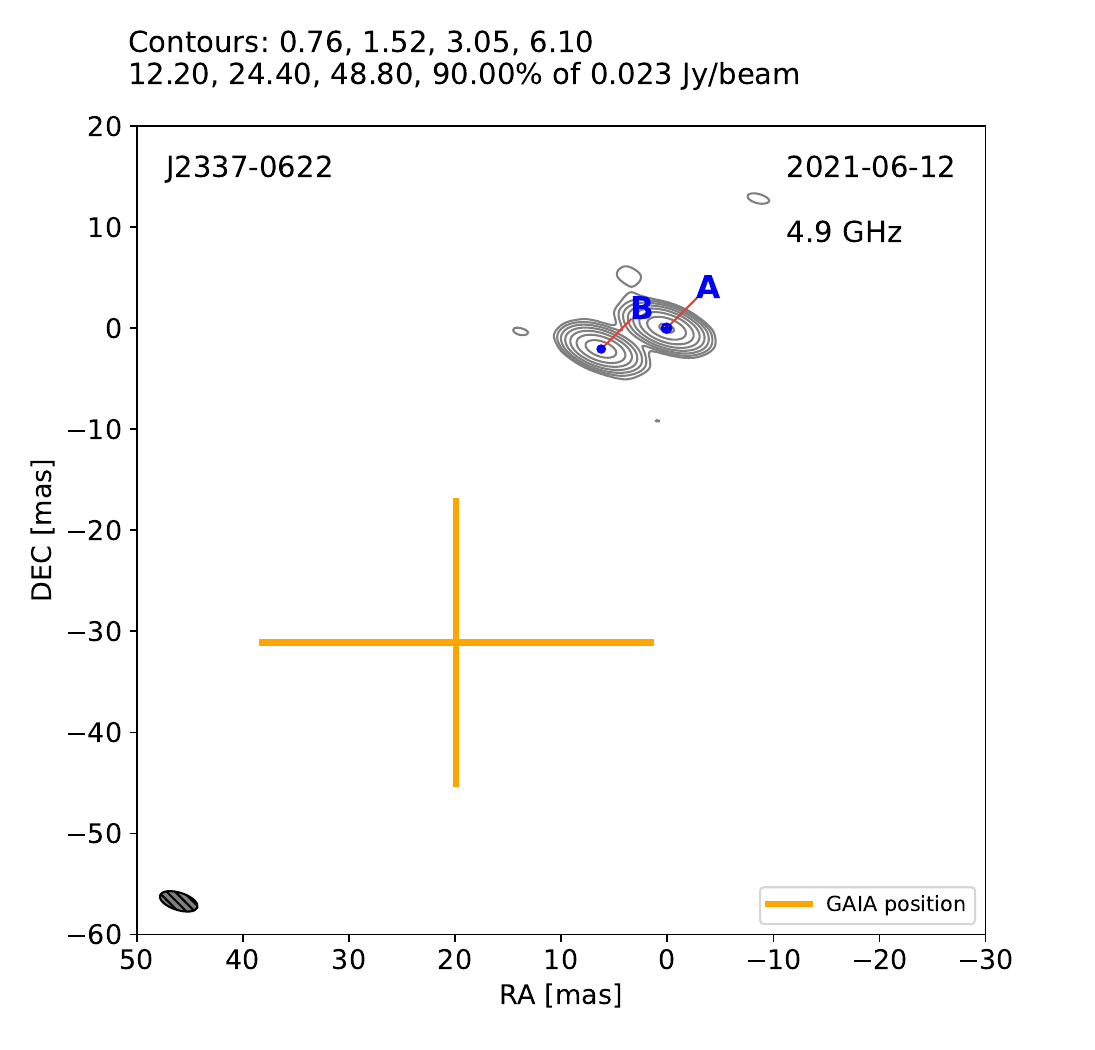}
    \caption{\textit{Left:} J2337-0622 EVN 4.9\,GHz data; \textit{Right:} EVN 4.9\,GHz data with GAIA position (orange cross). The contours start at four times the image rms noise of 0.430\,mJy/beam and increase by factors of two. The restoring beam size is 1.71$\times$3.72\,mas at $71.0$\,$^\circ$ PA. The source is not detected at 22.2\,GHz.}
    \label{fig:J2337}
\end{figure*}

\begin{table*}[h]
    \caption{Derived quantities for J2337-0622. See Table~\ref{tab:J2209} for a detailed description of the columns.}
    \vspace*{3mm}
    \adjustbox{width=1\textwidth}{%
    \label{tab:J2337}
    \centering
    \begin{tabular}{|| c c c c c | c c | c c | c c | c c | c c | c ||} 
        \hline
        Epochs & Frequency & Component & Flux  & Error  & Flux  & Error & FWHM  & Error & $\log(T_\mathrm{b,obs}/\mathrm{K})$ & Error  & SB    & Error & Distance & Error & Spectral Index \\ [0.5ex] 
               & [GHz]     &           & [mJy] & [mJy]  & Ratio &       & [mas] & [mas] &                                     &        & Ratio &       & [mas]    & [mas] &                \\
        \hline\hline
2015-12-11 & 4.3 & A & 23.4 & 3.5 & 1.90 & 0.37 & 1.20 & 0.10 & 9.08 & 0.10 & 2.75 & 1.00 & 6.75 & 0.46 & -0.34 \\
- & -   & B & 12.3 & 2.3 & -   & -   & 1.45 & 0.18 & 8.64 & 0.14 & -   & -   & -   & -   & -0.33 \\
        \hline
2015-12-11 & 7.6 & A & 19.3 & 3.0 & 1.89 & 0.40 & 0.73 & 0.06 & 8.94 & 0.10 & 16.74 & 6.43 & 6.23 & 0.65 & - \\
- & -   & B & 10.2 & 2.0 & -   & -   & 2.16 & 0.29 & 7.72 & 0.15 & -   & -   & -   & -   & - \\
        \hline\hline
2018-07-14 & 4.3 & A & 23.0 & 5.4 & 1.67 & 0.59 & 0.72 & 0.12 & 9.52 & 0.17 & 3.13 & 2.03 & 6.66 & 0.65 & -0.71 \\
- & -   & B & 13.8 & 4.1 & -   & -   & 0.98 & 0.21 & 9.02 & 0.23 & -   & -   & -   & -   & -1.59 \\
        \hline
2018-07-14 & 7.6 & A & 15.4 & 2.3 & 2.73 & 0.65 & 0.34 & 0.03 & 9.51 & 0.10 & 41.20 & 19.47 & 6.33 & 0.37 & - \\
- & -   & B & 5.6 & 1.3 & -   & -   & 1.31 & 0.25 & 7.90 & 0.19 & -   & -   & -   & -   & - \\
        \hline\hline
2020-11-20 & 4.3 & A & 38.8 & 7.6 & 2.95 & 1.00 & 1.28 & 0.16 & 9.24 & 0.14 & 2.06 & 1.25 & 7.14 & 1.06 & - \\
- & -   & B & 13.1 & 4.1 & -   & -   & 1.07 & 0.23 & 8.92 & 0.23 & -   & -   & -   & -   & - \\
        \hline\hline
2021-10-30 & 4.3 & A & 37.1 & 7.5 & 2.18 & 0.70 & 1.40 & 0.19 & 9.14 & 0.15 & 4.72 & 2.83 & 6.41 & 0.97 & -1.20 \\
- & -   & B & 17.0 & 4.8 & -   & -   & 2.05 & 0.44 & 8.47 & 0.22 & -   & -   & -   & -   & -2.41 \\
        \hline
2021-10-30 & 7.6 & A & 18.9 & 4.3 & 4.29 & 2.06 & 0.69 & 0.10 & 8.98 & 0.17 & $<$2.79 & - & 6.58 & 0.69 & - \\
- & -   & B & 4.4 & 2.0 & -   & -   & $<$0.56 & - & $>$8.53 & - & -   & -   & -   & -   & - \\
        \hline\hline
2021-06-12 & 4.9 & A & 26.6 & 3.1 & 1.38 & 0.12 & 0.86 & 0.04 & 9.31 & 0.06 & 0.79 & 0.12 & 6.49 & 0.17 & - \\
- & -   & B & 19.3 & 2.3 & -   & -   & 0.65 & 0.03 & 9.42 & 0.07 & -   & -   & -   & -   & - \\
        \hline\hline
2022-07-07 & 4.3 & A & 24.9 & 4.1 & 1.58 & 0.33 & 1.49 & 0.15 & 8.91 & 0.11 & 2.29 & 0.90 & 6.19 & 0.65 & 0.09 \\
- & -   & B & 15.7 & 3.0 & -   & -   & 1.80 & 0.24 & 8.55 & 0.14 & -   & -   & -   & -   & -0.63 \\
        \hline
2022-07-07 & 7.6 & A & 26.2 & 9.4 & 2.38 & 1.56 & $<$0.97 & - & $>$8.82 & - & $\sim$7.16 & 8.45 & 4.21 & 2.14 & - \\
- & -   & B & 11.0 & 6.3 & -   & -   & $<$1.69 & - & $>$7.96 & - & -   & -   & -   & -   & - \\
        \hline
    \end{tabular}
    }
\end{table*}



\section{Spectra}

\begin{figure*}[ht]
    \centering
    \caption{Radio spectra obtained from total flux densities in the VLBI clean maps from Astrogeo (black points) and the new EVN data (red points). We note that no correction has been made for different resolution or $(u,v)$-range in these observations and these plots should be understood to be qualitative in nature.}
    \includegraphics[width=0.9\textwidth]{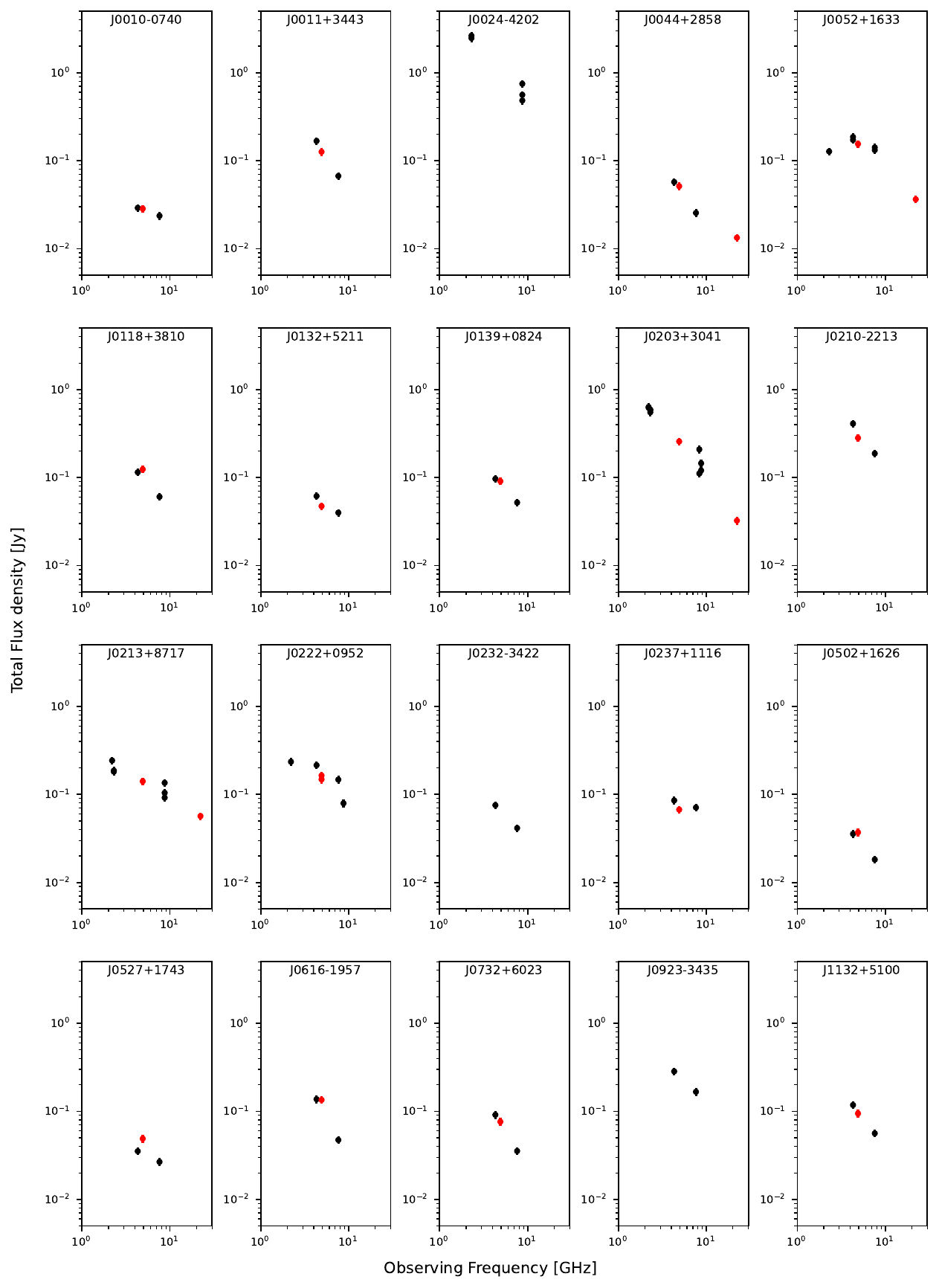}
    \label{fig:VLBI_spectra_1}
\end{figure*}
\newpage
\begin{figure*}[ht]
    \centering
    \caption{Fig.~\ref{fig:VLBI_spectra_1} continued.}
    \includegraphics[width=0.9\textwidth]{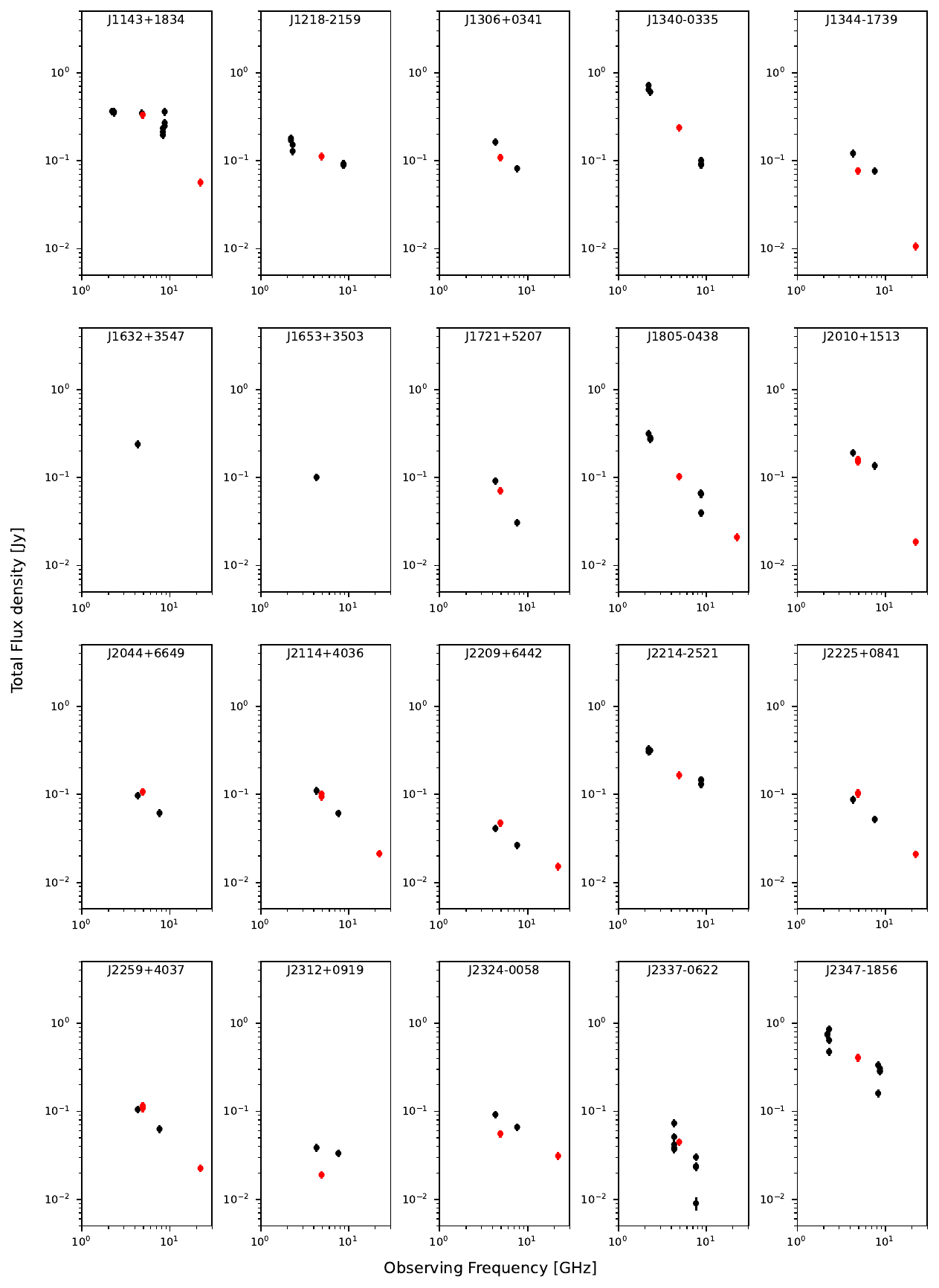}
    \label{fig:VLBI_spectra_2}
\end{figure*}

\section{Optical images}

\begin{figure*}[!htb]
    \caption{Optical images for the 40 initial milli-lens candidates, with the VLBI position indicated by the crosses.}
    \label{fig:optical1}
    \centering
    \vspace{0.14cm}
    \begin{subfigure}[t]{0.22\textwidth}
        \centering
        \includegraphics[width=0.95\textwidth]{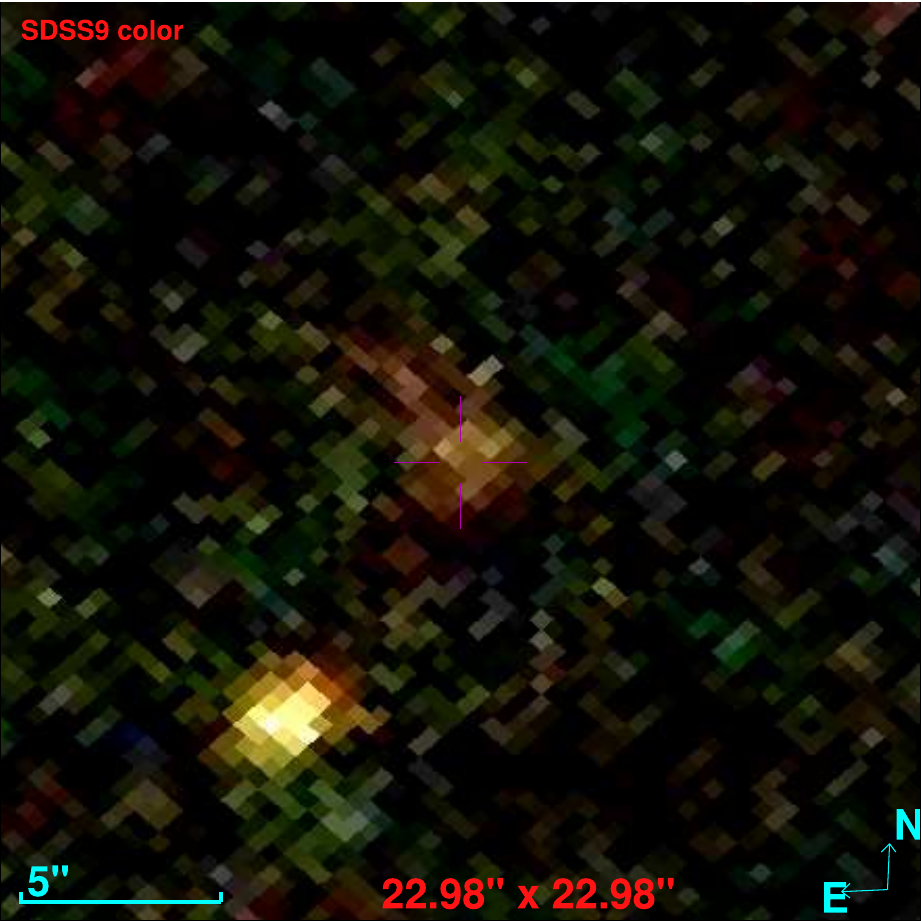}
        \caption{\tiny{J0010-0740 in SDSS DR9.}}
    \end{subfigure}%
    \hfill
    \begin{subfigure}[t]{0.22\textwidth}
        \centering
        \includegraphics[width=0.95\textwidth]{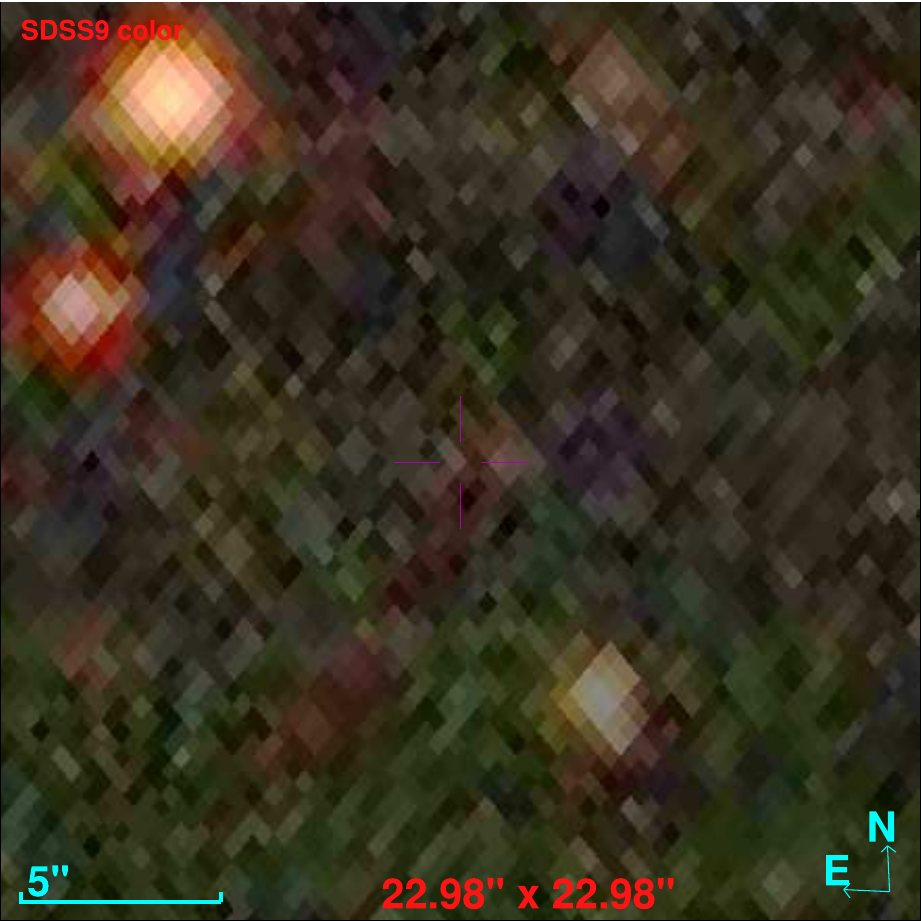}
        \caption{\tiny{J0011+3443 in SDSS DR9.}}
    \end{subfigure}%
    \hfill
    \begin{subfigure}[t]{0.22\textwidth}
        \centering
        \includegraphics[width=0.9\textwidth]{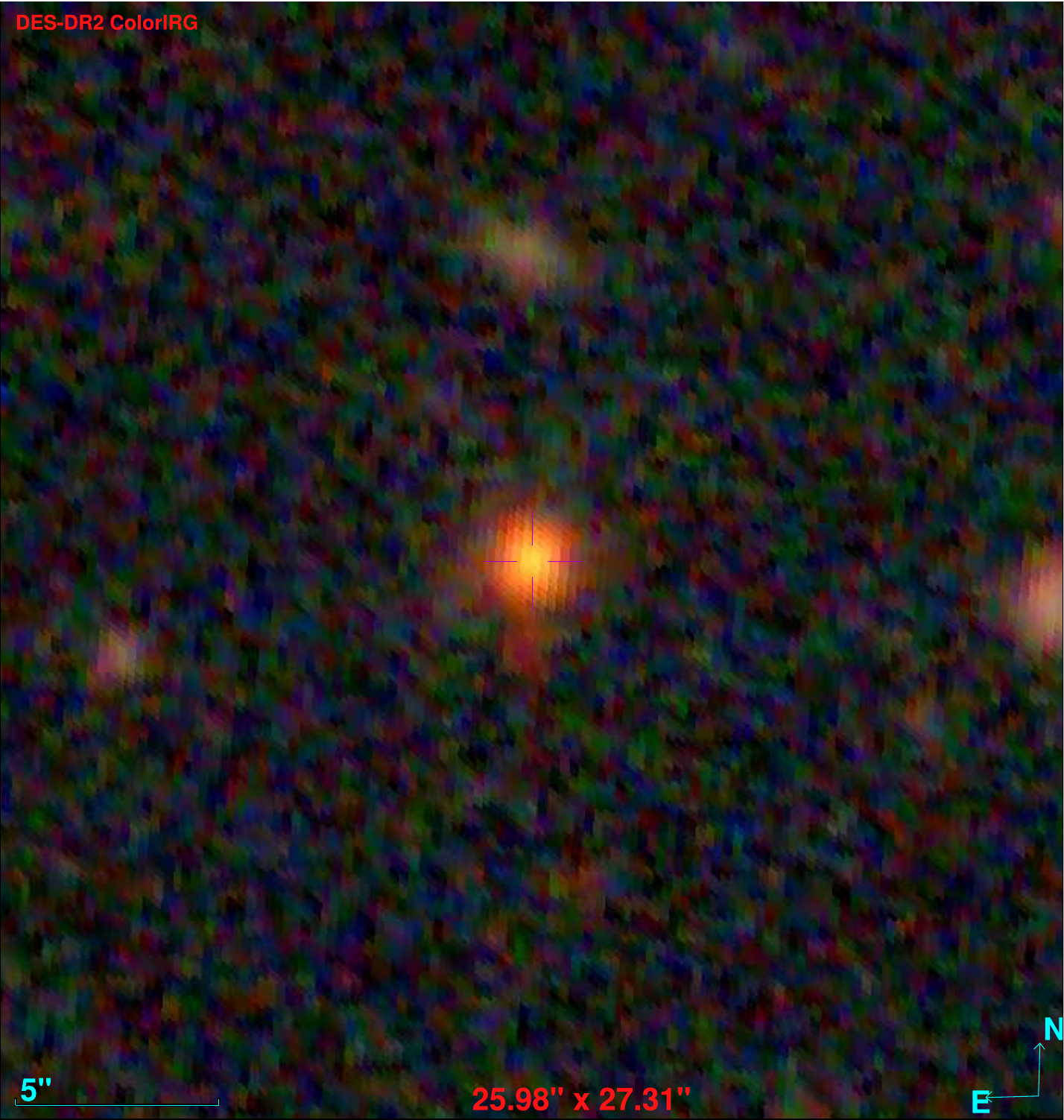}
        \caption{\tiny{J0024-4202 in DES DR2.}}
    \end{subfigure}%
    \hfill
    \begin{subfigure}[t]{0.22\textwidth}
        \centering
        \includegraphics[width=0.95\textwidth]{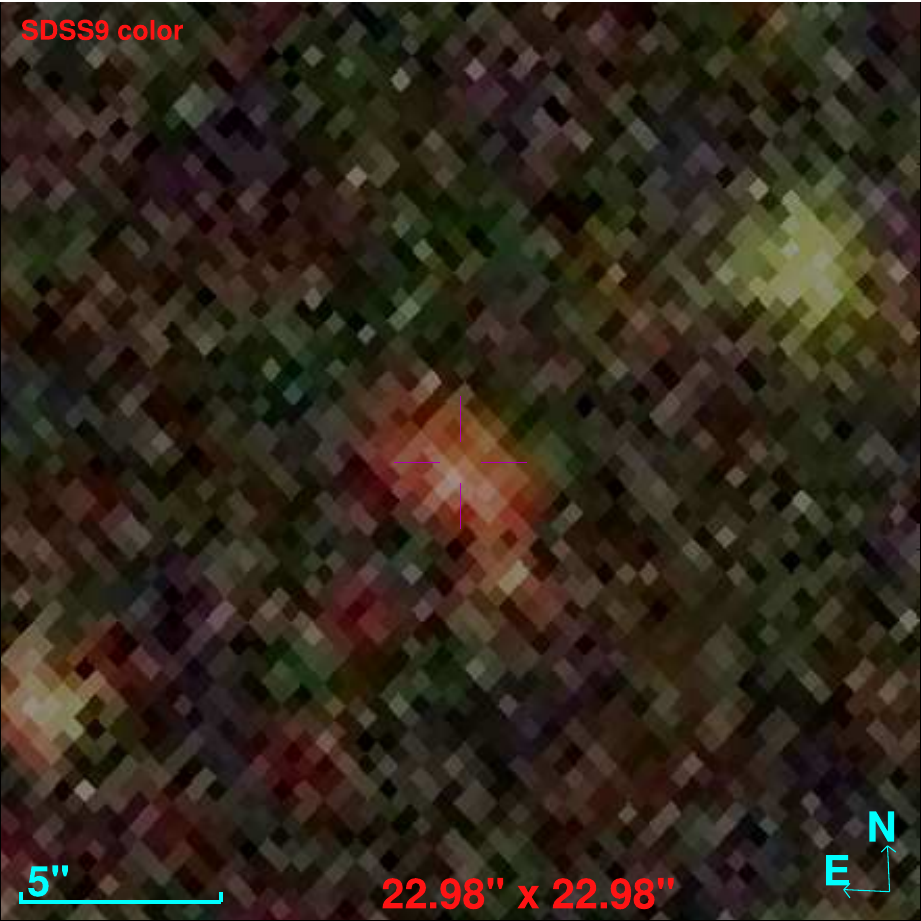}
        \caption{\tiny{J0044+2858 in SDSS DR9.}}
    \end{subfigure}
    
    \vspace{0.14cm}
    
    \begin{subfigure}[t]{0.22\textwidth}
        \centering
        \includegraphics[width=0.95\textwidth]{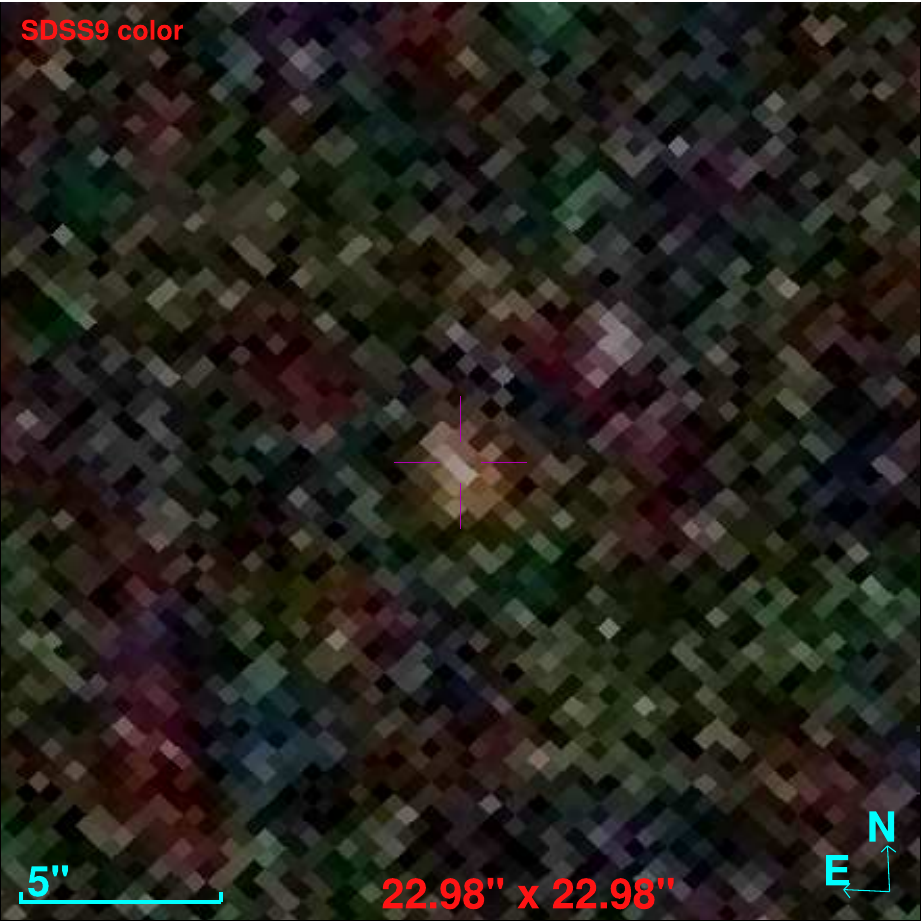}
        \caption{\tiny{J0052+1633 in SDSS DR9.}}
    \end{subfigure}%
    \hfill
    \begin{subfigure}[t]{0.22\textwidth}
        \centering
        \includegraphics[width=0.9\textwidth]{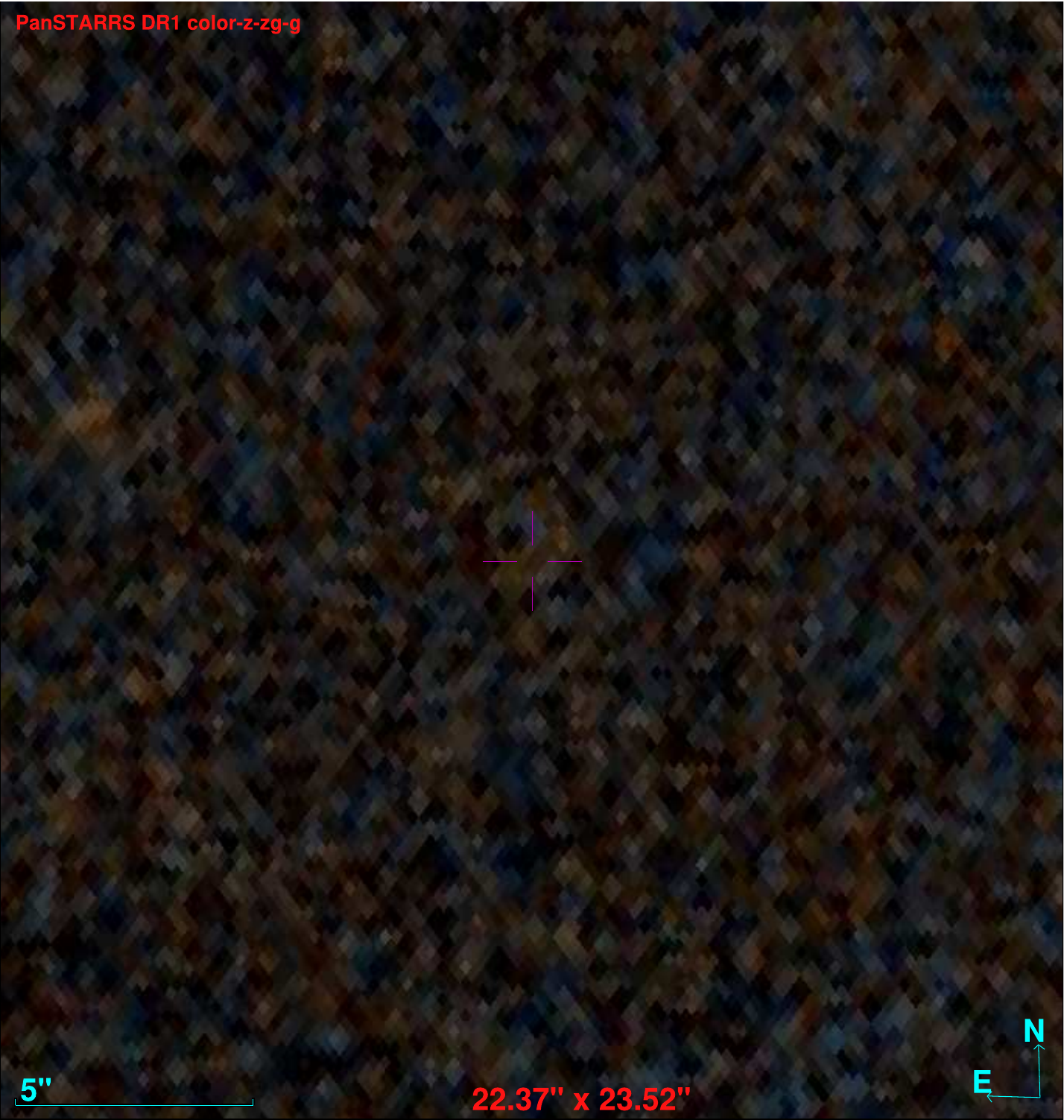}
        \caption{\tiny{J0118+3810 in PanSTARRS DR1 (no z).}}
    \end{subfigure}%
    \hfill
    \begin{subfigure}[t]{0.22\textwidth}
        \centering
        \includegraphics[width=0.95\textwidth]{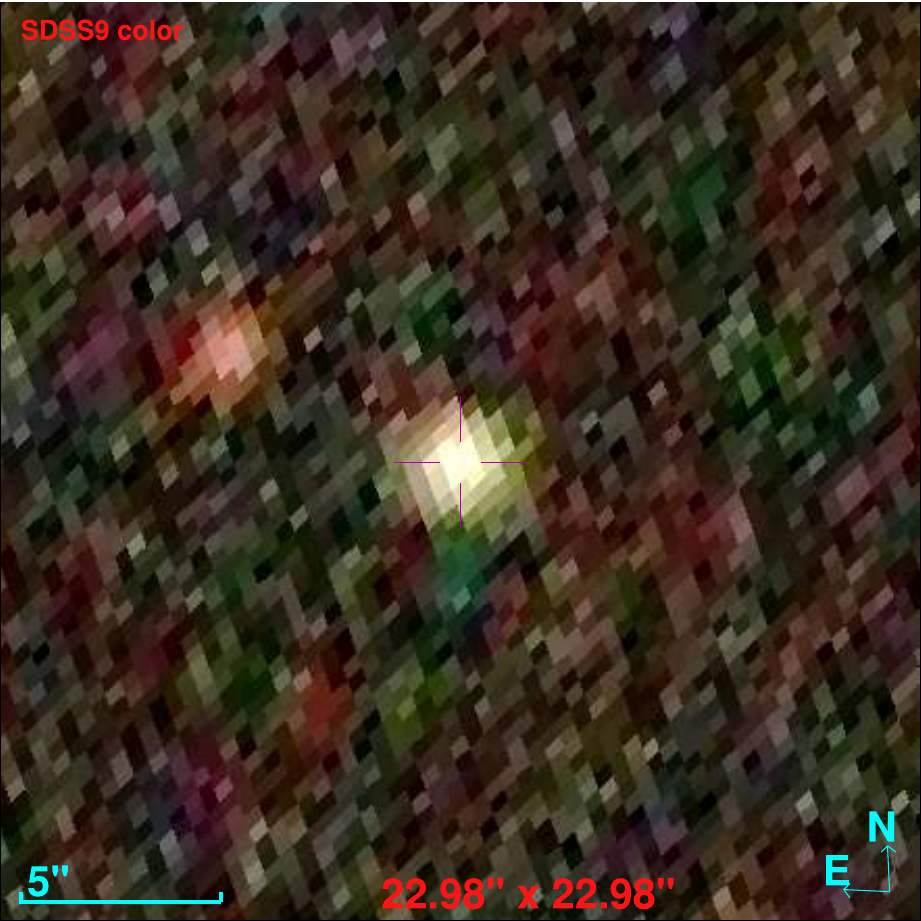}
        \caption{\tiny{J0132+5211 in SDSS DR9.}}
    \end{subfigure}%
    \hfill
    \begin{subfigure}[t]{0.22\textwidth}
        \centering
        \includegraphics[width=0.95\textwidth]{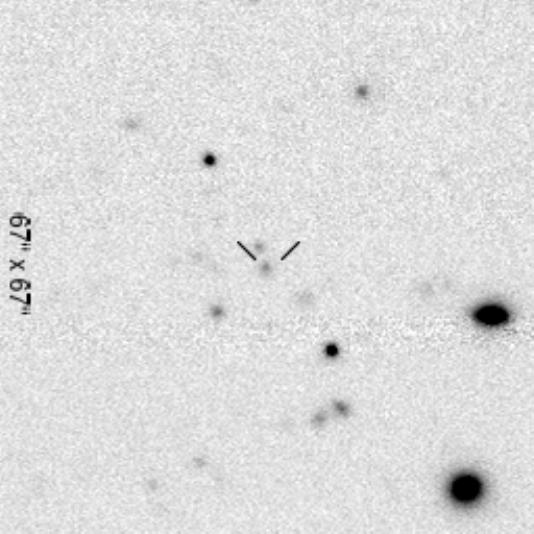}
        \caption{\tiny{J0139+0824 in DES DR8.}}
    \end{subfigure}
    
    \vspace{0.14cm}
    
    \begin{subfigure}[t]{0.22\textwidth}
        \centering
        \includegraphics[width=0.95\textwidth]{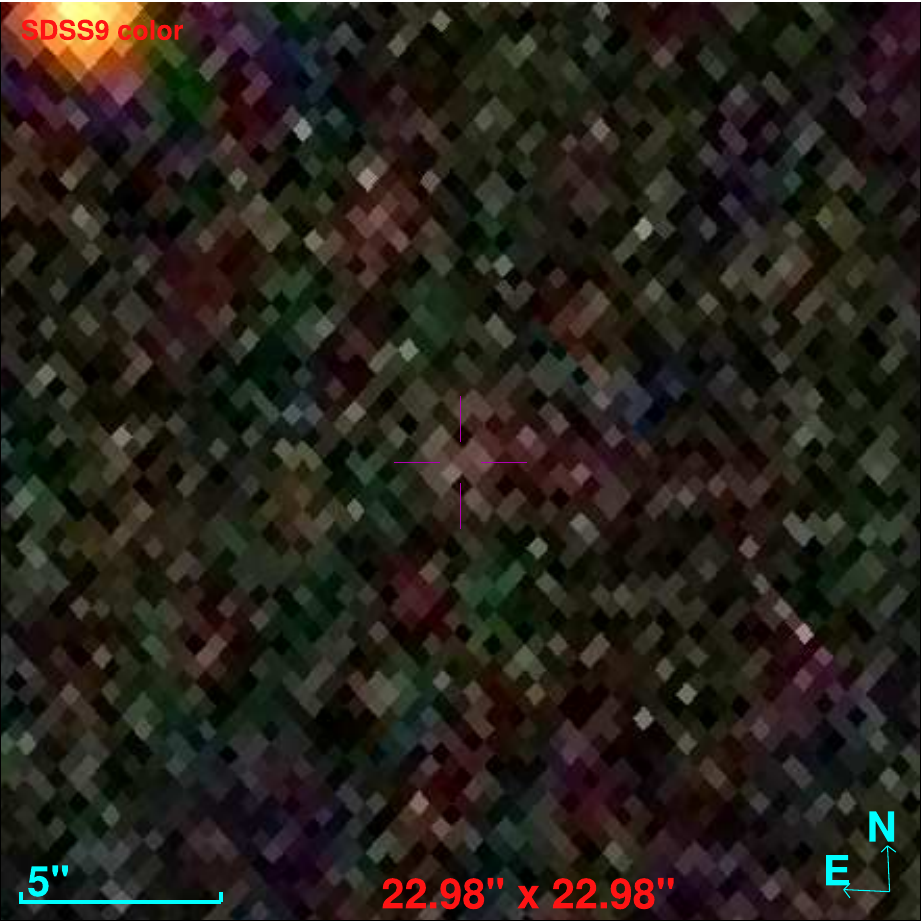}
        \caption{\tiny{J0203+3041 in SDSS DR9.}}
    \end{subfigure}%
    \hfill
    \begin{subfigure}[t]{0.22\textwidth}
        \centering
        \includegraphics[width=0.95\textwidth]{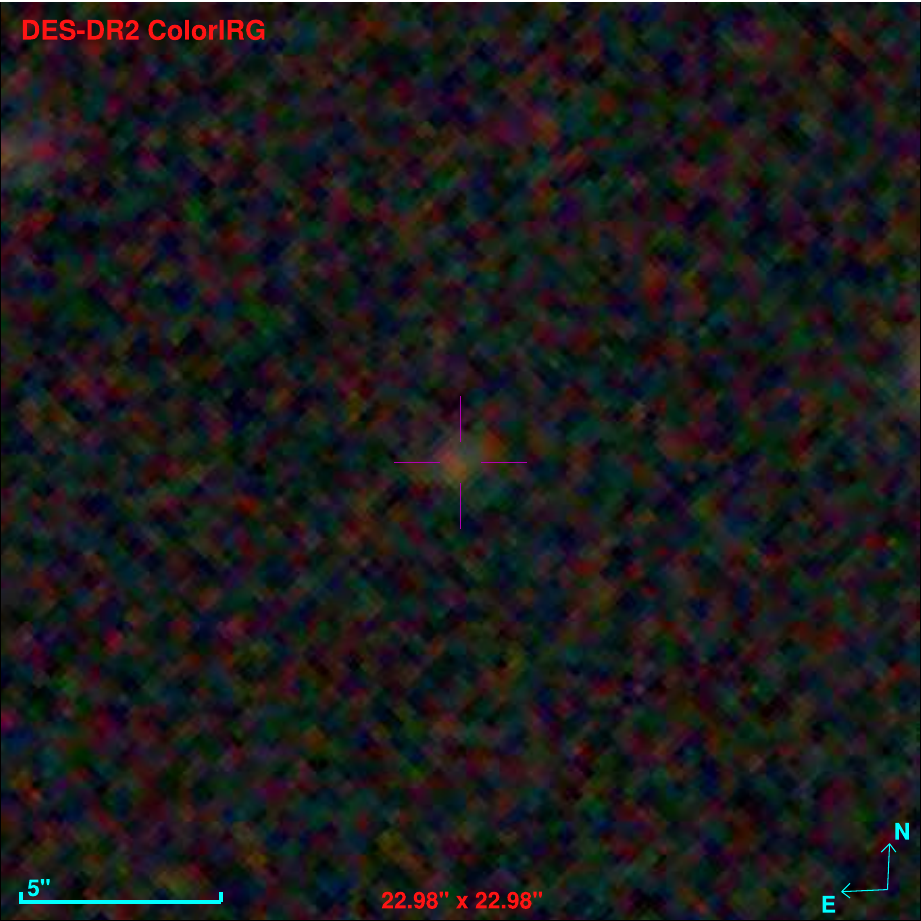}
        \caption{\tiny{J0210-2213 in SDSS DR9.}}
    \end{subfigure}%
    \hfill
    \begin{subfigure}[t]{0.22\textwidth}
        \centering
        \includegraphics[width=0.9\textwidth]{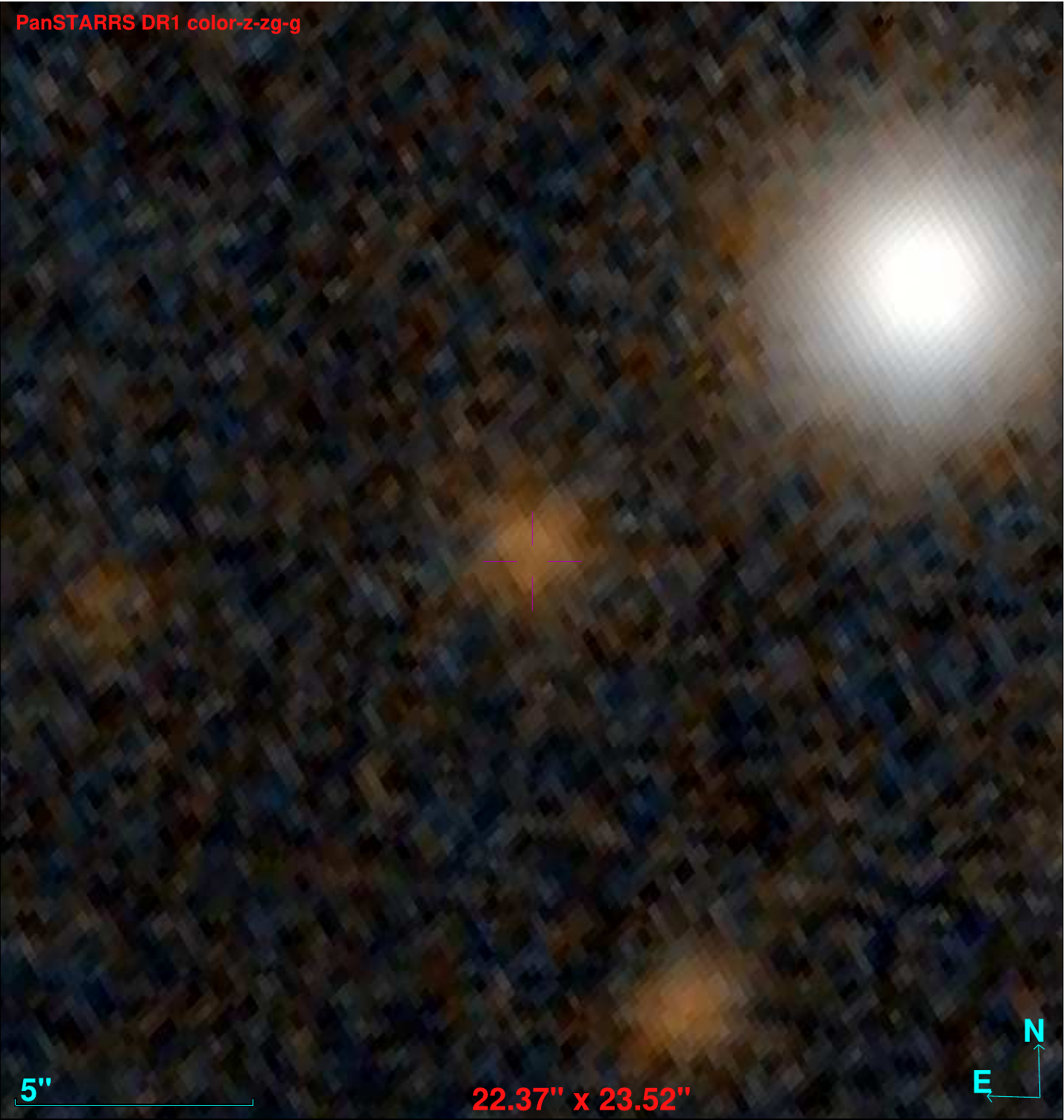}
        \caption{\tiny{J0213+8717 in PanSTARRS DR1 (no z).}}
    \end{subfigure}%
    \hfill
    \begin{subfigure}[t]{0.22\textwidth}
        \centering
        \includegraphics[width=0.9\textwidth]{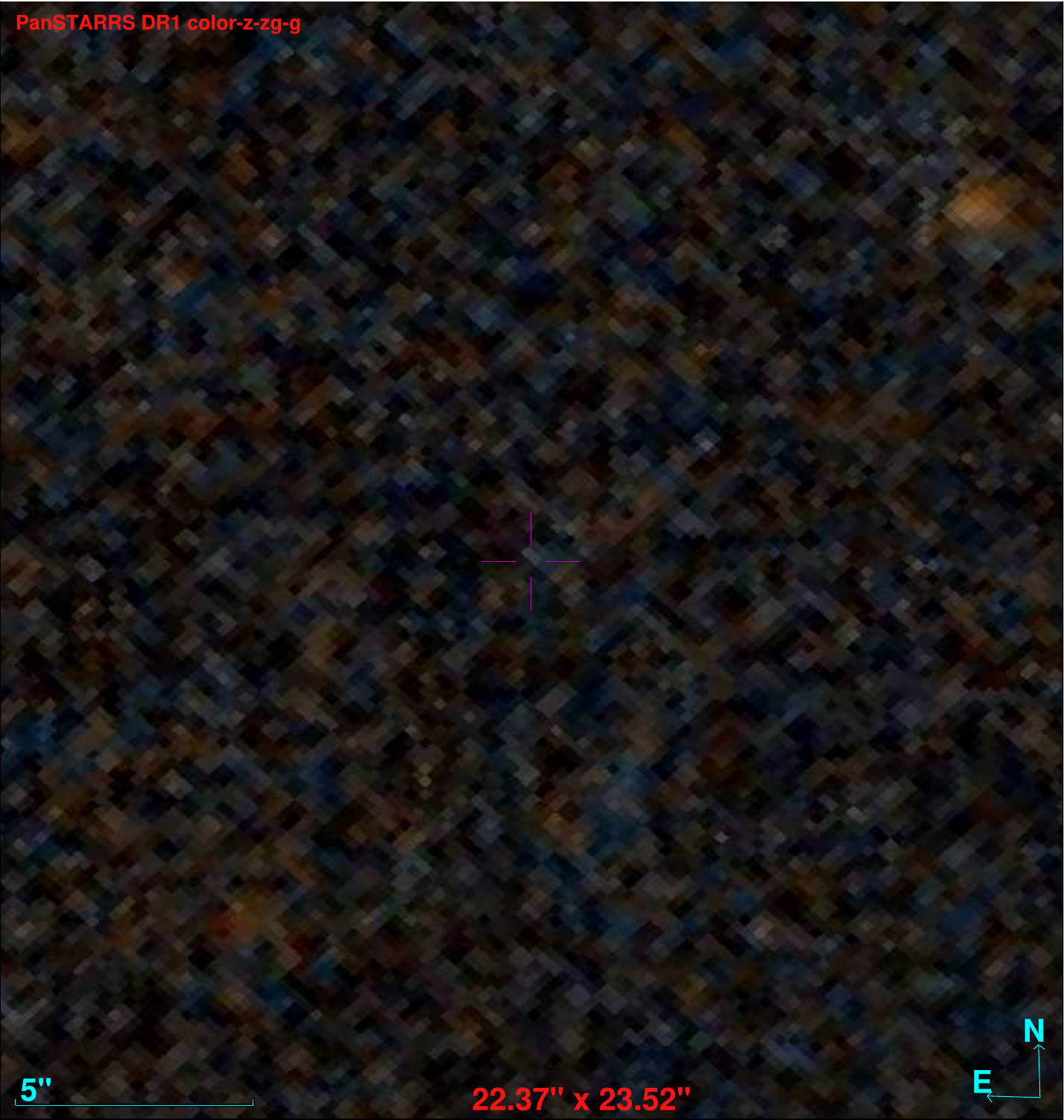}
        \caption{\tiny{J0222+0952 in PanSTARRS DR1 (no counterpart, no z).}}
    \end{subfigure}
    
    \vspace{0.14cm}
    
    \begin{subfigure}[t]{0.22\textwidth}
        \centering
        \includegraphics[width=0.95\textwidth]{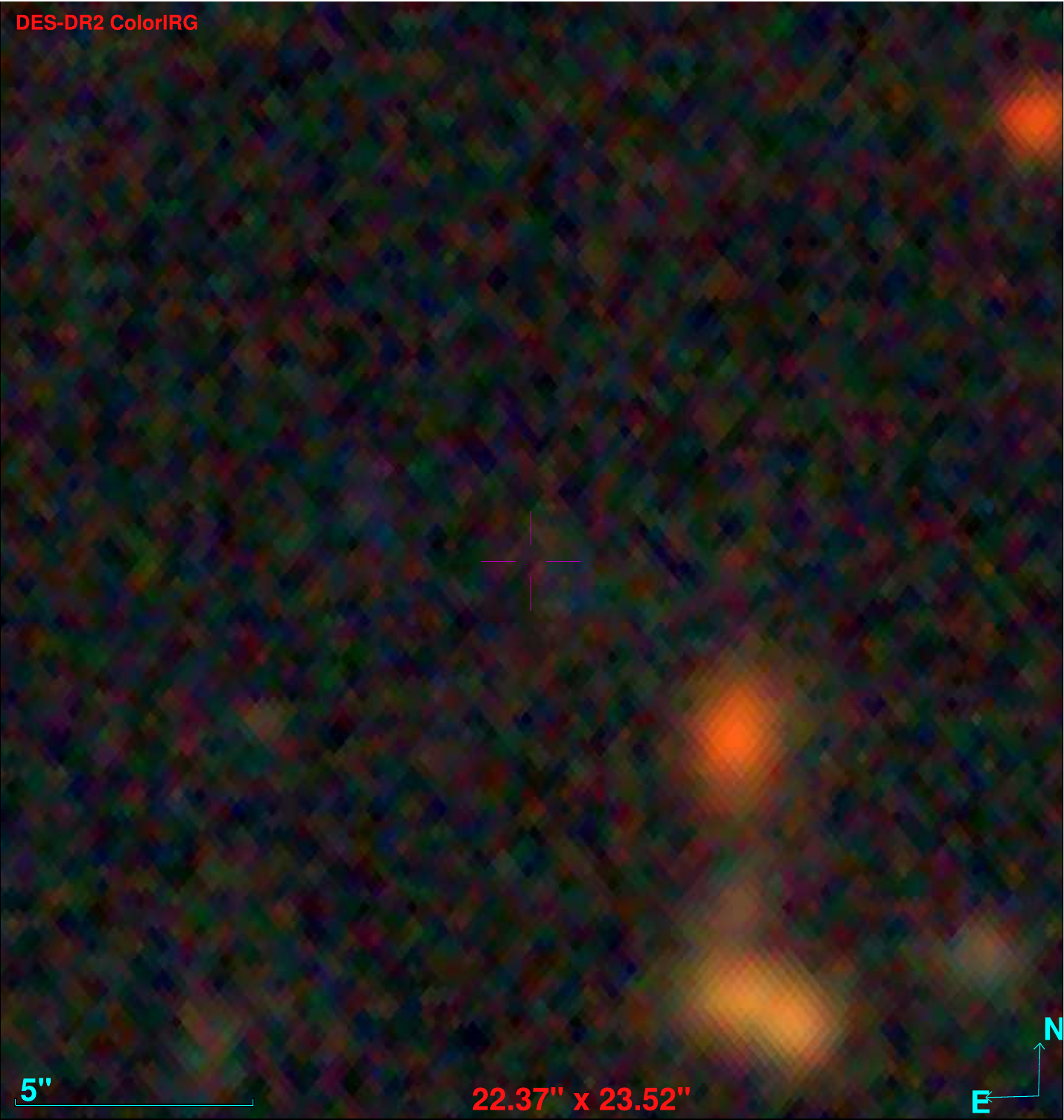}
        \caption{\tiny{J0232-3422 in DES DR2 (no z).}}
    \end{subfigure}%
    \hfill
    \begin{subfigure}[t]{0.22\textwidth}
        \centering
        \includegraphics[width=0.9\textwidth]{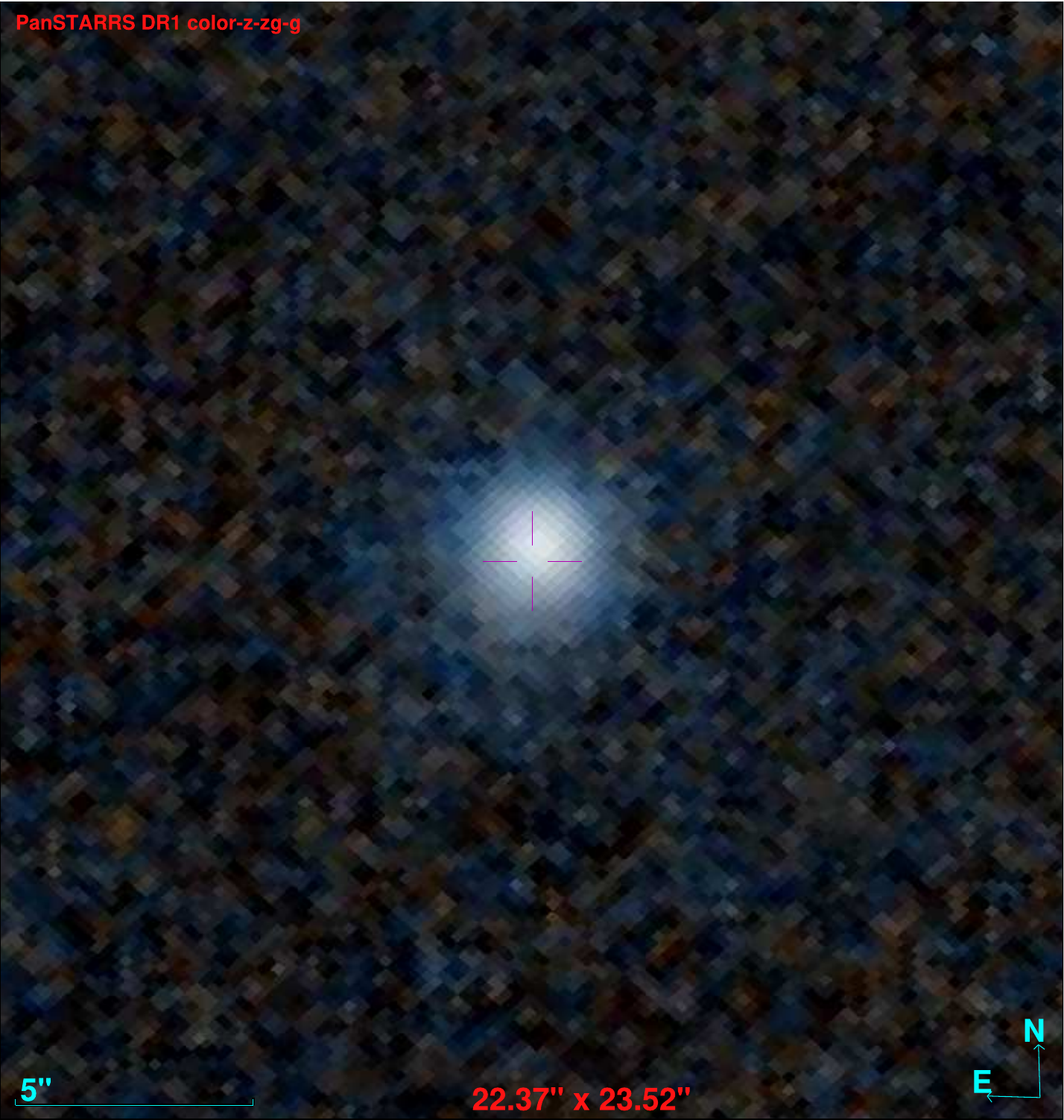}
        \caption{\tiny{J0237+1116 in PanSTARRS DR1.}}
    \end{subfigure}%
    \hfill
    \begin{subfigure}[t]{0.22\textwidth}
        \centering
        \includegraphics[width=0.9\textwidth]{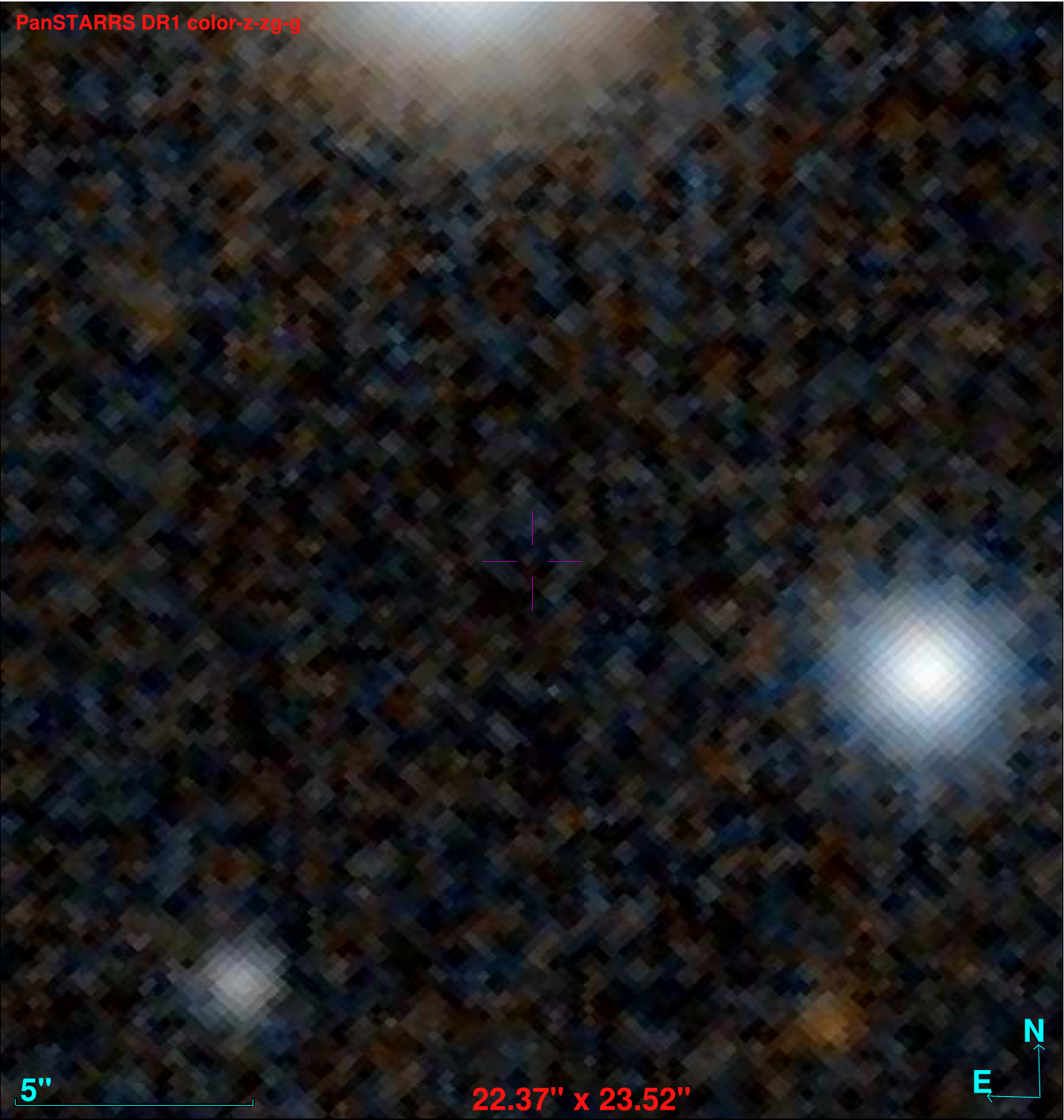}
        \caption{\tiny{J0502+1626 in PanSTARRS DR1 (no counterpart, no z).}}
    \end{subfigure}%
    \hfill
    \begin{subfigure}[t]{0.22\textwidth}
        \centering
        \includegraphics[width=0.95\textwidth]{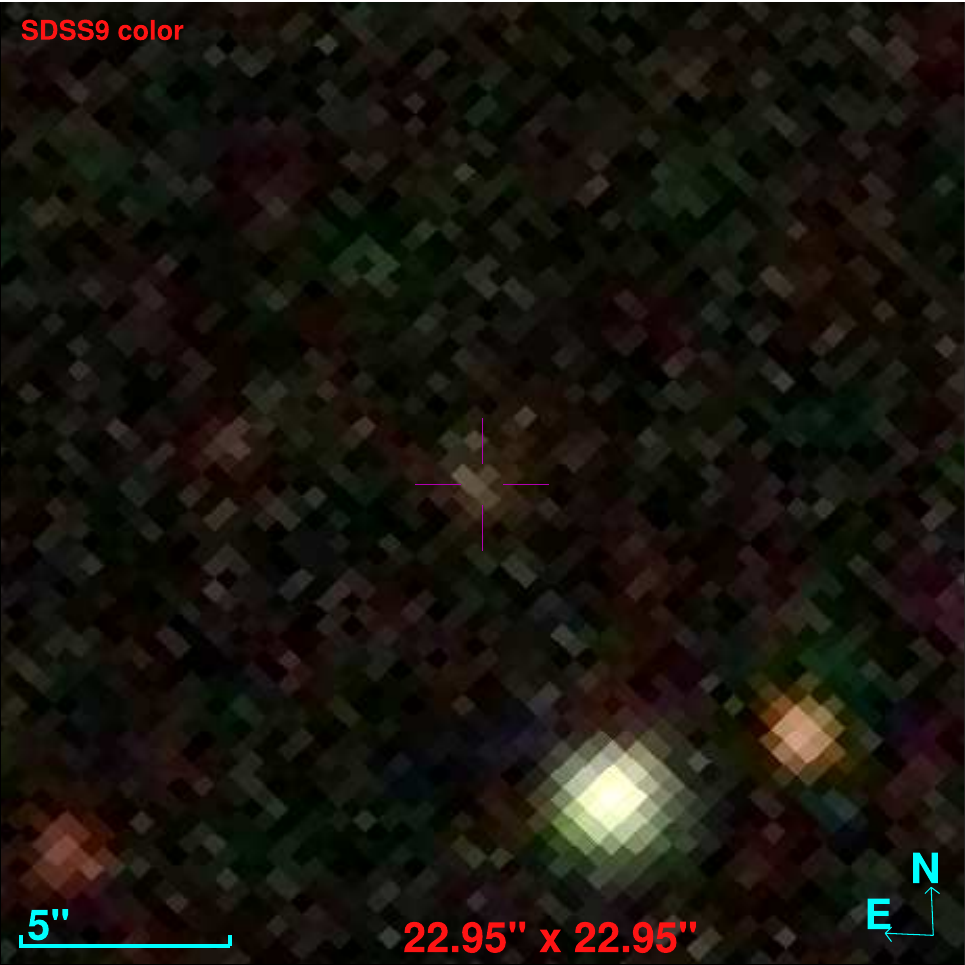}
        \caption{\tiny{J0527+1743 in SDSS DR9.}}
    \end{subfigure}
    
    \vspace{0.14cm}
    
    \begin{subfigure}[t]{0.22\textwidth}
        \centering
        \includegraphics[width=0.9\textwidth]{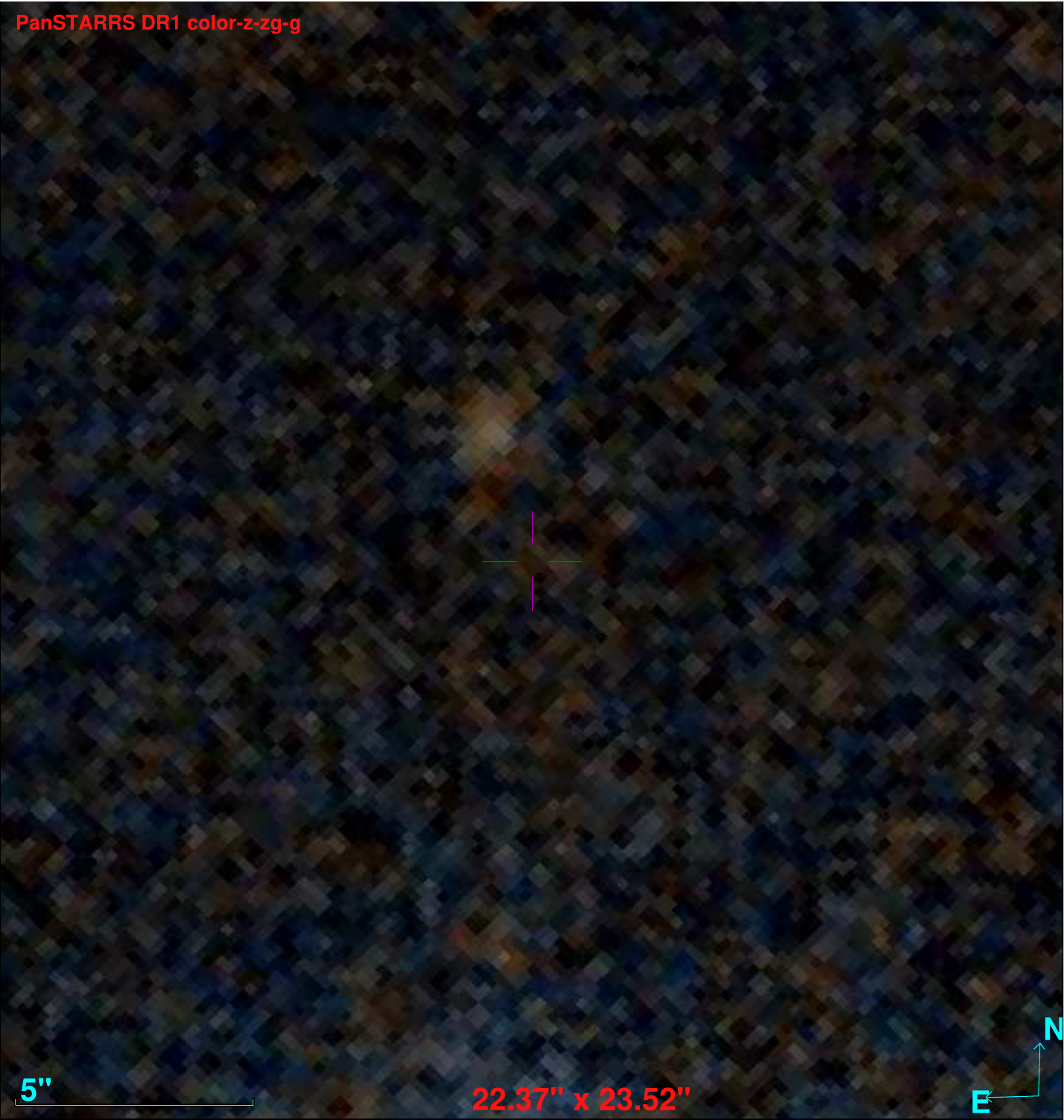}
        \caption{\tiny{J0616-1957 in PanSTARRS DR1 (no counterpart, no z).}}
    \end{subfigure}%
    \hfill
    \begin{subfigure}[t]{0.22\textwidth}
        \centering
        \includegraphics[width=0.95\textwidth]{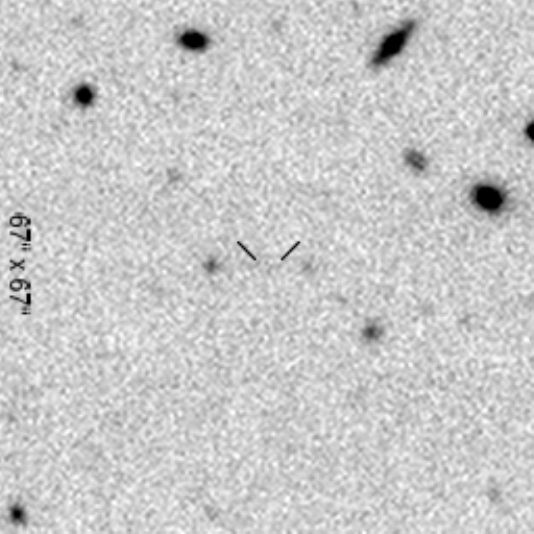}
        \caption{\tiny{J0732+6023 in DES DR8.}}
    \end{subfigure}%
    \hfill
    \begin{subfigure}[t]{0.22\textwidth}
        \centering
        \includegraphics[width=0.9\textwidth]{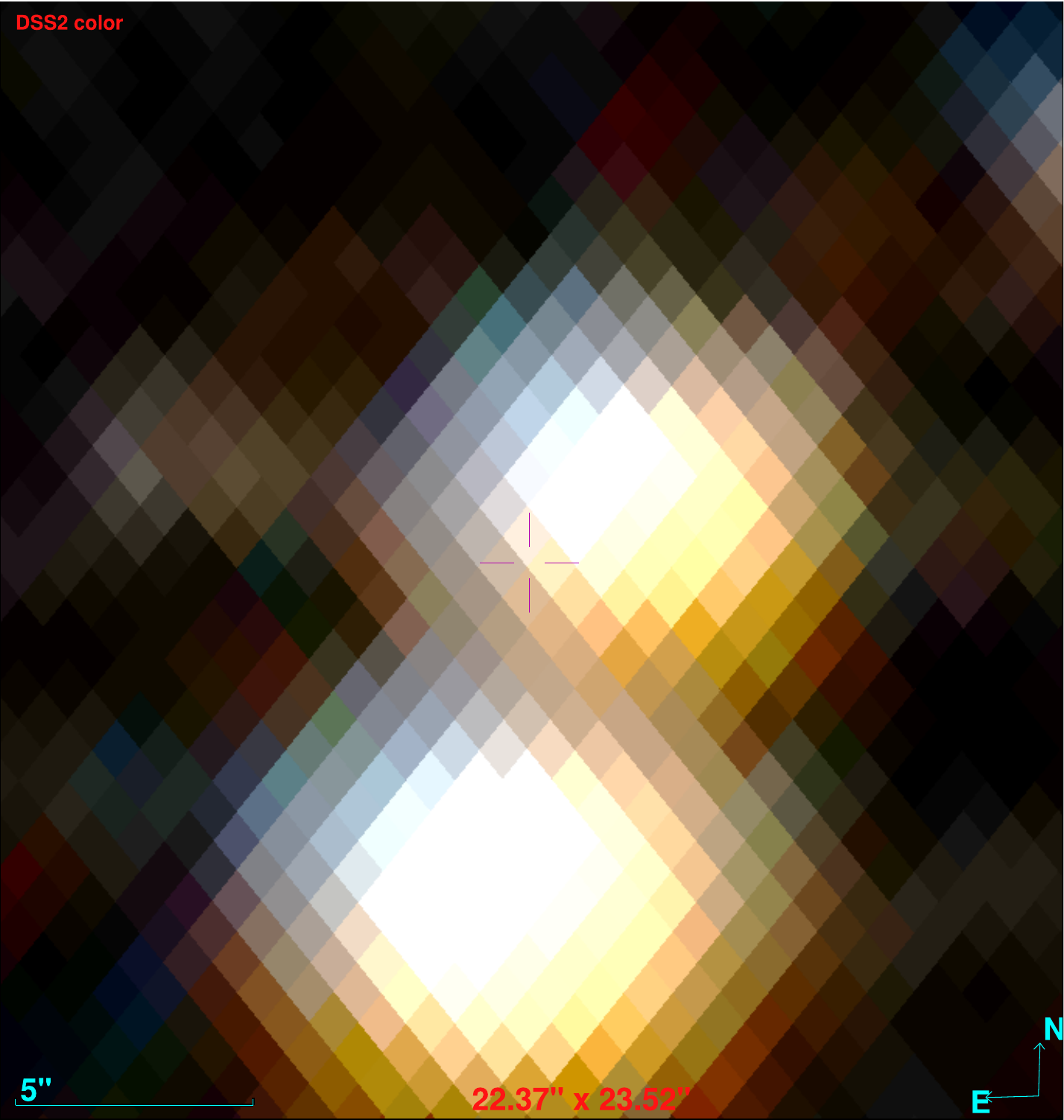}
        \caption{J0923-3435 in DSS2. Counterpart are two stars (no z).}
    \end{subfigure}%
    \hfill
    \begin{subfigure}[t]{0.22\textwidth}
        \centering
        \includegraphics[width=0.95\textwidth]{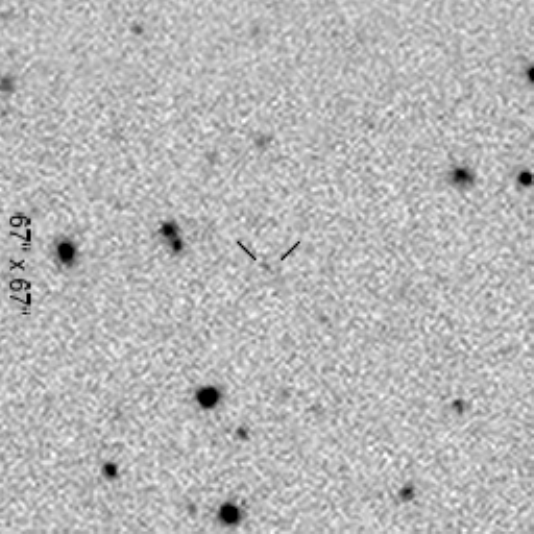}
        \caption{J1132+5100 in DES DR8.}
    \end{subfigure}
\end{figure*}

\newpage
\begin{figure*}[!htb]
    \caption{Fig.~\ref{fig:optical1} continued.}
    \label{fig:optical2}
    \centering
    \vspace{0.14cm}
    \begin{subfigure}[t]{0.22\textwidth}
        \centering
        \includegraphics[width=0.95\textwidth]{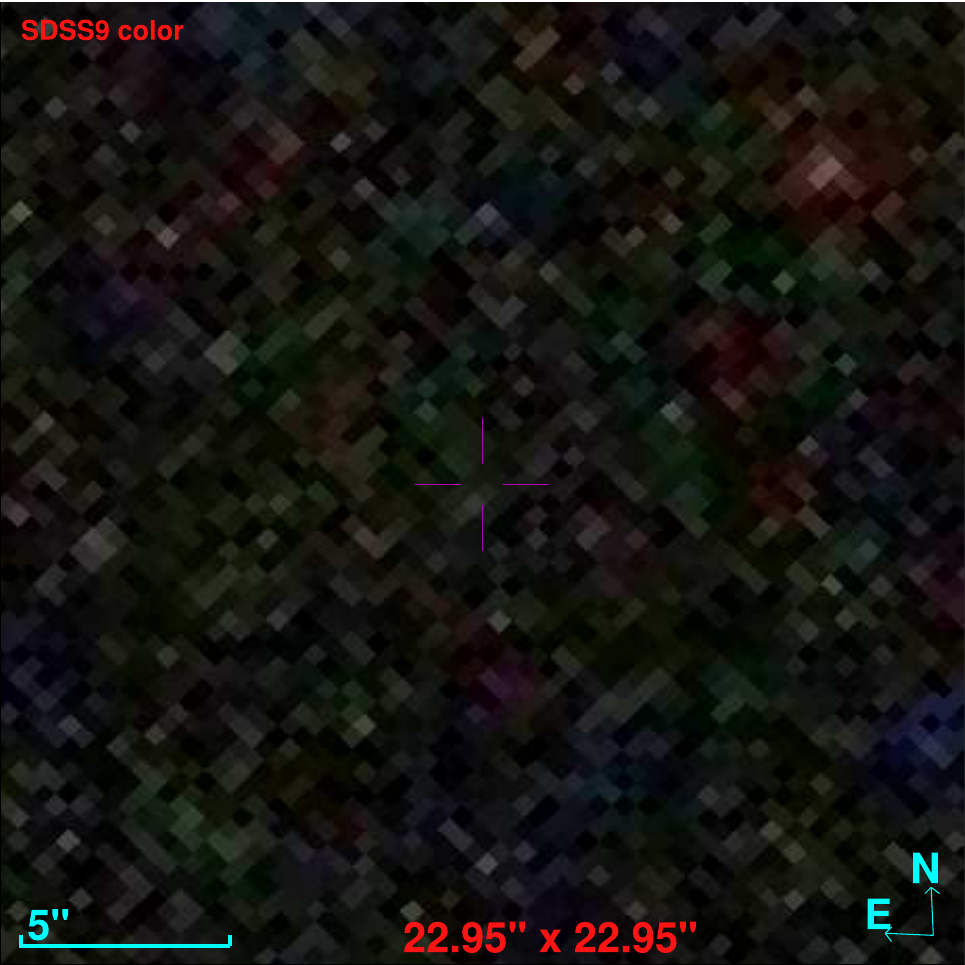}
        \caption{\tiny{J1143+1834 in SDSS DR9. Counterpart in DR 17 and DES DR10.}}
    \end{subfigure}%
    \hfill
    \begin{subfigure}[t]{0.22\textwidth}
        \centering
        \includegraphics[width=0.9\textwidth]{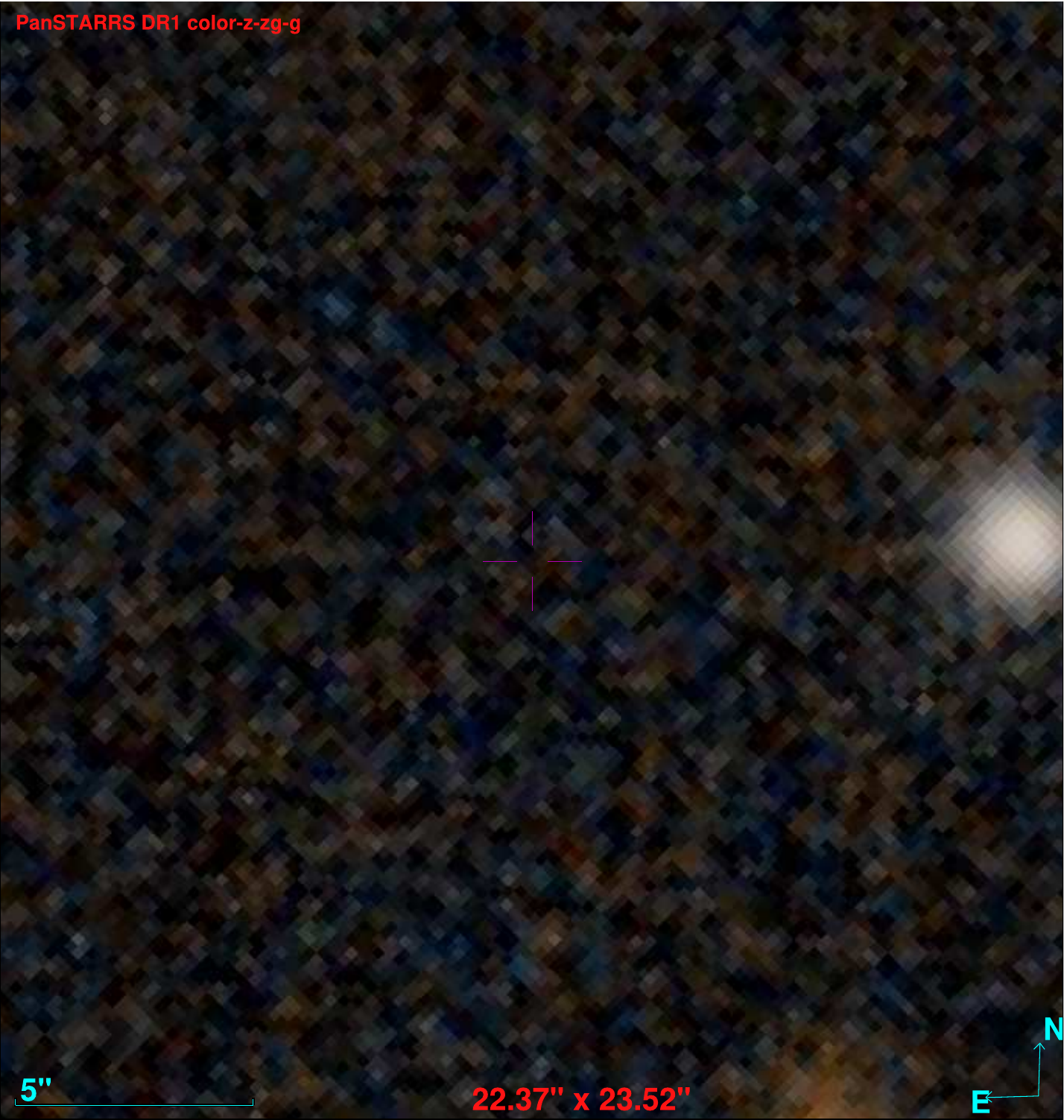}
        \caption{\tiny{J1218-2159 in PanSTARRS DR 1 (no counterpart, no z).}}
    \end{subfigure}%
    \hfill
    \begin{subfigure}[t]{0.22\textwidth}
        \centering
        \includegraphics[width=0.95\textwidth]{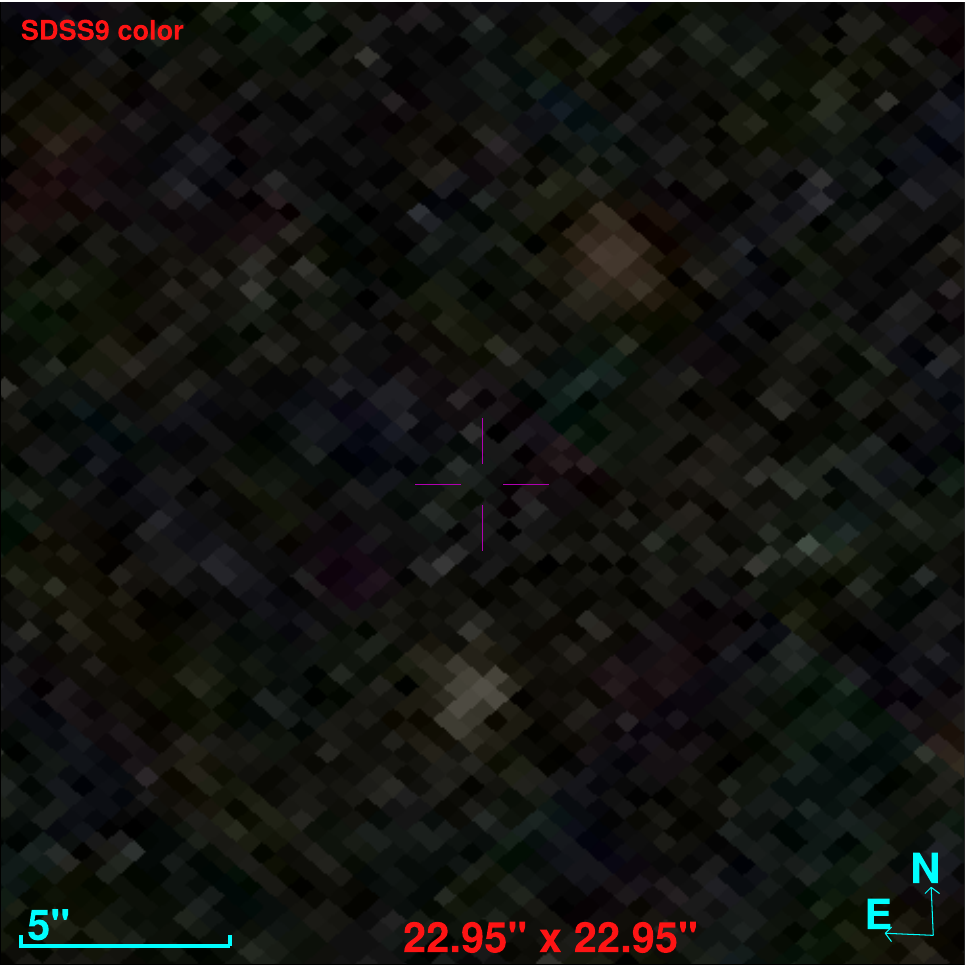}
        \caption{\tiny{J1306+0341 in SDSS DR9. (no counterpart, no z).}}
    \end{subfigure}%
    \hfill
    \begin{subfigure}[t]{0.22\textwidth}
        \centering
        \includegraphics[width=0.95\textwidth]{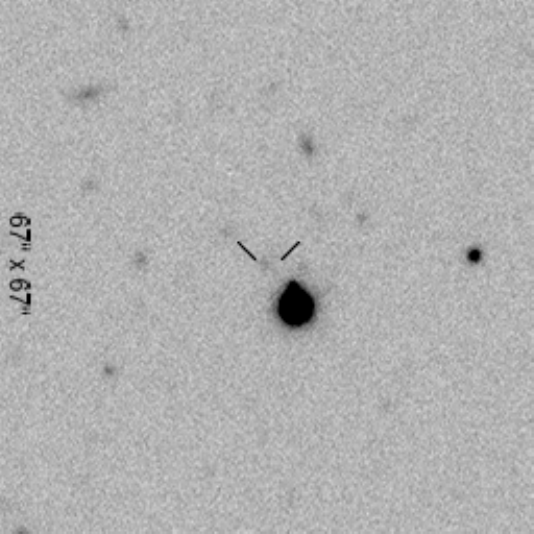}
        \caption{\tiny{J1340-0335 in DES DR8.}}
    \end{subfigure}
    
    \vspace{0.14cm}
    
    \begin{subfigure}[t]{0.22\textwidth}
        \centering
        \includegraphics[width=0.9\textwidth]{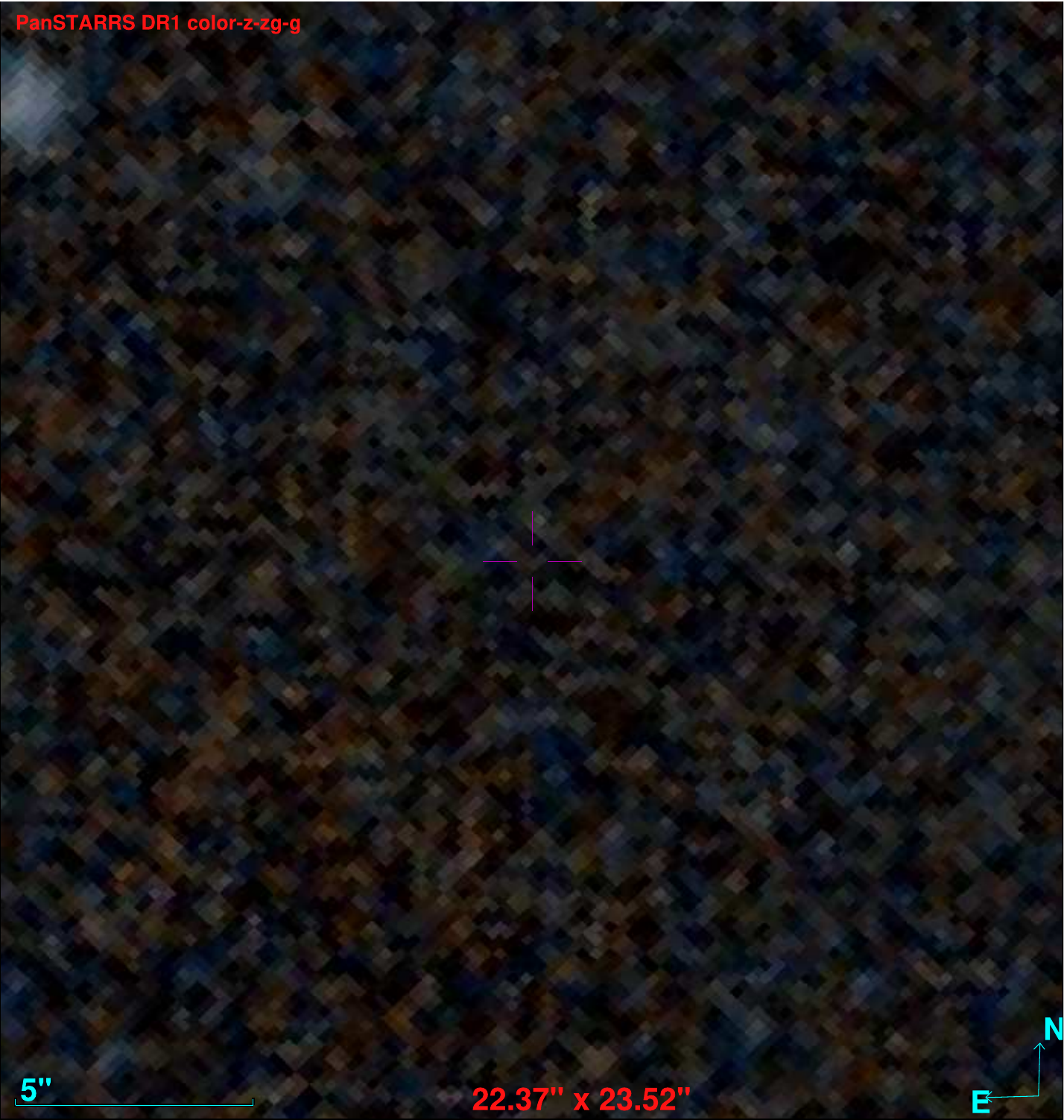}
        \caption{\tiny{J1344-1739 in PanSTARRS DR1 (no counterpart, no z).}}
    \end{subfigure}%
    \hfill
    \begin{subfigure}[t]{0.22\textwidth}
        \centering
        \includegraphics[width=0.95\textwidth]{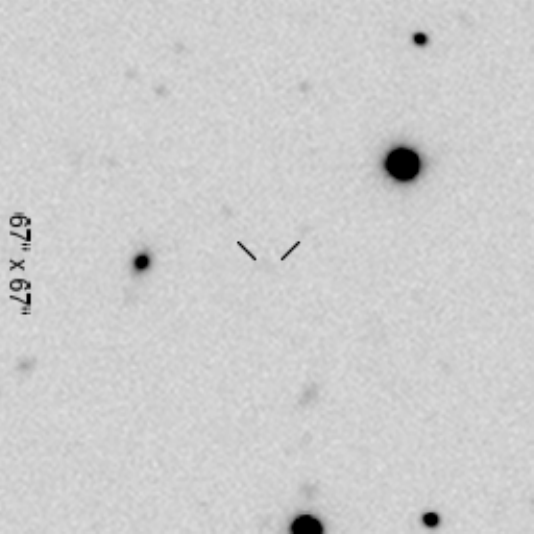}
        \caption{\tiny{J1632+3547 in DES DR8.}}
    \end{subfigure}%
    \hfill
    \begin{subfigure}[t]{0.22\textwidth}
        \centering
        \includegraphics[width=0.95\textwidth]{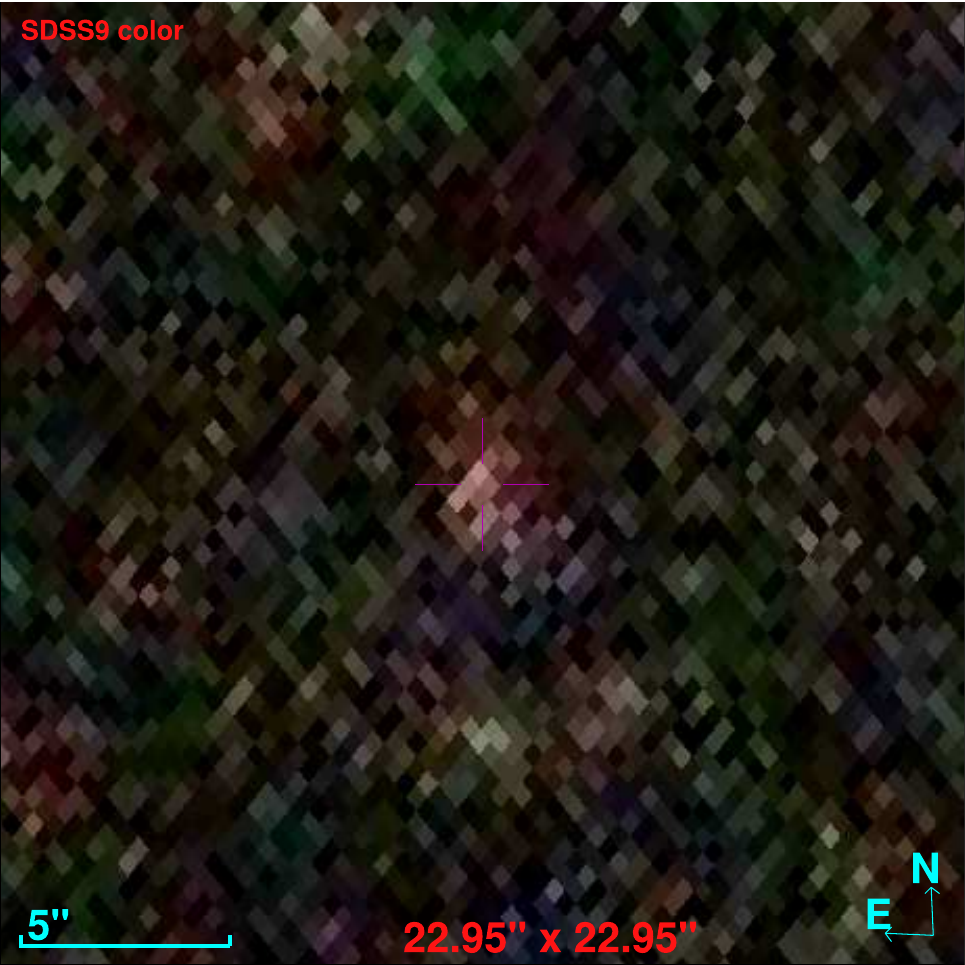}
        \caption{\tiny{J1653+3503 in SDSS DR9.}}
    \end{subfigure}%
    \hfill
    \begin{subfigure}[t]{0.22\textwidth}
        \centering
        \includegraphics[width=0.95\textwidth]{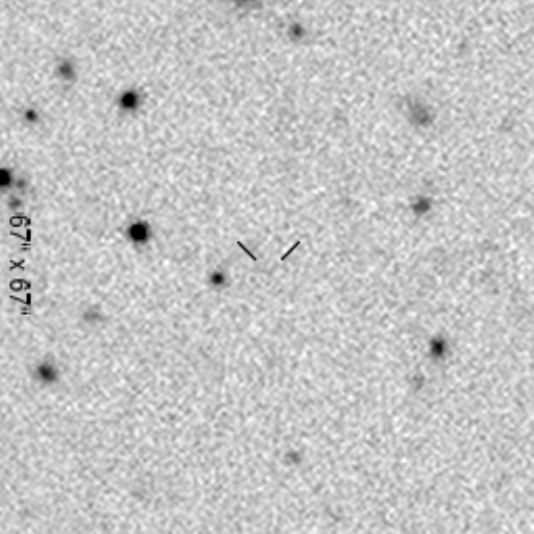}
        \caption{\tiny{J1721+5207 in DES DR8.}}
    \end{subfigure}
    
    \vspace{0.14cm}
    
    \begin{subfigure}[t]{0.22\textwidth}
        \centering
        \includegraphics[width=0.9\textwidth]{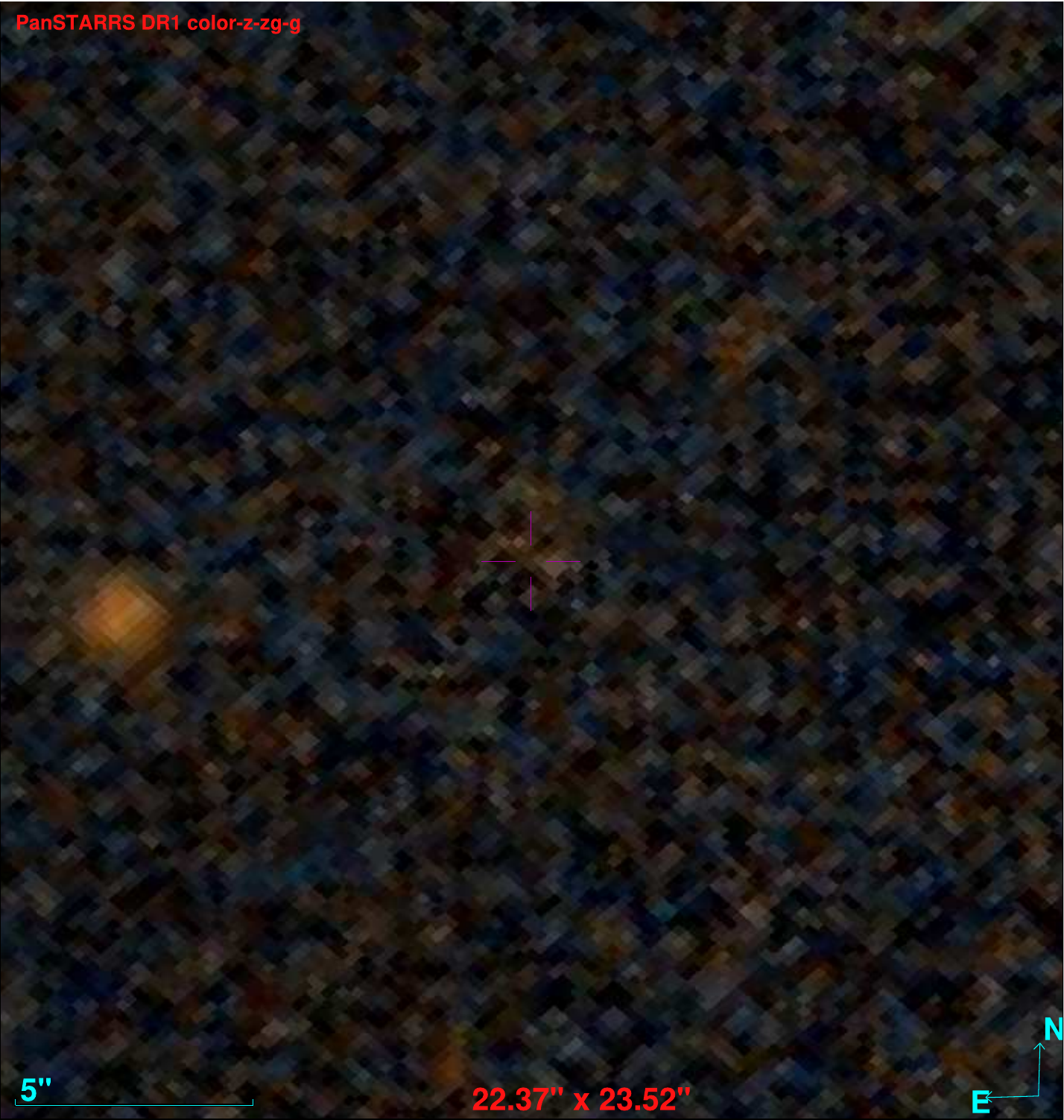}
        \caption{\tiny{J1805-0438 in PanSTARRS DR1 (no z).}}
    \end{subfigure}%
    \hfill
    \begin{subfigure}[t]{0.22\textwidth}
        \centering
        \includegraphics[width=0.9\textwidth]{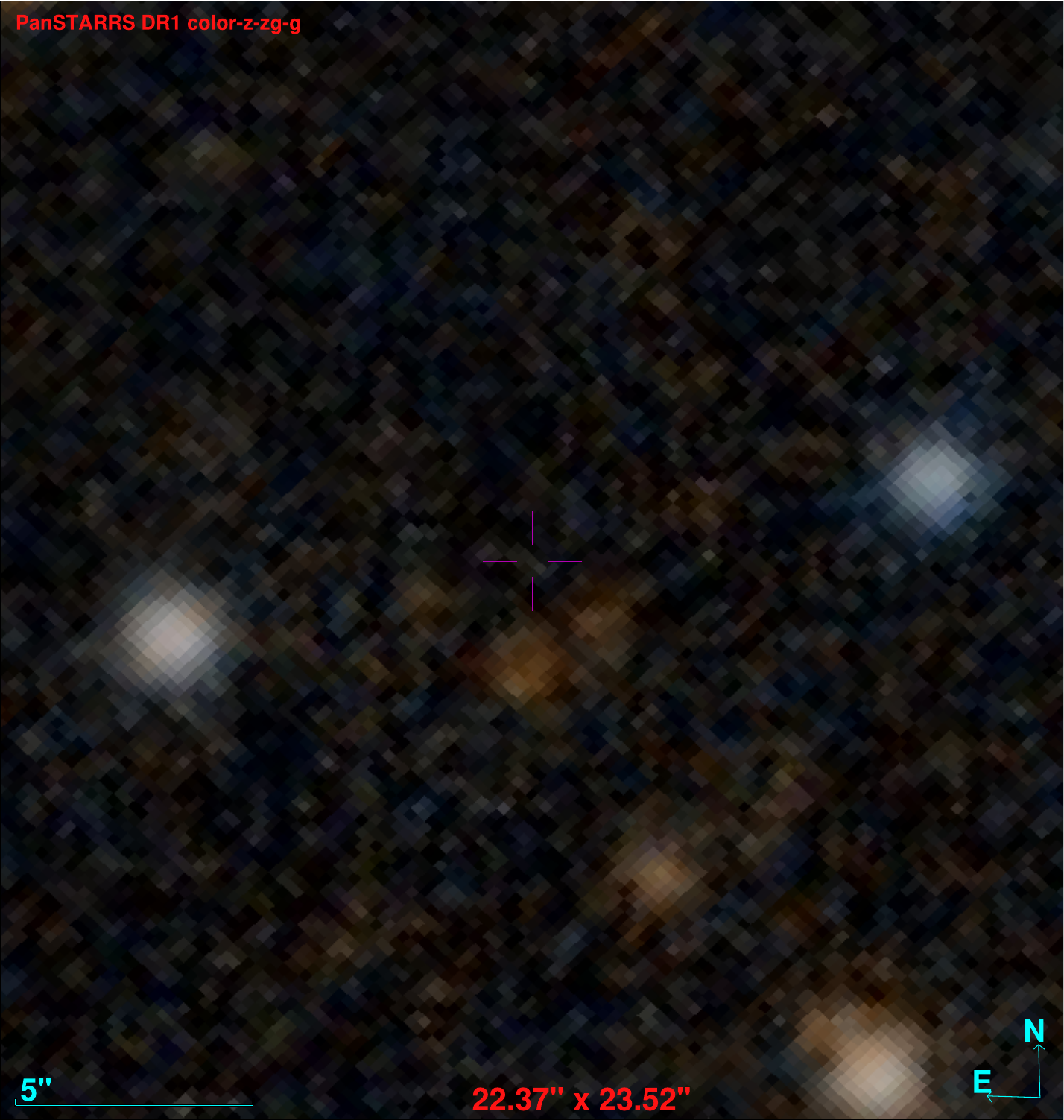}
        \caption{\tiny{J2010+1513 in PanSTARRS DR1 (no counterpart, no z).}}
    \end{subfigure}%
    \hfill
    \begin{subfigure}[t]{0.22\textwidth}
        \centering
        \includegraphics[width=0.9\textwidth]{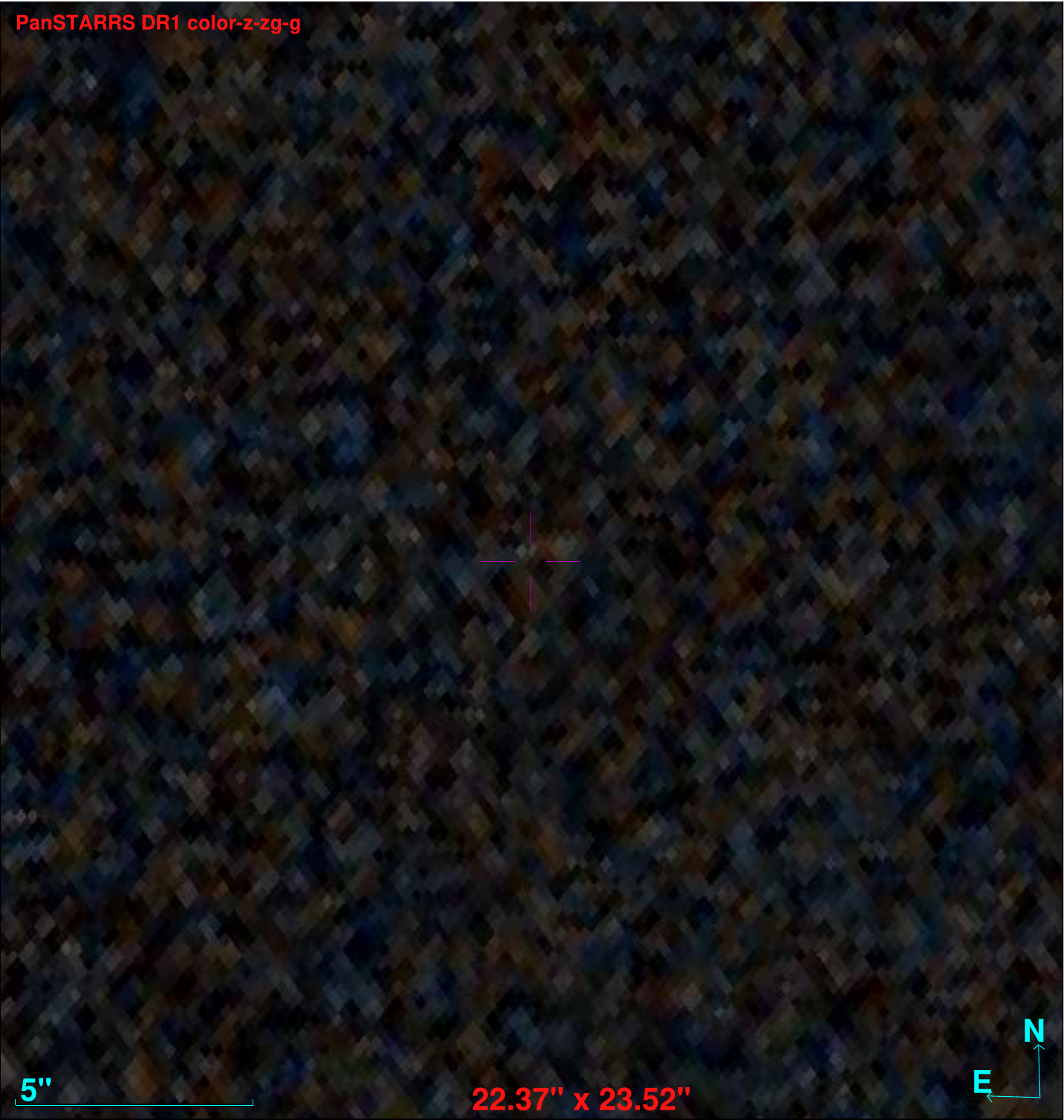}
        \caption{\tiny{J2044+6649 in PanSTARRS DR1 (no counterpart, no z).}}
    \end{subfigure}%
    \hfill
    \begin{subfigure}[t]{0.22\textwidth}
        \centering
        \includegraphics[width=0.9\textwidth]{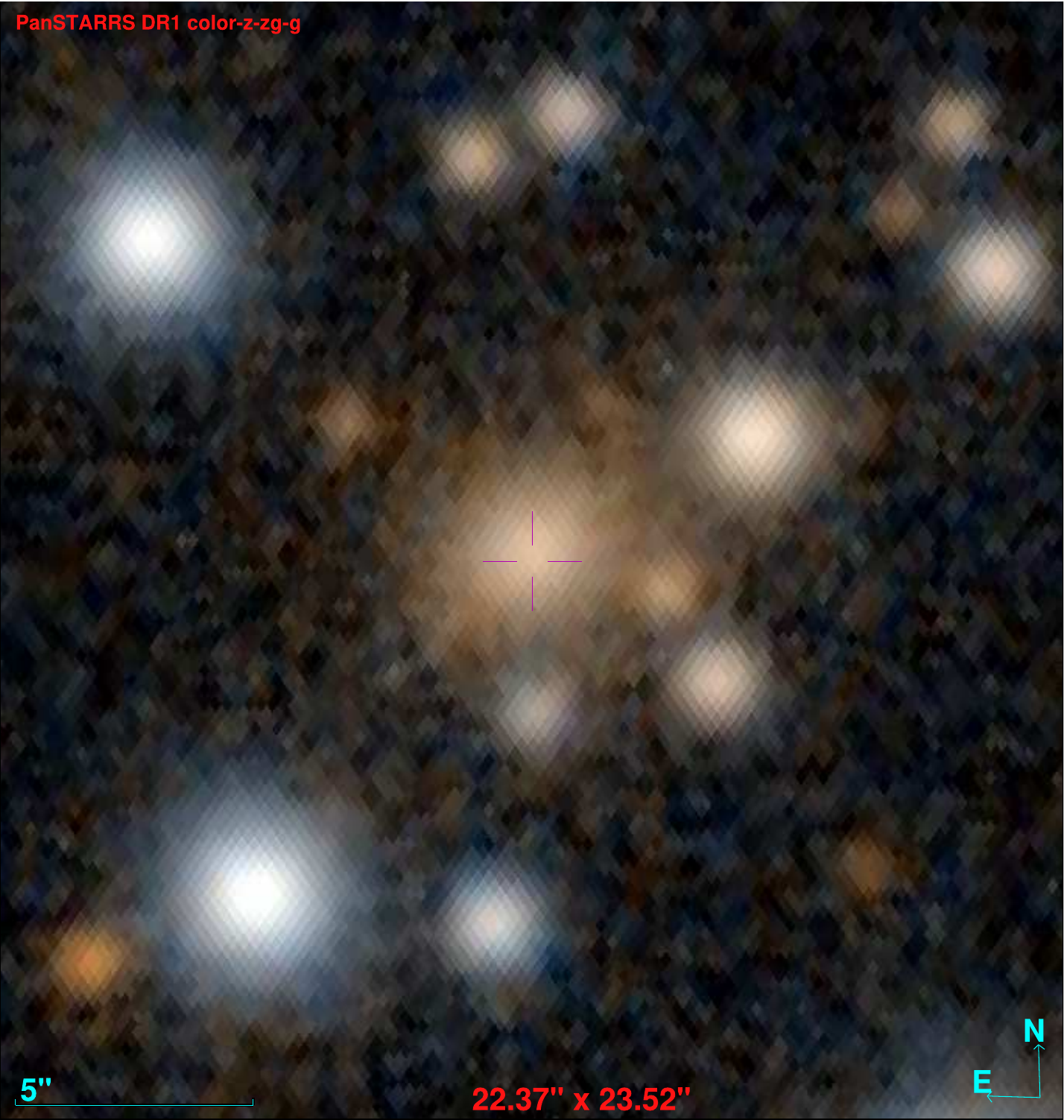}
        \caption{\tiny{J2114+4036 in PanSTARRS DR1.}}
    \end{subfigure}
    
    \vspace{0.14cm}
    
    \begin{subfigure}[t]{0.22\textwidth}
        \centering
        \includegraphics[width=0.9\textwidth]{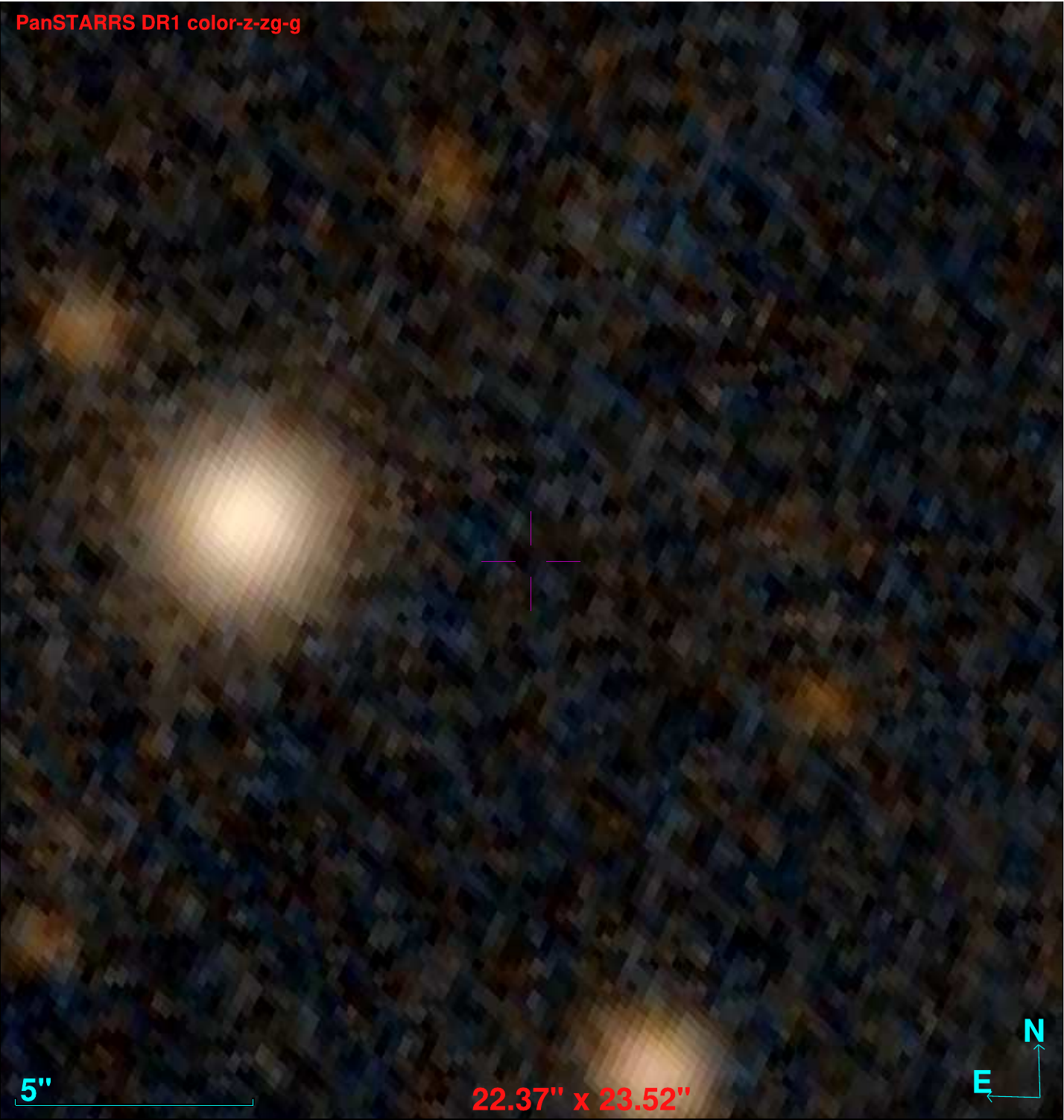}
        \caption{\tiny{J2209+6442 in PanSTARRS DR1 (no counterpart, no z).}}
    \end{subfigure}%
    \hfill
    \begin{subfigure}[t]{0.22\textwidth}
        \centering
        \includegraphics[width=0.9\textwidth]{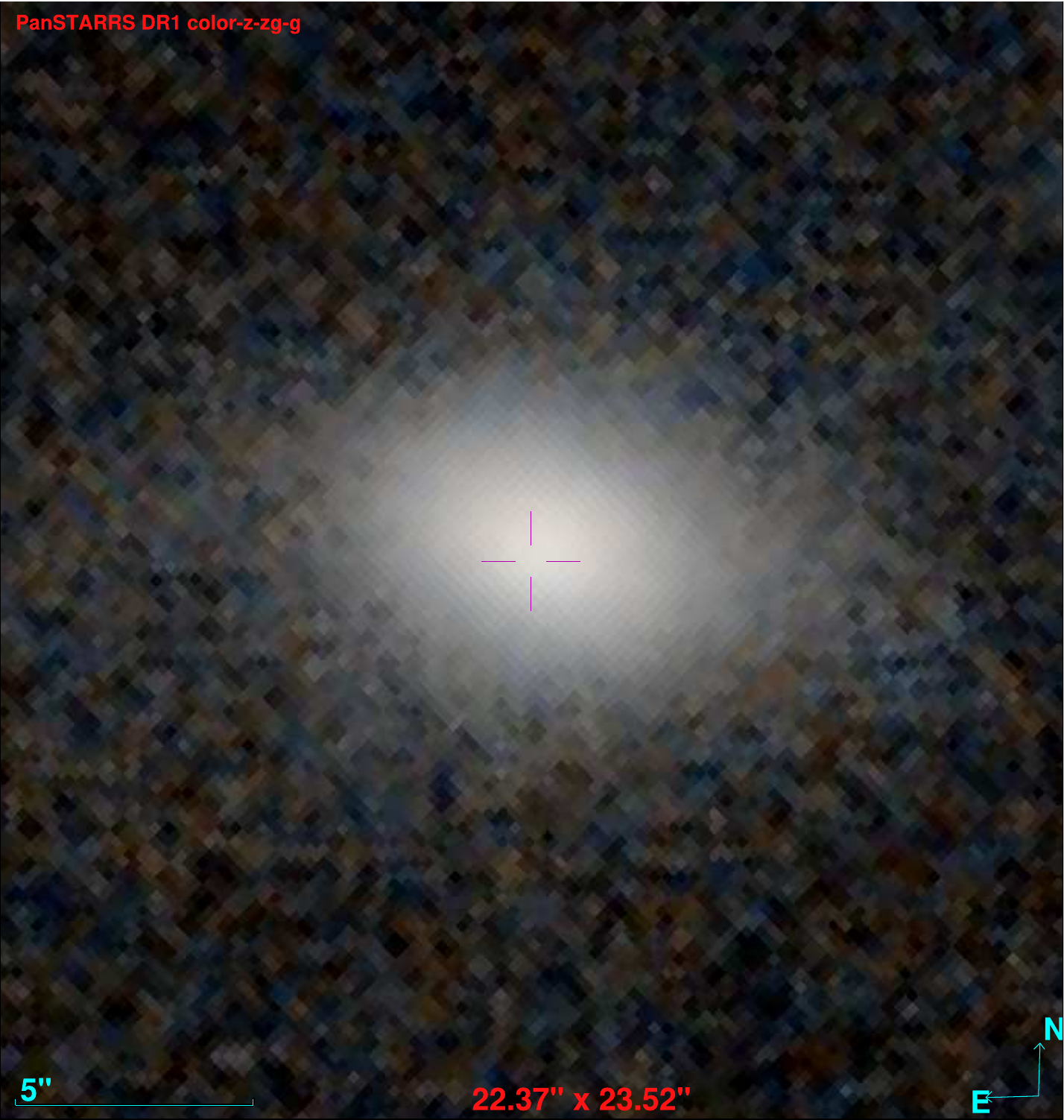}
        \caption{\tiny{J2214-2521 in PanSTARRS DR1.}}
    \end{subfigure}%
    \hfill
    \begin{subfigure}[t]{0.22\textwidth}
        \centering
        \includegraphics[width=0.95\textwidth]{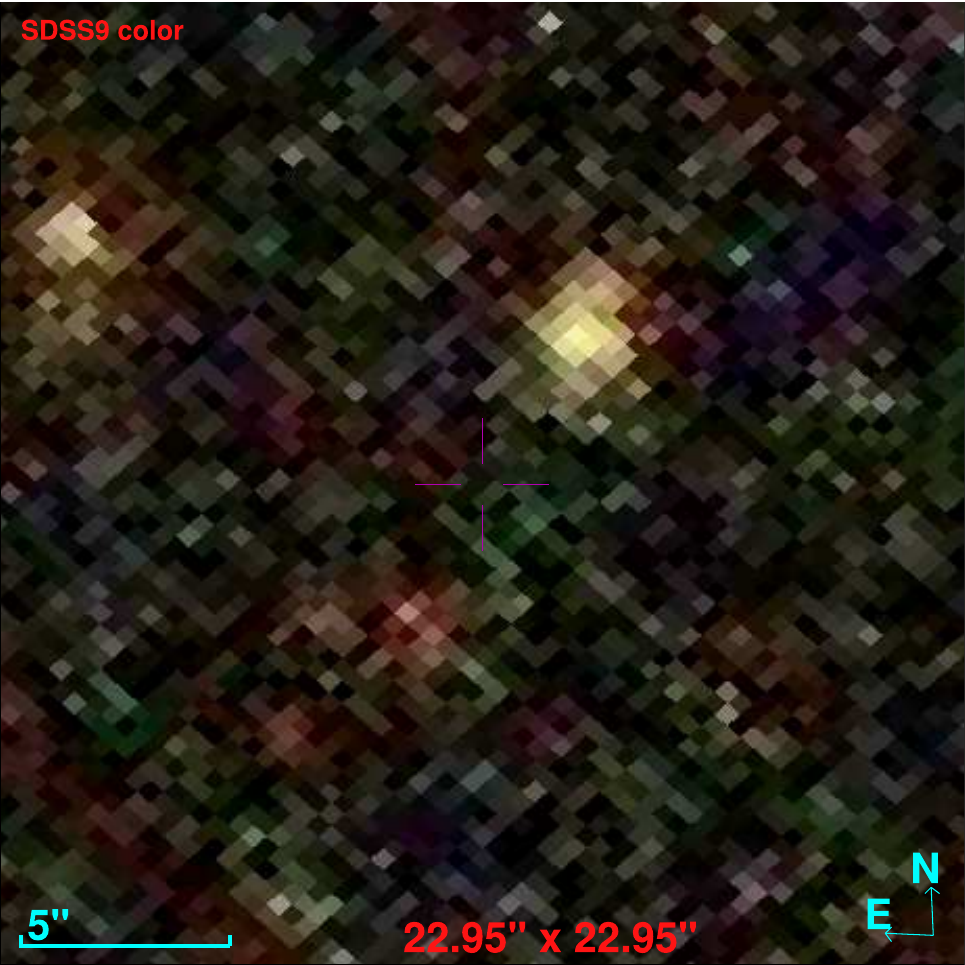}
        \caption{\tiny{J2225+0841 in SDSS DR9.}}
    \end{subfigure}%
    \hfill
    \begin{subfigure}[t]{0.22\textwidth}
        \centering
        \includegraphics[width=0.9\textwidth]{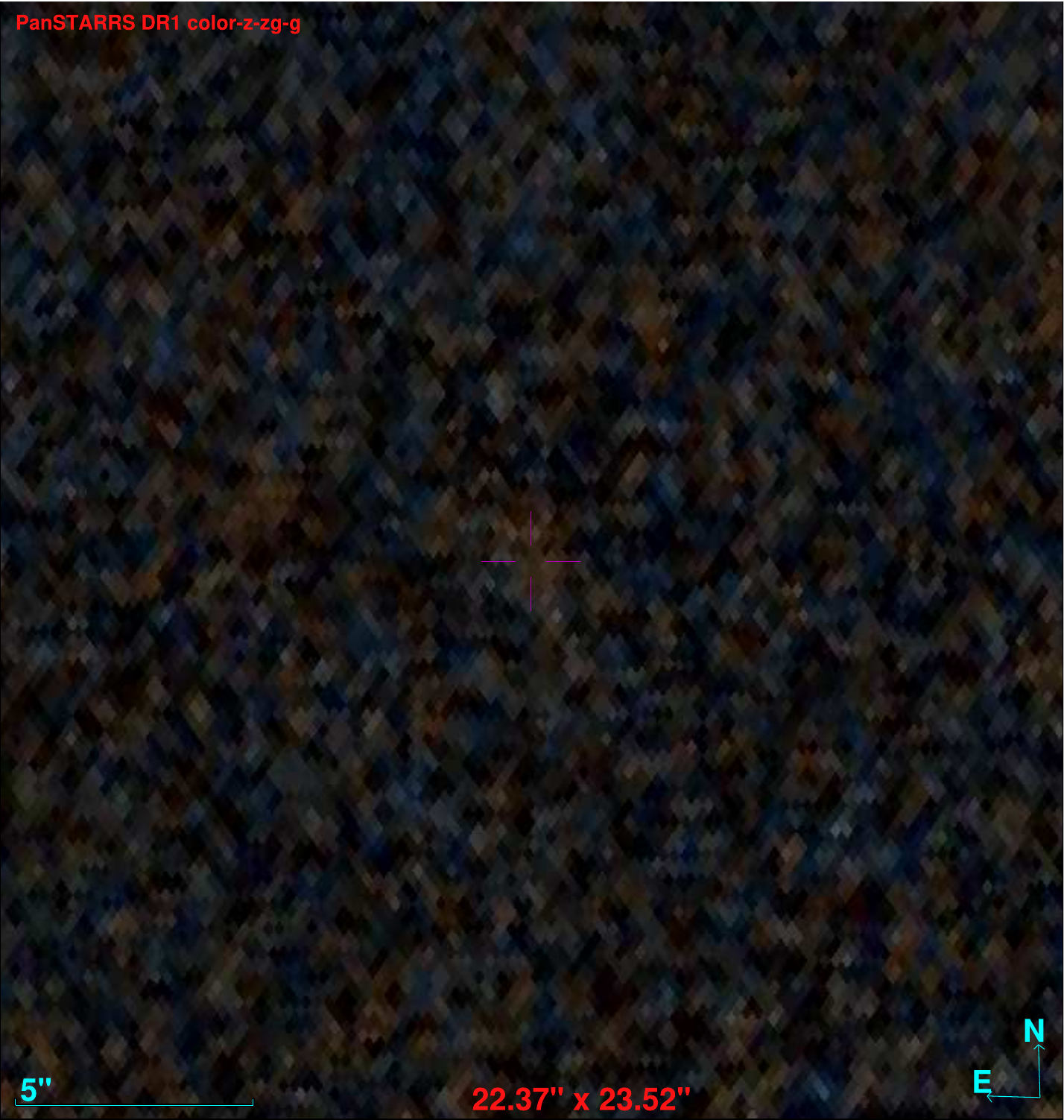}
        \caption{\tiny{J2259+4037 in PanSTARRS DR1 (no counterpart, no z).}}
    \end{subfigure}
    
    \vspace{0.14cm}
    
    \begin{subfigure}[t]{0.22\textwidth}
        \centering
        \includegraphics[width=0.95\textwidth]{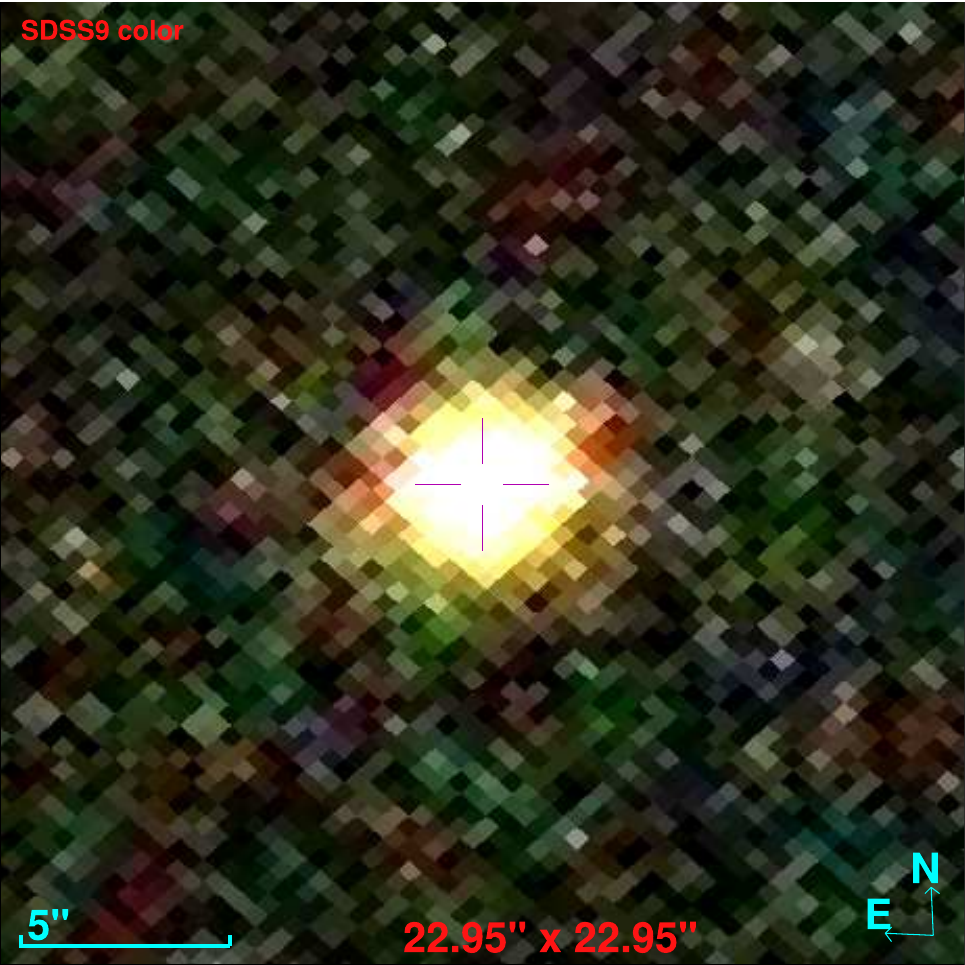}
        \caption{\tiny{J2312+0919 in SDSS DR9.}}
    \end{subfigure}%
    \hfill
    \begin{subfigure}[t]{0.22\textwidth}
        \centering
        \includegraphics[width=0.95\textwidth]{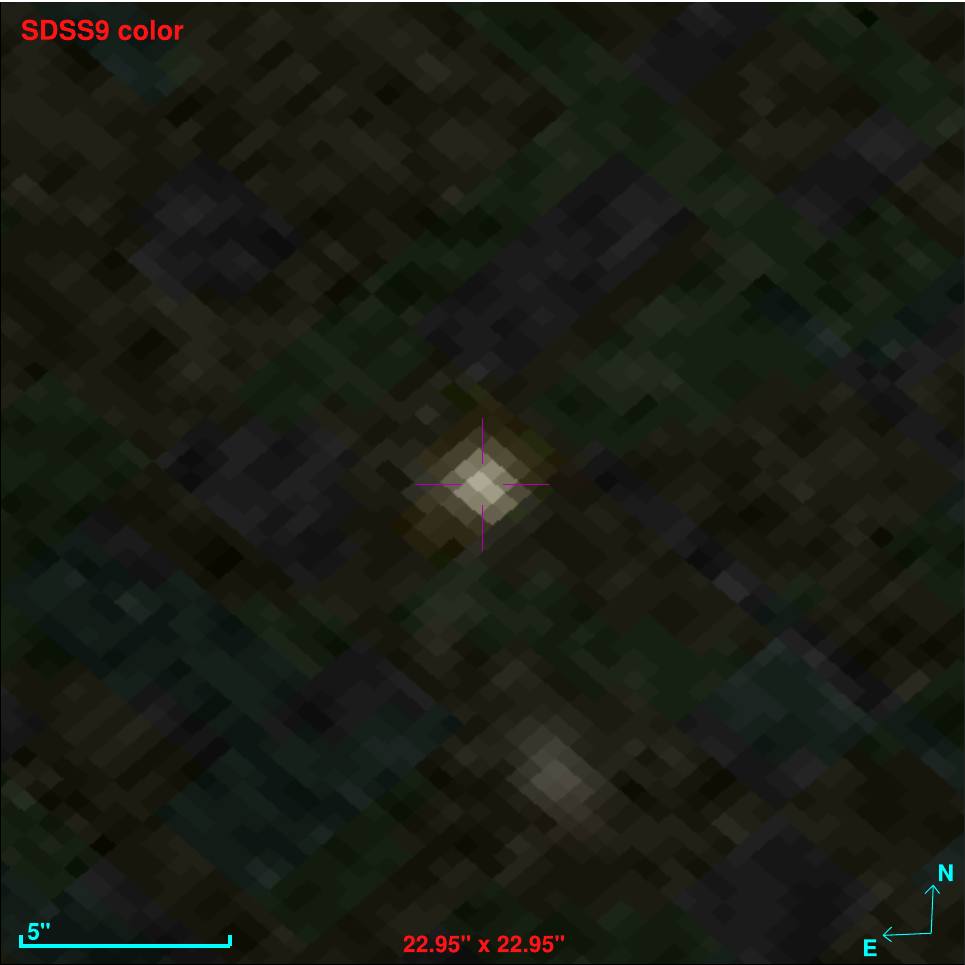}
        \caption{\tiny{J2324-0058 in SDSS DR9.}}
    \end{subfigure}%
    \hfill
    \begin{subfigure}[t]{0.22\textwidth}
        \centering
        \includegraphics[width=0.95\textwidth]{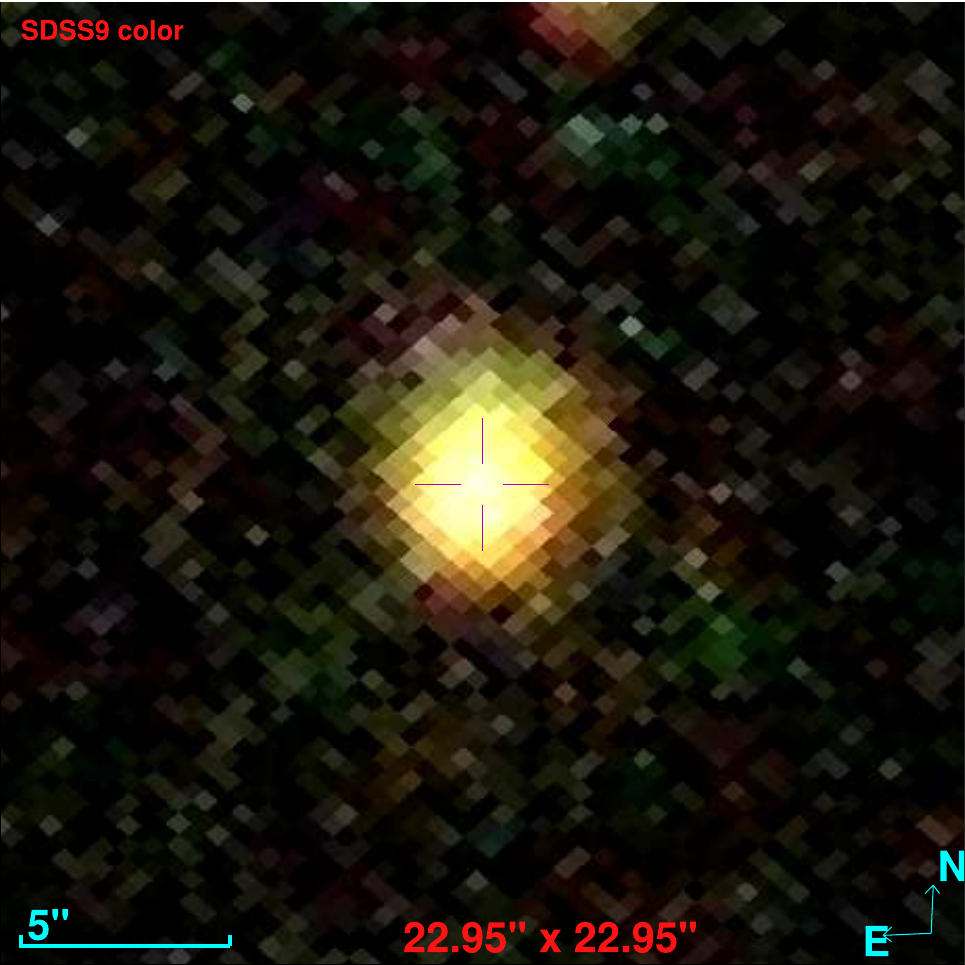}
        \caption{\tiny{J2337-0622 in SDSS DR9.}}
    \end{subfigure}%
    \hfill
    \begin{subfigure}[t]{0.22\textwidth}
        \centering
        \includegraphics[width=0.95\textwidth]{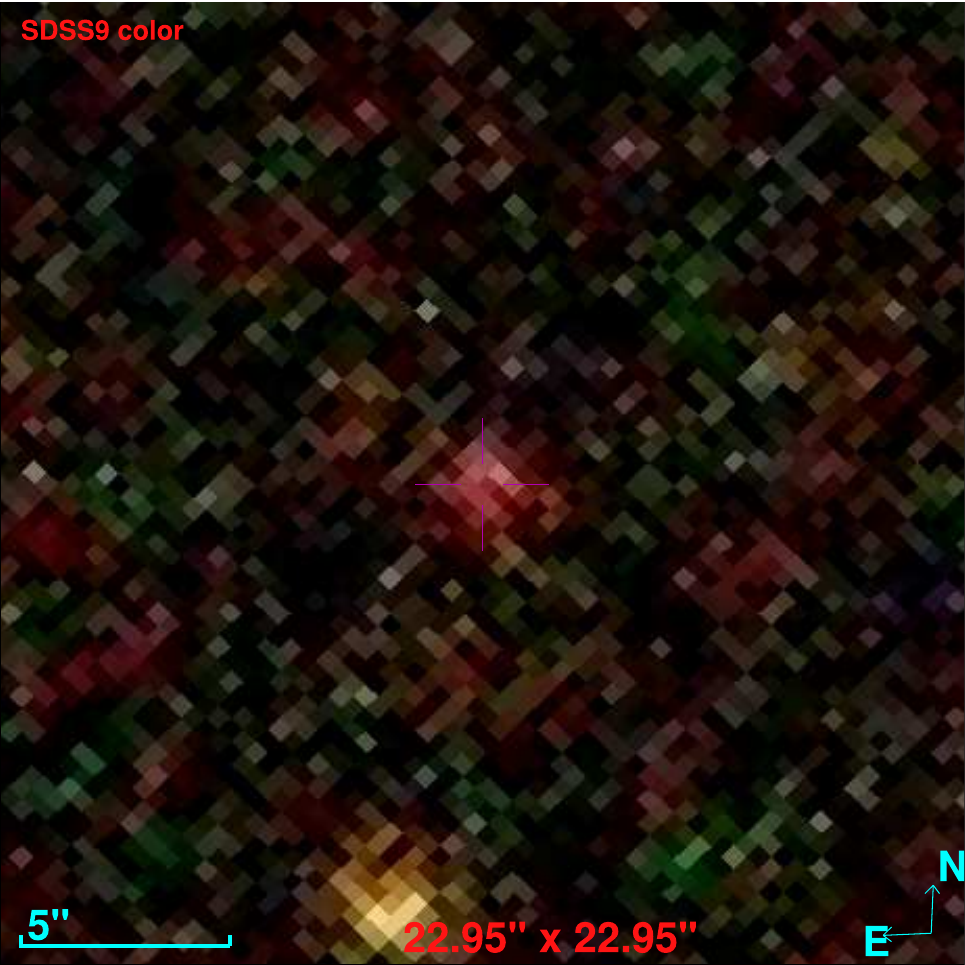}
        \caption{\tiny{J2347-1856 in SDSS DR9.}}
    \end{subfigure}
\end{figure*}

\end{document}